\documentclass[11pt,fleqn]{book} 

\usepackage[top=3cm,bottom=3cm,left=3.2cm,right=3.2cm,headsep=10pt,letterpaper]{geometry} 

\usepackage{xcolor} 
\definecolor{rubinblue}{RGB}{5,139,140} 
\definecolor{rubinlightblue}{RGB}{0,186,188} 

\usepackage{avant} 
\usepackage{mathptmx} 
\usepackage{bm}
\usepackage{microtype} 
\usepackage[utf8]{inputenc} 
\usepackage[T1]{fontenc} 

\usepackage{csquotes}
\usepackage{natbib}
\usepackage{aas_macros}
\usepackage{hyperref}

\usepackage{etoolbox}

\newcounter{caffil}
\newcommand{\caffil}[2]{%
    \refstepcounter{caffil}%
    \noindent%
    ${}^{\thecaffil}$\,#2 \label{#1}%
}

\newcommand{\cauthor}[2]{%
  \def\nextitem{\def\nextitem{,}}
  \renewcommand*{\do}[1]{\nextitem\ref{##1}}
  #1${}^{\docsvlist{#2}}$
}

\hypersetup{
  colorlinks,
  citecolor=rubinblue,
  linkcolor=rubinlightblue,
  urlcolor=rubinblue,
  linkbordercolor=rubinblue}
  
\usepackage{titlesec} 

\usepackage{graphicx} 
\graphicspath{{Pictures/}} 

\usepackage{lipsum} 

\usepackage{tikz} 

\usepackage[english]{babel} 

\usepackage{enumitem} 
\setlist{nolistsep} 

\usepackage{booktabs} 

\usepackage{eso-pic} 

\usepackage{titletoc} 

\contentsmargin{0cm} 
\titlecontents{chapter}[1.25cm] 
{\addvspace{15pt}\large\sffamily\bfseries} 
{\color{rubinblue!60}\contentslabel[\Large\thecontentslabel]{1.25cm}\color{rubinblue}} 
{}  
{\color{rubinblue!60}\normalsize\sffamily\bfseries\;\titlerule*[.5pc]{.}\;\thecontentspage} 
\titlecontents{section}[1.25cm] 
{\addvspace{5pt}\sffamily\bfseries} 
{\contentslabel[\thecontentslabel]{1.25cm}} 
{}
{\sffamily\hfill\color{black}\thecontentspage} 
[]
\titlecontents{subsection}[1.25cm] 
{\addvspace{1pt}\sffamily\small} 
{\contentslabel[\thecontentslabel]{1.25cm}} 
{}
{\sffamily\;\titlerule*[.5pc]{.}\;\thecontentspage} 
[] 

\titlecontents{lsection}[0em] 
{\footnotesize\sffamily} 
{}
{}
{}

\titlecontents{lsubsection}[.5em] 
{\normalfont\footnotesize\sffamily} 
{}
{}
{}
 
\usepackage{fancyhdr} 

\fancypagestyle{plain}{\fancyhead{}} 

\makeatletter
\renewcommand{\cleardoublepage}{
\clearpage\ifodd\c@page\else
\hbox{}
\vspace*{\fill}
\thispagestyle{empty}
\newpage
\fi}

\usepackage{amsmath,amsfonts,amssymb,amsthm} 

\newtheoremstyle{rubinbluenumbox}
{0pt}
{0pt}
{\normalfont}
{}
{\small\bf\sffamily\color{rubinblue}}
{\;}
{0.25em}
{\small\sffamily\color{rubinblue}\thmname{#1}\nobreakspace\thmnumber{\@ifnotempty{#1}{}\@upn{#2}}
\thmnote{\nobreakspace\the\thm@notefont\sffamily\bfseries\color{black}---\nobreakspace#3.}} 

\newtheoremstyle{blacknumex}
{5pt}
{5pt}
{\normalfont}
{} 
{\small\bf\sffamily}
{\;}
{0.25em}
{\small\sffamily{\tiny\ensuremath{\blacksquare}}\nobreakspace\thmname{#1}\nobreakspace\thmnumber{\@ifnotempty{#1}{}\@upn{#2}}
\thmnote{\nobreakspace\the\thm@notefont\sffamily\bfseries---\nobreakspace#3.}}

\newtheoremstyle{blacknumbox} 
{0pt}
{0pt}
{\normalfont}
{}
{\small\bf\sffamily}
{\;}
{0.25em}
{\small\sffamily\thmname{#1}\nobreakspace\thmnumber{\@ifnotempty{#1}{}\@upn{#2}}
\thmnote{\nobreakspace\the\thm@notefont\sffamily\bfseries---\nobreakspace#3.}}

\newtheoremstyle{rubinbluenum}
{5pt}
{5pt}
{\normalfont}
{}
{\small\bf\sffamily\color{rubinblue}}
{\;}
{0.25em}
{\small\sffamily\color{rubinblue}\thmname{#1}\nobreakspace\thmnumber{\@ifnotempty{#1}{}\@upn{#2}}
\thmnote{\nobreakspace\the\thm@notefont\sffamily\bfseries\color{black}---\nobreakspace#3.}} 
\makeatother

\newcounter{dummy} 
\numberwithin{dummy}{section}
\theoremstyle{rubinbluenumbox}
\newtheorem{theoremeT}[dummy]{Theorem}

\newtheorem{exerciseT}{Exercise}[chapter]
\theoremstyle{blacknumex}
\newtheorem{exampleT}{Example}[chapter]
\theoremstyle{blacknumbox}

\newtheorem{definitionT}{Definition}[section]
\newtheorem{corollaryT}[dummy]{Corollary}
\theoremstyle{rubinbluenum}

\RequirePackage[framemethod=default]{mdframed} 

\newmdenv[skipabove=7pt,
skipbelow=7pt,
backgroundcolor=black!5,
linecolor=rubinblue,
innerleftmargin=5pt,
innerrightmargin=5pt,
innertopmargin=5pt,
leftmargin=0cm,
rightmargin=0cm,
innerbottommargin=5pt]{tBox}

\newmdenv[skipabove=7pt,
skipbelow=7pt,
rightline=false,
leftline=true,
topline=false,
bottomline=false,
backgroundcolor=rubinblue!10,
linecolor=rubinblue,
innerleftmargin=5pt,
innerrightmargin=5pt,
innertopmargin=5pt,
innerbottommargin=5pt,
leftmargin=0cm,
rightmargin=0cm,
linewidth=4pt]{eBox}	

\newmdenv[skipabove=7pt,
skipbelow=7pt,
rightline=false,
leftline=true,
topline=false,
bottomline=false,
linecolor=rubinblue,
innerleftmargin=5pt,
innerrightmargin=5pt,
innertopmargin=0pt,
leftmargin=0cm,
rightmargin=0cm,
linewidth=4pt,
innerbottommargin=0pt]{dBox}	

\newmdenv[skipabove=7pt,
skipbelow=7pt,
rightline=false,
leftline=true,
topline=false,
bottomline=false,
linecolor=gray,
backgroundcolor=black!5,
innerleftmargin=5pt,
innerrightmargin=5pt,
innertopmargin=5pt,
leftmargin=0cm,
rightmargin=0cm,
linewidth=4pt,
innerbottommargin=5pt]{cBox}

\makeatletter
\renewcommand{\@seccntformat}[1]{\llap{\textcolor{rubinblue}{\csname the#1\endcsname}\hspace{1em}}}                    
\renewcommand{\section}{\@startsection{section}{1}{\z@}
{-4ex \@plus -1ex \@minus -.4ex}
{1ex \@plus.2ex }
{\normalfont\large\sffamily\bfseries}}
\renewcommand{\subsection}{\@startsection {subsection}{2}{\z@}
{-3ex \@plus -0.1ex \@minus -.4ex}
{0.5ex \@plus.2ex }
{\normalfont\sffamily\bfseries}}
\renewcommand{\subsubsection}{\@startsection {subsubsection}{3}{\z@}
{-2ex \@plus -0.1ex \@minus -.2ex}
{.2ex \@plus.2ex }
{\normalfont\small\sffamily\bfseries}}                        
\renewcommand\paragraph{\@startsection{paragraph}{4}{\z@}
{-2ex \@plus-.2ex \@minus .2ex}
{.1ex}
{\normalfont\small\sffamily\bfseries}}

\usepackage{hyperref}

\newcommand{\thechapterimage}{}
\newcommand{\chapterimage}[1]{\renewcommand{\thechapterimage}{#1}}

\def\thechapter{\arabic{chapter}}
\def\@makechapterhead#1{
\thispagestyle{empty}
{\centering \normalfont\sffamily
\ifnum \c@secnumdepth >\m@ne
\if@mainmatter
\startcontents
\begin{tikzpicture}[remember picture,overlay]
\node at (current page.north west)
{\begin{tikzpicture}[remember picture,overlay]
\node[anchor=north west,inner sep=0pt] at (0,0) {\includegraphics[width=\paperwidth]{\thechapterimage}};
\draw[anchor=west] (5cm,-9cm) node [rounded corners=20pt,fill=rubinblue!10!white,text opacity=1,draw=rubinblue,draw opacity=1,line width=1.5pt,fill opacity=.6,inner sep=12pt]{\huge\sffamily\bfseries\textcolor{black}{\thechapter. #1\strut\makebox[22cm]{}}};
\end{tikzpicture}};
\end{tikzpicture}}
\par\vspace*{230\p@}
\fi
\fi}

\def\@makeschapterhead#1{
\thispagestyle{empty}
{\centering \normalfont\sffamily
\ifnum \c@secnumdepth >\m@ne
\if@mainmatter
\begin{tikzpicture}[remember picture,overlay]
\node at (current page.north west)
{\begin{tikzpicture}[remember picture,overlay]
\node[anchor=north west,inner sep=0pt] at (0,0) {\includegraphics[width=\paperwidth]{\thechapterimage}};
\draw[anchor=west] (5cm,-9cm) node [rounded corners=20pt,fill=rubinblue!10!white,fill opacity=.6,inner sep=12pt,text opacity=1,draw=rubinblue,draw opacity=1,line width=1.5pt]{\huge\sffamily\bfseries\textcolor{black}{#1\strut\makebox[22cm]{}}};
\end{tikzpicture}};
\end{tikzpicture}}
\par\vspace*{230\p@}
\fi
\fi
}
\makeatother

\newcommand{\Mnomsun}{\hbox{$\mathcal{M}^{\rm N}_\odot$}}
\newcommand{\Rnomsun}{\hbox{$\mathcal{R}^{\rm N}_\odot$}}

\newcommand\blfootnote[1]{%
  \begingroup
  \renewcommand\thefootnote{}\footnote{#1}%
  \addtocounter{footnote}{-1}%
  \endgroup
}

\usepackage{etoolbox}
\makeatletter
\patchcmd\@makeschapterhead
  {\includegraphics[width=\paperwidth]{\thechapterimage}}
  {\includegraphics[width=\paperwidth,height=4.13333in]{\thechapterimage}}
  {}{}
  
\begin{document}
\title{Rubin Observatory LSST Transients and Variable Stars Roadmap}


\begingroup
\thispagestyle{empty}
\AddToShipoutPicture*{\put(0,0){\includegraphics[scale=1.25]{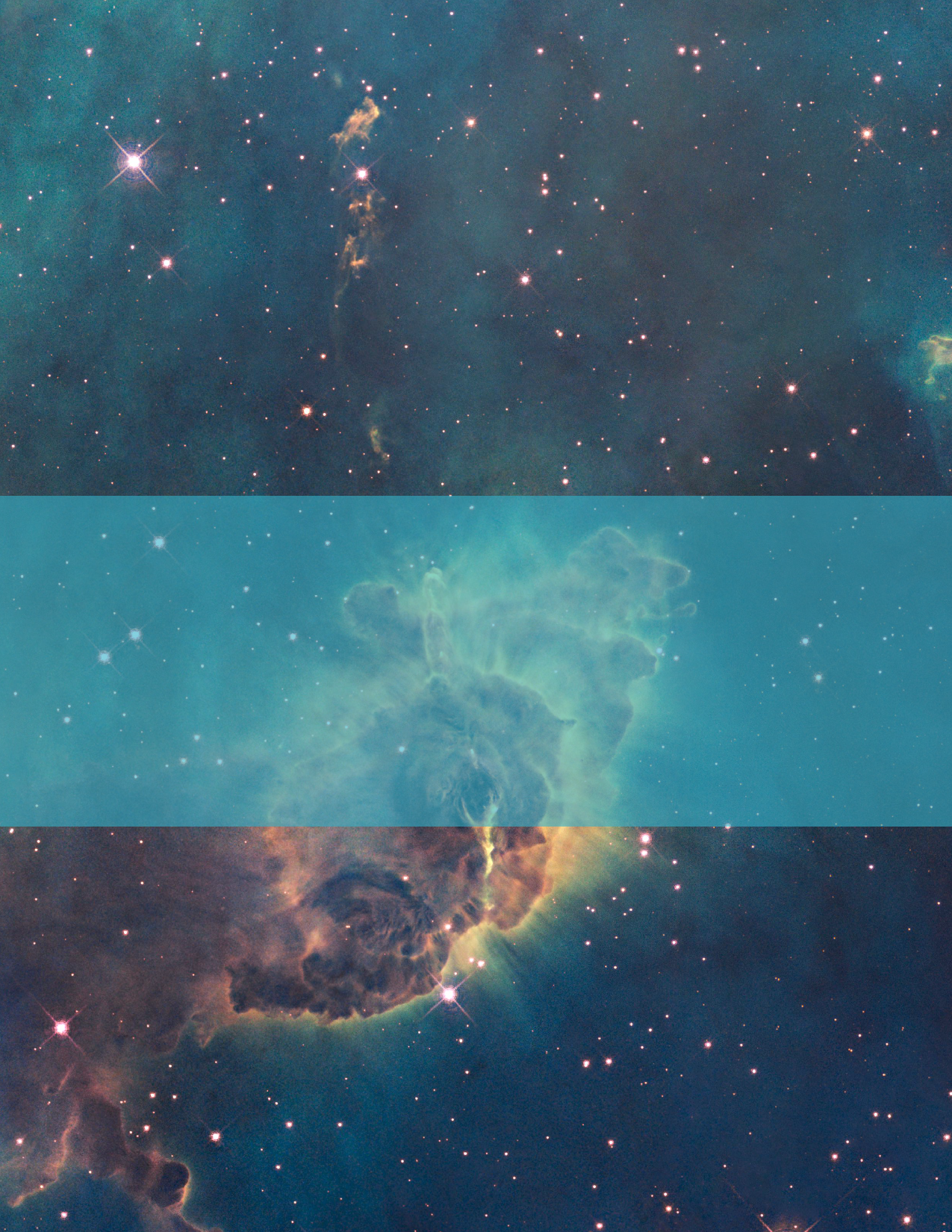}}} 
\centering
\vspace*{4.5cm}
\par\normalfont\fontsize{35}{35}\sffamily\selectfont
\textbf{Rubin Observatory LSST Transients and Variable Stars Roadmap}\\
\vspace*{0.5cm}
{\Huge The Rubin LSST TVS Science Collaboration}\par 

\vspace*{0.8cm}
{\Large Edited by Kelly Hambleton, Federica Bianco and Rachel Street}
\endgroup


\newpage
~\vfill
\thispagestyle{empty}

\noindent \textit{First release, \today; Cover image credit ESA/Hubble} 

\chapterimage{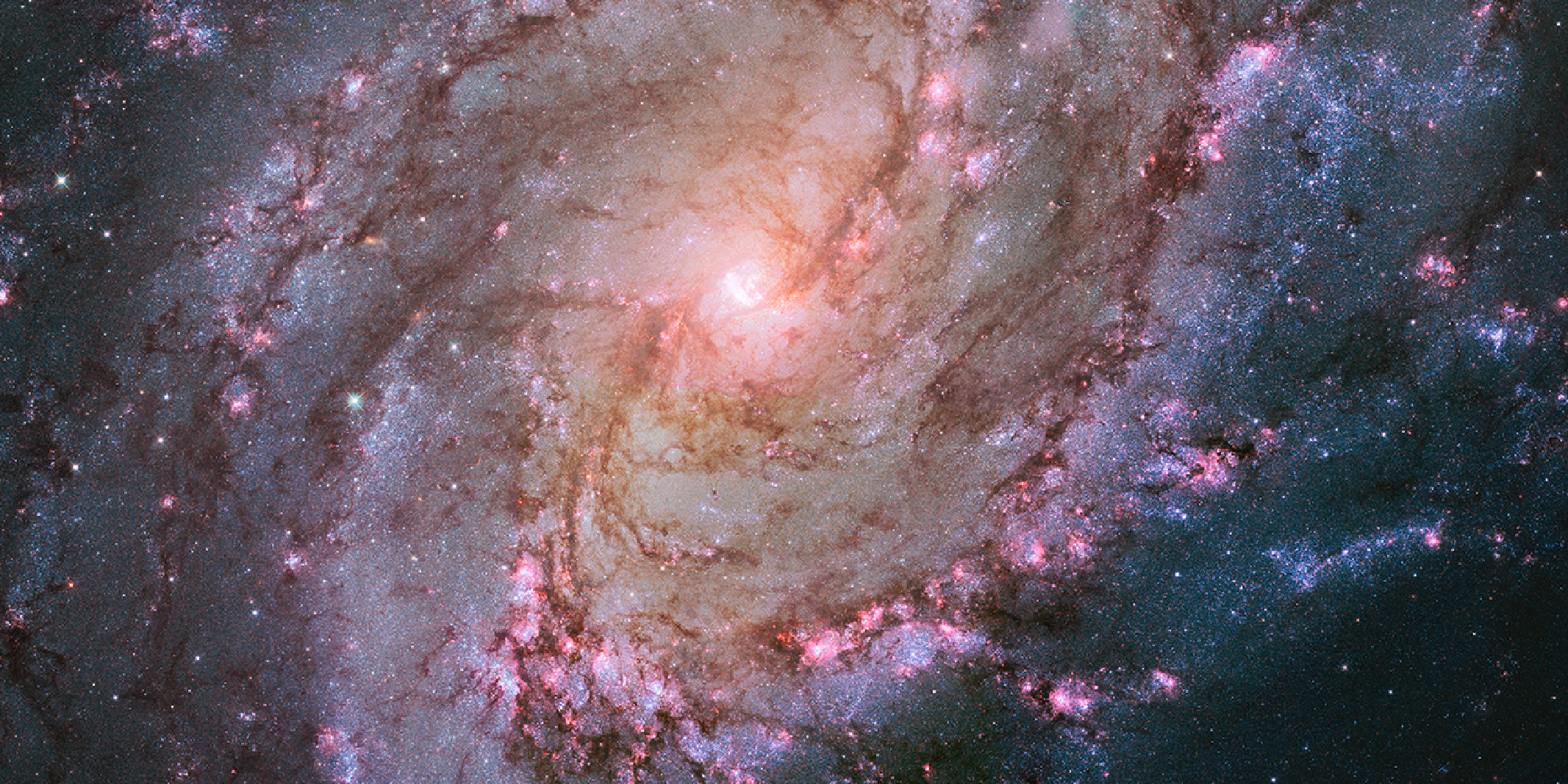} 

\pagestyle{empty} 
\hypersetup{linkcolor=black}
  \tableofcontents
  


\pagestyle{fancy} 

\chapterimage{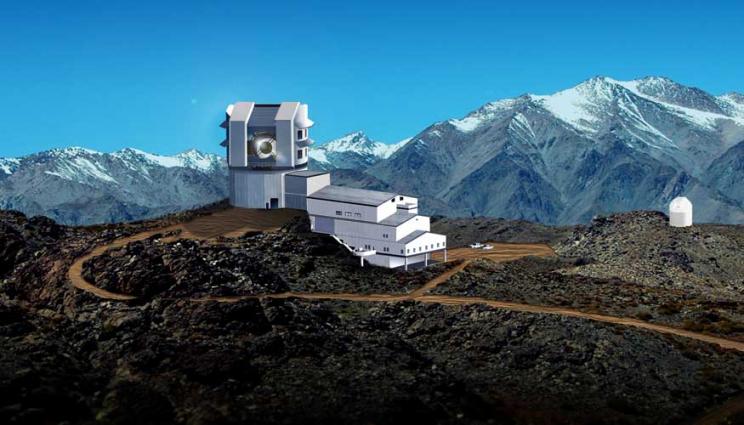} 
\chapter*{List of Authors}

\cauthor{Kelly M.~Hambleton,}{af:vu}
\cauthor{Federica B.~Bianco,}{af:del_phy, af:jrb, af:del_data, af:vro}
\cauthor{Rachel	Street,}{af:lco}
\cauthor{Keaton Bell,}{af:uow}
\cauthor{David Buckley,}{af:saao,af:uoc,af:uofs}
\cauthor{Melissa Graham,}{af:uow}
\cauthor{Nina Hernitschek,}{af:vandy}
\cauthor{Michael B.	Lund,}{af:nasa_esi}
\cauthor{Elena Mason,}{af:inaf_tri}
\cauthor{Joshua Pepper,}{af:lu}
\cauthor{Andrej Pr\v{s}a,}{af:vu}
\cauthor{Markus	Rabus,}{af:concep}
\cauthor{Claudia M. Raiteri,}{af:inaf_tori}
\cauthor{R\'{o}bert Szab\'{o},}{af:konk, af:kon_coe, af:lendu}
\cauthor{Paula Szkody,}{af:uow}
\cauthor{Igor Andreoni,}{af:jssi, af:mary, af:godda}
\cauthor{Simone Antoniucci,}{af:inaf}
\cauthor{Barbara Balmaverde,}{af:inaf}
\cauthor{Eric Bellm,}{af:uow}
\cauthor{Rosaria Bonito,}{af:inaf_pal}
\cauthor{Giuseppe Bono,}{af:inaf}
\cauthor{Maria Teresa Botticella,}{af:inaf}
\cauthor{Enzo Brocato,}{af:inaf}
\cauthor{Katja Bučar Bricman,}{af:ung}
\cauthor{Enrico Cappellaro,}{af:inaf_pad}
\cauthor{Maria Isabel Carnerero,}{af:inaf_tori}
\cauthor{Ryan Chornock,}{af:berk}
\cauthor{Riley Clarke,}{af:del_phy}
\cauthor{Phil Cowperthwaite,}{af:carne}
\cauthor{Antonino Cucchiara,}{af:nasa, af:marin}
\cauthor{Filippo D'Ammando,}{af:inaf_bol}
\cauthor{Kristen C. Dage,}{af:mcgil}
\cauthor{Massimo  Dall'Ora,}{af:inaf_tor}
\cauthor{James R. A. Davenport,}{af:uow}
\cauthor{Domitilla de Martino,}{af:inaf_capo}
\cauthor{Giulia de Somma,}{af:inaf_capo}
\cauthor{Marcella Di Criscienzo,}{af:inaf_roma}
\cauthor{Rosanne Di Stefano,}{af:cfa}
\cauthor{Maria Drout,}{af:toront}
\cauthor{Michele Fabrizio,}{af:inaf}
\cauthor{Giuliana Fiorentino,}{af:inaf}
\cauthor{Poshak Gandhi,}{af:south}
\cauthor{Alessia Garofalo,}{af:inaf}
\cauthor{Teresa Giannini,}{af:inaf}
\cauthor{Andreja Gomboc,}{af:ung}
\cauthor{Laura Greggio,}{af:inaf}
\cauthor{Patrick Hartigan,}{af:rice}
\cauthor{Markus Hundertmark,}{af:heidel}
\cauthor{Elizabeth Johnson,}{af:jh}
\cauthor{Michael Johnson,}{af:max}
\cauthor{Tomislav Jurkic,}{af:uor}
\cauthor{Somayeh Khakpash,}{af:del_phy, af:del_data}
\cauthor{Silvio Leccia,}{af:inaf_capo}
\cauthor{Xiaolong Li,}{af:del_phy, af:del_data}
\cauthor{Davide Magurno,}{af:inaf}
\cauthor{Konstantin Malanchev,}{af:uc,af:moscow}
\cauthor{Marcella Marconi,}{af:inaf_capo}
\cauthor{Raffaella Margutti,}{af:berk}
\cauthor{Silvia Marinoni,}{af:inaf}
\cauthor{Nicolas Mauron,}{af:mont}
\cauthor{Roberto Molinaro,}{af:inaf_capo}
\cauthor{Anais Möller,}{af:swin}
\cauthor{Marc Moniez,}{af:paris}
\cauthor{Tatiana Muraveva,}{af:inaf}
\cauthor{Ilaria Musella,}{af:inaf_capo}
\cauthor{Chow-Choong Ngeow,}{af:nci}
\cauthor{Andrea Pastorello,}{af:inaf_pad}
\cauthor{Vincenzo Petrecca,}{af:inaf}
\cauthor{Silvia Piranomonte,}{af:inaf_roma}
\cauthor{Fabio Ragosta,}{af:inaf_roma}
\cauthor{Andrea Reguitti,}{af:uab, af:mas, af:inaf_pad}
\cauthor{Chiara Righi,}{af:inaf}
\cauthor{Vincenzo Ripepi,}{af:inaf_capo}
\cauthor{Liliana Rivera Sandoval,}{af:tex}
\cauthor{Keivan G.~Stassun,}{af:vandy}
\cauthor{Michael Stroh,}{af:ciera}
\cauthor{Giacomo Terreran,}{af:lco}
\cauthor{Virginia Trimble,}{af:irvi}
\cauthor{Yiannis Tsapras,}{af:heidel}
\cauthor{Sjoert van Velze,n}{af:leid}
\cauthor{Laura Venuti}{af:seti}\&
\cauthor{Jorick S.~Vink.}{af:arma}

\bigskip

\caffil{af:vu}{Department of Astrophysics and Planetary Science, Villanova University, 800 Lancaster Ave, Villanova, 19085 PA}

\caffil{af:del_phy}{Department of Physics and Astronomy, University of Delaware, Newark, DE, 19716, USA}

\caffil{af:jrb}{Joseph R. Biden, Jr. School of Public Policy and Administration, University of Delaware, Newark, DE, 19716, USA} 

\caffil{af:del_data}{Data Science Institute, University of Delaware, Newark, DE, 19716, USA}

\caffil{af:vro}{Vera C. Rubin Observatory Construction Project} 

\caffil{af:lco}{Las Cumbres Observatory, 6740 Cortona Drive, Suite 102, Goleta, CA 93117-5575, USA}

\caffil{af:uow}{Astronomy Department, University of Washington, Box 351580, Seattle, WA 98195, USA}

\caffil{af:saao}{South African Astronomical Observatory, PO Box 9, Observatory Rd, Observatory 7935, South Africa}

\caffil{af:uoc}{Department of Astronomy, University of Cape Town, Private Bag X3, Rondebosch 7701, South Africa}

\caffil{af:uofs}{Department of Physics, University of the Free State, PO Box 339, Bloemfontein 9300, South Africa}

\caffil{af:vandy}{Department of Physics and Astronomy, Vanderbilt University, Nashville, TN 37235, USA}

\caffil{af:nasa_esi}{NASA Exoplanet Science Institute – Caltech/IPAC, 1200 E. California Blvd, Pasadena, CA 91125 USA}

\caffil{af:inaf_tri}{INAF Osservatorio Astronomico di Trieste (INAF-OATS), Via G.B. Tiepolo 11, 34143 Trieste (TS), Italy}

\caffil{af:lu}{Lehigh University, Department of Physics Deming Lewis Lab 16 Memorial Drive East}

\caffil{af:concep}{Departamento de Matem\'atica y F\'isica Aplicadas, Facultad de Ingenier\'ia, Universidad Cat\'olica de la Sant\'isima Concepci\'on, Alonso de Rivera 2850, Concepci\'on, Chile}

\caffil{af:inaf_tori}{INAF-Osservatorio Astrofisico di Torino, via Osservatorio 20, I-10025 Pino Torinese, Italy}

\caffil{af:konk}{Konkoly Observatory, Research Centre for Astronomy and Earth Sciences,  ELKH, Budapest, Konkoly-Thege Mikl\'os \'ut 15–17., H-1121, Hungary}

\caffil{af:kon_coe}{CSFK, MTA Centre of Excellence, Budapest, Konkoly Thege Mikl\'os \'ut 15-17., H-1121, Hungary}

\caffil{af:lendu}{MTA CSFK Lend\""ulet Near-Field Cosmology Research Group"}

\caffil{af:jssi}{Joint Space-Science Institute, University of Maryland, College Park, MD 20742, USA}

\caffil{af:mary}{Department of Astronomy, University of Maryland, College Park, MD 20742, USA}

\caffil{af:godda}{Astrophysics Science Division, NASA Goddard Space Flight Center, Mail Code 661, Greenbelt, MD 20771, USA}

\caffil{af:inaf}{INAF}

\caffil{af:inaf_pal}{INAF-Osservatorio Astronomico di Palermo, Piazza del Parlamento, 1 90134, Palermo, Italy}

\caffil{af:ung}{Center for Astrophysics and Cosmology, University of Nova Gorica, Vipavska 13, 5000 Nova Gorica, Slovenia}

\caffil{af:inaf_pad}{INAF-Osservatorio Astronomico di Padova, Vicolo dell'Osservatorio 5, 35122 Padova, Italy}

\caffil{af:berk}{Department of Astronomy, University of California, Berkeley, CA 94720, USA}

\caffil{af:carne}{Carnegie Observatories: Pasadena, California, US}

\caffil{af:nasa}{NASA Headquarter, 300 Hidden Figures Way SW, Washington D.C. 20546}

\caffil{af:marin}{College of Marin, 120 Kent Ave. , Kentfield CA 94904}

\caffil{af:inaf_bol}{INAF-IRA Bologna, Via P. Gobetti 101, I-40129 Bologna, Italy}

\caffil{af:mcgil}{Department of Physics, McGill University, 3600 University Street, Montreal, QC H3A 2T8, Canada}

\caffil{af:inaf_tor}{INAF-Universita degli Studi di Roma Tor Vergata, Università degli Studi di Napoli Federico II}

\caffil{af:inaf_capo}{INAF-Osservatorio Astronomico di Capodimonte (INAF-OANA), Salita Moiariello 16, 80131, Napoli (NA), Italy}

\caffil{af:inaf_roma}{INAF-Osservatorio Astronomico di Roma, via Frascati 33, I-00078 Monte Porzio Catone (RM), Italy}

\caffil{af:cfa}{Center for Astrophysics, Harvard, Observatory Building E, 60 Garden St, Cambridge, MA 02138, United States}

\caffil{af:toront}{Dunlap Institute for Astronomy \& Astrophysics, University of Toronto, 50 St. George Street Toronto, Ontario Canada M5S 3H4}

\caffil{af:south}{School of Physics \& Astronomy, University of Southampton, SO17 1BJ, UK}

\caffil{af:rice}{Physics and Astronomy, Rice University: Houston, Texas, USA}

\caffil{af:heidel}{Zentrum f{\"u}r Astronomie der Universit{\"a}t Heidelberg, Astronomisches Rechen-Institut, M{\"o}nchhofstr. 12-14, 69120 Heidelberg, Germany}

\caffil{af:jh}{Johns Hopkins Department of Physics \& Astronomy, 3400 N Charles St Baltimore, MD, US 21218}

\caffil{af:max}{Fundamental Physics in Radio Astronomy, Max Planck Institute for Radio Astronomy, 53121 Bonn, Germany \& DLR-Institute of Data Science, 07745 Jena, Germany}

\caffil{af:uor}{Faculty of Physics, University of Rijeka, R. Matejcic 2, 51000 Rijeka, Croatia}

\caffil{af:uc}{Department of Astronomy, University of Illinois at Urbana-Champaign, 1002 W Green Street, Urbana, IL 61801, USA}

\caffil{af:moscow}{Sternberg Astronomical Institute, Lomonosov Moscow State University, Moscow 119620, Russia}

\caffil{af:mont}{Université De Montpellier, 641 Av. du Doyen Gaston Giraud, 34000 Montpellier, France}

\caffil{af:swin}{Centre for Astrophysics and Supercomputing, Swinburne University of Technology, Mail Number H29, PO Box 218, 31122 Hawthorn, VIC, Australia.}

\caffil{af:paris}{Université Paris-Saclay, 339 Rue du Doyen André Guinier, 91440 Bures-sur-Yvette, France}

\caffil{af:nci}{National Central University Graduate Institute of Astronomy No. 300, Zhongda Rd, Jhongli City, Taiwan}

\caffil{af:uab}{Departamento de Ciencias Fisicas, Universidad Andres Bello, Fernandez Concha 700, Santiago, Chile}

\caffil{af:mas}{Millennium Institute of Astrophysics (MAS), Nuncio Monsenor Sòtero Sanz 100, Santiago, Chile}

\caffil{af:tex}{Department of Physics and Astronomy, University of Texas Rio Grande Valley, Brownsville, TX 78520, USA}

\caffil{af:ciera}{Center for Interdisciplinary Exploration and Research in Astrophysics (CIERA) and Department of Physics and Astronomy, Northwestern University, Evanston, IL 60201, USA}

\caffil{af:irvi}{University of California Irvine, Irvine, CA 92697, United States}

\caffil{af:leid}{Leiden Observatory, Niels Bohrweg 2, 2333 CA Leiden, Netherlands}

\caffil{af:seti}{SETI Institute, 339 Bernardo Ave, Suite 200, Mountain View, CA 94043, USA}

\caffil{af:arma}{Armagh Observatory and Planetarium, College Hill, BT61 9DG Armagh, Northern Ireland}

\caffil{af:inaf_oar}{INAF-Osservatorio Astronomico di Roma, via Frascati 33, I-00078 Monte Porzio Catone (RM), Italy}

\chapterimage{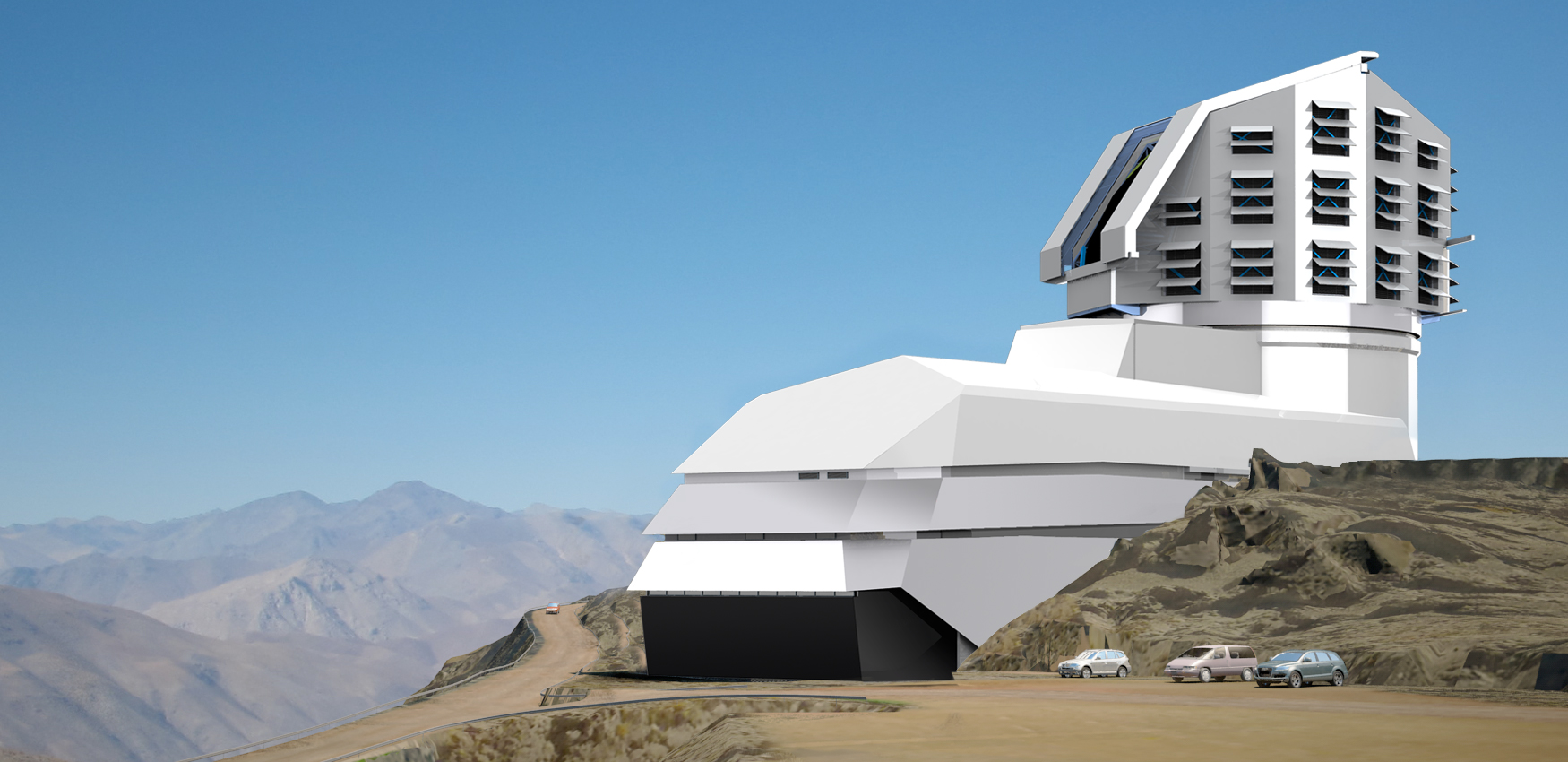} 

\chapter*{Abstract}
The\blfootnote{Image: The Vera C.~Rubin Observatory, \url{www.lsst.org/gallery}} Vera C.~Rubin Legacy Survey of Space and Time holds the potential to revolutionize time domain astrophysics, reaching completely unexplored areas of the Universe and mapping variability time scales from minutes to a decade. To prepare to maximize the potential of the Rubin LSST data for the exploration of the transient and variable Universe, one of the four pillars of Rubin LSST science, the Transient and Variable Stars Science Collaboration, one of the eight Rubin LSST Science Collaborations, has identified research areas of interest and requirements, and paths to enable them. While our roadmap is ever-evolving, this document represents a snapshot of our plans and preparatory work in the final years and months leading up to the survey's first light.

\chapterimage{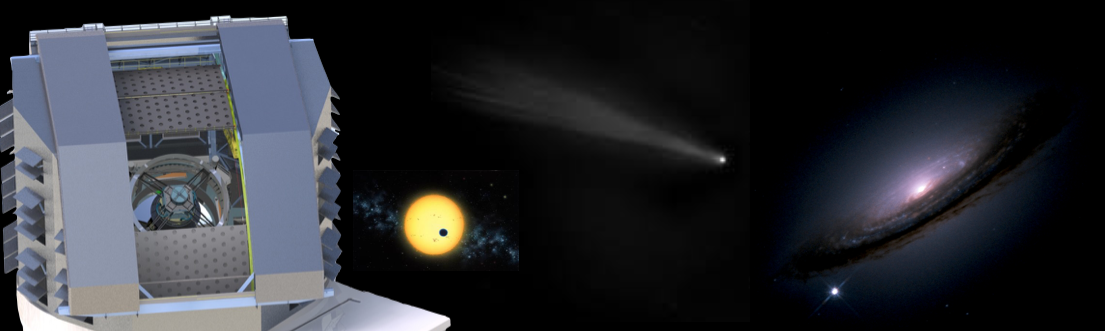} 

\chapter{Introduction} 

\noindent
\textsl{Authors: Rachel Street and Federica B. Bianco}

\bigskip

\noindent The Legacy Survey of Space and Time (LSST) at the NSF's Vera Rubin Observatory, due to begin science operations in 2023, will be a landmark project in astrophysics \citep{Ivezic2019}.  It will survey the entire southern sky every $\sim$3\,days in 6 filters, with a spatial resolution  of 0.2\," pix$^{-1}$ and to fainter limiting magnitudes ($g<25.0$\,mag in 30\,s)
than previously possible.  Crucially, Rubin LSST will continue for 10\,yrs, delivering an unprecedented catalog of long-baseline light curves.  

Recent improvements in data processing technology make it possible to not only reduce the survey data in real-time but also to disseminate alerts of new discoveries within seconds.  This raises the possibility of detecting and characterizing previously unexplored categories of astronomical variability for the first time, particularly transient phenomena which simply occurred too quickly to be studied by previous surveys.  

Fully exploiting this ground-breaking data presents some challenges, however.  The volume of data and it's rate of delivery: $\sim$20\,TB of data {\em per night}, producing $\sim$10\,million alerts {\em per night} - far exceed the scale of any previous survey.  Furthermore, some scientific goals will require the coordination of follow-up observations made in real-time response to LSST alerts.  Substantial preparation will be required in order to maximize the scientific return.  

The Transients and Variable Stars Science Collaboration (TVS) exists to bring together scientists interested in studying these classes of object from Rubin LSST data, with the goals of preparing for and facilitating their science.  The purpose of this document is to outline the areas of scientific study of interest to TVS members in sufficient detail to allow for the necessary preparations required to achieve those aims.  

The paper is organized into four main sections, as follows:

\begin{itemize}
\item Time sensitive science cases: these are science cases that require prompt identifications and triggers of follow-up resources.
\item Non-time sensitive science cases: these are science cases that can be developed on the annual data releases. Some of these science cases may assume that the sample is pure, i.e. that prompt identification or characterization has been achieved. 
\item Deep Drilling Fields

\item Mini- and Micro-Surveys
\end{itemize}
Each of these section is further subdivided into:

\begin{itemize}
\item Extrinsic transients and variability: including geometric transients such as microlensing, eclipsing binaries, and transiting planets or debris disks. 
\item Intrinsic galactic and Local Universe transients and Variables: including stellar eruptions, explosions, pulsations.
\item Extra-galactic transients: including supernovae, GRB, blazars. 
\end{itemize}

Each of these section includes multiple science cases. Each science case includes: 
\begin{itemize}
\item Low hanging fruits
\item Pie in the sky
\end{itemize}
Each section identifies tasks (including but not limited to): 

\begin{itemize}
    \item Observations needed ahead of Rubin LSST to narrow the space of relevant variables;
\item Theory development to generate predictions on Rubin LSST observables including data integration activities required to incorporate multiwavelength/context/other time scale data;
\item Computational advancements and infrastructure required to handle Rubin LSST data volume;
\item Facilities upgrades and development to support follow-up.
\end{itemize}
Additional chapters include a chapter on methodology and infrastructure, which includes discussion of the classification of transients and variable stars, and further, an overview of existing brokers. As it is a fundamental aspect of the TVS, there is also a chapter dedicated to equity and inclusively, which includes the TVS {\it Call for Action} and the steps the TVS are taking to be a diverse, inclusive and equitable platform, focused on the exploitation of Rubin LSST data.

\section{Rubin LSST Data Products and their access -- An Overview}
\textsl{Author and Editor: M.~L.~Graham}
\subsection{Rubin LSST Data Products and their access}

\bigskip

\noindent
The Rubin Observatory Legacy Survey of Space and Time (LSST) is an astronomical project designed to generate significant advances in four science areas: cosmology (dark matter and dark energy); the Solar System (with a focus on potentially hazardous asteroids); the Milky Way Galaxy; and transient phenomena. To do this, the Rubin LSST will deliver a deep survey that covers $\sim$18,000 square degrees in the southern sky and will detect $\sim40$ billion stars and galaxies \citep{2008arXiv0805.2366I}. A total of $\sim825$ visits to each part of the sky within this area will be made in six filters, {\it ugrizy}, over 10 years. About $10$\% of the observing time will be devoted to community-proposed special programs that extend the areal coverage, depth, and/or sampling cadence (e.g., mini-surveys, deep drilling fields). Rubin LSST currently estimates that full operations will begin in late 2024. 

Rubin LSST will acquire $\sim$20 terabytes of raw data each night and process it in real time, distributing alerts on objects that vary in brightness or position within 60 seconds, delivering processed images and updated object catalogs within 24 hours, and releasing a yearly reprocessed data set including deep image stacks (\citealt{LSE163}).
To enable science with this massive data set, the Rubin LSST Data Management System includes the Science Platform: a web-based service for data access, analysis, and processing that includes software tools and computational resources (\citealt{LSE319}).
We discuss the Prompt and Data Release data products in further detail in \autoref{sec:lsst_prompt_dp} and \ref{sec:lsst_dr_dp}, respectively.

\subsection{Remotely-Accessible Computing Platforms for LSST Science}

The size of the LSST data products, particularly the image data, is expected to break the traditional paradigm of astronomers downloading a dataset to their local machine for analysis.  For many large scale analyses, it will be unfeasible or prohibitively expensive in cost and time for many users to download sufficient data for their purposes.  Instead, Rubin science users are transitioning to a ``next-to-the-data`` paradigm, where vast data products are hosted at large, central Data Access Centers (DACs), and made accessible through online platforms that users can interact with through their browser. At the time of writing, the Rubin LSST Science Platform (RSP) is the most developed of these platforms, and the community are encouraged to gain experience with analysis using its tools through the Rubin Data Previews.  We note that other platforms are in the process of development and offer complementary and overlapping functionality, such as the LINCC Computing facilities\footnote{\url{https://lsst.dirac.dev/}} offered by the DIRAC Institute and the University of Washington.  

The RSP offers three main aspects: an interactive web portal by which the database contents can be searched and visualized, a Jupyter notebook aspect providing an interactive coding environment and Application Programmable Interfaces (APIs), to enable queries of the DAC by remote services.  In the expectation that different science use-cases might have different requirements for these aspects, we summarize the conclusions of the previous sections in \autoref{tab:LSP_aspects_usage}.  
\begin{table}
    \centering
    \begin{tabular}{l|c|c|c|c|c|c}
        Tool & Pulsating & EXor/FUor & Blazars & Microlensing & Brown  & Eclipsing \\
             &  Stars  &             &         &              & Dwarfs & Binaries \\
        \hline
       Image inspection & Yes & - & Yes & Yes & - & Yes \\
       Interactive plotting  & Yes & - & Yes & Yes & Yes & Yes \\
       Database search  & Yes & Yes & Yes & Yes & - & Yes \\
       Python notebooks & Yes & - & Yes & Yes & Yes & Yes \\
       APIs & Yes & Yes & - & Yes & - & Yes\\
    \end{tabular}
    \caption{Summary of data access functionality required by different TVS Science topics.}
    \label{tab:LSP_aspects_usage}
\end{table}

\section{Purpose of This Document}

The goals of this document are many-fold:
\begin{itemize}
\item Stimulate research and preparations for Rubin LSST 
\item Outline areas of necessary work (theoretical and practical)
\item Identify weaknesses and strengths of different Rubin LSST observing strategies for time domain science
\item Identify areas of research and infrastructure which are currently under prepared.
\item Facilitate the communication of ideas between TVS subgroups and with the other Science Collaborations, and reduce redundancy in our efforts.
\item Build a document that can be referenced in future grant applications.
\end{itemize}

\chapterimage{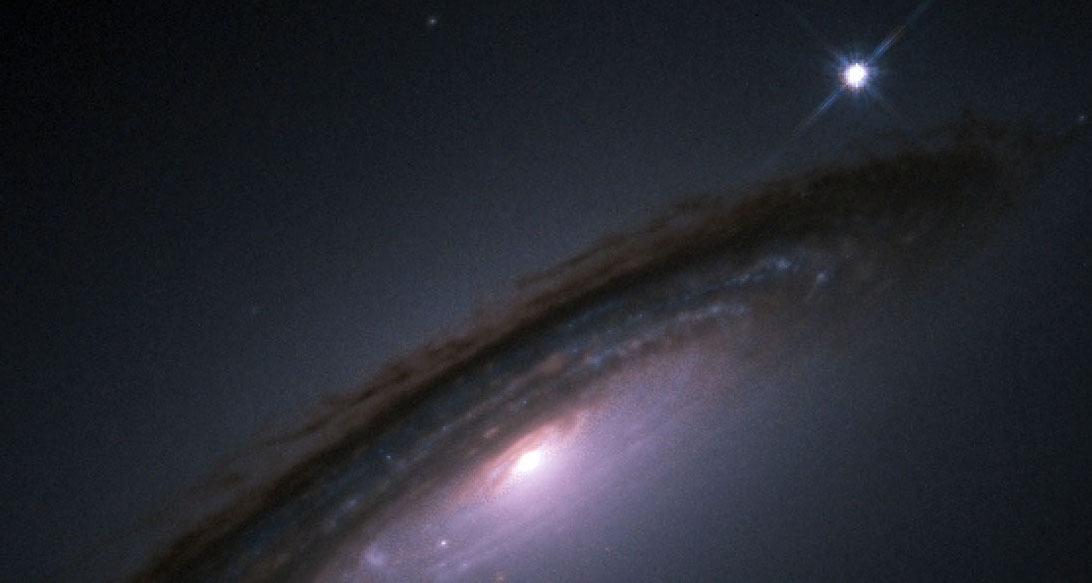}

\chapter{Time Critical science}

\section{Executive Summary}\label{sec:time_critical_science_summary}
\textsl{Authors: Rachel Street and Kelly Hambleton}

\bigskip

\noindent The\blfootnote{Image: Hubble Space Telescope-Image of Supernova 1994D (SN1994D) in galaxy NGC 4526.} nature and rapid delivery of Rubin LSST data prompt data products, combined with the groundbreaking survey volume, will open new areas of time critical science. \\
{\bf Microlensing:} Rubin LSST will detect statistically significant rates of microlensing events across the Galactic Plane and Magellanic Clouds.  The baseline survey duration and cadence is ideal to probe for lensing by isolated black holes, while stellar and planetary events are likely to need follow-up observations in order to properly characterise the events with sufficient cadence. \\
{\bf Eclipsing Binaries:} During the main survey, Rubin LSST will observe $\sim$1 millions contact binary stars. Of these, it is possible that, for the first time, a contact binary undergoing coalescence will be observed. The anticipated signature is the swift reduction of the orbital period.\\
{\bf Rare anomalies and SETI:} The consistency and duration of Rubin LSST also grants us the opportunity to search for deviations from understood physical behavior.  Anomalies may often reveal new physics, they can also be considered from the perspective of searching for optical signatures of extra-terrestrial civilizations. \\
{\bf Eruptions and Accretion Processes:} Rubin LSST will deliver statistically significant population samples of a wide range of intrinsically rare classes of object, and among other benefits this will provide insight into astrophysical accretion processes and outbursts occurring in different contexts.  These range from pre-main sequence stars to white dwarf, neutron star and black hole binaries, and in all cases the Rubin LSST data will highlight the impact of accretion on the evolution of these objects.  Uniquely, Rubin LSST will provide both the long-baseline time-series data necessary to quantify outburst cycles and object occurrence rates but also the rapid alerts necessary to enable outburst phases to be identified while in progress, triggering more detailed follow-up studies.\\
{\bf Explosive Transients:} Rubin LSST will discover an extensive sample of all classes of novae, intermediate-luminosity optical transients, supernovae, kilonovae, and gamma-ray bursts. These observations will reveal the mechanisms at work during and after the explosion or during mergers.\\
{\bf Active Galaxies: } Rubin LSST alerts will enable the swift multi-wavelength follow-up of blazar flux variations, including flairs. The optical counterparts of known blazars will also be observed and will reveal the nature of double black-hole binaries and the nature of neutron production in blazar jets.\\
{\bf Tidal Disruption Events: } Due to the depth and temporal sampling of Rubin LSST, the number of observed tidal disruptive events (TDEs) is expected to increase by an order of magnitude from $\sim$10/yr to $\sim$100/yr. This will enable studies of the supermassive black holes, host galaxies, the black hole occupation fraction and the central black hole occurrence as a function of redshift. Additionally, Rubin LSST will allow the probing of the optical emission, the cause of which is currently under dispute. \\
{\bf Gravitational Waves: } Primarily, Rubin LSST's target of opportunity program will be used to perform swift electromagnetic follow-up of gravitational waves produced by neutron star mergers, black hole mergers and black hole-neutron star mergers. Additionally, the early follow-up observations of kilonovae will be triggered by Rubin LSST's rapid alerts.\\
The software and technology development necessary to achieve these science goals are discussed. 

\section{Rubin LSST \emph{Prompt} data products}\label{sec:lsst_prompt_dp}
\textsl{Editor and Author: Melissa Graham}

Every standard visit image ($\sim30$s integration) acquired by the Rubin LSST will be immediately reduced, calibrated, and processed with Difference Image Analysis (DIA), wherein a template image (a deep stack of previously obtained images) is subtracted to generate a difference image. All sources detected in the difference image represent the time-variable components of transient phenomena (e.g., supernovae), variable stars (e.g., RR Lyrae), and moving objects (e.g., asteroids). For all difference-image sources detected with a signal-to-noise ratio (SNR) of at least 5, an alert packet containing information about the source (location, fluxes, derived parameters, and $\sim6^{\prime\prime}\times6^{\prime\prime}$ cutouts) will be generated and released within $60$ seconds of the end of image readout. Alerts are world public and can be shared with anyone, anywhere. Alerts on moving objects will be integrated into the Minor Planet Center. Due to the very high bandwidth of the Rubin LSST Alert Stream it will only be delivered in full, in real-time, to 7 Alert Brokers\footnote{Find the list of brokers at \url{https://www.lsst.org/scientists/alert-brokers}.} \citep{LDM-612}, who will serve them to their communities (some brokers plan to provide public access, e.g., \citealt{2018ApJS..236....9N}). All other prompt data products (e.g., visit, template, and difference images, catalogs of difference-image objects) will be made available to Rubin LSST members via the Science Platform within $24$ hours, but are subject to a two-year proprietary period, after which time they can be shared with anyone, anywhere, worldwide.

\subsection{Quantities of Particular Importance to Time-Domain Studies} 
There are four Prompt data types and quantities that we want to highlight which will be specifically useful for time-domain astronomy. First, all detected difference-image sources, for which there has been no variable source previously detected, will have forced photometry performed at their location using the last $\sim$30 days of difference images in order to, e.g., look for faint precursor events. This is commonly referred to as precovery photometry and will be available within 24 hours (as all Prompt data products). Second, all objects that were detected in difference images within the past $\sim$12 months will have forced photometry performed at their location in the new difference image in order to, e.g., continue to monitor known objects as they fade. This forced photometry will be available within 24 hours. Third, the catalog of difference-image sources will contain some light curve characterization parameters for, e.g., periodic and non-periodic features. The exact nature of these parameters is to be determined, but they will be updated to include new observations within 24 hours. Fourth, all difference-image sources will be associated with the nearest static-sky objects from the Data Release data products (see \autoref{sec:lsst_dr_dp}), so that the potential host galaxy and/or longer-term information about the variable star can be easily obtained. These associations will be included in the alerts and also the difference-image source catalogs. See also \cite{LSE163} for a full and complete description of the Rubin LSST data products.

\section{Extrinsic Transients} 
\textsl{Editor: Rachel Street}

\subsection{Microlensing}
\textsl{Authors: Rachel Street, Marc Moniez, Rosanne Di Stefano}
\label{sec:micro-time}

\bigskip

\begin{figure}[ht]
\begin{center}
\includegraphics[width=0.5\columnwidth]{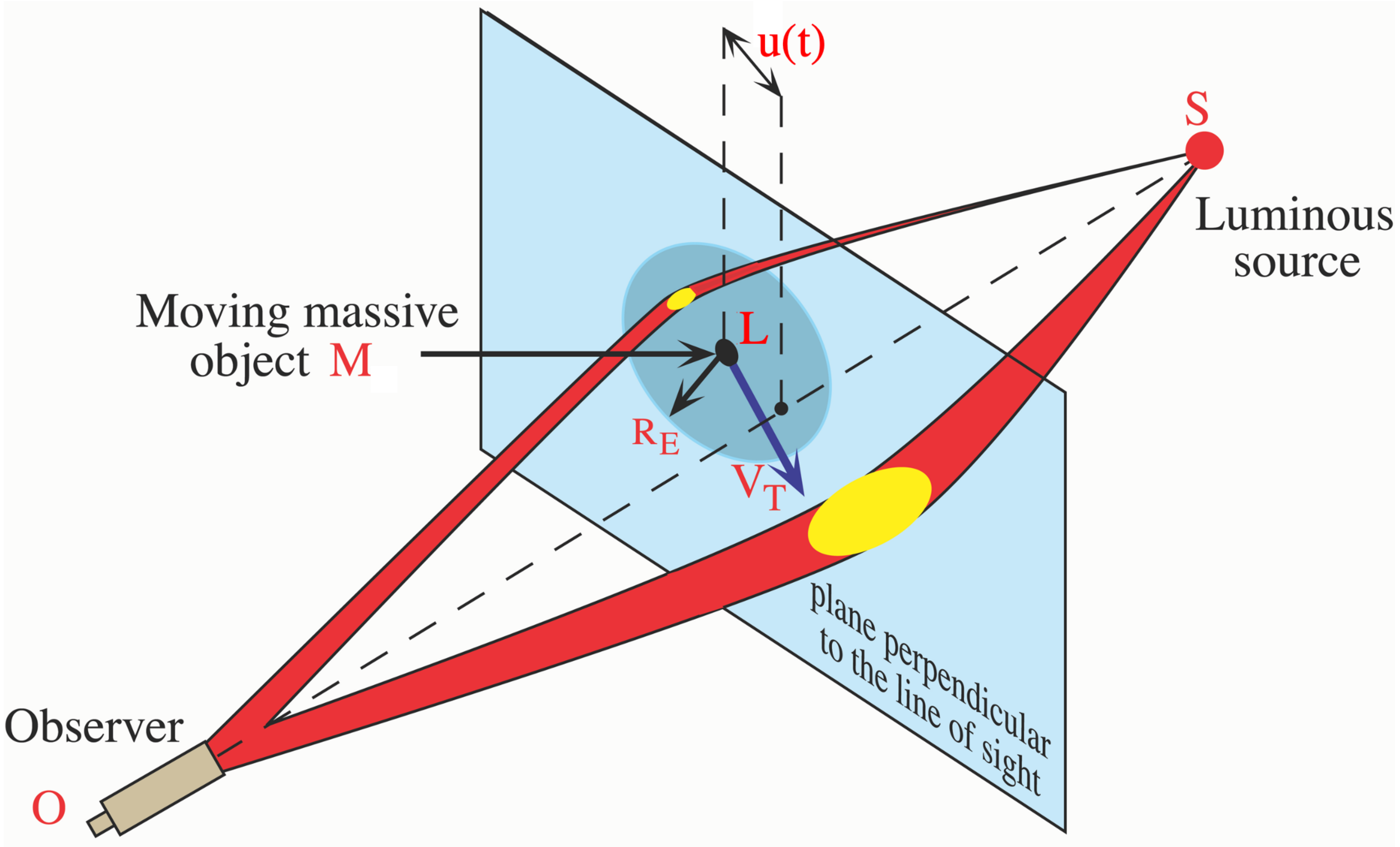}
\caption{ Geometry of a microlensing configuration.
{\label{principe-microlensing}}
}
\end{center}
\end{figure}

\noindent Microlensing occurs when a foreground object, lens $L$, passes directly between the observer, $O$, and a luminous background source, $S$ \citep{Paczynski_1986} (see \autoref{principe-microlensing}).
A review of the microlensing formalism can be found in \citet{Rahvar_2015}.  The gravity of the lens deflects light from the source with a characteristic radius, $R_{E}$, causing the source to brighten and fade as they move into and out of alignment, with a timescale, $t_{E}$, that scales with the square root of the lens mass (years for black holes (BHs), days to weeks for planets and stars). Assuming a single point-like lens of mass $M$ located at distance $D_L$ is deflecting the
light from a single point-like source located at distance $D_S$, the characteristic radius (Einstein radius) $R_{\mathrm{E}}$ is given by: 

\begin{equation}
R_{\mathrm{E}}\!\! =\!\! \sqrt{\frac{4GM}{c^2}D_S x(1-x)}
\simeq\! 4.54\ \mathrm{AU}\left[\frac{M}{M_\odot}\right]^{\frac{1}{2}}\!
\left[\frac{D_S}{10 kpc}\right]^{\frac{1}{2}}\!\!
\frac{\left[x(1-x)\right]^{\frac{1}{2}}}{0.5},
\end{equation}
where $G$ is the Newtonian gravitational constant, and $x = D_L/D_S$.
If the lens is moving at a constant relative transverse
velocity $v_T$, the characteristic lensing time scale is given by:
\begin{eqnarray}
t_{\mathrm{E}} \sim
79\ \mathrm{days} \times 
\left[\frac{v_T}{100\, km/s}\right]^{-1}
\left[\frac{M}{M_\odot}\right]^{\frac{1}{2}}
\left[\frac{D_S}{10\, kpc}\right]^{\frac{1}{2}}
\frac{[x(1-x)]^{\frac{1}{2}}}{0.5}\; . 
\end{eqnarray}
The so-called simple microlensing effect (point-like source and lens
with rectilinear motions) has several characteristic
features. Given the low probability of the alignment, the event should be singular in the history of the source (as well as of the deflector); the magnification due to lensing, independent of the color, is a simple function of time depending only on three parameters, with a symmetrical shape; as the source and the deflector are independent,
the prior distribution of the events' impact parameters $u_0$ (minimum distance, $u(t)$, see \autoref{principe-microlensing}) must be uniform; all stars at the same given distance and towards a given direction have the same probability of being lensed; therefore the sample of lensed stars should be representative of the monitored population at that distance, particularly with respect to the observed color and magnitude distributions.

Since this phenomenon doesn't require light to be detected from the lens itself, it is capable of exploring populations which are otherwise hidden from view due to their distance and/or intrinsic luminosity.  

Rubin LSST offers two complementary opportunities for microlensing discoveries: by extending the Wide-Fast-Deep survey to cover the Galactic Plane, and by conducting coordinated observations of a single Rubin pointing located on the Roman Space Telescope Bulge survey region.  Here we explore the scientific yield from both observing strategies. 

    \begin{itemize}
    \item Low hanging fruits
        \begin{enumerate}
        \item[a)] {\bf Galactic Plane Survey: Galactic population of single and binary black holes}\\
        Stellar evolution models imply that there should be millions of black holes residing in our galaxy \citep{Gould_2000}. 
        The past and present microlensing surveys all suffered from a drastic decline of the detection efficiency for events with durations $t_{\mathrm{E}}$ larger than a few years, which are expected from massive lenses ($M>20\,M_\odot$). This is the reason why the published limits on the contribution of compact objects to the Galactic dark matter were not very constraining beyond this mass.
        Recently, and following \citep{Mirhosseini_2018}, new limits have been published on the contribution of halo objects in the mass range $10\,M_\odot<M<1000\,M_\odot$ by combining EROS-2 and MACHO data \citep{Blaineau_2022}, that contain 14.1 million objects in common, consequently monitored for 10.6 years. This result illustrates the potential of 10 years of continuous observations. The light-curves measured by Rubin LSST, eventually completed by the aforementioned surveys, will allow one to reach the sensitivity either to detect heavy black holes up to a few thousands of $M_\odot$ and measure their Galactic density, or to exclude the possibility that they form a significant fraction of the Galactic hidden matter.
        
        The main condition to succeed in this task is to ensure a time-sampling of the Galactic fields and of LMC/SMC that spans the entire Rubin LSST survey duration and avoids very long gaps within the light-curves (apart from the inevitable inter-seasonal gaps). The final efficiency to long timescale events will not be sensitive to the details of the cadencing, so long as long gaps are avoided.  In this search, the past and present surveys databases will add precious information to confirm or not microlensing candidates found with Rubin LSST alone.

        Rubin LSST will complement the results from the gravitational wave observatories,
        since it is capable of directly measuring the statistical abundances of intermediate mass black holes (single or multiple), through their lensing effect on background sources. Rubin LSST will reach fainter magnitudes than OGLE ($\sim$23 mag vs $\sim$20 mag), so we expect that by monitoring a few billion stars in the Galactic Plane it will detect hundreds of microlensing events due to BHs.  

        \item[b)] {\bf Observing Self-lensing in the Magellanic Clouds}\\
        By including the Magellanic Clouds in the Wide-Fast-Deep survey, Rubin LSST will be able to detect microlensing events where both the lens and source lie in the Clouds. Hence Rubin LSST will explore stellar and stellar remnant populations in another galaxy.
        By conducting a long duration, self-consistent survey including the Milky Way Bulge, Galactic Disk and both Magellanic Clouds (LMC, SMC) Rubin LSST will compare the populations in different local environments with different star formation histories.  Since microlensing events are transients, new events reveal more of the underlying population each year.
    
        \item[c)] {\bf Understanding galactic and extra-galactic population of planets and low-mass stars}\\
        Despite outstanding discoveries from Kepler \citep{Borucki2010} and other surveys, the vast majority of known exoplanets lie within 1\,kpc of the Sun\footnote{\url{https://exoplanetarchive.ipac.caltech.edu/}}. Microlensing can probe to much greater distances ($\leqslant$8.5\,kpc) and allows sensitivity to planetary systems with planets of all masses  in orbits between $\sim$1--10\,AU \citep{Bond_2001, Bond_2004}.  The microlensing rate for surveys outside the Bulge has been measured in a few fields by \citet{Moniez_2017} and estimated for Rubin LSST by \citet{Sajadian_2019} based on the minion\_1016 OpSim {\it with minimal coverage of Plane fields}.  They found an average rate of 15 events deg$^{-2}$ year$^{-1}$ in the disk and 400 events deg$^{-2}$ year$^{-1}$ in the Bulge. This detection rate can be doubled if the cadence is increased from 6\,d to 2\,d.  The ``edge-on'' orientation of the SMC results in a higher probability of self-lensing (where both source and lens lie within the same galaxy), raising the possibility of detecting stellar and perhaps even planetary companions in a galaxy other than the Milky Way \citep{Di_Stefano_2000}.  Rubin LSST is predicted to detect 20--30 events year$^{-1}$ \citep{Mr_z_2018}, provided the SMC is monitored at least once every few days.  
        \end{enumerate}
        
    \item Pie in the sky
        \begin{enumerate}
        \item[a)]{\bf Galactic Plane Survey: Mesolensing} \\
        The rate of microlensing events scales as $\sim$\,${D_L}^{-3/2}$, so nearby objects traveling at relatively high angular velocities are more likely to lens background stars than similar more distant objects \citep{Di_Stefano_2008, Di_Stefano_2008a}. Rubin LSST will investigate the mass distribution of faint objects in the local neighborhood, such as low mass dwarfs, stellar remnants, and free-floating planets.  

        \item[b)] {\bf Galactic Plane Survey: Predicted lensing} \\
        Measurements of stellar proper motions with Rubin LSST will be valuable for predicting future microlensing events, as has been done from Gaia and Pan-STARRS 1 data \citep[to shallower limiting magnitudes than Rubin LSST,][]{bramich2018}.
        \end{enumerate}
        
    \end{itemize}

\subsubsection{Preparations for Microlensing Science}
Microlensing events are transient in the sense that they are non-repeating, so almost all observations must be obtained during the course of a given event. One specificity of Rubin LSST is worth highlighting for the search of microlensing effects: its excellent photometric accuracy. 

Until now, searches for microlensing effects by 1-2m class telescopes have mainly detected events with strong amplification, corresponding to an impact parameter $u_0<1$. The search should strongly benefit from the large diameter of Rubin LSST, by significantly improving the detection rate of weakly amplified events with impact parameter $u_0>1$.

The scientific goals described above necessitate the following capacities:

\begin{itemize}
    \item {\bf Identifying events from alerts}\\
    The first step to achieving the above science goals is thus to accurately identify microlensing events from Rubin LSST data, at as early a stage as possible.  Heavy use of the Rubin LSST prompt data products via one or more broker services is therefore anticipated. Recent work by \cite{Godines2019} has begun to apply machine learning techniques to identify microlensing events using alerts from the ZTF survey as a precursor; this is complemented by the development of metrics by \cite{khakpash2021classifying}.  Work is underway to apply Godines' software LIA\footnote\url{{https://github.com/rachel3834/LIA}} as a filter within the ANTARES\footnote{\url{https://antares.noao.edu/}} and the FINK\footnote{\url{https://fink-broker.org/}} brokers, thereby supplying a detection stream to the community.  We note that at the time of writing, a number of ANTARES and FINK filters include microlensing as one of their alert products, but more work is required to test and improve the performance of these filters in Rubin LSST-like data sampling with a realistic range of events, and to understand the selection biases. 
    It is worth mentioning that the Photometric Rubin LSST Astronomical Time-Series Classification Challenge (PLAsTiCC) \citep{kessler2019models} has provided a large sample of simulated Rubin LSST-like lightcurves including microlensing event that has initiated the development of machine learning algorithms to classify Rubin LSST lightcurves. We emphasize that in order for the classifiers to perform well, microlensing blending and second order effects like parallax and finite source effect should be simulated more accurately.
    
    \item {\bf Follow-up Observations}\\
    Long timescale (several years) Rubin LSST lightcurves for black hole candidates are likely to be of sufficiently high cadence that no additional timeseries photometry is required but this is not true for stellar lensing candidates, which can exhibit anomalous variations on timescales of days or shorter.  Timeseries imaging of selected targets will therefore need to be obtained in rapid response ($<$day) to alerts and to continue for $\sim$weeks, with the cadence dynamically scaled to match the evolving features of the targets.  
    
    Spectroscopic follow-up of all selected candidates will be highly desirable to help to characterize the source star and to derive an independent estimate of its angular size and distance.  This information is essential to resolve degeneracies in the microlensing model.  
    
    High spatial resolution imaging, both during and several years after the event, can also be used to constrain the properties and motion of the lens.  
    
    Although microlensing events can brighten by several magnitudes, a high fraction of Rubin LSST candidates are expected to be fainter than $i>$17.0\,mag.  Photometric follow-up will require narrow-field optical (and ideally NIR) imagers on 2--4\,m class telescopes, while spectroscopy will require high throughput spectrographs of R$\sim$few 1000 on 4--10\,m facilities.  AO imaging will be restricted to the brighter targets and will require 10\,m class facilities.  

    \item{{\bf{Facillities/Software Requirements}}\newline
    Depending on the survey strategy, Rubin LSST is likely to produce a few hundred microlensing alerts outside the Galactic Bulge per year and thousands of events within it, with a false alarm rate which has yet to be determined.  Since timeseries follow-up will be required as well as rapid-response observations, a Target and Observation Manager (TOM) system will offer great advantages to manage the overall program and coordinate observing efforts.}
\end{itemize}

\subsection{Eclipsing Binary Stars}
\textsl{Author: Andrej Pr\v sa} \label{sec:tcebs}

Binary systems are objects containing two or more stellar components that orbit around a common center of mass. When binary systems are oriented along a preferential line of sight, such that the stars pass in front of each other during their orbit, the binary system is classified as eclipsing. Eclipsing binary stars are typically considered non-transient: although extrinsically variable, most systems exhibit strictly periodic observables. Deviations from strict periodicity are, however, due to transient phenomena: spots, pulsations, gravitational and magnetic interactions between components, third body interactions, stellar evolution effects, etc. Thus, we consider these time-critical transient phenomena \emph{extraneous} to binarity: the time-sensitive triggers should be driven by the respective transient science rather than binarity. The benefit of finding transients in binaries, of course, is access to fundamental stellar parameters; we defer the discussion of this aspect to \autoref{sec:ntcebs}.

One exception to the above statement is coalescence. Merging of two compact objects (white dwarfs, neutron stars or black holes) is receiving all due attention thanks to the recent successes of the gravity wave physics \citep{Abbott_2016}, but coalescence of contact binary systems (two main sequence stars sharing a common envelope) is interesting in its own right. The wealth of physics that accompanies such a dramatic event spans plasma physics, equations of state, radiative properties, energy flows with mass transfer, all the way to outbursts and rapid rotation effects. Given that contact binaries are very common (about 15\% of all eclipsing binary systems), Rubin LSST is well positioned to catch such coalescence events over its nominal 10-yr survey.

\begin{itemize}

    \item Pie in the sky:
    \begin{enumerate}
        \item [a)] {\bf detecting a contact binary coalescence event}\\ 
        We have yet to observe such an event and the Rubin survey certainly holds the prospect of finding it. The 10-yr baseline of observing the expected ~1 million contact binaries \citep{prsa2011} will likely catch many candidates that exhibit a decreasing orbital period; of those, the prime candidates will feature a rapid, accelerated period decrease and those should be immediately followed up by point-and-stare instruments.
    \end{enumerate}

\end{itemize}

\subsubsection{Preparations for Eclipsing Binary Science}

As stated above, the vast majority of the eclipsing binary deliverables from Rubin LSST are non-time-critical, with the exception of coalescence. A coalescence event (the merging of two non-compact stars) will inevitably be a dramatic event: as the stars spiral inwards, the light curve periodicity will deteriorate and bursts of highly variable light are expected. We can bank on that variability to trigger follow-up observations of possible coalescence events.

\begin{itemize} 

\item{\bf {Identifying Events from Alerts}}\\
A binary star light curve that corresponds to a possible coalescence precursor will exhibit a significant period shortening, followed by a number of highly variable light events. These are likely to trigger an alert and, provided that the system was on our watch list, the alert should be quite straight-forward to identify and pass from the event brokers to follow-up facilities.

\item{\bf{Follow-up Observations}}\\
Once identified, a coalescence event should be followed up photometrically and spectroscopically as quickly and as completely as possible. The process of two stars coalescing into a single rapidly rotating star has yet to be observed, but the presence of blue stragglers and yellow giants clearly indicate that these events \emph{do} happen. While the timescales for coalescence are uncertain and thus far only theorized, an immediate follow-up campaign is crucial to catch and characterize a coalescence event as it is happening.

\item{\bf {Facilities/Software Requirements}}\\
Brokers are needed to properly flag a short-period (0.1-0.3 day) light curve with a shortening orbital period that starts to exhibit high amplitude light variations as these might be telltale signs of a coalescing binary system. Such events should trigger alerts for manual vetting and, if confirmed, immediate follow-up. Depending on the magnitude of the observed object, we need some of the largest observatories pointing to it as quickly as possible, and following up for as long as possible.

\end{itemize}

\subsection{The Search for Extraterrestrial Intelligence}
\textsl{Authors: James R. A. Davenport \& Federica B. Bianco}
\selectlanguage{english}

\bigskip

\begin{figure}[ht]
    \centering
    \includegraphics[width=0.70\columnwidth]{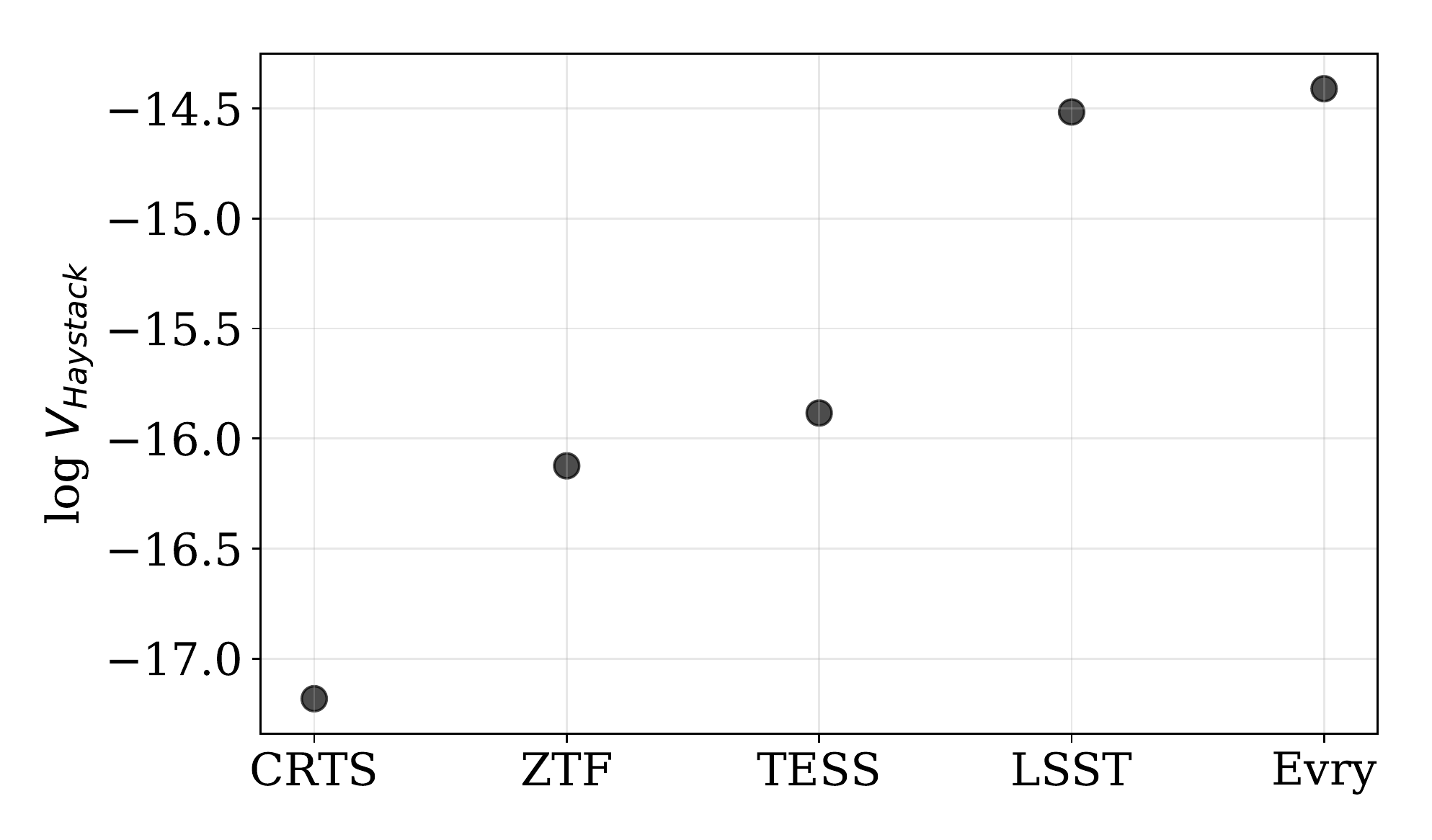}
    \caption{Comparison of the 8-D “Cosmic Haystack” SETI
search volume fraction, defined by \citep{wright18SETI}, computed using five optical surveys with varying designs. The Haystack fraction covered by these optical surveys is 1-2 dex larger than typical SETI programs conducted in the radio.
Evryscope \citep{Law15evryscope} is the best survey considered here for SETI work, narrowly beating Rubin LSST due to its wide field of view and very dense light curves. Adapted with permission from \citet{Davenport19}.}
    \label{fig:seti}
\end{figure}

\noindent Traditional Searches for Extraterrestrial Intelligence (SETI) and searches for ``technosignatures'' (signatures in the data unexplainable by natural phenomena) currently focus on dedicated observations of single stars or regions in the sky with the aim to detect excess or transient emission from intelligent sources. The latest generation of synoptic time domain surveys, for which Rubin LSST is the flagship, enable an entirely new approach: spatio–temporal SETI, where technosignatures may be discovered from spatially resolved sources or multiple stars over time (\autoref{fig:seti}).

\noindent Optical searches for SETI rely on four families of technosignatures:
\begin{itemize}
\item {Unnatural orbit alterations of objects in the Solar System (Loeb 2019)}.
\item Unnatural flux patterns of otherwise normal astrophysical variability (e.g. flares, pulsations, transits, etc), such as broadcasting a prime or Fibonacci sequence using transits \citep{arnold2005artificial, wright2016dysonian, Benford_2016, Guillochon_2015}.
\item Spatial correlations of events or phenomena: e.g. coordination of transiting systems, or rebroadcasting events such as novae along the ``SETI ellipse" \citep[the ellipsoid for receiving synchronized signals from multiple transmitters;][]{makovetskii1977nova,lemarchand1994passive, tarter2001seti, shostak2004scheme}.
\item{Spatial over-densities or unusual distributions: e.g. an over-density within the “Earth Transit Zone” band of stars that would see Earth as a transiting exoplanet \citep{heller2016search}, or in spatial clusters \citep{Davenport19}}.
\item{ Unusual variability profiles or statistical distributions of fluxes either on short timescales such as Boyajian’s Star \citep{marengo2015kic}, or long timescales such as disappearing stars \citep{villarroel2019vanishing} or occasionally missing transits \citep{kipping2016cloaking}. }
\end{itemize}

All of these signals are expected to be exceedingly rare, making the search for SETI a truly difficult problem of anomaly detection, a proverbial needle in the haystack. Specific anomaly detection algorithms for the technosignatures we listed above need to be designed to tap into the enormous discovery potential that Rubin LSST’s data rate and complexity uniquely enable.

\begin{itemize}
\item Low hanging fruits:
\begin{enumerate}
\item[a)] {\bf{Establishing a baseline for signals}}\\
    Spatiotemporal monitoring of the Rubin LSST sources, in conjunction with multiwavelength observatories, will enable the establishment of a baseline for signals at the relevant frequencies. The short time scale variations can be explored with a novel observing methodology in which the Rubin LSST telescope tracking is turned off during exposures in order to produce star trail images \citep{thomas2018searching}. These trailed images enable photometric time series to be measured for bright sources with time resolution of 14 milliseconds. The 9.62 square-degree Rubin LSST field-of-view dramatically extends the reach of this technique and enables the first large scale search of this kind. The combination of these high frequency observations with standard Rubin LSST observations and those of complimentary multiwavelength observatories, will enable the creation of a framework for the detection of anomalies.
\end{enumerate}

\item Pie in the sky:
\begin{enumerate}
    \item[a)] {\bf{Detection of anomalous signals}}\\
    The detection of anomalous signals or unknown-unknowns would have a profound impact. A credible technosignature detected by Rubin LSST would be an invaluable discovery. While contact is not something that could be attempted with Rubin, any credible detection of anomalies that remain unexplained by any physical process would be transformational with profound implication for our self-perception and our society. As such a discovery has not yet been made, the timeline for such a discovery would encompass the complete 10-yr survey.
\end{enumerate}
\end{itemize}

\subsubsection{Preparations for SETI}
\begin{itemize}
    \item{\bf{Identifying events from alerts}}\\
    As the unnatural nature of a SETI signal is expected to be recognised through the identification of unphysical patterns, it is unlikely that a single nightly alert would determine a SETI trigger (with the exception of interstellar objects transiting in our Solar System, a pur·view of the Solar System Science Collaboration).  However, a collection of nightly alerts may be identified as a potential technosignature, and in that case immediate follow up would be desired as, due to the possible deliberate nature of the signal, it could be turned off at any time. 
    
    \item{\bf{Follow-up observations}}\\
    All anomalous detections of this nature would require follow-up: intense, rapid follow-up in multiple wavelengths and at different time scales to ascertain the anomalous nature of the phenomenon.

    \item{\bf{Facilities/software requirements}}\\
    To fulfill the revolutionary discovery potential of the Rubin LSST project, new SETI algorithms and tools in the data-driven spatiotemporal domain are under development~\citet{Davenport19}, including a SETI framework to ingest these alerts and search for technosignatures in real time. \citet{Davenport19} provides a roadmap to the development of most of this software. A cumulative database can be re-analyzed, providing reproducibility and transparency that will increase the credibility of detections \citep{wright18SETI} Software for detection of anomalous sources in the Solar System would be a purview of the Solar System Science Collaboration \citep{Schwamb_2019}.
\end{itemize}

\section{Galactic and Local Universe transients and variables} 
\textsl{Editor: Paula Szkody}

\subsection{Young Eruptive Protostars} \label{sec:exorfuor}
\textsl{Authors: Teresa Giannini, Rosaria (Sara) Bonito, Simone Antoniucci}

\bigskip

\begin{figure}[ht!]
\begin{center}
\includegraphics[width=0.7\columnwidth]{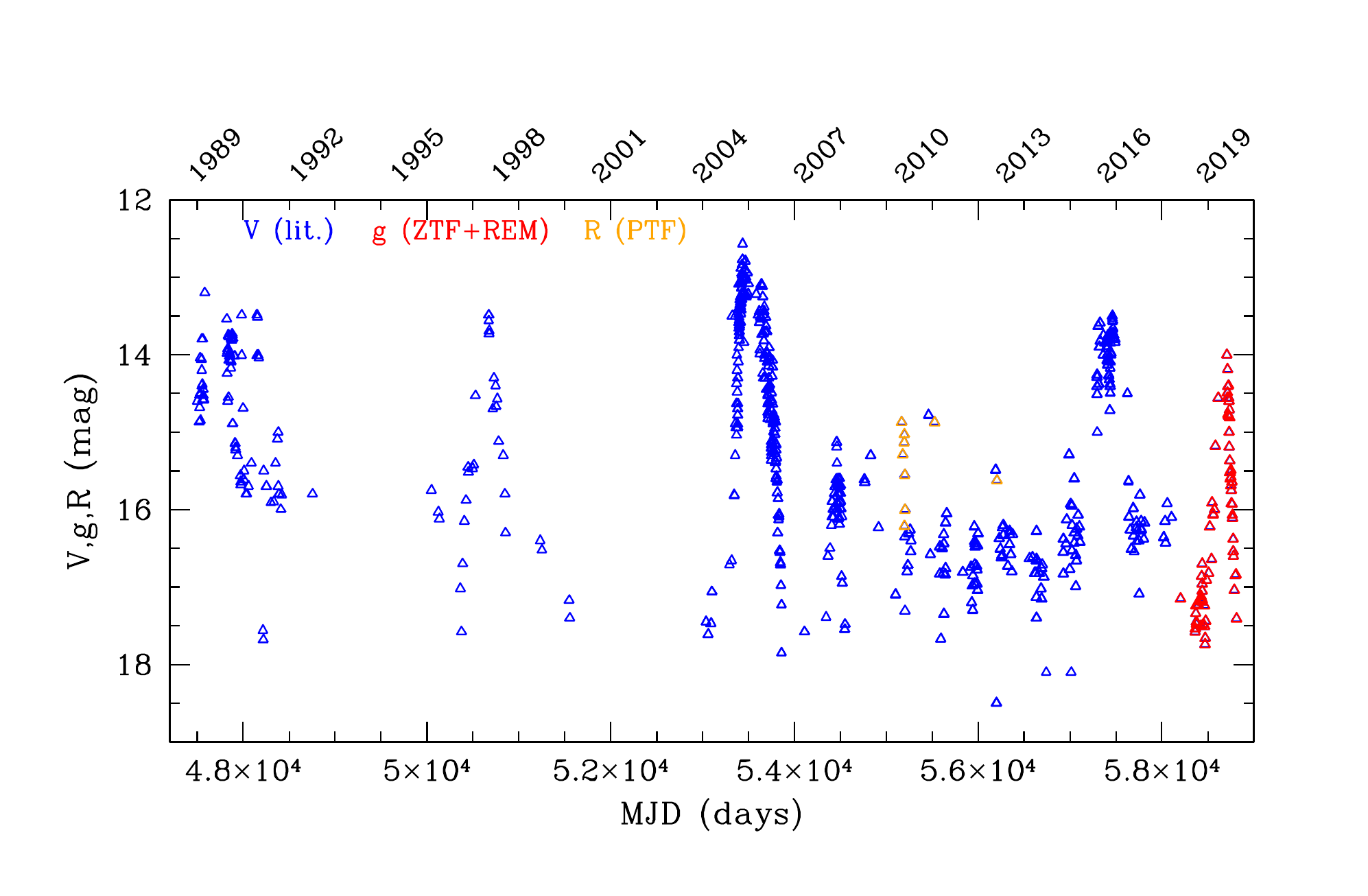}
\caption{A 30-year optical light curve of the EXor V1118 Ori \citep[adopted from][]{giannini_2020}.}
\label{fig:v1118ori_2}
\end{center}
\end{figure}

\noindent Intense accretion activity is a defining feature of the large majority of Pre-Main Sequence (PMS) stars. While small and irregular photometric variations (typically 0.2--1 mag) caused by disk accretion variability are commonly seen in the light curves of many classical T Tauri stars (CTTSs), some young sources display powerful UV-optical outbursts of much larger intensity (up to 4--6 mag, see \autoref{fig:v1118ori_2}). So far, only $\sim$ 20 objects have been recognized
as ``genuine'' eruptive protostars \citep{Audard_2014} and even
less have been monitored long-term. 

Depending on the different properties (burst duration, recurrence time between subsequent bursts, accretion
rate, presence of absorption or emission lines), young eruptive protostars are classified either as FUors \citep{Hartmann_1985} or EXors \citep{gh1989}
Observationally, FUors and EXors are very different objects. FUors are characterized by bursts of long duration (tens of years) with accretion rates 
of the order of 10$^{-4}$--10$^{-5}$ M$_\odot$/yr and spectra dominated by absorption lines, while EXors undergo shorter outbursts (months--one year) 
with a recurrence time of months to years with accretion rates of the order of 10$^{-6}$--10$^{-7}$ M$_\odot$/yr, and are characterized by emission line spectra. 

For both classes of objects it is believed that bursts are due to accretion of material that piles-up at the inner edge of the disk \citep{D_Angelo_2010}. However, the mechanism responsible 
for the burst triggering is not known: proposed scenarios involve gravitational or thermal instabilities inside the disk \citep{Audard_2014} or perturbation by an external body 
(orbiting planets or close encounters with nearby stars). At the moment, however, {\it none of the proposed models are able to provide a realistic 
view of the observed burst phenomenology} because the scarcity of the observations prevents putting tight constraints on the 
physical parameters involved.
 
The unprecedented sensitivity, spatial coverage and, even more importantly, observing cadence of Rubin LSST will allow for the first time a {\it statistical}
approach for the discovery and monitoring of eruptive protostars. In particular, both  the telescope lifetime and the sky coverage 
cadence, will permit  optimal monitoring of EXor-type variables.

\begin{itemize}
\item Low hanging fruits
\begin{enumerate}
\item[a)]{\bf Significantly increase the number of new EXor candidates in our Galaxy}\\
Approximately 20 EXors are known so far, mostly found serendipitous during observational campaigns dedicated to different scientific aims. With Rubin LSST we will have an unprecedented opportunity to significantly improve this number. Considering an r-band limiting magnitude of 24.7 for a single visit, we will observe all the stars with r-band $\sim$ 15, even in obscured regions (with A$_V$ $\le$ 10 mag).  Our selection criteria will be based on: the object's location in a star formation region; the shape of the Spectral Energy Distribution both in and out of the high brightness phase; a light-curve assessment (rise/decline timescale with a typical duration of months, a speed of about 0.05 mag/day and a burst amplitude between 2 and 4--5 mag in g-band); the burst duration (months/few years with 
recurrence expected at months/years); and Rubin LSST color-color analysis \citep[during the burst phase, a significant excess emission in the UV is expected, i.e.][]{Venuti_2015}. The Rubin LSST [$g-r$] vs. [$u-g$] color-color diagram represents a powerful diagnostic tool for selecting EXor candidates. Given the optimal sky coverage and cadence of Rubin LSST, we expect to increase the number of EXor candidates by about an order of magnitude during the first year of observations. 
\item[b)] {\bf Monitor known objects to identify and characterize both their low- and high-\newline brightness states}\\
The Rubin LSST main survey cadence of a couple of visits per month is ideal to properly sample the light curves of known objects, since it allows the monitoring of the rise/decline phases as well as the probing of the short amplitude variability characteristics in the quiescent phase. Our goals are: to construct a library of the light curves of known objects to be used as a reference for the identification 
of other members of the EXor class; and to  measure the physical parameters and the mass accretion rates during different phases of the source activity by means of optical/near-infrared spectroscopic 
follow-up. Since bursts occur typically every five-ten years, we expect to be able to observe at least one burst in a temporal range of 3-10 years.

\end{enumerate}

\item{Pie in the sky}
\begin{enumerate}
\item[a)] {\bf Discover new FUors and follow their rising phase}\\
With a survey lifetime of about a decade, it is reasonable to expect detections of new FUors. FUor candidates will be selected from the shape of the lightcurve (monotonic raising during years with brightness variations up to 5-7 magnitudes) and the location of the source in a star formation region. 
Prompt optical and near-IR spectroscopic as well as X-ray observations are needed to characterize the physical parameters 
and the  mass accretion rate during the rising and peak phases. FUor outbursts occur typically on time-scales of 100 years. Therefore, it is likely that the entire telescope lifetime will be needed to discover a reasonable number of FUor events.

\item[b)] {\bf Investigate the occurrence of EXor/FUor-like phenomena in evolutionary phases earlier than pre-main sequence}\\
The large majority of EXors/FUors are PMS stars with age about 10$^7$ yr. In principle, however, nothing prevents the existence of eruptive variables younger than PMS stars (the so-called Class I sources, with ages of about 10$^{5-6}$ yr). However, the detection of very young eruptive variables is challenging. Being  still immersed in their natal environment, they are heavily extincted in the UV and optical bands, i.e. in the photometric bands where the accretion burst is most intense. According to the model by \cite{Whitney_2003}, a 1 L$_\odot$ Class I source at a distance of 140 pc (Taurus, the closest star formation region) has $g$-band $\sim$ 16 mag. 
Considering the $g$-band limiting magnitude of 24.7 per single visit, we will be able to detect a burst of 6-7 mag of such sources provided that the local visual extinction does not exceed 15 mag.  Assuming that Class I sources undergo EXor events on timescales similar to pre-main sequence EXors, we expect to observe significant bursts in the first 1--3 years of the telescope's lifetime. 
\end{enumerate}
\end{itemize}

\subsubsection{Preparations for Young Eruptive Protostar Science}
Intense accretion activity is a common trait of young, eruptive protostars. The investigation of public surveys (ASAS-SN, Gaia, iPTF, ZTF), as well as the photometric monitoring (e.g.VST/OmegaCAM) of selected star forming regions is needed before Rubin LSST begins in order to refine the above  diagnostic tools. To further identify more of these interesting objects, we rely on broker alerts to identfy eruptive events, and further follow-up observations and facilities as detailed:
\begin{itemize}

\item{{\bf{Identifying events from alerts}}}\\
Candidate EXors/FUors can be identified on the basis of :
\begin{itemize}
    \item Location within a star-forming region;
    \item Quasi-monotonic increase of brightness in all bands, with a typical rising time of some hundredths of magnitude per day. A prompt alert is envisaged, because this could trigger a spectroscopic follow-up essential to characterize the onset of the outburst;
    \item Colors that become more and more blue with the increase in brightness.
\end{itemize}
At the time of writing, none of the available brokers perform color-color analysis and computation of the rising time.
\item{{\bf {Follow-up observations}}

Optical/near-infrared spectroscopic follow-up is needed to confirm the presence of emission lines in the spectrum and to measure the mass accretion rate. Depending on the source brightness and visibility, we will activate ToO observations at the ESO facilities (SoXS, X-Shooter), LBT (MODS, LUCI), and TNG (GIARPS, Dolores, NICS).
In case of detection of a Class I outburst, prompt follow-up 
with near-infrared imaging/spectroscopy will be activated. Such a discovery would also be a case of interest for observations with JWST, to study the accretion variability in the mid-infrared. Prompt follow-up in X-ray is also envisaged. In this respect, we remark that we will have access to GTO time for SoXS. Also, we have an ongoing program for observation with e-Rosita (Stelzer, Giannini, Bonito; 2018) to study the X-ray emission and its variability in accretion bursts. 
The eROSITA All-Sky Survey (currently in safe mode), started in 2020 and will operate for 4 yrs, which will allow for simultaneous observations with Rubin LSST.
Multi-epoch observations, when available, will be used to also explore the variability of these sources in X-rays, thus allowing us to perform a multi-band study of variability.}
\item{{\bf{Facilities/software requirements}}\\
EXor/FUor bursts occur on timescales of months. When an alert is received, we need to investigate the history of the source to reconstruct the light curve in the preceding quiescence phase.  We therefore need access to all the Rubin LSST photometric data, that must be stored in a dedicated repository. To store the light curves of all the known eruptive protostars for comparison purposes and to store all the light curves of perspective protostars, adequate computing storage is required. 
}
\end{itemize}

\subsection{Compact Binaries: Cataclysmic Variables (CVs)}
\textsl{Authors: Paula Szkody, Elena Mason and David Buckley \label{sec:CVs}}
 
Accretion onto compact objects allows a powerful probe of binary evolution and accretion physics, particularly for those systems in tight orbits with late-type donor stars \citep{Pala22}. Several types of systems comprise compact binaries. Cataclysmic variables (CVs) have accretion onto a white dwarf (WD) and include novae with 10--15 magnitude outbursts caused by thermonuclear runaways on the white dwarf; dwarf novae with 2--9 mag outbursts caused by disk instabilities; and nova-likes which have high and low states of accretion \citep{Warner95}. The AM CVn subclass of CVs consists of two white dwarfs, where one is accreting from a mass transferring evolved Helium WD \citep{Solheim2010}.  There also exist detached white dwarf binaries; pre-CVs consisting of a white dwarf and a low mass companion that have emerged from a common envelope but not yet begun mass transfer \citep[e.g.][]{2016MNRAS.458.3808R}; and short period detached binaries with a sub-stellar brown dwarf or planet companion \citep{2005MNRAS.357.1049D}. Low mass x-ray binaries (LMXBs) encompass accretion onto a neutron star (NS) or a black hole (BH) \citep{VanParadijs2001}. They experience rare outbursts due to disk instabilities, but also vary between intermediate and quiescent states. 
Longer orbital period high mass X-ray binaries (HMXBs) are also comprised of neutron stars or black holes accreting from early type companions \citep{2011Ap&SS.332....1R}. They also undergo outbursts due to enhanced accretion, either due to enhanced mass loss from the companion or periastron passage in an eccentric orbit. With a record number of different accretion states among these compact binaries, it has become critical to map the accretion history of each class of objects to correctly frame their evolutionary scenario.
 
Accretion onto compact objects is driven by angular momentum losses. Mechanisms of angular momentum loss have been identified as magnetic braking in the donor star and gravitational radiation losses. Which mechanism drives the angular momentum loss depends on the nature of the compact binary and its evolutionary phase \citep{Rappaport82}. It is known that MSPs, LMXBs and CVs can change their accretion states by stopping mass transfer, or switching from high to low accretion regimes and vice-versa. The timescale for large accretion state variations is usually of the order of days to months for LMXBs and MSPs, and from hours to years for CVs and HMXBs.
 \begin{itemize}
 \item Low hanging fruits:
 \begin{enumerate}
 \item[a)] {\bf Identify new novae} \\
 New novae can be identified by their outburst amplitudes through the alert system. Following the alert, the community should activate spectroscopic follow-up observation (especially UV and optical, but virtually across the whole electromagnetic spectrum since novae emit from gamma to radio wavelengths) to physically characterize the event (e.g. ejecta mass, filling factor, duration of the H burning phase, WD composition) and constrain the WD mass.  The best estimate currently available for the nova rate in the Galaxy is  $\sim$50 nova/yr, although with large uncertainty \citep[e.g.][]{shafter2017}.
 \item[b)] {\bf Identify recurrent novae outbursts} \\
Continuous all sky monitoring and monitoring of known novae potentially will allow the discovery of new recurrent novae (i.e. novae which show more than one outburst in the human time scale). Increasing the sample of known recurrent novae together with the ability of characterizing each of their eruptions in detail (through spectroscopic follow-up across the whole electromagnetic spectrum) will allow us to constrain their physical parameters (e.g. ejecta mass and filling factor, super soft source phase duration, WD mass) for a comparative analysis of novae and recurrent novae.  

 \item[c)] {\bf Analyze low states of known nova-likes} \\
 We will monitor the several dozen known novalike objects that enter low accretion states to determine when they enter these unpredictable low states. When in this state, spectroscopic follow-up will enable the identification of the underlying stars. These low states usually only last for weeks and are the only time when the accretion disk disappears and the stars can be seen. This enables the star masses to be measured in these high accretion rate systems from time-resolved spectra to determine the radial velocity curves. With the number of known systems, 1--2 low states each year are expected. As new systems are found during the 10 year survey, they can be added to the list.
 \item[d)] {\bf Find new high amplitude dwarf novae} \\
 Through Broker alerts, the identification of high amplitude dwarf novae (greater than 5 mag, known as superoutbursts) will allow for follow-up with high cadence photometry during their 2--3 week bright phases. This will allow the determination of superhump periods which will enable the determination of the mass ratio and clues to the white dwarf's evolutionary history. Some high amplitude dwarf novae have long spells between outbursts (years) and are intrinsically faint during quiescence, making them difficult to study. However, they represent an important tip of the iceberg of the low luminosity CV population, giving clues to their ultimate evolution.
 \end{enumerate}
 \item Pie in the sky
 \begin{enumerate}
 \item[a)] {\bf Observe new CV eruptive behavior} \\
 While CVs produce several different types of known outbursts, new types and forms of the outburst behavior are still possible. Since the rise times are short (less than 1--2 days) and the entire outburst might not last long, early identification through broker alerts can enable spectroscopic follow-up that ensures a correct classification.
 \end{enumerate}
 \end{itemize}

\subsubsection{Preparations for Cataclysmic Variable Science}
Cataclysmic variables produce different types of outbursts with varying amplitudes and durations. To fully take advantage of the Rubin LSST data products as discussed above, timely alerts, follow-up observations and specific software are required, which are detailed below:
\begin{itemize} 

\item{\bf {Identifying events from alerts}}\\
 All of the aforementioned science cases require rapid alerts to enable swift follow-up. Finding the outbursting objects or low states in the nova-likes will require specific filters in a Community Broker. The varying shape of a dwarf nova outburst will need particular attention in this regard to facilitate correct identification. Due to the high number of alerts expected, machine learning tools to identify this class of object, especially unusual outliers to the normal outburst behavior, will be essential. Enabling correct ML will require a good training set that is obtained during the first year to separate out normal vs unusual behavior.

 \item{\bf{Follow-up observations}}\\
 The correct identification of a nova versus a dwarf nova or AM CVn system will require spectroscopic follow-up with a 2--10\,m class telescope to identify Balmer emission (or absorption features for a dwarf nova outburst) and Helium lines for an AM CVn. Multiwavelength observations (X-ray, UV, IR and radio) are often needed to study the outburst once an unusual object is identified.

\end{itemize}

\subsection{Compact Binaries: Neutron Star Binaries }\label{sec:nsbinaries_tc}
\textsl{Authors: Elena Mason, Domitilla de Martino, Kristen Dage}
 
\bigskip

\noindent Neutron star (NS) binaries are binary systems with neutron star components. Some neutron star binaries contain milli-second pulsars (MSP) and can either be in the accretion state (AMSP, mostly discovered during outburst) or in the rotational powered state (RMSP, typically discovered in radio surveys and more recently also in the $\gamma$ energies by Fermi). According to the so-called {\it recycled scenario} \citep{alpar1982} MSPs are neutron stars that have been spun-up by a Gyrs-long accretion phase equivalent to Low Mass X-ray Binaries (LMXBs), once the mass transfer has dropped enough to allow the activation of the (radio) MSP powered by the rotation of the magnetic field. 

Transitional milli-second pulsars (tMSPs) are NS binaries that have been observed to switch from an accretion to a rotation powered state, or vice-versa, where mass can be ``propelled" out of the system on a time scale of months to decades \citep[e.g][]{papitto2013,archibald2009,stappers2014,patruno2014,demartino2010,demartino2010b,demartino2013,bassa2014}. Hence, they demonstrate, at one time, the connection between LMXBs and MSPs (confirming the recycled scenario), as well as supporting the idea that transition may be driven by mass transfer variations. 
Only a few transitional systems are currently known. They have been observed in outburst (only one case) and in an intermediate sub-luminous state, where an accretion disk is present \citep{archibald2015,papitto2015}. These are the only LMXBs that have been observed to be high energy Gamma-ray emitters (as revealed by the Fermi-LAT, \citealt{demartino2010}). Radio and X-ray observations performed during the intermediate states reveal an anti-correlated behavior between these two bands indicative of intermittent jet emission (e.g. \citealt{patruno2014,demartino2010,demartino2015}, see also \citealt{papitto2014},\citealt*{papittoetorres2015}). Additionally, the observation of optical pulses during the intermediate sub-luminous state \citep{ambrosino2017,papitto2019} poses and argument against accretion as it seems that the majority of the accretion disk material is expelled through this method.

\begin{itemize}
 \item Low Hanging fruits:
 \begin{enumerate}
 \item[a)]{\bf Observe changes of state for any MSP, LMXBs and tMSP, known and/or newly discovered}\\
Any change of state reported (and alerted) by Rubin LSST will trigger dedicated follow-up observations aimed at solving the binary system parameters. In particular a binary system entering into one or the other state will trigger time-resolved optical spectroscopy to determine the donor star and/or the disc parameters. If the period is unknown, with the help of optical photometry, it will be possible to trace the secondary and, in case of accretion disk, the primary orbital motions. If a LMXB-to-RMSP transition occurs, the observations will be crucial to trigger optical spectroscopy and multi-band optical photometry to derive the donor star's spectral type, the effects of pulsar irradiation and orbital parameters. These low-states will enable triggering of radio pulsar searches (eg MeerKAT and SKA).
The binary orbital parameters are crucial to  perform searches for pulses in the Gamma-ray and X-ray regimes. Hence, it is of utmost importance that we receive prompt alerts for any system entering into a sub-luminous state before it enters its radio state or resumes full accretion. Time resolved optical follow-up of systems in the MSP state will be needed to establish binarity and constrain the companion star and the mass function of the binary system.  
\item[b)]{\bf Super-orbital periodicities in X-ray binaries}\\
Some neutron star X-ray binaries (and at least one black hole binary) are known to exhibit super-orbital periods (i.e. periods greater than the orbital period of the system) in both the X-rays and optical \citep[see][and reference therein]{Hu19, Dage22, Thomas2022}. This is thought to be due to a warped, precessing accretion disc \citep[e.g.]{Brumback20}, with radiation-driven warping proposed as the lead scenario \citep{Ogilvie2001}.  While this has mostly been studied in the X-ray, with optical photometry and time domain coverage, it will be possible to identify even larger numbers of systems with super-orbital periods, which will expand our knowledge of accretion disc geometries with unusual configurations. 
\end{enumerate}
\item Pie in the sky:
\begin{enumerate}
\item[a)]{\bf{Identification of unexpected transitions or transition frequencies}}\\ 
By observing known binary systems with neutron star components (and those that are newly observed with Rubin LSST), we are well placed to identify new and unexpected transitions. If such a transition is identified, this will imply revision of current evolutionary scenarios. This science case will be ongoing over the 10-yr Rubin LSST survey mission. 
\end{enumerate}
\end{itemize}

\subsubsection{Preparations for Neutron Star Binary Science}

\begin{itemize}
    \item{\bf{Identifying events from alerts}}\\
    The prompt data alerts of Rubin LSST will enable an alert in real time about any change of state of known systems (whether MSP, LMXB or tMSP) as well as for newly discovered systems (that we anticipate to discover by combining Rubin LSST with other next generation sky multiwavelength surveys -- e.g. THESEUS, eROSITA and SKA). 
Obtaining notifications of new systems in real time will:
\begin{itemize}
\item Provide a census of the transitional systems among MSPs thus helping framing the evolutionary phase of each accreting NS, and 
\item Allow for dedicated follow-up that will allow the determination of the system parameters and/or characterize the physical mechanisms behind each observed phenomenology. 
\end{itemize}
    It is imperative that the alert brokers are prepared to identify and produce alerts for Neutron Star Binaries. This will need to include a list of all known Neutron Star Binaries so that they may be monitored for changes in their luminosity so that follow-up observations may be swiftly undertaken.
    \item{\bf{Follow-up observations}}\\
    By combining Rubin LSST with multiwavelength surveys such as THESEUS, eROSITA and SKA, we anticipate finding a significant number of new objects. If a LMXB-RMSP transition occurs, optical spectroscopy and multi-band optical photometry will be required. This will further trigger radio pulsar searches (eg MeerKAT and SKA). Follow-up X-ray and radio observations performed in the intermediate state can provide information about jet emission. 
\end{itemize}

\subsection{Compact Binaries: Black Hole X-ray Binaries (BHBs)}
\textsl{Authors: Michael Johnson, Poshak Gandhi}
\selectlanguage{english}

\bigskip

\noindent Black hole X-ray binaries (BHBs) are interacting binary systems consisting of a stellar-mass black hole and a donor star. Early spectral-type donor systems tend to be wind-fed, whereas in late spectral-type donors, the black hole will accrete matter from the donor via Roche lobe overflow and an accretion disc will form around it. As matter from this disc approaches the black hole, it releases gravitational potential energy in the form of broadband radiation. The radiation peak from the sites of maximal energy release in the innermost accretion zones occur in X-rays. Longer wavelength thermal radiation arises as one progressively moves away from the centre, with additional bright transient radiation arising in particular components such as non-thermal radiation from plasma jets \citep[e.g. ][]{charles06}. Finding and characterising Galactic BHBs has important implications for the end stages of stellar evolutionary population synthesis, as well as the progenitors and physics of gravitational wave binaries.

\begin{figure}[ht]
\begin{center}
\includegraphics[width=0.60\columnwidth]{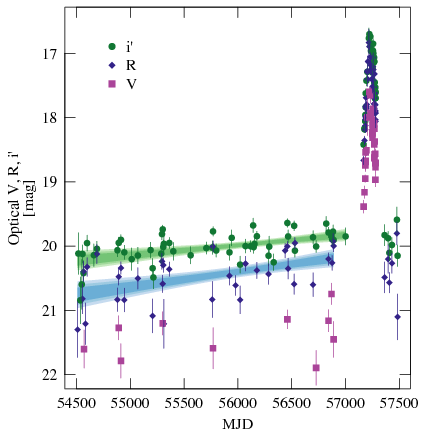}
\caption{{Long term optical lightcurve of an outburst from BHB GS\,1354--64. Note the subtle but definite rise in the $i'$ and $R$ filters over years, before the enormous rise associated with the outburst \citep{Koljonen_2016}. 
{\label{330425}}%
}}
\end{center}
\end{figure}

BHBs spend their lives predominantly in states of either quiescence or outburst. The explanation for the latter state was outlined by, \cite[e.g. ][]{lasota2001disc} to be a trigger associated with accumulated matter on the inner radius of the accretion disc reaching a critical limit, causing the disc to undergo thermal runaway and initiating a shockwave that propagates away from the inner radius. As the fronts from this shockwave propagate outwards, the matter behind the fronts becomes hot and ionised, beginning to diffuse inwards and eventually triggering the outburst. These series of events cause the optical signature of the BHB to steadily rise over the course of years whilst the matter is building up and to sharply increase in the weeks building up to the outburst due to the critical limit being reached. This behavior has been observed directly within only a select few BHBs, e.g. GS\,1354--64 by \cite{Koljonen_2016} and V404\,Cyg by \cite{Bernardini_2016}. The optical lightcurve for GS\,1354--64 is shown in \autoref{330425}. There is enormous interest in being able to identify trends leading up to an outburst, not only to learn about the physical processes at play, but to make the advanced coordination of  multiwavelength observations easier. 

The number of BHBs expected to be observed by Rubin LSST was explored by \cite{Johnson_2018} where they estimated the fractional Galactic low mass X-ray binary (LMXB) population in which accurate recovery of the orbital period is likely to be possible. Furthermore, \cite{wiktorowicz2021predicting} calculated the number of X-ray binaries expected to be detectable with Rubin LSST via self-lensing.

 \begin{itemize}
 \item Low hanging fruits:
 \begin{enumerate}
 \item [a)]{\bf{Observing the increase in luminosity of known black hole binaries (BHBs)}}\\ By observing the increase in luminosity that occurs prior to the outburst, we are able to trigger dedicated multi-wavelength follow-up in order to characterise the event. In the X-ray, BHBs can be distinguished as they tend to exhibit characteristic evolutionary paths in X-ray flux/spectral slope (‘q’) diagrams, together with strong radio emission from jets (e.g. \citealt{fender2004towards}). They also tend to show strong stochastic correlated multiwavelength flux variability down to milli-second order timescales (e.g. \citealt{gandhi2017elevation}). BHB masses can be estimated via dynamical radial velocity variations (\citealt{casares2014mass}). So far, only the brightest tip of the population has been followed up in detail in the optical band \citep[e.g., ][]{Corral_Santana_2016} and only $\sim30\%$ of detected BHBs have sufficient observations to dynamically confirm the presence of a black hole. Therefore increasing these statistics is vital to improve our understanding of these outbursts. Currently, there are roughly 5--10 X-ray outbursts detected per year, only 2--3 of which are bright enough to be followed up in the optical.  The deeper mag limits of Rubin LSST monitoring, as well as the possibility to access redder filters, should substantially enhance the detection of fainter and heavily extincted optical counterparts.
 \end{enumerate}
 
 \item Pie in the sky: 
  \begin{enumerate}
 \item [a)]{\bf{Characterising the long term trends of BHB outbursts}}\\
 With a larger sample of better characterized BHBs, we will be able to better understand the long-term trends of outbursts. This could provide the key to predicting the occurrence of BHB outbursts, allowing us to trigger follow-up {\it before} the event. Additionally, the long term outburst trend is what distinguishes the optical signature of BHBs from other systems that also exhibit ellipsoidal modulation, such as cataclysmic variables. Characterisation of this trend may also allow for the classification of BHBs, without the need for X-ray follow-up. However, this characterisation would require additional detailed observations and analysis of BHBs throughout their outburst phases. 
 \end{enumerate}
 \end{itemize}

\subsubsection{Preparations for Black Hole Binary Science}
\begin{itemize}
    \item{\bf{Identifying events from alerts}}\\
    It is imperative that the alert brokers are prepared to identify and produce alerts for Black Hole Binaries. The catalogue matching service from AMPEL \footnote{\url{https://ampelproject.github.io/}} will be used for multiwavelength confirmation for potential BHBs, utilising archival data from sources such as eROSITA, Spitzer and ZTF. For known BHB systems, the watchlist from Lasair\footnote{\url{https://lasair.roe.ac.uk/}} will be used to monitor for rises in the optical. As many may only be visible with the current generation of telescopes during outburst, follow-up must be triggered promptly. Further, prompt alerts will enable prompt follow-up, which is necessary to characterize the outburst events.
    \item{\bf{Follow-up observations}}\\
    The long term rise expected in the years prior to the outburst will likely be sufficiently sampled with the regular Rubin LSST cadence. A higher frequency of observations will however be required to characterise the sharp rise directly preceding the outburst. Suitable telescopes to perform these observations range from the LCOGT (0.8\,m) to the New Technology Telescope (358cm), depending on the brightness of the target. 
    
    \item{\bf{Facilities/software requirements}}\\
    Some BHBs systems will only be observable after the brightness has begun to rise. Therefore we require an algorithm capable of classifying the outburst from data that only cover the brightest observable section of the observable signature. 
\end{itemize}

\section{Extra Galactic Transients} 
\textsl{Editor: Federica B. Bianco}

\subsection{Supernovae}
\textsl{Authors: Elizabeth Johnson \& Melissa Graham \& Virginia Trimble \& Anais M\"oller \& Fabio Ragosta \& Federica B. Bianco}

\bigskip

\noindent Supernovae (SNe) are powerful stellar explosions, known for their extremely bright luminosity, where the relevant star is essentially destroyed through one of two physical processes: gravitational collapse of an iron core that has reached the Chandrasekhar limit inside a more massive star or detonation of a white dwarf driven over the Chandrasekhar limit by mass transfer from a companion. The latter are called Type Ia SNe (or thermonuclear), the former, Type Ib, Ic, and II (core-collapse), from a traditional classification scheme based on whether hydrogen features appear in the spectra (Type II) or not (Type I). It is the Type Ia's that have most often been used as distance indicators to probe cosmic expansion and show that this expansion is currently accelerating.

In addition to being exquisite an cosmological tool (an area of specialization of the Dark Energy Science Collaboration\footnote{{\url{https://lsstdesc.org/}}}), SNe are important phenomena in our Universe for two main reasons. The first reason lies in chemical evolution of galaxies, because they both synthesize additional heavy elements and distribute the products of nuclear reactions in the pre-explosion stars. Type Ia's provide more than half of the iron-peak elements and some intermediate mass (carbon-sulfur) material \citep{Guo_2008, Matteucci_2010, Simionescu_2015, Jimenez_2015}. The core collapse events generate everything from helium to iron, and also elements beyond iron from slow neutron capture in the parent stars and probably from rapid neutron capture in the event itself. The second reason is that the kinetic energy of expanding SNe remnants is a major source of heating and stirring of interstellar gas, which prevents the dust and gas clouds collapsing down to stars quickly, allowing stars and solar systems, such as our own, to form. Thus we are made (partly) of "star dust," and live in a habitable solar system owing to past supernovae; indeed, one that occurred just less than 5.5 Gyr ago may have triggered the formation of our own Solar System \citep{Loewenstein_2006, Gatto_2017}.   

Despite SNe having significant scientific value, their explosion mechanisms remain somewhat mysterious. They have well-speculated models and progenitor theories, but those progenitors are still controversial. For example, SNe Ia have evidence for both a single-degenerate explosion scenario, where the white dwarf is accreting material from a non-degenerate companion, and a double-degenerate explosion scenario that involves two white dwarfs. Additionally, SNe classes are not completely defined and with recent surveys, more objects exhibiting behaviours of two different SNe classes have been found. It is thus important to investigate whether supernova types are isolated families or a continuum and how to define them \citep{2003F}.

The advances we would like to make with supernova science in the era of Rubin LSST are as follows:

\begin{itemize}
\item Low Hanging Fruit:
\begin{enumerate}
    \item [a)] \textbf{Improved classification schemes}\\ 
    Ultimately, observational studies of individual and samples of SNe rest on a robust classification. Large numbers of Rubin LSST light curves, combined with the addition of photometric follow-up to provide improved light-curve coverage, would improve photometric classification. 
    Expanding the supernova dataset would bring some classes of unusual, rare, difficult to observe supernovae~\citep[for example fast evolving or Pair-Instability supernovae,][]{2021arXiv210508811H, 2019ApJ...881...87G} into a statistical domain, enabling a better understanding of the boundaries (or lack thereof) between classes.
    \item [b)]{\bf {Refined theoretical predictions and classification methods}}\\ 
    An overarching goal of SNe science is the identification of the SNe progenitor (which type of star exploded, in which environment and what triggered the explosion) and explosion mechanisms. Observations of a significant number of supernovae would allow for a more in-depth understanding of theory through the measurement of observables. For example the signatures of nickel distribution {\it vs.} shock interaction in early SNe\,Ia light curves and spectra \citep[see][and reference therein]{C2015}. While largely our characterization of SNe and physical insight come from spectra, the size of the Rubin LSST sample of SNe, combined with follow-up in multiple spectral bands, would enable the spectral feature of the event to be constrained photometrically. 
    \item [c)] \textbf{Direct identification of progenitors}\\ 
    By employing the use of follow-up facilities, i.e. {\it HST, JWST}, the direct identification of the progenitor stars (and/or their binary companions) of future nearby supernovae will be possible \cite[see][and reference therein]{smartt2009death}.
    \item [d)] {\bf{Association of supernovae with host galaxies}}\\
    With the large number of supernovae expected to be observed with Rubin LSST, studies of event rates, environmental dependence, progenitor stellar populations, host-galaxy associations are all possible, with the potential to reveal physical insight unobtainable from direct observations~\citep[see, for example,][and references therein]{2020ApJ...892..153M}. The high number of SNe and the detailed study of explosion mechanisms and progenitor models will help the disentangle to luminosity-metallicity-distance degeneracy. This will allow a better host-galaxy association for all the SNe detected. For this science case, however, photometric or spectroscopic redshifts are required to calibrate the photometric distance, both of which require follow-up observations.
\end{enumerate}
\item Pie in the sky:

\begin{enumerate}
    \item [a)] {\bf{Moving SN studies into the multi messenger astronomy realm}}\\  
    The prompt follow-up of SN candidates will allow the cross correlation of events detected with other facilities and also through other messengers.
    The possibility of surveying wide areas of the sky provides an opportunity to learn about the electromagnetic (EM) signatures of fast phenomena known in other regions of the Energy-frequency spectrum whose EM counterparts have been missed until to now. 
    \item [b)]{\bf{ Discovering new, unknown kinds of stellar explosions}}\\
    As we open up new regions of the parameter space with synoptic observations of an unprecedented volume of space we stand a chance to discover new phenomena, as recently happened for SNe Icn \citep{Fraser2021,GalYam2022,Perley2022,2022arXiv220507894P}.

\end{enumerate}
\end{itemize}

\subsubsection{Preparations for Supernovae Science}
\begin{itemize}

\item{\bf{Identifying events from alerts}}\\
It is imperative that the brokers \autoref{ssec:brokers} are prepared to produce swift alerts for supernovae. To ensure this, the community needs to continue working toward classification schemes and effective queries to the brokers to identify SNe (and more general time-domain objects) of interest and trigger follow up. This process must be largely, if not entirely, automated due to the large data volume delivered by Rubin. Early classification models for SNe that are capable to identify a transient with only a few observations are still rare \citep[see, for example,][]{Muthukrishna:2019, qu2021photometric}, especially when the desired classification goes beyond ``SNIa or not'' and into classification of SNe subtypes (see Model and Software requirements).
\item{\bf{Follow-up observations/archival data}}\\
To classify and study new SNe detected by Rubin LSST (especially early on in the LSST survey, while our photometric classification schemes have time to take advantage of the large data volume) we will need to leverage dens(er) coverage of the light curves and spectral observations from ancillary data from other facilities. These observations will include observations in bands and wavelength regimes complementary to Rubin's. To do so, it is essential to have a reliable network of observatories which can follow-up an object once an alert is distributed, and, ideally, an intelligent network system connecting follow-up facilities and monitoring and deploying observations to avoid duplication and collect the most relevant data.

Moreover, synergies, such as Rubin-Roman, or Rubin-Euclid, or Rubin-4MOST, can enable ancillary data collection, which would greatly enhance our understanding of Rubin discoveries. A collection of spectra, even for a relatively small subset of the detected SNe, enable spectro-photometric analyses, which improve the understanding of the composition of the source and the interstellar medium around it, and the physics mechanisms the source underwent to produce the detected event. This will further improve our knowledge of the parameter space we use to characterize the detected transient.  With the high pressure that Rubin LSST will put on follow-up facilities, selecting the most promising SNe for spectroscopic follow-up will be critical to improve our insight into stellar explosions.

\item{\bf{Facilities/software requirements}}
\begin{itemize}
\item{\bf{Precursor datasets}}\\
Algorithms for photometric classification are crucial for both (a) early light-curve classification from spectroscopic follow-up, and (b) for complete light-curve classification. For this, we need to obtain a more diverse, less biased sample of supernovae to constrain all relevant astrophysical processes and their observational manifestations. Efforts are underway, with recent algorithms providing the fast and accurate classification of supernovae in both early and late stages of formation \citep{Muthukrishna:2019,Moller:2020}. The current Extended LSST Astronomical Time-Series Classification Challenge (ELAsTiCC) challenge\footnote{\url{https://project.lsst.org/meetings/rubin2022/agenda/extended-lsst-astronomical-time-series-classification-challenge-elasticc}, an upgrade of the past PLAsTiCC challenge \url{https://plasticc.org/}} provides an ideal venue for the development of further models.

\item{\bf{Automation of the discovery and study chain}}\\
After detection and selection of interesting SNe, enabled by expert models interfacing with alert brokers, the triggering of follow-up resources should be enabled by software packages, like AEON~\citep{10.1117/12.2559986}, so that we may act promptly to classify young supernovae.
Our ability to make discoveries in this current and future era of time-domain astrophysics is limited by a two-pronged challenge: (i) it will be necessary to identify specific targets of interest amidst millions of alerts each night and; (ii) we will need the very fast identification of notable alerts, ideally within hours of detection, to enable prompt follow-up observations (spectroscopy and/or observations across the spectrum). Toolkits such as Target and Observation Manager systems~\citep[TOM][]{Street2018b} allow the exploration and especially prioritization of targets in an observing run. This will help scheduling the follow-up and allow the science community to have a fast first look at the phenomena. Luckily, tremendous advances have been made in the automation of the discovery chain in recent years, with major facilities (e.g. the Gemini telescopes) fully endorsing and embracing software like AEON to enable follow up of Rubin SNe. Many other facilities, especially those connected to Rubin through international in-kind data right agreements,\footnote{\url{https://project.lsst.org/groups/cec/node/5}} are working toward the integration of TOM, AEON, and similar software packages. Furthermore, the use of TOMs will be pivotal to our success with handling large data sets. For example, when deciding which object parameters to pre-calculate, store, and make available to queries {\it vs.} which to compute-on demand. 

\item {\bf{Model development}}\\
 To effectively deal with the 6-band sparse (2-3 images per week, intranight gap $\sim3$ days) we need more methodological development and software tools that can deal with multi-filter light curves, e.g., developing parametric models and methods for correctly treating sparsely sampled data.
 
Algorithms to associate supernovae with host galaxies are further needed. Efforts to organize galaxy catalogs and to create SNe matching software are underway \citep{Gupta16}. Algorithms to select follow-up targets would also be highly beneficial. These algorithms would maximize the use of our spectroscopic resources. Concurrently, these algorithms would fine-tune our training sets for photometric classification - one of our main limitations. Efforts have started using Active Learning algorithms \citep{Ishida:2019}. To then compare our different photometric classification algorithms, we require a sample of benchmark systems.
\end{itemize}
\end{itemize}

\subsection{Intermediate-Luminosity Optical Transients (ILOTs)\label{sec:ilot-time}}
\textsl{Authors: Andrea Pastorello \& Elena  Mason \& Federica B. Bianco \& Andrea Reguitti}

\bigskip

\noindent Intermediate-Luminosity Optical Transients \citep[ILOTS; e.g.][]{Berger2009} form a class of astrophysical objects identified through their relatively faint intrinsic luminosity.
Their absolute magnitudes are intermediate, between those of core-collapse supernovae (SNe) and classical novae ($-10>M_V>-15$), and for this reason they are frequently named as "Gap Transients" \citep[e.g.,][]{Kasliwal2012,Pastorello2019a}. They have been studied in depth only in recent years, hence our knowledge on their nature is still incomplete. This is even more  true since only a limited number of ILOTs have high-quality and well-sampled data sets, and while the early-time spectra and the luminosity of different species of ILOTs are very similar, they can be produced by a wide variety of physical mechanisms and progenitor stars. Although a fraction of SNe are in the same magnitude range as ILOTs, including low-luminosity stripped-envelope SNe, such as faint SNe Iax, Ca-rich transients, Ia candidates, or other faint SNe I \citep{Kasliwal2013, Valenti2009}, and low-luminosity Type II-P events \citep{Pastorello2004, Spiro2014}, our team is mostly interested in non-terminal stellar transients. These include:
\begin{itemize}

\item {\bf Giant eruptions of massive stars, in particular luminous blue variables (LBVs)}. LBVs are very bright sources, and among the most massive stars detectable in galaxies. Famous examples in our Galaxy are AG Car and HR Car. During canonical S Dor-like outbursts, LBVs experience erratic brightness variability over timescales of several months to a few years, with $\Delta$M of a couple of magnitudes, but without a significant change in the bolometric luminosity \citep{Humpherys1994, Smith2011, Humphreys2017}. In this phase, LBVs move to the right of the Hertzprung-Russell diagram, becoming redder and cooler, coming back to the left side in quiescence. 
The situation dramatically changes during a giant eruption, when they become the most luminous stars in their host galaxies. $\eta$ Car, in the Milky Way (MW), experienced a giant eruption in the mid-19th Century. During the giant eruption, $\eta$ Car reached an apparent magnitude of -0.7 (absolute magnitude $M_V$ of -14, \citealt{smith2011revised}), while in a second obscured little eruption it reached a magnitude of 6.2. With the single-band magnitude limit of 24--25 for Rubin LSST, not only would a little eruption be visible in the Local Group (distance modulus up to 27), but a giant eruption would be visible well beyond it. Current models propose that brightness changes in $\eta$ Car are due to an unstable multiple system, where the giant eruption was possibly triggered by a dynamically induced merger. However, other mechanisms may explain giant eruptive mass-loss events without invoking close stellar interactions in binary systems \citep[e.g. instabilities triggered by explosive shell burnings, or pulsational pair-instability; see][and references therein]{smith2014}.

During a giant eruption (which lasts years to decades), LBVs experience multiple outbursts whose individual peaks reach $M_V \approx -14$ mag \citep{Wagner2004, Smith2010b, Smith2017, Pastorello2010, Pastorello2013}. However, massive hypergiants may also have a single short-duration outburst with $M_V$ similar to that of giant eruptions; \citep{Smith2011, Tartaglia2015, Tartaglia2016}. Although their massive progenitors survive the outbursts, these transients may resemble (as energetics and spectral appearance) true SNe explosions. For this reason, extragalactic LBV-like outbursts are also dubbed SN impostors \citep{VanDyk2000}.

\item {\bf Red novae and non-degenerate stellar mergers}\\
Red Novae (RNe) are created by the successful ejection of a common envelope in a binary system that eventually lead to the coalescence of the two stars. The nature of (RNe), such as V838 Mon, V4332 Sgr and M31-RV, was debated until V1309 Sco was discovered in 2008. A combination of spectroscopic data and a well-sampled pre-outburst light-curve proved it to be the stellar merger of a common envelope binary \citep{Tylenda2011, Mason2010}. Recent discoveries of extra-galactic counterparts, the so-called Luminous Red Novae, LRNe \citep{Blagorodnova2017, Goranskij2016, Smith2016b} have extended the RNe zoo to much higher luminosities and masses \citep{Blagorodnova2017, Pastorello2019b, Cai2019, Pastorello2021, Blagorodnova2021, Cai2022}, and are expected to dominate the Rubin LSST sample \citep[e.g.][]{Howitt2020}. RNe/LRNe show multi-peaked light curves, and spectra that progressively transition from intermediate-type stars to K- and then M-type stars. TiO and VO molecules are usually detected in very late spectra of stellar mergers.

\item {\bf Intermediate-luminosity red transients (ILRTs)}\\ ILRTS show spectra that are initially quite blue, but become redder with time \citep{cai2021}. They show prominent H and Ca II lines, including the typical [Ca II] 7291, 7323 \AA \ doublet in emission. In ILRTs, the doublet is detected at all phases. Their light curves are reminiscent of those of sub-luminous Type II-L SNe \citep[e.g. SNe 2008S and NGC300-2008OT1;][]{Prieto2008, Bond2009, Botticella2009, Smith2009, Berger2009, Humphreys2011} or even Type II-P SNe \citep[e.g. M85-2006OT1 and PTF10fqs;][]{Kalkarni2007, Pastorello2007, Kasliwal2011}, and -- when observed -- the late-time decline rate is roughly consistent with the $^{56}$Co decay. In quiescence, the progenitors of ILRTs usually remain undetected in the optical and near-infrared bands, while they are fairly luminous in the mid-infrared domain. The progenitors of ILRTs are moderately massive stars (8-15 M$_\odot$), enshrouded in dusty cocoons. ILRTs are proposed to be electron-capture SNe from Super-Asymptotic Giant Branch (S-AGB) stars \citep[e.g.][]{Botticella2009, Pumo2009, Thompson2009}. Although this interpretation is disputed (see, e.g., \citealt{Andrews2021}), recent observational arguments seem to favor the terminal SNe explosion scenario for ILRTs \citep{Adams2016}. 
\end{itemize}
With the extensive deep data from Rubin LSST, we expect significant improvements in our understanding of ILOTs. Rubin LSST will facilitate the following science cases:

    \begin{itemize}
    \item {Low hanging fruit:}
    \begin{enumerate} 

\item [a)]{\bf Observations of known ordinary luminous blue variable (LBV) outbursts (S Dor-like) and the detection of new LBVs}\\
The full range of variability of a classical LBV can be comfortably monitored with single shot Rubin LSST observations up to about 30 Mpc, but with periodic stacks we can largely exceed this distance limit. For known LBVs, using also literature, data available in the public telescope archives and those publicly released by the surveys, we aim to obtain light curves with baselines of many decades. Multiple survey strategies may reveal different types of variability, which develop on different timescales. Well-studied LBVs will become reference objects. With Rubin LSST, we expect to find new LBV candidates in outburst and determine their occurrence rates.\\
On the other hand, the study of the canonical variability (the S Dor phase) of known luminous blue variables in the Milky Way, Large Magellanic Cloud (LMC) and Small Magelanic Cloud (SMC), and nearby Galaxies is a key instrument to unveil the role of binary interaction. The moderate-cadence of the main survey (with one observation every a few days and in different filters) allows us to investigate modulations and/or periodicity features in the light curves. Through the light and color curves, with supporting spectroscopic information, we can determine correlations among the observable parameters.

\item [b)]{\bf Observing giant eruptions/outbursts of LBVs and other massive stars}\\
Timescales of giant stellar eruptions (the so-called SN impostors) may range from a few weeks to decades (in the case of Giant LBV eruptions). As these event are mostly produced by very massive stars, they are likely rare events, but can still be monitored for many years. 
We aim at monitoring SN impostor light curves with Rubin LSST, as well as discovering evidence of pre-SNe outbursts prior to ejecta-CSM interacting SNe. Understanding if pre-SNe outburst events are common in the pre-SNe stages (as claimed by \citealt{Ofek2014} and observed by \citealt{Strotjohann2021}) is a major goal of our research. In addition, a wide database of objects well-followed until the very late phases is an essential tool to observationally discriminate SN impostors from faint ejecta-CSM interacting SNe. The progenitors of these supernova impostors in a relatively quiescent state  can be detected in earlier deep stacked Rubin LSST images.

\item [c)]{\bf (Luminous) Red novae and non-degenerate stellar mergers}\\
Stellar mergers are common \citep[e.g.][]{deMink2014}, with the rates tightly depending on their luminosity: \citet{Kochanek2014} estimated a Galactic rate of a few $\times 10^{-1}$~yr$^{-1}$ for V1309 Sco-like events ($M_V \approx -4$ mag); $\sim0.03$~yr$^{-1}$ for V838 Mon-like events (M$_V \approx -10$ mag)and; $\sim 10^{-3}$~ yr$^{-1}$ for NGC~4490-2011OT-like events (M$_V \approx -14$ mag). Although low-luminosity RNe are quite frequent, only about 25 RNe/LRNe have been observed in the Galaxy and in the Local Universe so far. Rubin LSST is expected to find many more candidates than are currently known. A large sample of objects with high-cadence and good signal-to-noise photometric and spectroscopic data are required for reliable comparisons with theoretical models \citep[e.g.][]{Matsumoto2022}. They will be a key tool to unveil the mechanisms triggering the RN outbursts, the fate of the binary systems (coalescence or not), and will finally allow us to provide more robust occurrence rate estimates. This will be an ongoing project over the lifetime of the Rubin LSST survey.

\item [d)]{\bf Increasing the number of Intermediate-luminosity red transients (ILRTs)}\\
The increasing number of ILRTs discovered in nearby galaxies suggests that these transients are relatively frequent, being about 8$\%$ of core-collapse SNe according to \citet{cai2021}. Rubin LSST will greatly increase the number of new discoveries, providing light curves from the explosion to very late phases. All of this is essential to determine the presence of classical SNe signatures, such as the shock breakout and the light curve $^{56}$Co tail. In addition, stack frames collected before and years after the explosion will provide high-quality images of the ILRT site which (in combination with deep infrared images collected with other facilities) are fundamental to the characterization of the progenitors and their final fates.

\end{enumerate}
\newpage
\item Pie in the sky
\begin{enumerate}
\item [a)]{\bf{Detection prior to outburst}}\\ 
The magnitude and colour information inferred from the inspection of Rubin LSST archive images will help constrain the progenitor parameters of new ILOTs, and allow us to detect variability patterns before the onset of the main transient event \citep[e.g.,][]{Tylenda2011,Pastorello2010,Pastorello2021a}. When stellar counterparts are not obviously found in single-visit images, the periodic stacks will allow us to go much deeper in magnitude, greatly increasing the probability to detect the quiescent progenitor.

\item [b)] {\bf{Understanding the fate of ILOTs}}\\ 
From the observations, the interpretation of the real nature of many ILOT species is still controversial. For ILRTs and some LBV-like eruptions, for example, it is still debated whether the star survives the eruption or not. For some LRNe, the fate of the binary system after the common envelope ejection is still controversial (final merger vs. surviving binary). The existence of a large image database will allow us to create deep stacks in order to monitor the source in the optical bands, up to very late phases. This is essential to follow the decline to below the luminosity threshold of the quiescent progenitor, eventually observing the complete disappearance of the source. Supporting IR observations, however, are necessary to constrain the dust formation affecting the optical photometry.

\item [c)]{\bf{Obtaining the underlying statistics of ILOTs}}\\ 
Increasing the statistics of well-sampled ILOTs is a key goal to (1) determine possible correlations among observed parameters, such as the peak absolute magnitude, plus the temperature evolution and the velocity of the ejected material; (2) identify correlations between observables and other physical parameters such as the energy released in the outburst and/or the progenitor mass; and (3) to correlate the physical parameters with the properties of the environments (stellar population, metallicity).
    \end{enumerate}

\end{itemize}

\subsubsection{Preparations for ILOT Science}
\begin{itemize}

\item {\bf{Identifying events from alerts}}\\
The planned Rubin LSST cadence will allow for a collection of homogeneously sampled light curves and colors of different ILOT sub-types. The availability of well-sampled templates will greatly favor the identification of ILOTs for the purposes of alerts. With this in mind, our team has a dedicated observational program, and is working on the preparation of a number of reference light curve templates for the different species of ILOTs. These templates will further enable the classification of new objects, as well as the comparisons and interpretation of the various transient species. 

\item {\bf{Follow-up observations}}\\
Each transient will have a well sampled light curve and color evolution, which will allow us to discover, classify and characterize new transients of the ILOT type. Through multi-wavelength mid- to high-resolution spectroscopic follow-up (in the UV, optical and NIR) it will be possible to understand the ejecta kinematics, density, ionization structure, geometry and, ultimately, the dynamics of our sample. Hence, upon receiving a Rubin LSST alert, it will be mandatory to proceed with dedicated spectroscopic follow-up. Only the brightest ILOTs with extended visibility will be monitored using mid to large size telescopes equipped with the best multi-wavelength spectrographs (such as SoXS@NTT). Rubin LSST multi-color photometry will also enable us to detect short-duration features in the light curves (e.g., humps due to shell-shell collisions or to reflection nebulae), and to sample the spectra spectral energy distribution (SED). 

Complementary UV photometry (in absence of UV spectroscopy)  will also be important as well as possibly X-ray follow-up (information is expected in the X-ray but has not yet beed observed). Photometric follow-up will be also requested in the optical and IR domains, to extend the SED sampling and to fill observational gaps due to the large gaps in the LSST mian-survey cadence.

\end{itemize}

\subsection{Gamma-ray Bursts}
\textsl{Authors: Maria Drout, Eric Bellm, Antonino Cucchiara}

\bigskip

\noindent Gamma-ray Bursts (GRBs) have been identified in the late 60's by the Vela spy satellites \citep{Klebesadel67}. After more than sixty years we know now that some GRBs are among the most luminous cosmic explosions in the Universe. Two classes of GRBs are currently known: the \emph{long} GRBs (LGRBs)  and \emph{short} GRBs (SGRBs), based on the duration of the gamma-ray emission. LGRBs have a duration equal or longer than two seconds, while SGRBs emission last less than two seconds \citep{Fishman95,Kouveliotou93}.

LGRBs are thought to be produced by the explosion of single massive ($\gtrsim 10-30 M_{\odot}$) stars \citep{Paczynski_1986,Fryer99,Woosley93}, while SGRBs are the byproduct of the merging process of two compact objects like neutron-star binaries (NSB), black hole binaries (BHB), or Neutron star - white dwarf pairs \citep{Eichler89,Goodman86,Meszaros97}.
SGRBs are also connected with the production of Gravitational Wave (GW) signals, as recently discovered by the LIGO-VIRGO observatories \citep{Abbott17A}.

LGRBs are likely connected with the earliest generation of massive stars produced in the Universe. A few GRBs (GRB\,090423, GRB\,090429B, GRB\,120923) have been either spectrocopically or photometrically confirmed at redshifts $z\gtrsim 9.4$ \citep{Tanvir09,cucchiara11, Tanvir18, Cucchiara13}, representing the furthest stellar objects ever discovered. These events, not only represent a unique laboratory for stellar evolution, but also pinpoint host high redshift galaxies irrespective of their intrinsic mass or luminosity. Such objects become complementary to, e.g., Hubble Ultra Deep Field samples and represent unique targets for  high-redshift Universe explorers like JWST and the Nancy Grace Roman Telescope.

The majority of GRB emission occurs in the high-energy regimes (Gamma rays), while a low-energy multiwavelength emission (the \emph{afterglow}), can be produced in the aftermath of the GRB explosions. While early GRB studies focus only on the properties of the gamma-ray prompt emission, the launch of the Neil Gehrels Swift Observatory (formerly known as {\emph Swift}) has shifted the attention to the early (first minutes to few hours) afterglow emission, in particular from X-ray to Optical, \citep{Racusin08,Racusin09, Kann10, Kann12}. With this new focus on the optical region for GRBs, we are hoping to probe the following science goals with Rubin LSST:
    \begin{itemize}
    \item Low hanging fruits
    \begin{enumerate}
\item [a)]{\bf{Detection of GRB emission}}\\
LGRBs and SGRBs produce, after an initial gamma-ray radiation, a jetted emission that interacts with the surrounding inter-stellar medium (ISM). The afterglow is generated by the synchrotron emission produced by the interaction of the ultra-relativistic blastwave with the ISM \citep{Panaitescu02,Meszaros97}. Optical/infrared rapid spectroscopy of GRB afterglow reveals the properties of the GRB host ISM as well as the presence of intervening systems. Absorption spectra have been key to investigate the cosmological metal content up to redshift $z\approx 7$ \citep{DeCia12, Chornock13, Cucchiara15}.
These also  can constrain the neutral hydrogen fraction, and the dynamics of re-ionization.

Also, rapid (within minutes of the GRB discovery) and timely spaced spectroscopic observations have revealed the presence of varying fine-structure transitions, which hint at the interaction of the surrounding medium with the GRB emission and/or with ISM particles ionized by the surrounding stellar UV background \citep{Vreswijk11}. Rubin sensitivity and survey strategies will be capable of providing alerts and triggered observations for newly discovered GRBs beyond what current facilities can do. Despite the degrading of space-based Gamma-ray detectors and the aging of the Neil Gehrels Swift Observatory and the Fermi satellite, the community should expect a constant rate of 100 GRBs/year, with at least one third of them being observable by Rubin.

\item [b)]{\bf{Detection of LGRB orphan afterglows and population studies}}\\
\noindent
The {\it Swift} satellites detects on average 100 GRB y$^{-1}$, but the number of actual GRB event is uncertain, due to the uncertainty in the jet opening angle of the initial gamma-ray emission ($\theta_{\rm jet}$). If the viewing angle (the angle between the jet axis and the observer's line of sight)  $\theta_{\rm view }>\theta_{\rm jet}$ we will not be able to detect the prompt emission, but, thanks to the deceleration of the blastwave and the subsequent decrease in the
Lorentz factor, $\Gamma$, we will be able to to detect the low-energy afterglow emission \citep{Dermer99}. These orphan afterglows arenproportional to $(1-{\rm cos} \theta_{\rm jet})^{-1}$, suggesting that we should identify $\sim$ 200 GRB y$^{-1}$. The peak of the detectable emission, occurring likely days post-burst is in the MHz-GHz regime, but the Rubin LSST single visit depth ($r\approx 24.7$) enables the detection of such an event. Assuming typical microphysical parameters for the GRB emission, Rubin LSST should be able to identify roughly 50 orphan afterglow GRBs per year \citep{Cenko15, Bhalerao17}. 

\item [c)]{\bf{The detection of GRBs as gravitational wave electromagnetic counterparts}}\\
\noindent
Short GRBs are due to the merger of two compact objects, which also produce gravitational waves. At a 200 Mpc distance, double neutron-star mergers and subsequent Kilonova emission can be detected from a the single visit in the i-band for up to 1 week.
Depending on the GW trigger facilities, the observing strategy, and the intrinsic properties of the GW progenitor, we expect between 5 and 20 SGRB/GW events to be observable by Rubin in at least one band \citep{Andreoni22}.

\end{enumerate}
    
    \end{itemize}

\subsubsection{Preparations for GRB Science}
\begin{itemize}
\item {\bf{Identifying events from alerts}}\\
Swift alerts are necessary for the timely follow-up of long and short GRBs. While broker-specific capabilities are not yet developed for GRB follow-up, we expect initially to take advantage of brokers focused on GW events, Kilonovae, and other fast transients. The capabilitied offered by Swift, Fermi and other smaller high-energy space based observatoris (e.g. BurstCube) will provide adequate constrain on the nature of these phenomena (SGRB/LGRBs/ultra-Long GRBs) thorugh the study of the X-ray/Gamma-ray counterparts.
Finally, Rubin alerts will provide localization of possible GRB to other multi-messenger facilities: e.g. IceCube neutrino detectors, radio survey (SKA, VLA), as well as Cherenkov Telescope Arrays.

\item {\bf{Follow-up observations}}\\
Absorption spectra have been key to investigating the cosmological metal content up to redshift $z\approx 6$ and to further constraining the neutral hydrogen fraction, and the dynamics of re-ionization. Further, spectroscopic observations have revealed the presence of varying fine-structure transitions. We plan swift follow-up observations using a plethora of facilities both on the ground and in space. These include target of opportunity observations at Gemini, Keck, the Very Large Telescopes, Telescopio Gran Canaria. Similarly, rapid response programs exist with the Hubble Space Telescope and JWST. The key role played by robotic telescopes around the World cannot be understated as they guaranteed continuous observations of GRBs past the Rubin survey sequence.

\item {\bf{Facilities/software requirements}}\\ 
We expect to access the Rubin data via the alerts produced by brokers, while at the same time we intend to compare the candidates with existing transients catalogs available through the Rubin LSST Science Platform. The need of building nightly lightcurve of these fast-evolving objects is imperative, especially when considering the magnitude limited capabilities for follow-up observations from the ground (${\rm m}_{AB}\approx22$ is a reasonable limit to securely detect weak metal absorption lines with a 8-m class telescope).
A direct link to a GRB-dedicated alert system (GCN Circular Network or the newly developed GCN Kafka Broker\footnote\url{{https://gcn.nasa.gov/}}) will guarantee the rapid identification and dissemination of GRB location, type, and temporal behavior information.

\end{itemize}

\subsection{Blazars}\label{blz1}
\textsl{Authors: Claudia M. Raiteri, Barbara Balmaverde, Maria Isabel Carnerero, Filippo D'Ammando, Chiara Righi}

\bigskip

\noindent Active galactic nuclei (AGNs) include a broad variety of sources that share the common property of emitting persistent huge luminosities from a very compact region ($\sim 1$ pc). Their power is thought to come from the accretion of matter onto a supermassive black hole ($10^6$--$10^{10}$ M$_\odot$). Some AGNs are very powerful radio sources, with twin jets of plasma extending up to Mpc distances from the central engine, as in the radio galaxies 3C 236 and Centaurus A. When one of the jets is oriented close to the line of sight, its emission is amplified by relativistic effects and these beamed sources are called "blazars".  Therefore, blazars are the most suitable objects to investigate the physics of the inner parts of extragalactic jets. 
    
    Blazars include flat spectrum radio quasars (FSRQs) and BL Lac objects (BL Lacs). The classical separation between the two classes depends on the spectroscopic properties, BL Lac showing only weak emission lines, if any \citep{stickel1991,stocke1991}.
    
    Blazars emit at all wavelengths, from the radio to the $\gamma$-ray band, and their flux is variable on all the observable time scales, from minutes to years. They also show spectral changes.
    The variability can be due to both geometric effects \citep[i.e.\ variations of the viewing angle of the emitting regions, see e.g.][]{raiteri2017_nature} and intrinsic (i.e.\ energetic) processes. Orientation changes can be produced by magnetohydrodynamic instabilities developing inside the jet, by orbital motion in a binary black hole system or by a precessing jet. Energetic processes include the injection and acceleration of particles, formation of shock waves and magnetic reconnection.
   
    Blazar radiation is polarized and both the polarization degree and angle are variable too. Indeed the lower energy radiation, from radio to UV and in some cases up to X-rays, is well explained as synchrotron radiation produced by relativistic electrons in the magnetized jet, while the origin of the high-energy radiation (X and $\gamma$ rays) is still debated. According to leptonic models, it is produced by inverse-Compton scattering of soft photons from the same relativistic electrons \citep{konigl1981}. Alternatively, hadronic models suggest that the high-energy emission is caused by synchrotron radiation produced by protons and muons and by particle cascades \citep{boettcher2013}. Hadronic models also predict the production of neutrinos. In this respect, the recent detection by IceCube of ultra-high-energy neutrinos that can possibly be associated with blazars opens an exciting new observing window on these multimessenger sources \citep{aartsen2019}. 
    
    Radio-loud AGN (and in particular blazars) are among the best sources to search for black hole binaries (BHBs) as they are hosted in giant elliptical galaxies that are thought to result from galaxy mergers. According to the binary black hole  models of \citet{lehto1996} and  \citet{sundelius1997}, when a secondary black hole impacts the accretion disk of the primary black hole, a periodic optical outburst signal is produced. The outburst timing gives strong constraints on the two BH masses involved and a measurement of the spin of the primary black hole. This model has been successfully applied to the quasi-periodic light curve of the BL Lac object OJ~287, which shows double-peaked outbursts every $\sim 12$ years \citep{valtonen2016}. In contrast, \cite{villata1998} proposed a scenario where the two jets launched by the two SMBHs of the binary system are bent by the interaction with the ambient medium and their axes undergo long-term precession, so that the orientation of the outflow varies quasi-periodically in time. More recently, the blazar PG~1553+113 has been found to exhibit a two-year quasi-periodic behaviour at $\gamma$ rays \citep{ackermann2015} that is also recognizable in the optical light curves. Radio observations have revealed precessing/wobbling motion of the jet, which suggests that geometrical effects must play a major role \citep{caproni2017,lico2020}.

\begin{figure}[ht]
\begin{center}
\includegraphics[width=0.7\columnwidth]{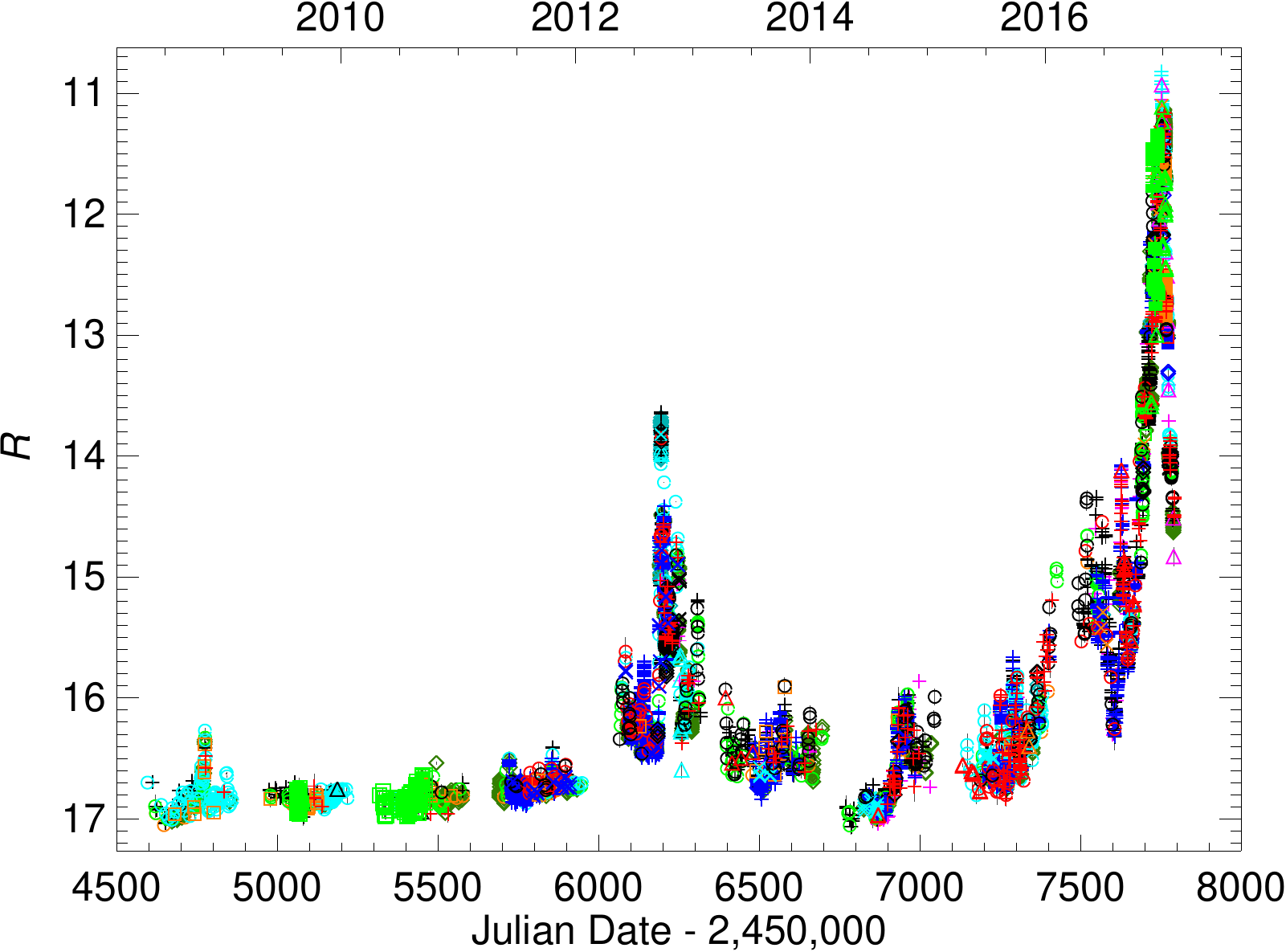}
\caption{The long-term $R$-band light curve of the blazar CTA~102 that has been obtained by the Whole Earth Blazar Telescope (WEBT; https://www.oato.inaf.it/blazars/webt/) Collaboration. \citep{raiteri2017_nature}}
\label{cta102_rband}
\end{center}
\end{figure}
\newpage
The availability of Rubin LSST data in real time will allow us to:

    \begin{itemize}
    \item Low hanging fruits
    \begin{enumerate}
     \item [a)]{\bf Trigger multi-wavelength observations of flares and other interesting events}\\ 
     By monitoring blazar flux variations, we are able to trigger follow-up observations every time an interesting event (usually a flare) is observed. This will enable immediate observations using multi-wavelength facilities. Follow-up observations may include also optical observations in polarimetric and spectroscopic mode. Multi-frequency light curves will allow us to study cross-correlations and time delays between flux changes in the optical band and those in other bands with time series analysis methods.  This in turn will shed light on the emission and variability mechanisms (particle acceleration/cooling, shock waves, orientation effects), and on the location of the various emitting regions in the jet. This requires the comparison with theoretical models for blazar emission and variability.
     
     \item [b)]{\bf Search for optical counterparts of unidentified $\gamma$-ray sources}\\ 
     Rubin LSST will observe known blazars that have previously been observed by the Fermi Gamma-ray Space Telescope satellite (and in the future by the Cherenkov Telescope Array-CTA). The Fermi Large Area Telescope Fourth Source Catalog \citep[4FGL,][]{4FGL} Data Release 3 (DR3) that has recently been published includes 6658 sources, 68\% of which are identified or associated with blazars. There are 2157 sources that still lack a secure identification with counterparts at other wavelengths and most of them are expected to be blazars, too. The Rubin LSST discovery of optical flaring activity in sources lying in the positional uncertainty region of these $\gamma$-ray sources would lead to a robust identification. Some identifications are expected already after the first year of the survey.

     \end{enumerate}
     \item Pie in the sky
    \begin{enumerate}

    \item [a)]{\bf Understand the production of neutrinos by blazar jets}\\ 
    By looking for active blazars in the sky region from which the ultra-high-energy events revealed by IceCube (and in the future KM3NeT) are coming, we plan to verify the power of blazars as cosmic accelerators. The first robust association between a flaring blazar and a high-energy IceCube neutrino was established in 2017 \citep{ice2018a}. \citet{righi2017} estimated that the future neutrino experiment KM3Net will be able to detect the neutrino signal of several BL Lacs in a few years. In this context, the Rubin LSST continuous mapping of the sky will be a formidable tool to establish such connections. The comparison between the source behaviour and theoretical predictions will clarify the framework in which neutrinos are produced in blazar jets.

    \item [b)]{\bf Reveal periodic flux changes as a signature of binary black hole binary systems}\\ 
    Geometrical versus intrinsic scenarios that describe the periodic flux changes in BHBs may be discerned through polarimetric measurements during the (quasi) periodic outbursts. Since Rubin LSST will monitor a large portion of the sky for $\sim 10$ years, we expect that it will be able to either confirm or reject BHB candidates and detect new ones, with important implications also for gravitational wave astronomy. 
    \end{enumerate}

    \end{itemize}

\subsubsection{Preparations for Blazar Science}

\begin{itemize}
\item{\bf{Identifying events from alerts}}\\
Identifying interesting events, such as flares, in time to conduct follow-up observations will necessarily rely on the Rubin LSST Prompt data products delivered by a community Broker.  More work is required to establish suitable filters within the available broker systems and to measure their performance.  The nature of the phenomenon makes it particularly important that Rubin LSST alerts be cross-matched against multiwavelength information at all frequencies, from radio to $\gamma$ rays in real-time, including FIRST, NVSS, VLASS, WISE, 2MASS, SDSS, Gaia, ROSAT, XMM-Newton, Swift and Fermi.  Furthermore, it would be highly beneficial to cross-match alerts against catalogs of blazars and blazar candidates, such as BZCAT \citep{massaro2009}, CRATES \citep{healey2007}, 3HSP \citep{chang2019}, WIBRaLS and KDEBLLACS \citep{dabrusco2019}, BROS \citep{itoh2020} and ABC \citep{paggi2020}.

\item{\bf{Follow-up observations}}\\
The blazar science case would highly benefit from a support telescope of the meter class equipped with a polarimeter to monitor the polarimetric behaviour of selected objects, especially during active states. Polarimetry can distinguish between non-thermal and thermal phenomena and is expected to give information on the jet magnetic field. Spectropolarimetry on larger telescopes would also be advantageous. 

\item{\bf{Facilities/software requirements}}\\
In addition to accessing Rubin LSST data via the brokers, it will be necessary to explore the Data Release products, using the database search and visualization tools provided by a data access centre (DAC). Interactive image and data plotting tools will be required, and overlaying images from surveys spanning the full wavelength range will be essential in assessing target behavior. Some very useful tools are available at the ASI Space Science Data Center (https://www.ssdc.asi.it/), such as the Sky Explorer, SED builder, and Multi Catalog Search. Implementation of tools like those in the LSST Science Platform (RSP) would improve Rubin LSST data exploitation. 
Moreover, the ability to cut-out images of interesting objects is important, in particular to analyze low-luminosity BL Lac objects, where the jet point-like emission is drowned into the emission of the extended host-galaxy. Such a problem is common to other AGN studies in general.

Software tools to analyse the timeseries photometry will be required to accomplish the science goals above. As with other science cases, while some code already exists, it is likely to require expanding and updating to handle the Rubin LSST data products, and to operate within the RSP environment. Methods to perform time series analysis are further detailed in \autoref{blz2}.

One critical point for blazar studies is the saturation limit of Rubin LSST. Blazar flaring states can involve flux changes up to six magnitudes \citep[][see \autoref{cta102_rband}]{raiteri2017_nature}, so there is an implicit danger of loosing information during the most interesting phases. Therefore, methods to avoid or at least mitigate saturation are strongly needed and can involve either software or observing solutions that should be tested during the commissioning phase. An analysis of the impact of saturation on blazar variability studies with Rubin LSST was performed by \citet{raiteri2022}. The possibility to avoid saturation by adopting short exposures or the star trail observing techniques \citep{thomas2018} was discussed in \citet{raiteri2018}.
    
\end{itemize}


\newpage
\subsection{Tidal Disruption Events}
\textsl{Authors: Katja Bricman, Sjoert van Velzen}

\bigskip

\noindent Tidal Disruption Events (TDEs) occur when a star passes within the Roche radius of the supermassive black hole (SMBH) in a center of a galaxy, where it is torn apart by strong tidal forces \citep{Rees:1988bf,Evans:1989qe}. After the disruption, approximately one half of the stellar debris escapes the gravitational pull of the black hole, while the other half remains bound and eventually returns to the SMBH. The fallback rate of the bound material follows a  $t^{-5/3}$ decline.

The disruption can be observed as a bright flare of light originating from the center of a non-active galaxy. The observed optical emission depends on various parameters relating to the objects and dynamics involved, such as, for example, the black hole mass, the pericenter radius, and the stellar mass, radius, and composition \citep[e.g.][]{Kochanek:1993cm, Gomboc:2005wu, Lodato:2008fr, Guillochon:2012uc, Mockler:2018xne}. Thus, optical light curves of TDEs provide a unique opportunity for detecting dormant SMBHs and measuring their masses. TDEs enable the study of black holes with masses up to $10^8 \, M_\odot$, since at larger SMBH masses the (classical) Roche radius for a Solar type star lies inside the black hole event horizon. However, if a heavier SMBH is rotating rapidly, we can expect to detect tidal disruptions of Solar type stars from black holes with $M_{BH} > 10^8 M_\odot$ \citep{Kesden:2012b}, making measurements of black hole spins using TDEs a possibility.

TDEs are rare events, with the rate between $10^{-4}$ and $10^{-5}$ per galaxy per year \citep[e.g.][]{Magorrian:1999vm, vanVelzen:2014dna}. To date the sample of optically detected TDEs consists of a couple of dozen events \citep[e.g.][]{vanVelzen:2010jp, Gezari:2012sa, Arcavi:2014iha, Chornock:2013jta, Holoien:2014jha, Holoien:2016uaf, Leloudas:2016rmh, Wyrzykowski:2016acu, Blagorodnova:2017gzv, Holoien:2018oby, vanVelzen:2018dwv, Holoien:2019zry}, mostly discovered in the last 15 years with optical surveys such as SDSS, Pan-STARRS, PTF, iPTF, ASASSN and ZTF. The majority of these events have follow-up spectroscopic observations and/or UV observations. Currently, we are detecting $\sim$10 TDEs per year. 

The observed sample shows a variety of light curves \citep{vanVelzen:2018dwv}, however the majority of them exhibit a steep decay consistent with $t^{-5/3}$ in early times with a late time accretion disk emission consistent with a near constant power law decline. They evolve slowly, on time scales from months to years, and usually show no AGN variability in the host prior to the transient phase. The peak absolute magnitudes of TDEs are around $-20$, while their spectra are blue with broad He and/or H lines. Throughout their evolution they remain at approximately a constant temperature with $T_{BB} \approx 2 \times 10^4$ K and show little to no color evolution. TDEs tend to show a preference towards post-starburst E+A host galaxies \citep{Arcavi:2014iha, 2016ApJ...818L..21F, Law-Smith:2017zne, Graur:2017vjj}, though they also appear in other types of galaxies.

Rubin LSST is expected to detect $\sim 1000$ well sampled TDEs per year \citep{vanVelzen:2010jp, Bricman:2019mcg}. A large sample of successfully identified objects with frequent temporal sampling is essential in order to unveil several properties of TDEs that are still not well understood. For example, our knowledge about the origin of the optical emission is incomplete, new observations are needed to determine whether the emission is due to the reprocessing of the inner disk (\citealt[e.g.][]{Loeb:1997dv, Guillochon:2013jda}) or due to shocks caused by stream-stream collisions (\citealt[e.g.][]{Piran:2015gha, Krolik:2016grf, Bonnerot:2016cob}). Furthermore, discovering TDEs pre-peak allows photometric and spectroscopic studies throughout the evolution of the event, thus probing accretion physics and the environments of dormant SMBHs. 

The availability of Rubin LSST data in real time will allow exploring science cases described in detail below. All science cases depend on our ability to identify TDEs from the large number of transients Rubin LSST will discover each night. Therefore, an automated algorithm for early-time photometric identification of TDEs from other transients (especially SNe) is required.

\begin{itemize}
    \item Low hanging fruits
    \begin{enumerate}
        \item [a)]{\bf{Probing the origin of the optical emission}}\\
        The properties of early-time, pre-peak light curves of low-redshift TDEs can be determined through consideration of the optical emission. Thanks to the depth of Rubin LSST we can measure the temperature (from the $u-r$ color) and the rise of the earliest emission from low-redshift TDEs (z$\sim$0.05), which will provide new insight into the origin of their optical emission. If the emission is due to reprocessing of the inner disk, we should expect a smooth rise to peak and a large blackbody radius. While emission due to shocks from intersecting streams should yield a smaller blackbody radius and may show larger rms variability.
        
        \item [b)]{\bf{Increasing the number of well-sampled TDEs}}\\ 
        Rubin LSST is expected to detect $\sim 1000$ well sampled TDEs per year \citep{vanVelzen:2010jp, Bricman:2019mcg}, which will allow statistical studies of SMBHs. TDEs discovered with Rubin LSST will enlarge the current observed sample by a factor of $\sim$10. This will allow searching for correlations between observed properties of TDEs, for example between galaxy mass and peak luminosity, peak time, color at peak. A larger observed sample will enable measurements of the luminosity function of TDEs as a function of galaxy mass or galaxy type. 
        
        \item [c)]{\bf{Determination of the masses of quiescent supermassive black holes}}\\ 
        Frequently sampled light curves of TDEs observed with Rubin LSST will enable fitting numerically predicted models to observations and the determination of SMBH masses.
        To achieve this science goal, we need accurate models that can describe the optical light curves of TDEs and can provide SMBH mass estimation. An example of such models include \texttt{MOSFiT} \citep{Guillochon:2017bmg}, which does a great job describing "normal" events see \citep{Mockler:2018xne}. However it struggles to reproduce TDEs with ``unusual" properties (e.g. ASASSN-15lh \citealt{Dong:2016, Leloudas:2016} or AT2018fyk \citealt{Wevers:2019}).
        
        \item [d)]{\bf{ The late-time observations of accretion disks}}\\ 
        The depth and frequent temporal sampling of Rubin LSST will enable late-time ($>1 $~yr post peak) observations of the brightest events. The detection of these plateaus \citep{vanVelzen:2018fuv} will enable back hole mass measurements.
        
        \item [e)]{\bf{Identifying potential host galaxies of TDEs by color}}\\ 
        With a sample of $\sim$1000 TDEs and information on their host galaxies, we will be able to investigate whether the color of the host galaxy is correlated with the probability of a TDE going off in the galaxy. Since TDEs seem to favour post-starburst hosts, with a clear preference towards E+A galaxies \citep{Arcavi:2014iha}, information on the type of galaxies in Rubin LSST's field of view will be useful for faster classification of the events \citep{2018ApJ...868...99F} - a nuclear transient in E+A/post-starburst galaxy is more likely to be a TDE than a SNe.
    \end{enumerate}
    \item Pie in the sky
    \begin{enumerate}
        \item [a)]{\bf{Redshift evolution of the TDE rate}}\\
        After a few years of the Rubin LSST operations we should have a sample of $\sim$1000 photometric TDE candidates with photometric redshift from their host galaxies. After turning this into a volume-limited sample, the number in different redshift bins is directly proportional to the disruption rate at each redshift. If the TDE rate is proportional to the  galaxy merger rate, we can expect a strong increase of the disruption rate from z=0.1 to z=1. On the other hand, the decreasing density of $M\sim 10^6 M_\odot$ black holes with redshift may lead to a decrease of rate \citep[e.g.][]{Kochanek:2016zzg}.
    
        \item [b)]{\bf{Black hole occupation fraction}}\\ 
        After a few years of Rubin operations, using our sample of $\sim 1000$ photometric TDEs, we can measure the rate as a function of galaxy mass (and galaxy type). Comparing this to the predicted rate as a function of mass, we obtain the fraction of galaxies that host black holes, i.e., the black hole occupation fraction.
        
        \item [c)]{\bf{Off-nuclear TDEs observed indirectly through recoiling black holes or stripped satellite galaxies}}\\ 
        After we have established the photometric properties of nuclear TDEs as seen by Rubin LSST, we can relax our requirement for the location of the transient in the host galaxy and search for these more rare events. A measurement of the fraction of non-nuclear massive black holes provides constrains on their seed formation mechanism in the early universe \citep{Greene:2019G}.
    \end{enumerate}
 \end{itemize}

\begin{figure}[ht]
\centering
\includegraphics[width=.6\textwidth]{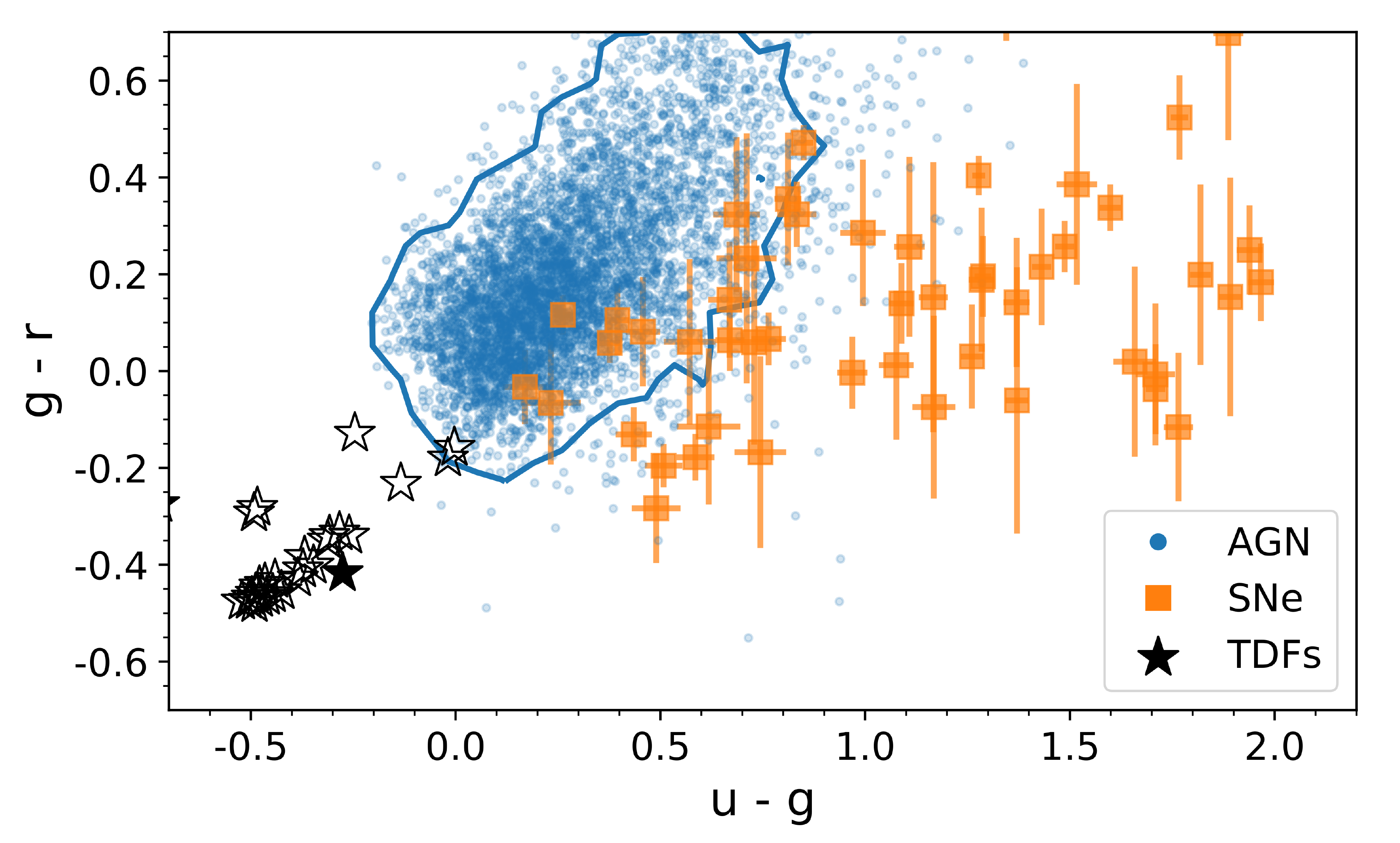}
\caption{Color-color diagram of nuclear transients: mean $u - g$ color vs. mean $g - r$ color. TDEs lie in the lower left part of the diagram, with blue mean colors $u - g \sim - 0.5$ and $g - r \sim -0.4$, making them clearly recognizable from SNe and AGN. Adapted from \cite{vanVelzen:2010jp}.}
\label{fig:tdes_uband}
\end{figure}

\subsubsection{Preparations for TDE Science}

\begin{itemize}

\item {\bf{Identifying events from alerts}}\\
The targets from Rubin LSST will be selected in real time, via filters in brokers. A filter to select TDE candidates based on light curve properties (e.g., temperature, decay time, temperature evolution) is being developed in the Ampel \citep{Nordin:2019} Broker. We expect the parent sample will be $\sim$10 000 per year and the filtered sample $\sim$100 pear year at magnitude $\sim$20. 

\item {\bf{Follow-up observations}}\\
Simultaneous observations in UV and X-ray, together with spectroscopic observations of the brightest events, are required to constrain the emission mechanism. As Rubin LSST is a photometric survey, which will achieve unprecedented depths, there will be no spectra for the majority of the candidates and we will need to rely on the photometric identification of TDEs from a large sample of transients. It is essential that TDEs are discovered pre-peak and that the observed light curves have sufficient multi-band/mulit-wavelngth coverage with frequent color measurements, especially in the bluer bands (\emph{u, g, r}). We emphasize observations in \emph{u} band are crucial to discern between SNe and TDEs \citep{vanVelzen:2010jp, Hung:2017lxm, vanVelzen:2018dwv}, see also \autoref{fig:tdes_uband}.

\begin{figure}[ht!]
\begin{center}
\includegraphics[width=0.9\columnwidth]{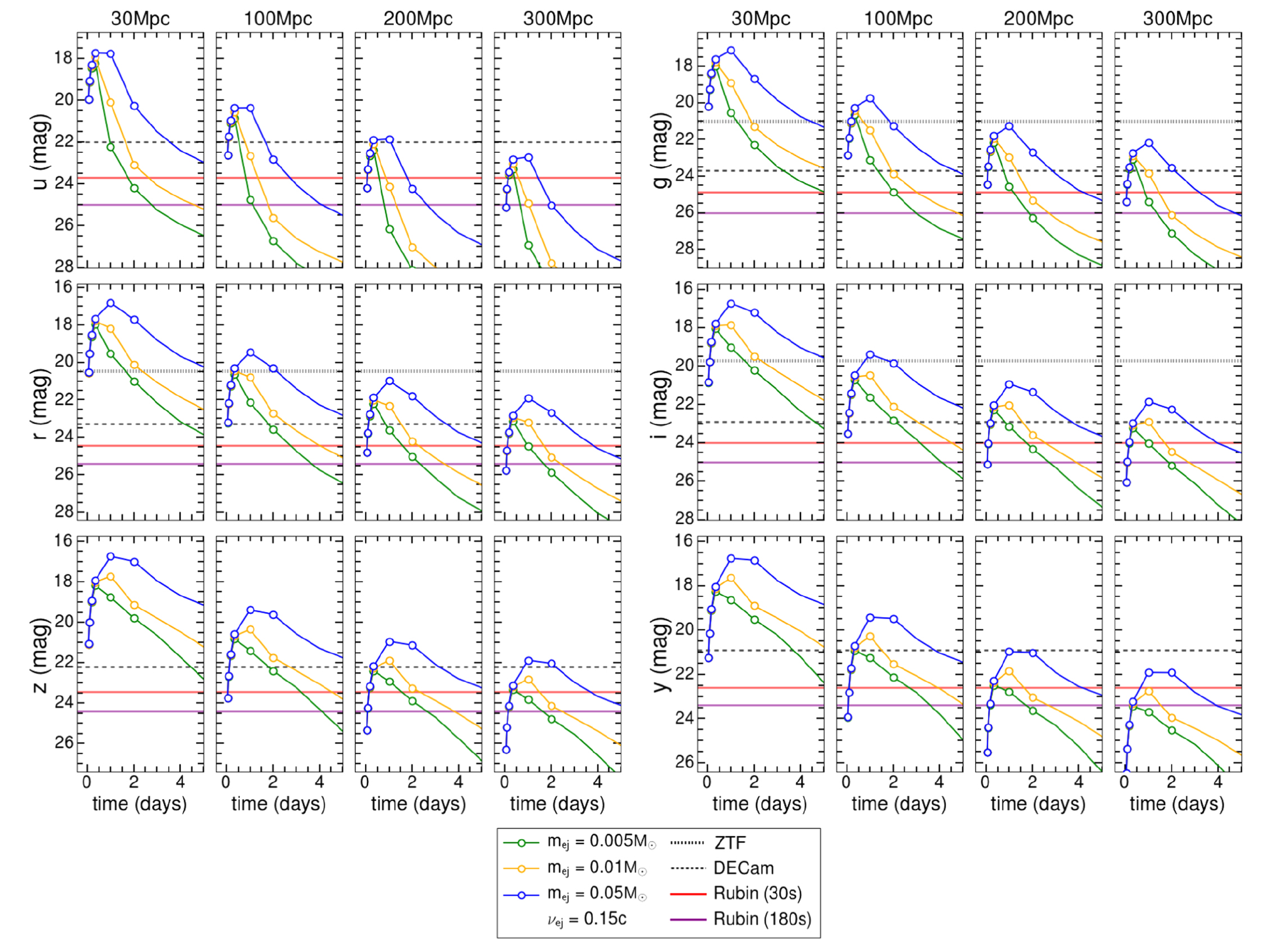}
\caption{Simulated kilonova light-curves in the six Rubin LSST filters for different properties of the ejecta (mass and velocity) at four representative distances (30, 100, 200, and 300 Mpc).Dotted and dotted–dashed horizontal lines mark typical 5$\sigma$ detection thresholds of ZTF and DECam, respectively, assuming 30~s exposure times. Red and purple solid lines: Rubin LSST 5$\sigma$ detection thresholds for exposure times of 30~s and 180~s under ideal observing conditions. The superior sensitivity of the Rubin Observatory is essential to detect the multi-color emission from kilonovae.  From \cite{Andreoni22}.}
\label{fig:KNe}
\end{center}
\end{figure}

\item{\bf{Facilities/software requirements}}\\
\texttt{emcee} or other MCMC fitting code for light curve feature extraction will be required (fitting can take $\sim$ 10 seconds per light curve). This will be important to estimate colors from observations in different filters from different nights. 

For TDE identification, photometric and spectroscopic follow-up in UV, X-ray and optical is possible for the brightest events. Additional data on galaxy types in the Rubin LSST field of view will also be required. Population studies can be done with Rubin LSST data only. For follow-up observations, Swift-like observations for UV/X-ray are required for events at $z<0.2$. Optical spectra from 2-8 meter class telescopes are also important.\\
Archival data products from eROSITA, Gaia, final co-add of WISE, and BlackGEM will be essential.

\end{itemize}


\subsection{Electromagnetic (EM) Counterparts of Gravitational Wave (GW) Events}
\textsl{Authors: Raffaella Margutti, Phil Cowperthwaite, Igor Andreoni, Michael Stroh, Giacomo Terreran, Silvia Piranomonte, Enzo Brocato, Ryan Chornock} 

\bigskip

\noindent The discovery of the electromagnetic counterparts to the binary neutron star (NS) merger GW170817 at 40 Mpc has opened the era of GW+EM multi-messenger astronomy \citep{AbbottNSdiscovery}. The true power of GW detections becomes apparent when they are paired with  electromagnetic (EM) data. The identification of an EM counterpart provides numerous benefits including: improved localization leading to host-galaxy identification; determination of the source's distance and energy scales; characterization of the progenitor's local environment; the ability to break the modeling degeneracies between distance and inclination; insights into the hydrodynamics of the merger and the physics of the jet launching mechanism; and information about quantities of heavy elements synthesized in such events. Furthermore, the identification of the EM counterpart facilitates other fields of study such as determining the primary sites of heavy r-process element production, placing limits on the neutron star (NS) equation of state and making independent measurements of the local Hubble constant (see e.g., \citealt{Margutti2021} for a recent review). 

GWs have now been detected from many BH-BH mergers \citep{Abbott_2016}, from binary NS mergers \citep{Abbott_2017,Abbott_2020}, and most recently from the merger of a black hole with a neutron star \citep{Abbott_2020}. 

In this rapidly evolving field, the frontier is now to characterize the diversity of the EM counterparts to compact-object mergers. Additionally, new sources of detectable GW emission might be revealed soon, e.g. in the form of a GW-burst from a highly asymmetric stellar explosion or more exotic event.

\begin{figure}[h!]
\begin{center}
\includegraphics[width=0.8\columnwidth]{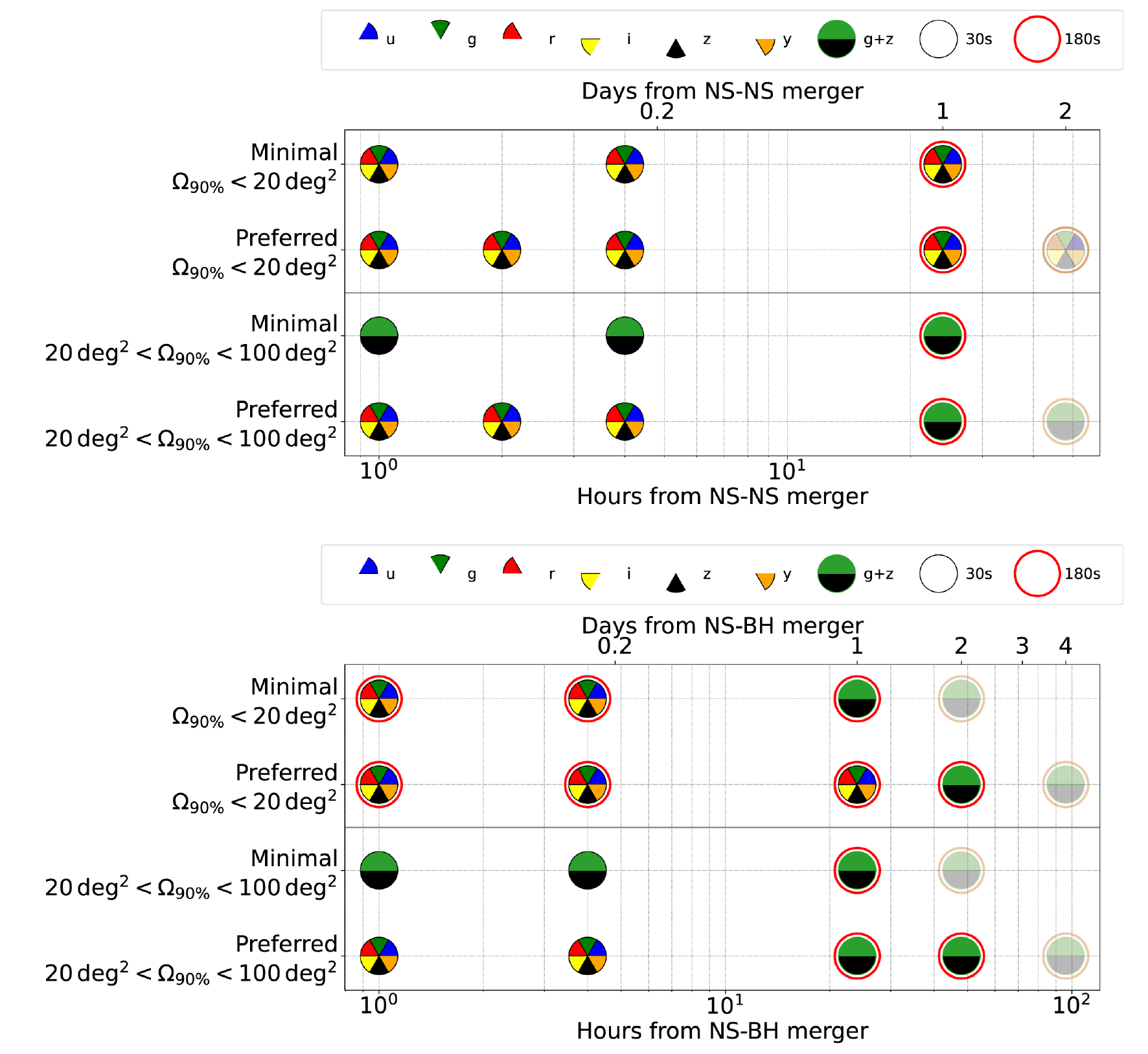}
\caption{Rubin LSST observational follow-up strategy for NS–NS (top) and NS–BH (bottom) mergers for different GW localization regions $\Omega$. Solid markers indicate planned observations over the entire localization area, while semitransparent markers indicate possible extra observations to be carried out if the optical counterpart has not yet been identified. Additional details can be found in \cite{Andreoni22}.}
\label{fig:NSstrategy}
\end{center}
\end{figure}

\begin{figure}[h!]
\begin{center}
\includegraphics[width=0.8\columnwidth]{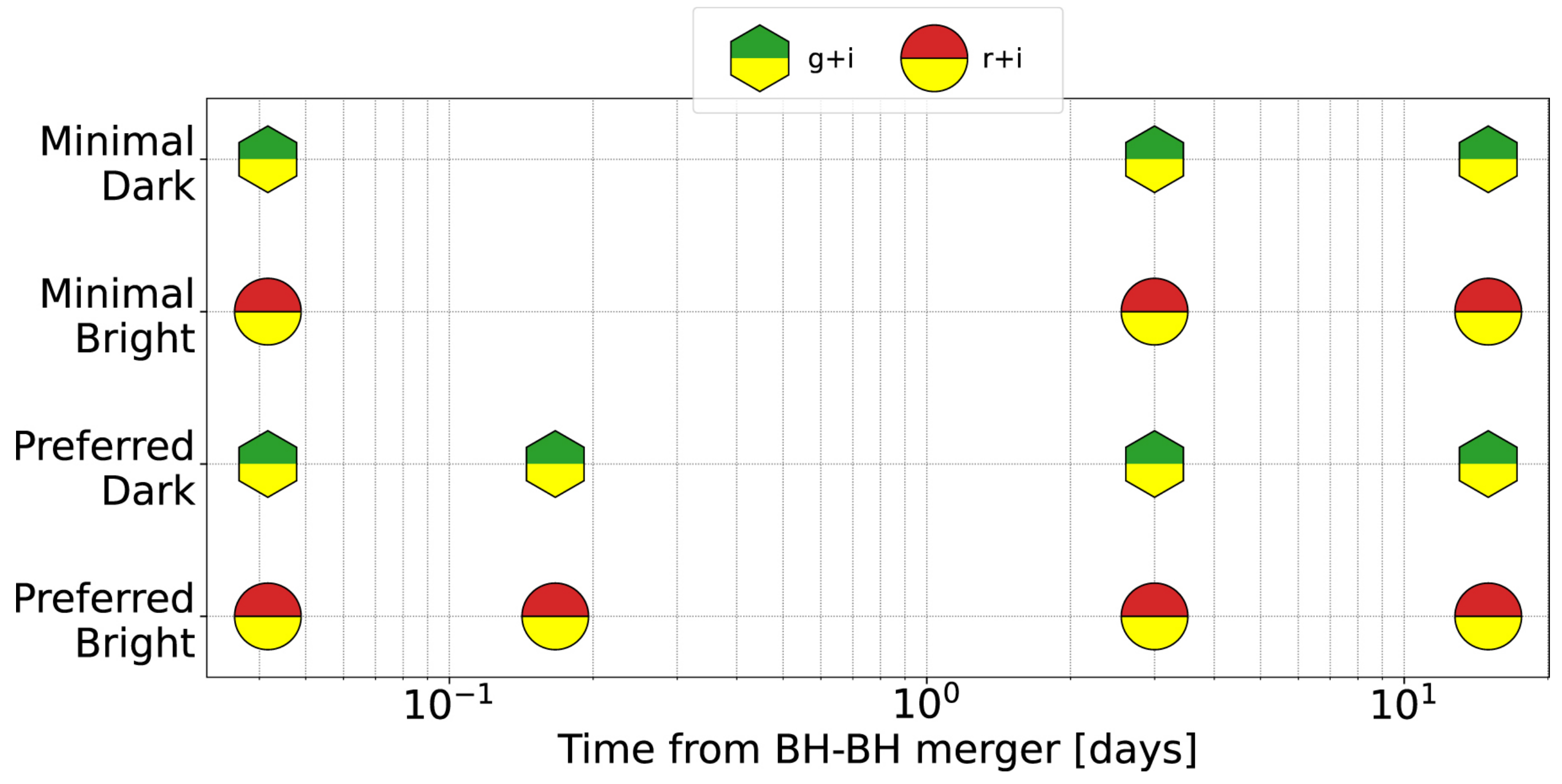}
\caption{Rubin LSST observational follow-up strategy for BH-BH mergers for different GW localization regions $\Omega$. Additional details in \cite{Andreoni22}.}
\label{fig:BHstrategy}
\end{center}
\end{figure}

\begin{table*}[h!]
    \centering
    \begin{tabular}{@{\extracolsep{4pt}}ccccccc}
    \hline\hline
    & \multicolumn{3}{c}{O5} \\
\cline{2-4} \cline{5-7}
         & Total & $20<\Omega_{\rm{90\%}}\le100$ (Rubin) & $\Omega_{\rm{90\%}}\le20$ (Rubin) \\ 
    \hline
   NS--NS  & $190^{+410}_{-130}$ & $22^{+49}_{-15}$ ($12^{+8}_{-3}$) & $13^{+29}_{-9.1}$ $(7^{+15}_{-5})$  \\
  NS--BH  & $360^{+360}_{-180}$ & $45^{+45}_{-23}$ $(24^{+24}_{-12})$ & $23^{+23}_{-12}$ $(12^{+12}_{-6})$ \\
 BH--BH  & $480^{+280}_{-180}$ & $104^{+61}_{-39}$ & $70^{+41}_{-26}$ \\
    \hline
    \end{tabular}
    \caption{Predicted number of NS--NS, NS--BH and BH--BH merger GW detections during the fifth  LIGO, Virgo, and KAGRA (LVK) observing run (O5), with which Rubin LSST science operations will likely overlap, assuming a duration of one calendar year for the run. Numbers within brackets indicate the events that will be accessible to Rubin LSST. Modified from \cite{Andreoni22}. Based on results by \cite{Petrov_2022}.}
    \label{tab:GWmergers}
\end{table*}

Rubin LSST can play a critical role in this nascent field of multi-messenger astronomy in the 2020s, when the gravitational wave detector network is expected to detect higher rates of merger events involving NSs ($\sim$10 per year) out to distances of several hundred Mpc. Rubin LSST, equipped with target-of-opportunity (ToO) capabilities and the optimal strategies for the follow-up of gravitational-wave sources, will be the premiere machine for the discovery and early characterization for NS mergers and other gravitational-wave sources. Specifically, Rubin LSST ToOs of optical counterparts of GW sources, which can be both thermal KNe and GRB afterglows, will uniquely enable the following science, if equipped with the ToO observing strategy as in \autoref{fig:NSstrategy}:
\begin{itemize}
\item Low hanging fruits
\begin{enumerate}
\item [a)]{\bf{Population studies of EM counterparts of NS mergers}}\\  The primary science goal for studies of EM transients from GW sources in the 2020s will be growing the sample size of known events, with a strong focus on finding the faintest events at the edge of the detection horizon of GW detectors. Targeted follow-up (see \autoref{fig:KNe}-\ref{fig:NSstrategy} and \autoref{tab:GWmergers}) will be much more efficient at achieving this goal compared to waiting for serendipitous discoveries from the Rubin LSST Wide-Fast-Deep (WFD) survey  \citep[e.g.][]{Cowperthwaite2019, Setzer2019, Andreoni2022serendip} and a combined multi-messenger analysis will be of much higher scientific value \citep[e.g.][]{Dietrich2020Sci}. 
Using Rubin LSST for targeted follow-up will build a large sample of EM counterparts, which are essential for conducting statistically rigorous systematic studies that will allow us to understand the diversity of EM behaviors, their host environments, the nature of merger remnants, and their contribution to the chemical enrichment of the universe through cosmic $r$-process production, which shapes the light-curves and colors of ``kilonovae'' (KNe) associated with GW events \citep[e.g.][]{Metzger2015}    

\item [b)]{\bf{Very early observations of KNe (e.g., $\lesssim10$~hr post-merger)}}\\ Despite the fact that the optical counterpart of GW170817 was discovered less than 11 hours post-merger \citep[e.g.][]{Andreoni17, Arcavi17,Coulter17,Cowperthwaite17,Drout17,Kasliwal17,Lipunov17,Smartt17,Soares17,Tanvir17,Valenti17,Villar17}, these observations were still unable to definitively determine the nature of the early time emission. Understanding this early-time emission is crucial for identifying emission mechanisms beyond the KNe (e.g., neutron precursor, shock-cooling, e.g., \cite{PiroKollmeier18}. In particular, mapping the rapid broad-band SED evolution will allow us to separate these components, and also to distinguish KNe from most other astrophysical transients. Rubin LSST's prompt data alerts will enable the rapid follow-up of KNe for the purpose of identifying their emission mechanisms (see \autoref{fig:KNe}-\ref{fig:NSstrategy} and \autoref{tab:GWmergers} for follow-up and detection details).  
\item [c)]{\bf{Discovery of the EM counterparts of NS-BH mergers}}\\ 
NS-BH mergers can produce KNe under some physical conditions. However, depending on the mass ratio of the binary and the NS equation of state, there may be less or more material ejected \citep[e.g.][]{Foucart2018}, and hence a brighter or fainter EM counterpart. It is also unclear if NS-BH mergers will be able to produce the bright early-time blue emission seen in GW170817 \citep{Metzger2015}. Furthermore, these systems will be gravitationally louder and thus GW-detections will, on average, be at greater distances. This combination of increased luminosity-distance and potentially fainter counterpart means that Rubin LSST will be an essential tool for discovering their EM counterparts (see \autoref{fig:KNe}-\ref{fig:NSstrategy} and \autoref{tab:GWmergers} for follow-up and detection details).
\end{enumerate}
\item Pie in the sky
\begin{enumerate}
\item [a)]{\bf{Discovery of EM counterparts of binary black hole (BH-BH) mergers}}\\ 
There are numerous speculative mechanisms for the production of an optical counterpart to a BH-BH merger \citep[e.g.][]{Perna,Loeb,Stone,deMink,McKernan}. Yet, none have been unambiguously observed. Rubin LSST will be able to place the deepest limits on the optical emission from BH-BH mergers, with a high statistical confidence in the case of non-detections, or might be able to discover the first EM counterpart to BH-BH mergers. In general, Rubin LSST has the best capabilities to explore the currently uncharted
territory of EM counterparts of unidentified GW sources. See \autoref{fig:BHstrategy} for a diagram describing the follow-up procedures and \autoref{tab:GWmergers} for the probability of detection with Rubin LSST.
\end{enumerate}
\end{itemize}

\subsubsection{Preparations for EM Counterpart Science}
\begin{itemize}

\item {\bf{Identifying events from gamma-ray burst triggers}}\\
While, for the most part, Rubin LSST will be used for the purpose of follow-up in the context of GW science, triggers from gamma-ray space observatories will be used to identify and to prompt expeditious follow-up observations of GRBs and KNe.
Working with current and future X-ray/radio/UV facilities interfacing with brokers to best reduce sky localization areas to search, filter out ``undesired transients”, and to ensure that the relevant information is shared will allow us to make important decisions about which Rubin LSST sources to observe.

\item {\bf{Follow-up observations}}\\
Once the GW counterpart is identified, a deep pointed multi-band follow-up is required to extract as much information as possible about the physics governing the thermal and non-thermal emissions associated with the GW sources.
Given the current estimates of KNe rates, Rubin LSST will be able to detect $\sim 10^2-10^3$ events within z = 0.25, but their classification might be challenging. \cite{Andreoni2022serendip} using several simulated cadence strategies for Rubin LSST found that currently available cadences will be  able to identify more than 300 KNe out to $\sim$1400 Mpc over a ten-year survey. Among those, we expect about 3--32 KNe that are recognizable as fast-evolving transients similar to the GW170817.
The samples of candidate counterparts coming from the Rubin LSST search will be photometrically and spectroscopically followed up by larger telescopes to determine their nature, removing the numerous expected contaminants. 
Once the most promising counterpart candidates are selected among the yet unclassified sources, deep photometry and spectroscopy are required to study all the properties of the emission.
While the selection of the most promising candidates can be done using 1- to 3-meter telescopes (such as the Palomar 200 inch Hale telescope, the Liverpool
Telescope, the Telescopio Nazionale Galileo and the Nordic Optical Telescope), to capture detailed features of the KN spectrum and its evolution larger telescopes of 4–-10 meters are required. Some examples of the required instruments are the X-shooter spectrograph on the ESO Very Large Telescope, the EFOSC2 instrument in spectroscopic mode at the ESO New Technology Telescope, the Goodman Spectrograph on the 4\,m SOAR telescope and the FLAMINGOS2 near-infrared spectrograph at Gemini-South. The Hubble Space Telescope and James Webb Telescope are additional key instruments that avoid challenging atmospheric absorption in the case of infrared spectroscopy.

\item {\bf{Facilities/software requirements}}\\ 
Considering that only through real-time follow-up we can learn about the nature of these fast transients, once we will have access to the Rubin data via the alerts produced by the brokers, we will need fast transient discovery algorithms \citep[e.g.][]{Andreoni2022serendip} and software able to classify the light curves of these  fast-evolving candidates every night. This requirement is mandatory and it will help us confirm the discovery of a GW counterpart to subsequently proceed to deep multi-wavelength follow-up with others observing facilities.

\end{itemize}


\chapterimage{head1.png}
\chapter{Non Time-Critical Science}

\section{Executive Summary}\label{sec:non_time_critical_science_summary}
\textsl{Authors: Kelly Hambleton}

\bigskip

\noindent The\blfootnote{Hubble Space Telescope image of M83, Image Credit: Hubble/Space Telescope Science Institute} wide-fast-deep 10-yr long survey will provide a plethora of data spanning the entire southern sky. This unparalleled volume of data will provide extensive new insights into non-time critical transient and variable star science.\\
\noindent{\bf {Transiting Exoplanets:}} It is anticipated that hot Jupiters around Sun-like stars within distances of approximately 0.1\,AU from their host stars will be prime targets for Rubin LSST. Other missions, such as TESS, will be used to calibrate Rubin LSST's planet detection methods.\\
{\bf {Eclipsing Binary Stars:}} A full census of short-period and contact binary stars will be obtained, which can further be used as a population probe of the Galaxy. Color information will provide temperature estimates for all known eclipsing binaries in the Southern sky and binary stars will further provide calibration for trigonometric parallaxes. Rubin LSST will enable a deeper understanding of contact binaries. \\
{\bf {Microlensing:}} The microlensing optical depth, event rate and event duration distributions will be analysed, which will provide information about the mass density distribution and kinematics of compact objects towards a given direction. Through injection of simulated events into the data from the Annual Data Releases, the selection bias for microlensing events can be analysed.\\
{\bf {Pulsating Stars (General):}} The classification and subsequent creation of a pulsational H-R diagram form the primary goals for general pulsating star science with Rubin LSST. Due to the relatively large gaps between data, the initial focus will be on long period pulsating stars, with lower amplitude short-period pulsators and multi-periodic pulsators becoming more accessible, at least from a statistical perspective, later on in the survey.\\
{\bf {Pulsating Stars (Cepheid and RR Lyrae Stars):}} Cepheids and RR Lyrae stars are widely revered as exceptional standard candles. The depth and breadth of Rubin LSST observations will enable the 3D structure of many systems, including the local group galaxies, to be studied in detail using these bright pulsating stars. Rubin LSST will further enable the period-luminosity and period-luminosity-color relations to be calibrated (including the metallicity term). These improvements will, in turn, improve estimates of H$_o$, especially with additional observations of Ultra Long Period (bright) Cepheids.\\
{\bf {Pulsating Stars (Long Period Variables):}} The primary goal for Long Period Variables (LPVs) is to further understand them in the context of standard candles. In the infrared bands, long period variables have been shown to have scatter in their period-luminosity relation similar to that of Cepheid variables. By collecting data in the reddest bands with Rubin LSST, LPVs could provide new additional evidence towards the Hubble tension.\\
{\bf {Galactic Globular Clusters}} The stellar populations within globular clusters will be probed, including the cataclysmic variables, exotic binary stars (i.e. containing neutron stars and black holes) and pulsating stars. The distribution of globular clusters in the galaxy will also be analysed to understand the effects of tides.\\
{\bf {Brown Dwarfs:}} The known sample of brown dwarfs has previously been hampered by their low luminosities and obtaining extensive light-curve coverage has such proved difficult. Rubin LSST will generate a large census of brown dwarf light curves to significantly greater distances than before. This will further enable the study of brown dwarf variability, which is currently attributed to weather in the atmospheres of the brown dwarfs. \\
{\bf {Young Erruptive Variables:}} Rubin LSST will enable the statistical study of accreating pre-main sequence objects that undergo outbursts (EXor), which is currently only thought to be 2\%. Further, the mechanism that triggers the observed outburst events will be analysed through comparison of the available light curves.\\
{\bf Compact Binaries (Cataclysmic Variables):} All the known Cataclysmic Variables (CVs) will be continuously monitored and new CVs will be discovered and classified. Additionally, CVs in eclipsing binaries and polars will be observed, which will enable a deeper look at the mass distribution of CVs and the ultimate evolution of high magnetic field systems after the common envelope phase.   \\
{\bf Compact Binaries (Neutron Star Binaries):} A census of transitional milli-second pulsars (tMSPs), milli-second pulsars (MSPs) and low mass x-ray binaries LMXBs will be obtained. The various types will be analysed to answer important questions about NS state changes.\\
{\bf Compact Binaries (Black Hole Binaries, BHBs):} Long term observations of a large number of BHBs in quiescence will be undertaken for the first time. This will enable a deeper understanding of the companions in these systems and further will enable period determination. In the optimal cases, constraints can be placed on the binary component masses.\\
{\bf Luminous Blue Variables:} Extended observations of LBVs will enable a significant number to be observed during outburst; will enable the association of the LBV identified in its eruptive state with its non-eruptive counterpart; and further enable the identification of LBVs that are supernova precursors.  \\
{\bf Light Echoes of eruptions and explosions:} Light echoes can be studied to understand the dust distribution local to stellar explosions and the explosion itself. Using Rubin LSST, known light echoes will be studied to eventually form a training set to identify new light echoes. Light echoes can set constraints on galactic explosion history and lead to the discovery of unknown supernovae.\\
{\bf Blazars:} Long term observations of blazars will enable the statistical study of their behaviour including periodicities and chromatic flairs. The environment of the blazar host galaxies can be better understood and the blazar population as a function of redshift can be explored.\\
{\bf Supernove:} The long-term monitoring of supernovae will enable a deeper understanding of their rates and progenitors. A comparison of supernova rates with models will be undertaken and the intrinsic properties of supernovae as a function of redshift will also be measured. Light curves of all types of supernovae will be obtained, which will enable a more complete understanding of star formation history.  \\
The follow-up and archival data, alongside any software and hardware necessary to complete the outlined objectives are discussed.

\section{Rubin LSST Data Release products (for time domain)}\label{sec:lsst_dr_dp}

\textsl{Editor and Author: Melissa Graham}

After the first half-year of operations, and on an annual basis thereafter, Rubin LSST will process and release all of its data via the Science Platform. This will include all of the data products associated with difference imaging analysis (DIA; very similar to the Prompt data products discussed in \autoref{sec:lsst_prompt_dp}), the raw and calibration images, deeply coadded all-sky image mosaics in each filter, and object catalogs with measurements and parameters derived from both the visit and coadded images (including catalogs of moving-object orbits based on Rubin LSST data alone). Object catalogs of forced photometry in all direct images for the union set of all objects detected in any image (or image stack) will also be provided. In general, for time-domain studies the annual data release will be the most highly {\it characterized} set of data products, and will be best suited for e.g., population studies and events rates analyses. See also \cite{LSE163} for a full and complete description of the Rubin LSST data products.

\section{Extrinsic transients and variables} 
\textsl{Editor: Michael Lund, Joshua Pepper}

\subsection{Transiting Exoplanets}\label{subsection:Transiting exoplanets}
\textsl{Authors: Michael Lund, Joshua Pepper, Keivan Stassun}

\bigskip

\noindent It is known that nearly all stars are orbited by exoplanets, and nearly all known exoplanets with precisely measured physical properties come from transit surveys, such as Kepler/K2 \citep{Borucki2010, Howell2014} and TESS \citep{ricker2015}.  For a transiting planet that can also be dynamically measured (such as with radial velocity (RV) measurements), one can determine the planet mass, radius, orbital period, orbital eccentricity, and bulk density. Even if the mass of the planet cannot be measured via RV, due to the small RV signal or faintness of the host star, knowing the radius and orbital period of large numbers of planets can enable demographic studies of exoplanetary populations, as with the Kepler mission \citep[][and references therein]{Bryson2020}.

\begin{figure}[ht]
   \begin{center}
   \includegraphics[width=0.70\columnwidth]{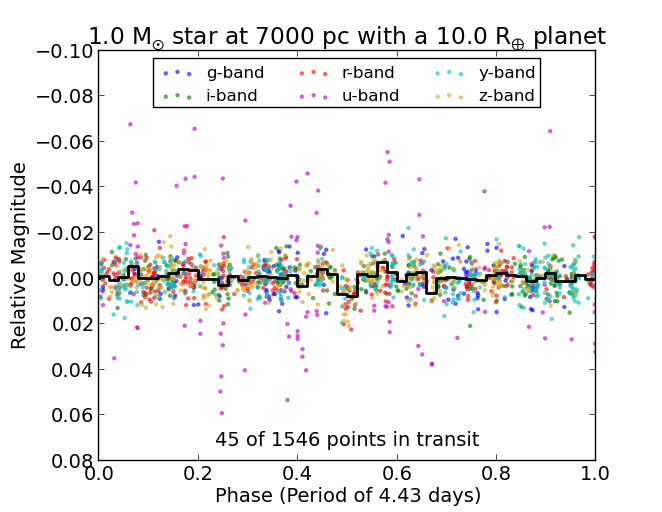}
   \caption{Phase-folded simulated light curve of a transiting Hot Jupiter observed in all 6 bands of a Wide-Fast-Deep field, first published in \cite{Lund_2015}.}
   \label{fig:6bandtransit}
   \end{center}
\end{figure}

Surveys searching for transiting planets generally are only able to probe a limited parameter space of host stars. Most searches that observe large areas of the sky are constrained to search for transits of relatively nearby stars, as these surveys tend to focus on brighter stars. Some searches probe more distant stars, but only in a single region of sky, such as the Kepler field. This trade-off in survey design is a result of the two ways that planet yields can be increased, either through higher cadence observing a narrow region, or by observing more stars down to a limiting magnitude with a wider field. Rubin LSST is not designed with transiting exoplanets in mind, and so provides a different sort of challenge, as well as opportunity. The deep-drilling fields constitute a small fraction of the Rubin LSST survey, but have a comparable cadence to ground-based planet searches. The wide-fast-deep fields observe at a lower cadence, but will cover half the sky much fainter than surveys that are focused on exoplanet detection. Exoplanet transits will be detectable in these light curves, as shown in \autoref{fig:6bandtransit}. The result is that for these fields the detection efficiency will be lower, but the range of stars being searched will include populations not normally prioritized in transiting planet searches, such as very late type stars, white dwarfs, stars in the galactic bulge, and stars in clusters. This will enable Rubin LSST to provide insight into planet occurrence and formation rates around stars of varying mass, metallicity, and stage of stellar evolution.

    \begin{itemize}
    \item Low hanging fruit
    
    \begin{enumerate}
       \item [a)]{\bf{The Detection of Hot Jupiters}}\\
       We expect to observe hot jupiters, planets with the approximate mass of Jupiter found within 0.1\,AU of their host star, around a range of main sequence stars (with smaller planets being detectable when orbiting smaller host stars). Standard exoplanet transit detection methods include the search for periodic transit signals in light curves with algorithms like BLS \citep{Kovacs2002}, coupled with careful exclusion of likely false positives, such as various types of eclipsing binary stars (there may also be a significant benefit to newer transit detection methods that have been proposed such as SBLS \citep{Panahi2021}, a recent effort to optimize the BLS search algorithm for sparsely-sampled light curves). These tools, when applied to Rubin LSST light curves of all colors, should provide candidate exoplanets that would not be found using any other current or previous planet searches \citep{Lund_2015, Jacklin_2015, Jacklin_2017}. This will be ongoing through the main survey.
       \item [b)]{\bf{The calibration of Rubin LSST's planet detection through other missions}}\\ 
       Rubin LSST's detection of planets may be compared to pre-existing planet searches. The faint end of TESS host stars may overlap with the bright end of stars observed by Rubin LSST. With TESS optimized for detecting short-period planets, Rubin LSST's detection efficiency could be calibrated by trying to recover exoplanets that are found by TESS around stars in the magnitude overlap.
    \end{enumerate}
    \item Pie in the sky
    \begin{enumerate}
       \item [a)]{\bf{The detection of planets orbiting white dwarfs}}\\ 
       It is possible that planets will be detected around white dwarfs. Such detections would provide insight into the formulation and structure of planetary systems. The occurrence rate of planets around white dwarfs currently only has upper limits.  Current transit surveys have been unable to survey large numbers of white dwarfs due to their intrinsic faintness and roughly even distribution across the sky. Since even an Earth-sized planet transiting a white dwarf will block a significant amount of light, these events will be detectable with only a few points in transit. The ability of Rubin LSST to detect transiting planets around white dwarfs has been preliminarly explored in both \cite{Cortes_2019} and \cite{Lund_2018}.
    \end{enumerate}
    \end{itemize}

\subsubsection{Preparations for Transiting Exoplanet Science}
\begin{itemize}

\item {\bf{Follow-up observations/archival data}}\\
After transit surveys identify candidate transit signals, there are a series of additional observations and analyses required to verify that the signals are indeed arising from exoplanets, and also to measure the properties of the planets. The full process typically involves multiple rounds of follow-up spectroscopy and/or photometry (e.g. \citealt{Collins2018}) to identify particular types of false positives \citep{Brown2003, Sullivan2015}. Some such observations will be possible for particular transit candidates identified by Rubin LSST. However the faintness of the typical star observed by Rubin LSST compared to the bulk of transit host stars ($V < 16$), let alone transit hosts with dynamically confirmed planets ($V < 15$) will mean the majority will not be bright enough for follow-up observations.

In general, such observations include moderate-precision RV observations to rule out certain false positives caused by EBs and to better characterize the stellar properties, high-spatial-resolution imaging (via AO or Speckle) to identify nearby unseen luminous neighbours, and then precision RV observations to dynamically confirm the planet and measure its mass. 

\item {\bf{Facilities/software requirements}}\\ 
There are two main types of software that are required for transiting exoplanet science. We will need the ability to search for transit signals in Rubin LSST light curves using BLS or other transit search tools, and potentially also detrending software. Future signal detection techniques, such as those based on machine learning algorithms, might be necessary given the the number of stars observed by Rubin LSST.  Current software does not exist for such large scales and will likely be created using the Rubin Science Platform\footnote{\url{https://data.lsst.cloud}}.
\end{itemize}

\subsection{Eclipsing binary stars} \label{sec:ntcebs}
\textsl{Authors: Andrej Pr\v sa, Keivan Stassun, Chow-Choong Ngeow}

\bigskip

\noindent Eclipsing binary stars (EBs) play an important role in stellar astrophysics primarily because they allow us to determine stellar masses and radii to an unparalleled accuracy that often exceeds 1--3\% \citep{torres2010}. The run-of-the-mill process of determining these fundamental parameters involves acquiring photometric and spectroscopic data, fitting the data by a sophisticated eclipsing binary model such as WD \citep{wilson2014} or PHOEBE \citep{prsa2018}, critically evaluating the solution uniqueness and, finally, deriving fundamental parameters.

To derive \emph{absolute} parameters (i.e.~masses in kg or $\Mnomsun$, radii in m or $\Rnomsun$), we need to be able to convert angular dimensions to absolute dimensions, and photometry gives us only the former. Thus, without spectroscopy or another means of determining absolute dimensions, the path to absolute dimensions from eclipsing binary modeling remains elusive. This implies that, with Rubin LSST data alone, we will only be able to obtain \emph{relative} sizes of stars; to transition to absolute sizes, we will need additional data or additional assumptions.

\begin{figure}[ht]
    \centering
    \includegraphics[width=0.49\textwidth]{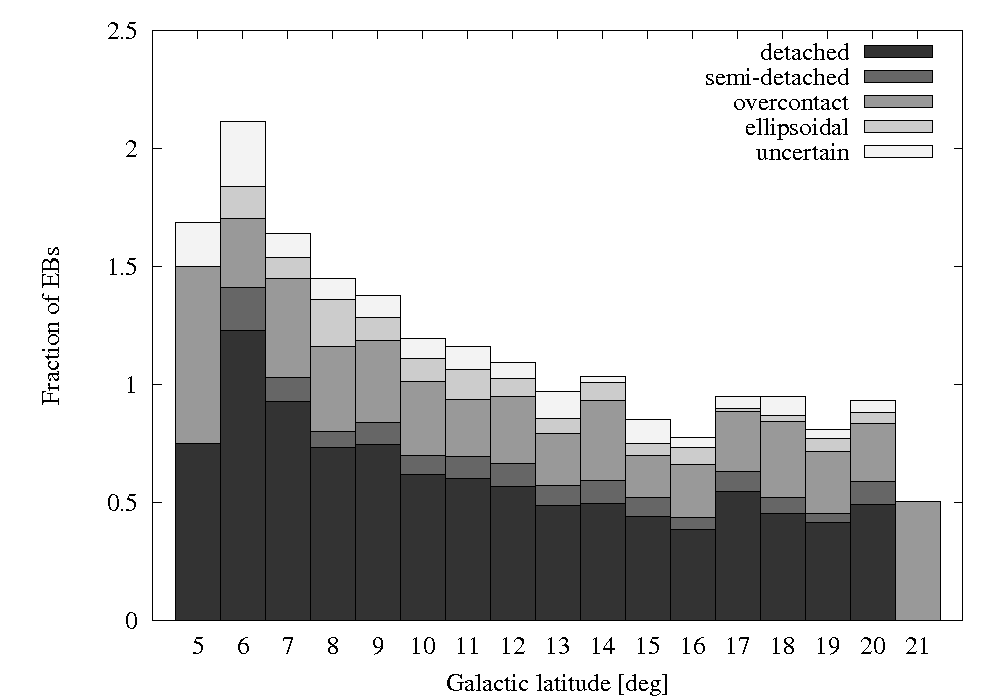}
    \includegraphics[width=0.49\textwidth]{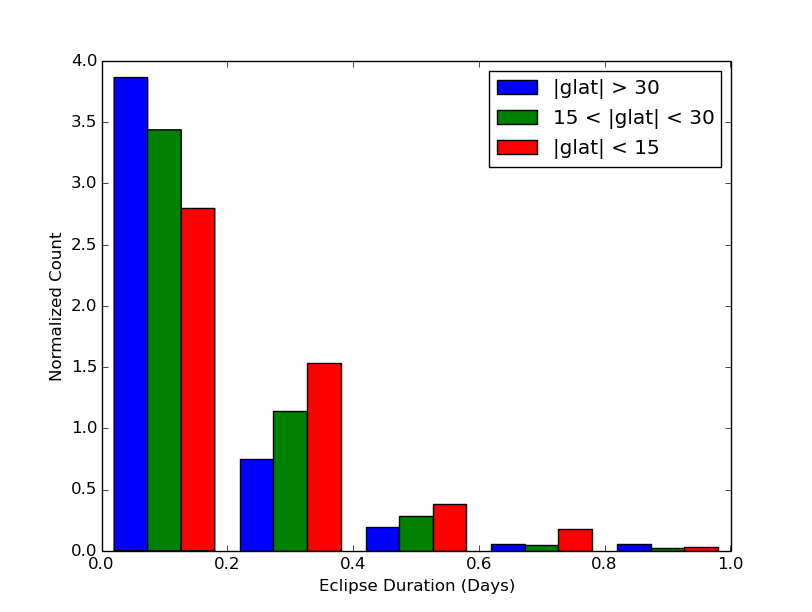} \\
    \caption{Density function of eclipsing binary stars as a function of galactic latitude, observed by the Kepler mission. The left panel depicts the area- and number-normalized count of eclipsing binaries as a function of the inferred morphology type. The right panel depicts the distribution of eclipse durations as a function of galactic latitude. Both panels provide strong evidence that the increasing occurrence rates towards the galactic plane are \emph{not} a consequence of the increased number of stars, but of the genuine differences in the sizes of stars belonging to different stellar populations.}
    \label{fig:eclipse_durations}
\end{figure}

Even relative sizes provide us with a wealth of information about stellar populations. \autoref{fig:eclipse_durations} depicts a density distribution of EBs as a function of galactic latitude: the left panel provides the area- and number-normalized fraction of EBs as a function of latitude, and the right panel compares eclipse durations for three groups of EBs delineated by latitude. If the number of EBs were uniform across the galaxy, then the left panel distribution would be flat, and the three groups in the right panel would exhibit the same behavior. The observed occurrence rate dependence on galactic latitude clearly indicates that there is an intrinsically higher probability for binaries to eclipse closer to the galactic disk, implying that the stellar population of stars closer to the galactic plane features comparatively larger stars. This conclusion can be drawn even though the absolute scales of these binaries are not known.

\begin{figure}[ht!]
    \centering
    \includegraphics[width=0.8\textwidth]{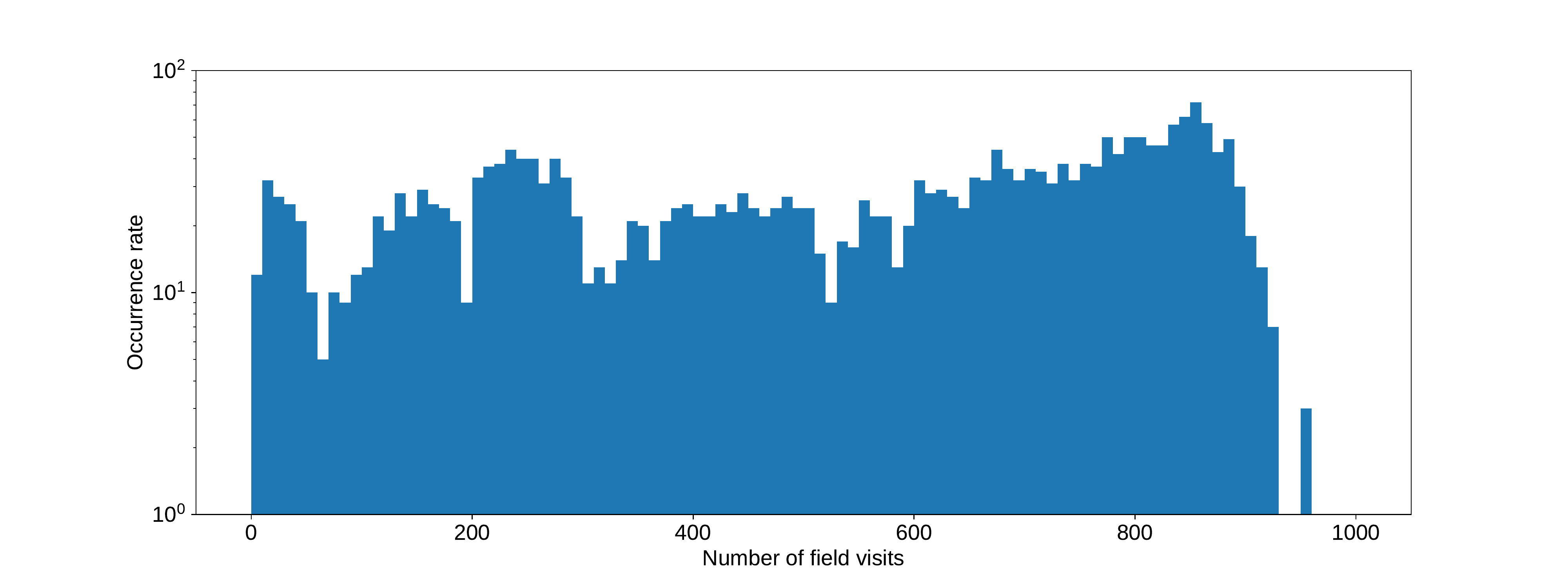} \\
    \includegraphics[width=0.8\textwidth]{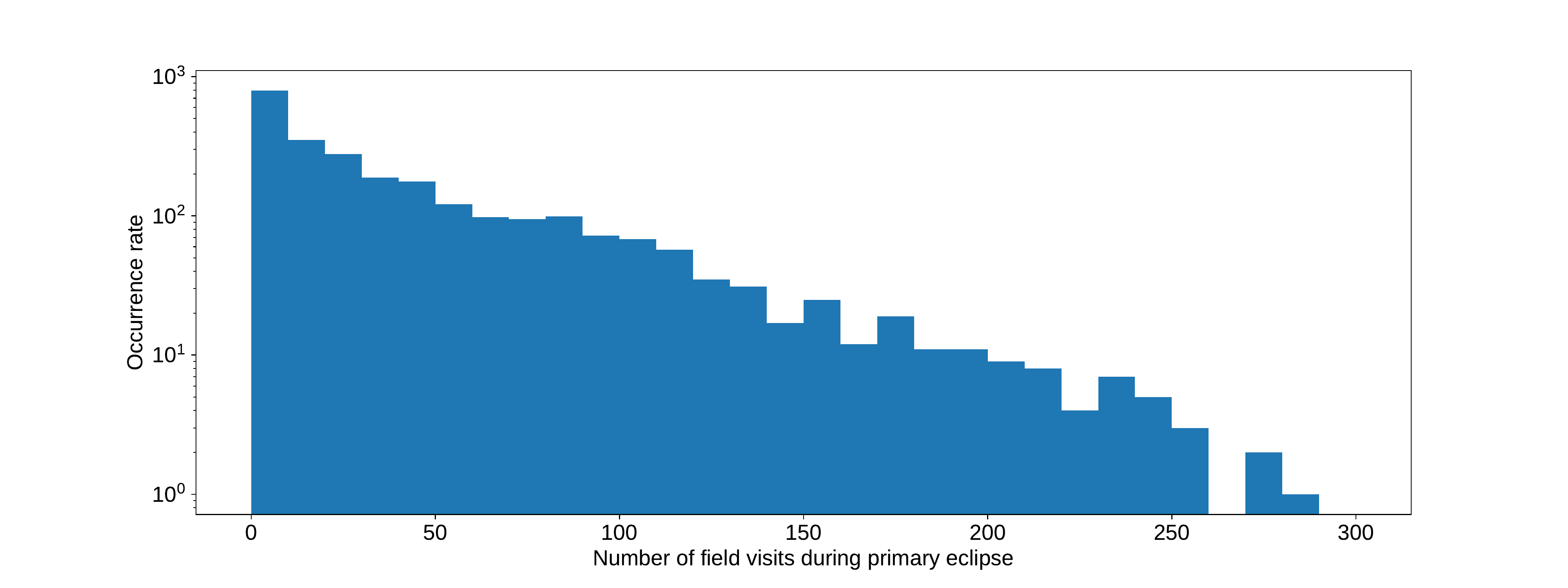} \\
    \includegraphics[width=0.8\textwidth]{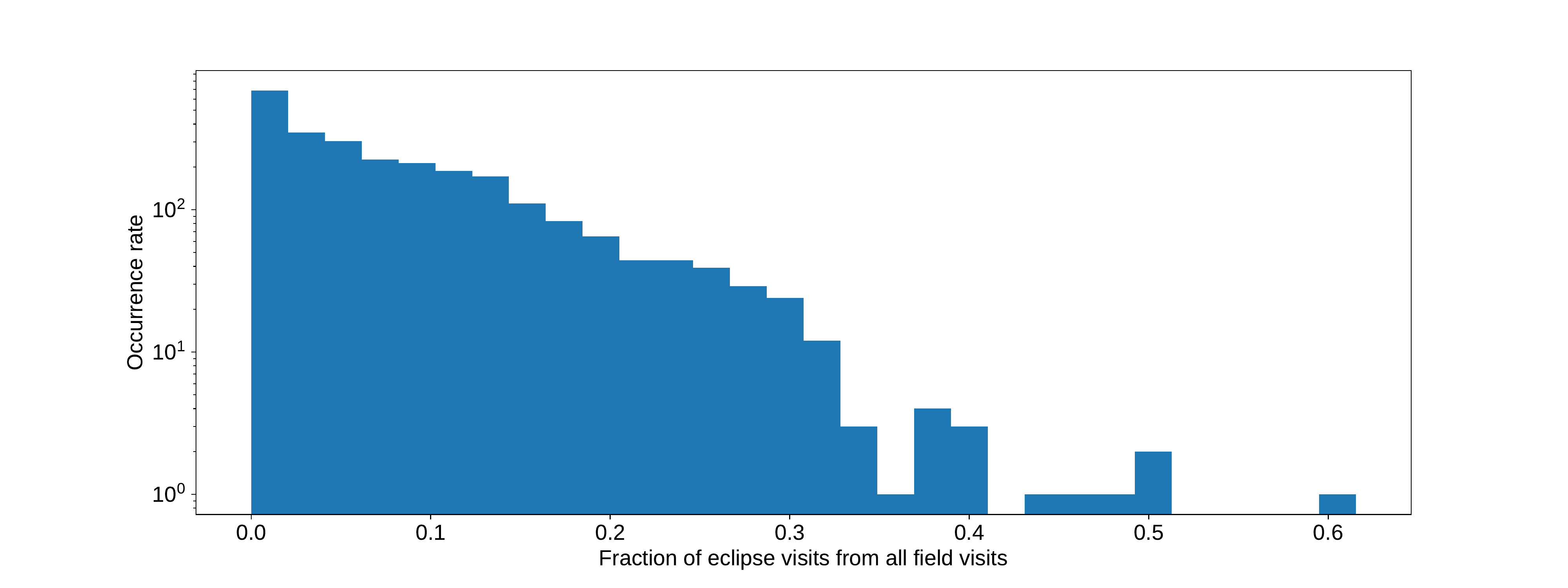} \\
    \caption{Field visit distributions for southern TESS eclipsing binary stars. Top: the overall distribution of field visits over LSST's 10-yr span. Middle: the distribution of primary eclipse visits. Bottom: the fraction of primary eclipse visits to all field visits. This measures the likelihood that an eclipsing binary will be detected: the higher the fraction, the likelier the detection (assuming other variables such as eclipse SNR being equal). Note the log-scale of occurrence rates in all three panels. The relationship between occurrence rates and primary eclipse visits, and the relationship between occurrence rates and fractions, are remarkably log-linear.}
    \label{fig:eb_visits}
\end{figure}

To evaluate the effectiveness of the Rubin LSST survey, we simulated Rubin LSST observations of all southern sky binaries discovered by the Transiting Exoplanet Survey Satellite (TESS; \citealt{ricker2015}) during their first 26 sectors of observations. Rubin LSST timestamps were simulated by the opsim 2.1 run \citep{bianco2022}; they covered the main survey, deep drilling fields, and mini-surveys. For each survey mode we counted the number of field visits and compared them to the number of visits while binaries are in eclipse. This fraction is used as a metric to determine the likelihood of the survey to discover EBs. \autoref{fig:eb_visits} depicts the results for the main survey: top panel shows the distribution of field visits that contain TESS EBs; the middle panel shows the distribution of field visits while the EBs are in primary eclipse; and the bottom panel shows the ratio between the two. The larger the number of in-eclipse visits, the better the odds of detecting EB variability, determining the correct orbital period and classifying it as an EB.

    \begin{itemize}
    \item Low hanging fruits
    \begin{enumerate}
    \item [a)]{\bf Observing known southern sky EBs}\\ 
    Rubin LSST will, as part of the main survey, observe the fields in 6 passbands up to $\sim$900 times. The use of passbands will provide us with color information on these targets, which in turn will yield effective temperature estimates. As we know that the objects are EBs ahead of time, this science goal can be met as soon as the survey starts. Given the survey's bright magnitude limit, the obvious sources of known EBs are OGLE, Gaia and, to a lesser extent, TESS.
    
    \item[b)] {\bf Near-complete census of short-period EBs}\\ 
    Thanks to the increased probability of eclipses, short-period EBs are easy targets for long baseline surveys. Additionally, because of tidal and rotational distortion, the amount of received light varies even when there are no eclipses. Given the 10-year span of the main survey and the large number of in-eclipse observations, we expect the census of eclipsing binaries with periods shorter than $\sim$1 day to be near-complete. Another added benefit is that, while we need to wait 10 years to maximize fidelity and ephemeral accuracy, discoveries and reliable classification can happen much sooner, likely within the first 2 years for the bulk of the sample. Typical candidates are contact binaries (W UMa-type systems), semi-detached binaries (Algol-type systems) and close detached binaries.
    
    \item[c)] {\bf Detect non-conservative mass loss in contact systems} \\
    Contact systems share a common envelope; if that envelope becomes too large, either because of evolutionary expansion of one of the components or because of mass transfer within the system via Roche lobe overflow, mass can be lost through Lagrangian points L2 or L3. When that happens, mass loss is non-conservative: there is no mechanism to reclaim that mass. This stage of stellar evolution is fleeting but it can be detected through theoretically driven cross-cuts in the parameter space of the contact binary, most notably mass ratio and period change. The main survey will provide us with an unprecedented number of contact binaries (cf.~previous point) and is thus quite likely to provide us with the discovery of non-conservative mass loss candidates. Because of the sensitivity to period changes, a full 10-yr survey will likely be needed for this task.
    
    \item[d)] {\bf Calibration and data integrity validation using high signal-to-noise ratio EBs}\\ 
    While not strictly astrophysical in nature, this project allows us to better qualify the survey output. EB light curves are specific in their shape and, when signal-to-noise ratio (SNR) is high, difficult to confuse with other variability types. Given that a bright, high-SNR EB is in the field, looking for ``ghost'' signals elsewhere in the field will provide us with insight into potential instrumental artifacts such as blooming, electronic cross-talk, saturation artifacts, etc.
    
    \item[e)] {\bf Calibration of trigonometric parallaxes and the near-field distance scale}\\
    Because the luminosities of EBs can be determined without regard to knowledge of the distance to the EB, EBs can be used as standard candles to test trigonometric parallaxes and to calibrate the distance scale (see, e.g., \citealt{Stassun:2016,Stassun:2018,Stassun:2021}). For example, \citet{Stassun:2021} have demonstrated the ability with a sample of $\sim$150 benchmark-grade EBs to constrain global systematics in the Gaia parallaxes down to $\sim$20~$\mu$as, for EBs as far as the Magellanic Clouds.
    
    \item [f)] {\bf Application of period-luminosity relation for (late-type) contact binaries}\\ 
    The empirical $gr$-band period-luminosity (PL), period-Wesenheit (PW) and period-luminosity-color (PLC) relations have been derived by \citet{Ngeow:2021} for late-type contact binaries, which could be used for distance measurements as an alternative to pulsating stars. Even though these $gr$-band PL/PW/PLC relations are in the Pan-STARRS1 system, they can be used to cross-check distance measurements from using pulsating stars (e.g. RR Lyrae) for nearby (dwarf) galaxies in the early stage of Rubin LSST before these PL/PW/PLC relations can be derived in native Rubin LSST filters.

    \end{enumerate}
    \item Pie in the sky
    \begin{enumerate}
    \item [a)] {\bf Full binary population probe}\\ 
    Given the survey's 6 passbands, we will get effective temperature estimates for \emph{millions} of EBs \citep{prsa2011}. That will allow us to study the distributions of binary stars as a function of galactic latitude (perhaps even longitude), multiplicity rates as a function of spectral type, individual stellar populations across the galaxy and in neighboring galaxies, etc.
    
    \item[b)] {\bf Characterize contact binary stars}\\ 
    The main benefit of contact binaries in terms of modeling is that their tidal deformation depends on the mass ratio between the two components. That implies that mass ratios can be recovered photometrically \citep{terrell2005}. For those stars where Gaia distances are also available, we will have the absolute scale of the systems and, thus, a full set of fundamental parameters. To further understand contact binary systems, this is imperative because, to date, we do not understand how energy and heat are transferred in common envelopes even though we have seen it observationally for decades. Further, low mass ratio binary components exhibit similar surface brightnesses, which is only possible when there is a strong amount of mixing in the envelope. Through knowledge of their fundamental parameters, we will be able to probe the envelope mixing of contact binary stars.
    
    \item[c)] {\bf Discover EB components in fleeting stages of stellar evolution}\\ 
    This project, an ``umbrella'' for the non-conservative mass transfer task described above, includes all non-time-critical transient phenomena related to binary stars: mass transfer and mass loss, tidal and magnetic coupling, coalescence and outbursts, and other time-variable phenomena. While ``fleeting,'' in astronomical terms these processes take a considerable amount of time in contrast to, say, nova outbursts, supernovae and other time-critical phenomena. The one concern here is a low number of observations over 10 years that may be prohibitive for \emph{recognizing} these fleeting stages.
    
    \item[d)] {\bf Discover EBs in the fundamentally new regime}\\ 
    With the survey's $r \sim 24.7$ magnitude breakdown, we will for the first time probe regimes that have been hither-to unexplored. This is predominantly the case for low mass components in tight binaries, i.e.~M-M pairs. These systems are intrinsically too faint for current surveys and the binarity rates are uncertain. In addition, there is a significant discrepancy between observed radii and theoretical predictions: observed radii seem to be larger by as much as $\sim10-20$\%. Rubin LSST will provide us with the means to probe this regime in detail and address both multiplicity rates and obtain, with the help of Gaia distances, fundamental parameters that can further calibrate models of stellar structure and evolution.
    \end{enumerate}
    \end{itemize}

\subsubsection{Preparations for Eclipsing Binary Science}
\begin{itemize}
\item {\bf{Follow-up observations/archival data}}

There are several surveys and noteworthy catalogs of eclipsing binary stars that can be used to augment Rubin LSST's dataset: OGLE \citep{udalski2015}, Gaia \citep{gaiadr32022}, and TESS \citep{ricker2015}. We do not anticipate a strong need for follow-up observations mostly because of the faint regime of Rubin LSST: most of the targets will be out of reach for spectroscopic follow-up, and the bright end of Rubin LSST targets will overlap with Gaia parallaxes, thus providing us with their absolute scales.

\item {\bf{Facilities/software requirements}}

Eclipsing binary targets may be identified using a data from prompt processing (24\,hrs) or the annual releases, but it is likely that more work is necessary in order to ensure that suitable filters are available. In order to successfully identify eclipsing binary stars, cross-matching Rubin LSST detections against the Gaia and OGLE catalogs will be mandatory, and other catalogs are highly desirable, such as VMC, VVV, ZTF and POSS.

The VaST tool (http://scan.sai.msu.ru/vast/) will be used to search for variable stars, and a Jupyter notebook will be written as a demonstration for the whole community.  Necessary analysis steps will consist of algorithms to search for variables and for period finding; some software exists for this purpose but is likely to need expanding, and adapting to work within the Rubin LSST (Data Access Center) DAC environment.  For example, current software requires ASCII or FITS files containing the lightcurve data whereas this information may be provided via database query within a DAC interface.  

Target assessment will require the functionality of image and lightcurve interactive inspection tools, as well as database search functions that can be customized to use our selection criteria.  A Jupyter notebook environment is considered to be a flexible means to achieve these analyses (making use of matplotlib for instance) within a DAC but interfacing with the VaST Tool will also require the use of APIs.

The overall scale of user-generated data products can be estimated from ongoing work by the TVS Crowded Field Photometry Task Force.  Preliminary estimates suggest that $\sim$100\,KB of output will be produced for each confirmed variable, including text and plotted figures.  

\end{itemize}
 
\subsection{Microlensing}\label{sec:micro-non-time}
\textsl{Authors: Marc Moniez, Rachel Street}

\bigskip

\noindent A detailed introduction for Microlensing can be found in \autoref{sec:micro-time}. Here we discuss the science objectives for microlensing from a non-time critical perspective, which focus on the distribution of microlensing events.
    
The microlensing optical depth towards a given source is the instantaneous probability for that source to lie behind the Einstein ring of a lens. It only depends on the mass density along the line of sight, from the observer to distance $D_S$ of that source:
       \begin{equation}
\tau(D_S)=\frac{4 \pi G D_S^2}{c^2}\int_{0}^{D_{S}} x(1-x)\rho(x) d x\, ,
\end{equation}

where $\rho(x)$ is the mass density of deflectors at distance $x D_S$. When considering a sample of $N_{obs}$ sources observed during a duration $T_{obs}$, the probability for any source to lie behind a lens has to be averaged over the $D_S$ distribution to be connected with the event frequency and duration estimates through the expression:
\begin{equation}
\tau =\frac{1}{N_{obs}T_{obs}}\frac{\pi}{2}\sum_{events}
\frac{t_{\mathrm{E}}}{\epsilon (t_{\mathrm{E}})},
\label{taumeas}
\end{equation}
where $\epsilon (t_{\mathrm{E}})$ is the average detection efficiency
of microlensing events with a time scale $t_{\mathrm{E}}$.
This number is conventionally defined as the ratio of the number of detected microlensing events with duration $t_{\mathrm{E}}$,
to the expected number of events where the source gets behind the Einstein ring of a lens during $T_{exp}$.
The optical depth can be considered as a probe that brings information about the mass density distribution of compact objects towards a given direction.

\begin{itemize}
\item Low hanging fruits
\begin{enumerate}
    \item [a)] {\bf{Optical depth, event rate and {$\bm{t_E}$} distribution}}\\
The distribution of the microlensing durations $t_E$ allows us to probe the kinematics of the population of compact objects.
The optical depth determination and the best use of the microlensing durations both require a careful estimate of the
detection efficiency $\epsilon (t_{\mathrm{E}})$.
This estimate requires the simulation of events in the domain of the parameter space that exceeds, by a large amount,
the expected domain of sensitivity, and averages are computed over all the parameters other than $t_E$ (including the impact parameter, source luminosity, colour and distance, and time of maximum magnification). The estimate of this efficiency is a difficult aspect of the microlensing interpretation, since identified sources of bias (like blending, parallax of finite source effects) may be complicated to take into account.
When analyzing a set of past data, these biases are untargeted, in the sense that they do not vary during the event progress,
and the efficiency studies are similar for all microlensing surveys.

Since the Rubin LSST survey strategy will not vary depending on what events are discovered, the simplest approach would be to exploit the Rubin LSST data in isolation, excluding follow-up.  Simulated events could be injected into real Rubin LSST time series photometry following a Data Release, the same detection algorithms applied to the full lightcurves available up to that point - meaning that events would be detected after the fact from largely "complete" lightcurves, allowing the algorithm's selection bias to be evaluated.  
\end{enumerate}
\end{itemize}
\subsubsection{Preparations for Microlensing Science}
\begin{itemize}
\item {\bf{Follow-up observations/archival data}}\\
The cross-matching of Rubin LSST difference imaging analysis (DIA) Sources against other catalogs would be beneficial,but not essential, to accurately estimate the stellar properties and spectral type.  For example, the Gaia catalog would provide distance information and spectral type in some cases.  

\item {\bf{Facilities/software requirements}}\\ 
Since this work will require access to the full lightcurve data for tens to hundreds of thousands of stars, at least some of the work is envisioned to be conducted through the Rubin LSST Science Platform (RSP).  Selecting suitable DIA Sources will require access to variability statistics computed on the Rubin LSST timeseries photometry, and a large catalog search.  The event simulation could also be conducted through the RSP, or on the downloaded timeseries.  
\end{itemize}

\section{Intrinsic Galactic and Local Universe Transients and Variables} 
\textsl{Editor: Kelly Hambleton}

\subsection{Pulsating stars: General \label{sec:pulse}}
\textsl{Authors: Kelly Hambleton, Keaton Bell, Massimo Dall'Ora, Ilaria Musella, Marcella Marconi, Marcella Di Criscienzo}

\bigskip

\noindent Rubin LSST observations will significantly improve our knowledge of pulsating stars. Stars in many stages of evolution experience global pulsations, these oscillations propagate through, and are affected by, the structure of the stellar interiors. They manifest as photometric variations in the time domain with frequencies equal to the stellar eigenfrequencies. By measuring these frequencies, we can constrain the interior stellar structure and fundamental parameters, which are otherwise not accessible in non-pulsating stars \citep[e.g.][]{Yu2018}. This is one of the most powerful methods for probing stellar structure, which would be otherwise impossible. Global properties (e.g., luminosities, radii, masses) obtained from the analysis of pulsation properties also enable many types of science, such as standardizing candles for distance determination, tracing stellar populations and revealing the absolute sizes/masses of exoplanets and binary companions \citep{Beck2014}. The detection of stellar pulsations helps us to understand driving mechanisms, making the pulsation frequencies more readily interpretable.
\begin{figure}[ht]
    \centering
    \includegraphics[width=0.65\textwidth]{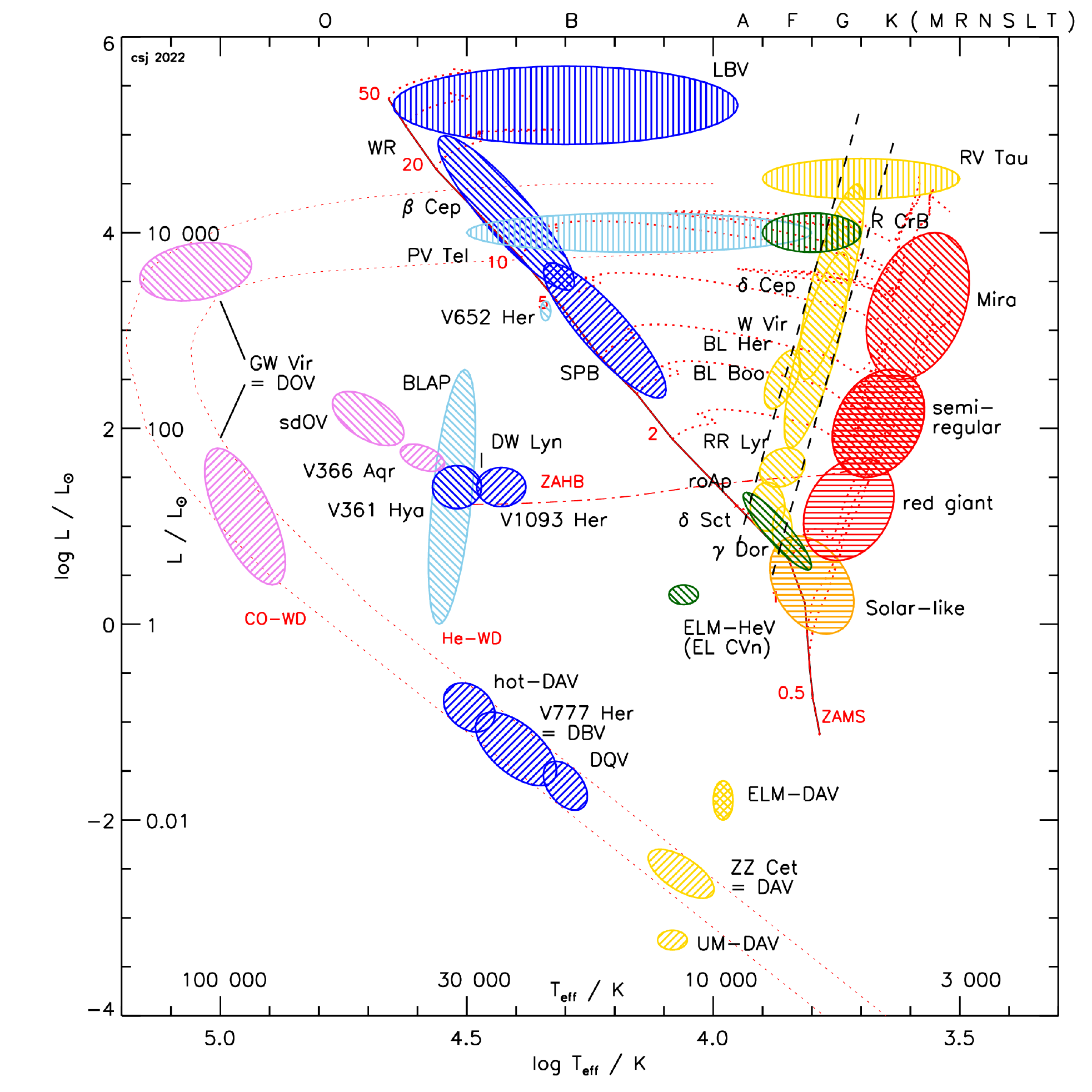}
    \caption{A pulsational H-R diagram that provides a schematic of the regions where various pulsating stars reside. Each known type of pulsating star is depicted. The solid red line represents the zero-age main sequence, the dotted red lines are evolutionary tracks for various masses, where the masses are the numbers along the ZAMS line (provided in solar masses). The cross-hatchings represent the primary mode types: acoustic p (pressure) modes (\textbackslash\textbackslash\textbackslash); gravity (buoyancy)
g modes (///); stochastically driven pulsators ($\equiv$); and strange modes (|||). This figure is courtesy
of Simon Jeffery, and is based on Figure 1 of \citet{Jeffery2015}. }
    \label{fig:pulse-hr}
\end{figure}  

    \begin{itemize}
    \item Low hanging fruits 
    \begin{enumerate}
    \item [a)]{\bf{The classification of an extreme number of pulsating stars}}\\ 
    Classification is one of the most prominent goals for pulsating stars science with Rubin LSST. While it is not expected that this will be complete, in the first years it will include long period variables such as Mira variables, Cepheids and RR Lyraes. On longer timescales, the long timebase of observations will enable stable multi-periodic signals to be resolved, such as subdwarf B stars, gamma Doradus stars, Delta Scuti stars and pulsating white dwarf stars (especially for brighter targets with high signal-to-noise). The catalog will grow incrementally with each additional Annual Data Release. 

    The minisurveys and deep drilling fields can also provide additional support for this goal, as the reliable classification of pulsating stars in a smaller field with a higher cadence could, in principle, provide a training set for machine learning algorithms to classify further pulsating stars in the main survey.

    \item [b)] {\bf{The generation of a Rubin LSST pulsational H-R diagram}}\\
    The different colors provided by Rubin LSST will enable the temperature determinations of all identified objects through the consideration of the average color of each object in several different bands. Further, when we combine Rubin LSST parallaxes and magnitudes, it will be possible to obtain the luminosities of all our objects. Thus Rubin LSST will provide the opportunity to generate a pulsational H-R diagram of an extreme number of pulsating stars in the Southern hemisphere (\autoref{fig:pulse-hr} is an example of a current pulsational H-R diagram). Upon commissioning, the first H-R diagram will be created, which can be built upon with subsequent data releases.
    
    \item [c)]{\bf{Population studies of pulsating stars}}\\
    The large spatial coverage of Rubin LSST will allow for population studies where positional variations can be taken into account and considerations, for example, the association with the thin and thick disk, can be statistically analyzed. For the shorter-period variables, typical pulsation timescales can be measured from computing structure functions (record of typical magnitude differences at different time lags, as often employed in quasar analyses, see e.g. \citealt{Hughes1992}). Again, the likelihood of identifying short period variables will increase with time and so this project will be ongoing through the 10-yr survey.
    \end{enumerate}

    \item Pie in the sky
    \begin{enumerate}
    \item [a)] {\bf{The classification and period determination of multi-periodic variables and variables with short periods}}\\ 
    The 3--5\,d cadence of Rubin LSST is well suited to studies of long period pulsating variables, such as Cepheid variables. The study of short period multi-periodic variables, such as white dwarfs and delta Scuti stars, for the purpose of asteroseismology will likely be more complicated and require the complete survey mission. Comparing the amplitudes of variations in specific frequency ranges in different filters can help to constrain (at least statistically) the geometries (spherical degree) or types of pulsation modes (gravity modes and pressure modes), however, this is better suited to deep drilling fields.
    
    \item [b)] {\bf{Identifying new types of pulsating stars}}\\
    In the last 40 years, since the advent of space missions, tens of new types of pulsating stars have been found. While Rubin LSST will is better suited to detect long period pulsators, the depth and coverage of observations do not exclude it from identifying new types of pulsating star. Sparse ground-based survey photometry from the Zwicky Transient Facility, for example, recently revealed a new class of blue large-amplitude pulsators \citep[BLAPs;][]{2019ApJ...878L..35K}.

    \end{enumerate}
    \end{itemize}

\newpage     
\subsubsection{Preparations for Pulsating Star (General) Science}

\begin{itemize}
\item {\bf{Follow-up observations/archival data}}\\
The most scientifically compelling pulsating star discoveries made from the Rubin LSST data (e.g. new variable classes, planet hosts, those with extreme pulsational characteristics or that belong to important stellar populations) can be targeted for follow-up with high-speed cameras on large telescopes to obtain data sets suitable for full asteroseismic analyses. SCORPIO on Gemini-S is one example of an instrument with high-speed capabilities that will be able to probe the Rubin LSST survey volume.
    
Additionally, representative samples of each class of pulsating variable star should be identified in existing survey data (e.g., stars with time domain photometry from PTF/ZTF, colors from SDSS, and distances from Gaia), to define initial classification algorithms. Different summary statistics (rms scatter, structure function turnover time lag, etc.) from other time domain surveys (PTF, ZTF, Pan-STARRS, ASAS-SN, etc.) should be considered in order to identify the features that best discern the variable subclasses and relate to physical properties of interest. 

\item {\bf{Facilities/software requirements}}\\ 
To store the catalog data, computing facilities are required with large amounts of storage ($\sim$Tb). Additionally, a server and the infrastructure to host an online version of the catalog would be a long term service goal.
    
The tools to determine the temperature and luminosity of the pulsating stars can undergo development prior to data acquisition. Furthermore, an analysis of the limitations placed by the spectral window of the main survey strategy is necessary to understand where asteroseismic results may be reliable. The development of tools that enable the separation of frequency aliases from the asteroseismic solution may revive the potential to seismically constrain stellar interiors with Rubin LSST data, at least at the population level. Since the majority of identification will be undertaken by assessing the scatter in the data (combined with color and temperature information) it is unlikely that significant headway will be made until the nominal mission is well underway due to the large number of data points needed.

Additional tools include period finding tools. Current tools, such as Period04 \citep{Period04}, are not well suited for bulk operations. Work is currently underway to generate such tools that will be applicable to the Rubin LSST extensive data set, i.e Gatspy \citep{Vanderplas2014} and Pyriod\footnote{\url{https://github.com/keatonb/Pyriod}} software package.
\end{itemize}

\subsection{Pulsating Stars: Cepheids and RR Lyrae Stars}  \label{sec:cepheids}
\textsl{Authors: Ilaria Musella, Marcella Marconi, Silvio Leccia, Roberto Molinaro, Vincenzo Ripepi, Giulia de Somma, Marcella Di Criscienzo}

\bigskip

\noindent Cepheids and RR Lyrae stars are bright pulsating stars that expand and contract due to the Kappa mechanism, which acts like a heat engine. They pulsate in the first overtone and fundamental modes and have periods on the order of days (RR Lyrae stars) to hundreds of days (Cepheids). Pulsating stars play a fundamental role both as distance indicators and stellar population tracers. Indeed, they can be used to obtain individual and mean distances and to calibrate secondary distance indicators that allow us to reach cosmological distances (> 100 Mpc), where the Hubble flow is undisturbed and it is possible to determine the Hubble constant $H_0$.  In addition, different types of variables, depending on their masses, belong to different evolutionary phases and, on this basis, they can be used to get information about the age and chemical composition of their host stellar population and hence to define the 3D distribution of stellar systems. As a consequence, pulsating stars can be also trace the possible presence of radial trends, halos and streams, providing important clues on the star formation history (SFH) of the host galaxy.
   
   As distance probes, the most prominent classes of pulsating stars are the Classical Cepheids (CCs), the RR Lyrae (RRLs), the Population II Cepheids (P2C), the Anomalous Cepheids (ACs) and the SX Phoenicis (SXPhs). Classical Cepheids, as population I stars, and RRs, as population II stars, represent the first rung of the cosmic distance ladder. The role of CCs and RRLs as distance indicators is based on the period-luminosity (PL) and period-luminosity-color (PLC) relations for the former and on the period-luminosity-metallicity (PLM) relation in the $B$ and $V$ optical bands, and on a PL in the NIR bands for the latter.
   
   In spite of the large number of observational and theoretical studies, a general consensus on the coefficients of these relations has not been reached. In particular, many uncertainties remain regarding the dependence on the host galaxy's chemical composition for the PL relation (for both for CCs and RRLs). While the Gaia mission is providing the most accurate distance determinations obtained so far for more than 1\% of the Milky Way stellar content \citep[see e.g.][]{Gaia+21}, the largely discussed problem of the $H_0$ tension \citep{Verde+19,Riess+21} remains unresolved. Indeed, the Hubble constant value based on the extragalactic distance scale, which is based on CCs is $H_0=73.04\pm1.04$ km s$^{-1}$  Mpc$^{-1}$ \citep{Riess+21} whereas the Planck Cosmic Microwave Background measurements adopting a $\Lambda$ Cold Dark Matter Model has obtained $H_0=67.4\pm0.5$ km s$^{-1}$ Mpc$^{-1}$ \citep{aghanim2020planck}. 
   
    Multi-band time series data collected by Rubin LSST, combined with its very accurate parallaxes and proper-motion measurements, will provide a fundamental benchmark, extending Gaia's capabilities to five magnitudes fainter, thus allowing us to observe variable stars not only in the Milky Way, but also in local group galaxies \citep[see e.g.][]{Oluseyi+12}. 
    
    The possibility to improve the knowledge of the ultra long period Cepheids (ULPs) is also a very exciting prospect. Currently, the number of know ULPs is small \citep[72;][]{Musella+21,Musella+22} often lacking homogeneous, accurate photometry. These variables, characterized by periods longer than about 80 days, are hypothesised to be the extension at higher period and mass of the Classical Cepheids. Thank to their luminosity, ULPs are observable up to cosmological distances (larger than 100 Mpc) allowing us to measure the Hubble constant without the need for secondary distance indicators. This occurrence would significantly reduce the error budget in the final Hubble constant value. In spite of this challenging property, the ULPs are not completely understood from the theoretical point of view, indeed, current stellar evolution and pulsation models do not predict such long periods in the corresponding color-magnitude diagrams, in particular in the lowest metal regime. In this context, to better understand and interpret observational data, we also need a solid theoretical scenario. 
    
    During the last two decades, an extensive and detailed theoretical scenario for several classes of pulsating stars, including Cepheids and RR Lyrae stars, has been built on the basis of nonlinear convective pulsation models \citep[see e.g.][]{Marconi_2005,Fiorentino_2007,Bono_2008,Bono_2010,Marconi_2010,Marconi_2011,Marconi_2012,Marconi_2015,DeSomma+20}. These models allow us to derive all observables, for example, the boundaries of the instability strip, light curves, periods, amplitudes and mean magnitudes. They cover a large range of masses, luminosities, metallicities and helium contents, and thus represent a robust and unique theoretical tool to interpret the observed behavior of different classes of pulsating stars and to fully exploit their crucial role as distance indicators and stellar population tracers.
    
    \begin{itemize}
    \item Low hanging fruits
    \begin{enumerate}
    \item [a)] {\bf Determine the 3D structure of the studied systems}\\ 
    Using the aforementioned PL/PLC/PLM relations, distance measurements and 3D distributions of variables in stellar systems will be obtained in the first three years after the commissioning phase through the comparison of theoretical models and the observed light curves. The Rubin LSST data will be combined with all the other public multiwavelength data sets including PTF/ZTF, SDSS and Gaia to achieve this objective. As additional data are received, we will continue to update our findings.
    
    \item [b)]{\bf Constraining the coefficients of period-luminosity and period-luminosity-color relations}\\ 
    Using Rubin LSST data, we will constrain the coefficients of the PL and PLC relations of Cepheid and RR Lyrae stars to an unprecedented level of accuracy. This result will be obtained after the 10-year survey. The calibration of these relations for Classical Cepheids and RR Lyrae stars and the adoption of Rubin LSST photometric and astrometric results will include the debated metallicity term \citep[see][]{Ripepi+21}. The collected data will allow us to determine the affect of metallicity on these relations. 

    \item [c)]{\bf Very accurate estimates for the stellar masses of RR Lyrae and Cepheid stars}\\ 
    This will be obtained through the comparison of the very extensive and accurate Rubin LSST light curves and theoretical light curves. Stellar masses are crucial to derive firm constraints on the efficiency of non-canonical phenomena (such as overshooting and mass loss in stellar evolution models), on pulsation-convection coupling and on very debated (but fundamental) parameters such as the helium to metal enrichment ratio. Indeed, this ratio has a key role in several fields of stellar and galactic astrophysics and has been shown to affect the predicted metallicity dependence of Cepheid PL relations \citep[see e.g.][and references therein]{Marconi_2005}.
    
    \item [d)] {\bf An extended census of ULPs}\\ The purpose of such a large census would be to achieve a photometrically homogeneous sample with accurate period determinations. This is possible with Rubin LSST thanks to the long observational baseline. This will allow us to test their reliability as standard candles able to directly reach cosmological distances without the adoption of secondary distance indicators.
    
    \end{enumerate}
    \item Pie in the sky
    \begin{enumerate}    
    \item [a)]{\bf{RR Lyrae and Cepheid stars as stellar population tracers}}\\ 
    The pulsational characteristics of RR Lyrae and Cephied stars depend on their physical and chemical properties. Therefore, in principle they can be (and have indeed been) used to identify the different stellar populations of the Milky Way and to characterize the properties of the Galactic components (e.g.the so-called Oosterhoff dichotomy that separately characterizes RR Lyrae in Galactic globular clusters and in the field of the Galactic Halo). The increasing quantity and quality of photometric data in different bands, together with the availability of the kinematic information from Gaia, provides us the opportunity to test a potentially robust tool to unveil both the formation and accretion history of the Milky Way. On this aspect, methods have been proposed in the literature to gain accurate distances, reddening and metallicity by using optical-NIR photometry \citep{Karczmarek2017}. In particular, these methods make use of the \textit{BVIJHK} passbands. While the use of the Johnson filters can be easily replaced with the Rubin LSST \textit{gri} filters, \textit{JHK} wavelengths are not covered by Rubin LSST. This opens the question of whether the reddest Rubin LSST passbands, namely \textit{zy}, can be effectively used for this method. Moreover, the availability of a large number of homogeneous \textit{u-}band light curves opens the possibility to use this band in this context.

    \item [b)] {\bf Firm constraints on the physical and numerical assumptions adopted in the pulsation models for RR Lyrae and Cepheid stars}\\
    As a final and very ambitious goal, we aim at putting firm constraints on the physical and numerical assumptions adopted in the stellar pulsation hydrodynamical models (e.g. the adopted opacity tables or the treatment of convection and overshooting) and to provide a sound calibration of the extragalactic distance scale thus casting light on  the debated tension in the H$_0$ determination between the values derived by Planck and those obtained using distance indicators.
    
    \end{enumerate}
    
    \end{itemize}

\subsubsection{Preparations for Pulsating Star (Cepheids and RR Lyrae Stars) Science}

\begin{itemize}
\item {\bf{Follow-up observations/archival data}}\\
    This project can be developed in synergy with all the existing and coming variability surveys that provide extensive data sets for pulsating stars such as RR Lyrae and Cepheids. The comparison with accurate and extensive observations is mandatory to test and improve pulsation models. The main projects are:
    \begin{itemize}
    \item {\bf Gaia DR3} for the comparison between theoretical and observational light curves and between individual distances obtained through theoretical tools and the ones inferred from Gaia parallaxes. These kinds of comparison are crucial to constrain the physical and numerical assumptions of pulsation models and in turn to improve our knowledge of stellar physics \citep[see e.g.][]{GaiaCOClememntini+17,DeSomma+20}.
    \item {\bf The VISTA near-infrared $YJK$s survey of the Magellanic System (VMC)} for testing the predictive capabilities of pulsation models in the near-infrared filters through the comparison of predicted pulsation properties with VMC@VISTA time-series data for Cepheids and RR Lyrae in the Magellanic Clouds \citep[see e.g.][]{Marconi+17,Ragosta+19}
    \item {\bf Optical Gravitational Lensing Experiment (OGLE)} for testing the predictive capabilities of pulsation models in the optical filters through the comparison of predicted pulsation properties with OGLE time-series data for Cepheids and RR Lyrae in the Magellanic Clouds \citep[see e.g.][]{Marconi+13}
\end{itemize}

\item {\bf{Facilities/software requirements}}\\ 
In the next year, before Rubin LSST's commissioning phase, we will build a complete theoretical scenario for Cepheids and RR Lyrae stars in the Rubin LSST bands to compare the observed pulsation properties with our models. In order to do this, we plan to enlarge the already computed extensive sets of nonlinear convective pulsation models for Cepheids and RR Lyrae stars, for varying chemical compositions ($Z$ and $Y$), masses and luminosities \citep[see e.g.][for pulsational models in the Johnson-Cousins filters]{Bono_1999,Fiorentino_2002,Marconi_2003,Di_Criscienzo_2004,Marconi_2005,Marconi_2010,DeSomma+20}. We will then transform these pulsational models into the Rubin LSST photometric $ugrizy$ bands.  
    
The $ugrizy$ light curves, mean magnitudes and colors, pulsation amplitudes and color--color loops will be derived, together with periods and analytical relations connecting pulsational to intrinsic stellar parameters (e.g. PL, PLC and Wesenheit relations). We will also probe their possible dependence on the metal content. Analogous works have already been developed to study the theoretical intrinsic properties of RR Lyrae and Cepheid stars in the SDSS bands \citep{Marconi_2006,Criscienzo_2012} and for Cepheid stars in the HST/WFC3 filters, which are typically used to study these variables \citep[F555W, F606W, F814W and F160W;][]{Fiorentino_2013}. First results have also been very recently obtained for the new theoretical period-luminosity-metallicity relations for RR Lyrae in the Rubin-LSST filters \citep{Marconi+22}. Rubin LSST observations will provide a very large database of pulsators hosted in different environments and characterized by different chemical compositions, allowing us to test the accuracy and reliability of our models and also to modify and/or refine the input physical parameters to get very good agreement between theoretical and observational properties. 

In order to host our theoretical templates, we need to have a dedicated powerful ($\sim$32 core) server. For the data analysis, we require software for period finding and for the classification and characterization (amplitude, mean magnitude, Fourier parameters) of pulsating stars, as described in \S\ref{sec:pulse}.
\end{itemize}

\subsection{Pulsating Stars: Long Period Variables}  \label{sec:LPV}
\textsl{Authors: Nicolas Mauron, Marcella Di Criscienzo, Marcella Marconi}

\bigskip

\noindent Long period variables (LPVs) are cool giants with periods or timescale of $100-3000$\,d, and $V$-band peak-to-peak amplitudes of $\sim 0.1-4.0$\,mag. Being on the asymptotic giant branch (AGB), they comprise a degenerate core composed of oxygen and carbon (or neon), a triple $\alpha$ burning zone and a region with CNO nuclear activity. This compact core (size 1/100 \mbox{R$_{\odot}$}) is surrounded by the convective envelope (size $100-500$ \mbox{R$_{\odot}$}, surface temperature $\sim 3000-4000$\,K). The surface chemistry of these stars may be altered by Hot Bottom Burning (HBB) and the Third Dredge-Up (TDU), whose efficiency depends on initial mass and metallicity. In addition to convection and chemistry, pulsation and mass-loss are critical. Intense mass ejection creates an expanding, dust-producing low-density envelope.
Despite recent progress on AGB evolution \citep{H_fner_2018}, it still remains a challenge to establish a consistent link between evolution, convection, chemistry, pulsation, shocks, dust formation, and mass loss. Rubin LSST is in a prime position to provide multi-color light curves, which will constrain pulsation models.\\
We also note that calibration of PL and PW relations of Mira stars and Semi-Regular Variables (SVRs) as well as ultra long period Cepheids (ULPs) will be crucial to provide alternative standard candles to Classical Cepheids with the power of reaching still farther distances.\\

    \begin{itemize}
    \item Low hanging fruits
    \begin{enumerate}
\item [a)]{\bf Understanding pulsations}\\ 
Pulsations have been observationally identified but the physical understanding is in an embryonic state \citep[despite its tremendous importance for Miras as standard candles,][]{Huang2018}. OGLE and MACHO experiments have shown that in the Period-Luminosity diagram, LPVs form six distinct parallel sequences. We know that membership in each of the sequences depends on the pulsation mode responsible for their variability \citep{Wood_2015}, but understanding which of these relationships is important for standard candles is not a simple question. However, LPVs are multiperiodic and normally only the primary period (the higher amplitude pulsation) is used in the period-luminosity (P-L) relation. \cite{Trabucchi_2019} showed that, as the star evolves, specific overtone modes gradually become stable and the primary mode shifts towards lower radial orders. In particular, the star initially rises on the P-L diagram while traversing these sequences from left to right(see \autoref{fig:trabucchi}).

Since Rubin LSST will measure thousands of LPV light curves, either for Galactic disc stars or stars in external galaxies (with a range of metallicities and ages), studies comparable to those of OGLE and MACHO will be achievable, with the advantage of extending the analysis to different Local Group environments.\\ 
Theoretical predictions of pulsation properties as a function of stellar parameter along the AGB need to be refined. While the commissioning and first year should be enough to detect most LPVs within a few Mpc, $5-10$ years will be necessary to firmly establish the longest periods (mostly $\sim2$\,yr) and eventually their long secondary periods.

\item [b)] {\bf To Learn about the dust production of LPVs in galaxies}\\ 
LPVs in galaxies are a dominant cause of dust enrichment (silicates and carbon dust) and the enrichment of some elements like carbon and lithium. Due to their large mass loss rates and cool temperatures, the AGB winds are a favorable site for dust production, via the condensation of gas molecules into solid grains. Recent investigations by \cite{Ventura_2012,Ventura_2014,Di_Criscienzo_2013,Dell_Agli_2019} and \cite{Nanni_2014} set the theoretical framework to model dust formation in expanding AGB winds. Rubin LSST will measure the pulsation periods or multi-period character of AGBs. It will also provide multi-color time-dependent information helping us to understand optical absorption/scattering by grains. The extensive knowledge of circumstellar dust at all metallicities is imperative for the SED fitting of LPVs undergoing huge mass-loss because they dominate the dust production rate of the host galaxy, with a special importance at high redshifts. 
    \end{enumerate}

\begin{figure}[ht]
    \centering
    \includegraphics[width=0.7\textwidth]{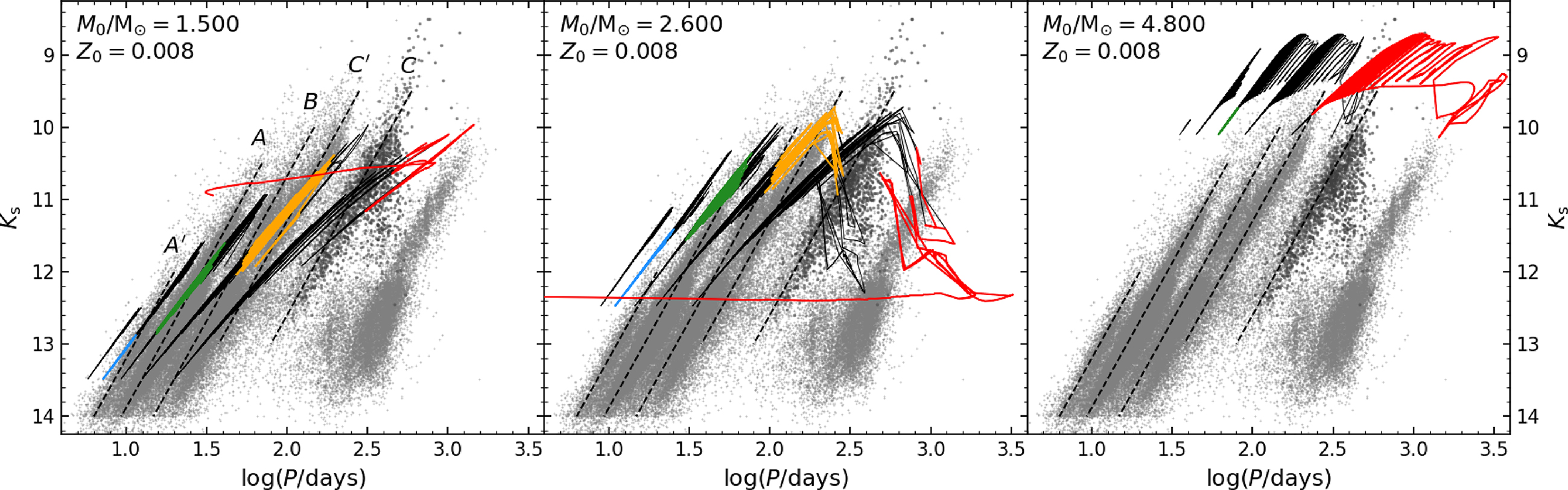}
    \caption{Evolutionary tracks  with $M=1.5M_\odot$ and Z = 0.008 at the beginning of the AGB superimposed to the observed $Ks$–$\log P$ diagram of observed LPVs in the LMC, from the OGLE-3 Catalogue \citep{Soszynski_2009}. Dashed lines identify the position of observed PL sequences. Observed stars classified as Miras are shown with darker colours on sequence C. For each evolutionary track, the periods of five modes are shown with coloured solid lines where dominant, and black lines otherwise.This figure is courtesy of Michele Trabucchi and is based on Figure 23  of \cite{Trabucchi_2019}.}
    \label{fig:trabucchi}
\end{figure}

    \item Pie in the Sky
    \begin{enumerate}
 \item [a)]{\bf Understanding long secondary periods (LSPs)}\\ LSPs happen for  about $25-30$\% of LPVs, as shown by the LMC/SMC surveys. The LSP is the reddest sequence in the P-L diagram (called D sequence), and has an unknown origin. It cannot be a radial fundamental pulsation since the period is $\sim4$ times longer than the fundamental period. The most favored explanations are binarity and non-radial g modes, but not without significant problems. The link between LSPs and mass loss via dust driven winds was raised by \cite{Wood_2009}. They showed that LSPs display some mid-IR excess compared to stars without LSPs.  LSPs cause mass ejection from giant stars. This mass and accompanying circumstellar dust is most likely in either a clumpy configuration or disk configuration. The 10 years of Rubin LSST observations will be fundamental to understanding how the presence of LSPs depend on environment or binarity, and for quantifying the percentage of LPVs that show this type of secondary variability.
 \item [b)]{\bf{Searching for symbiotic stars (SySt)}}\\
 SySt are composed of a giant and an accreting white dwarf or neutron star. They can play a role as progenitors of supernovae type Ia. Symbiotic binaries are considered among the widest interacting binary stars, which allows the onset of different phenomena such as the formation of accretion disks, Roche-lobe overflow, thermonuclear runaway, slow and recurrent nova outbursts on the surface of the compact component. Rubin LSST will help detect them through their $ugrizy$ colors \citep{Lucy_2018}. Rubin LSST will enable us to measure their AGB period and/or provide typical light curves, such as those shown by \cite{svui2j}. By surveying part of the Galactic disk, the halo and nearby dwarfs, Rubin LSST will bring complementary information on SySt populations and evolution. Note that only $\sim250$ Galactic and $\sim70$ extragalactic SySt are known so far \citep{Merc_2019}. Rubin LSST will contribute by increasing the number of known SySt in the Galaxy. A systematic search of light curves typical of SySt in available databases (OGLE, Catalina, ATLAS, ZTF) should be carried out to prepare Rubin LSST for characterization studies.
 In order to test different theoretical models of interacting phenomena such as nova and recurrent nova outbursts, determination of mass loss rates from the cold components, as well as the mass transfer mechanisms, is of key importance. Rubin LSST photometry will greatly contribute in classification of SySt, determination of the physical characteristics of the cold component  \citep{Akras_19}, dust characterization and mass loss rates of cold domponent in D-type SySt, especially when combined with near- and mid-IR surveys and observations (2MASS, WISE). Comparison between light curves in different bands could lead to detection of ellipsoidal variability  \citep{Mikolajewska_02}, which in turn could contribute to better understanding of SySt with Roche-lobe overflow as a mass transfer mechanism.

\item  [c)]{\bf The refinement of pulsation and shockwave models}\\
This will be done using the multicolor light curves obtained by Rubin LSST for thousands of LPVs, since each star will have observations in the $ugrizy$ bands. This science has already commenced with Gaia 2-filter photometry. But Rubin LSST will surpass Gaia because it will last approximately twice as long and because its data will have higher cadence and are observed at multiple wavelengths. To our knowledge, there are no theoretical models to mirror Rubin LSST's observations, although they would have the potential to identify dependencies on metallicity, luminosity, dust production, etc. We will additionally need high resolution spectral monitoring of a few dozen periodic LPVs (of diverse origin) to help constrain gas hydrodynamics and shocks and to complement Rubin LSST's photometry; the feasibility of this task is proven by \cite{Alvarez_2001}. To obtain this goal, the complete 10-yr survey is required. However, the 1st year of data will provide $2-3$ periods for pulsators with $P \sim 100-200$\,d.
\newpage
\item  [d)]{\bf The fine tuning of LPVs as standard candles}\\ This goal is fundamental to Rubin LSST, and it is imperative that we optimize the choice of Miras suitable for this science.

The Mira infrared P-L relation shows scatter that is comparable with that of Cepheids, but documentation on their infrared ($\sim 1-4\,\mu$) light curves is scarce. In addition, dust can be involved either in emission or in absorption in IR light curves. For this reason, we require deep-learning software to search for subgroups of OGLE and Rubin LSST Miras that would decrease the P-L relation scatter. A theoretical/empirical link between Rubin LSST optical light curves and 2\,$\mu$ light curves must be achieved. It is only after $\sim5-10$\,yr of operation that we shall have this information, provided a significant number of $\sim 2\,\mu$ light curves are obtained with additional ground based-telescopes.

    \end{enumerate}
    \end{itemize}
    
\subsubsection{Preparations for Pulsating Star (Long Period Variables) Science}
\begin{itemize}
\item {\bf{Follow-up observations/archival data}}\\
LPVs are important for stellar evolution and stellar population studies and could be decisive in the discussion  on the local Ho measurement and to solve the old tension on its measure. In fact LPVs can be used in place of local Type Ia supernovae (SNe Ia), whose sample size limits the precision of Ho.

Unfortunately, until now these type of studies have been very limited \citep{huang_2020} due to the lack of observations with a long (years) baseline, which is necessary to  derive periods and  the shape of the light curve from which accurate infrared mean magnitudes can be derived.
Rubin LSST will partially fill this void but new follow-up observations will be necessary after or during the survey.
In fact, Rubin LPV light curves will be mainly used to derive  accurate period measurements and derive the shape of the light curve but still need IR follow-up observations in order to produce useful PLRs (unless the PLR in the reddest Rubin bands can be derived). Infrared surveys will also be necessary to discriminate between C/O chemistry that is important to study the effect of metallicity on PLR and, especially, in the case of carbon stars, to correct the magnitudes from circumstellar reddening.
Ongoing survey such as the DUSTINGS program \citep{Boyer_2014} (started with Spitzer) must be pursued to obtain (at least) some temporal information in phase with Rubin LSST.

Last but not least, Rubin will never reach the Hubble (or better JWST) resolution, so for distant galaxies (>1.5Mpc), there may be a problem of
angular confusion (crowding) with neighboring stars which will impead the detection of many of Miras. However, if HST, JWST or adaptive optics in the next years will provide a list of potential Miras in distant galaxies, perhaps
Rubin LSST could monitor the pixel containing each of them and provide accurate periods.

\item{\bf{Software Requirements}}\\
Periodic and pulsating variables will primarily be identified from Rubin LSST light curves, once the baseline extends beyond a few times the main period of the object in question.  For most of the variables in this section, this means that targets will be identified from the Year 1 Rubin LSST Annual Data Release, with the assistance of selection cuts in photometric color and cross-matching with a number of external catalogs, particularly OGLE, Gaia, 2MASS, TESS and Spitzer.  This can be achieved using Rubin LSST Science Platform.  

Prerequisites for accomplishing these goals will be the extraction of good quality photometry from the crowded star fields in the Galactic Plane, as well as the availability within the Rubin LSST Science Platform of tools to analyse photometric timeseries, i.e. Period04 \citep{Lentz2014}.

The task of identifying candidates will be considerably enhanced if tables of basic variability statistics (including period, amplitude) are made available, and cross-matching of Rubin LSST stars against existing catalogs is pre-computed.  This is particularly important for classical radial pulsators.  For LPVs, a crossmatch with NIR and IR surveys will be desirable, such as 2MASS, WISE, DUSTINGS etc., in order to search for optical counterparts of dusty objects. Some work remains to extend existing software tools (designed for single light curve analysis) and adapt them to handle Rubin LSST's multi-filter light curves in an optimal way. 

\end{itemize}

\subsection{Galactic Globular Clusters}
\textsl{Authors: Liliana Rivera Sandoval}

\bigskip

Galactic globular clusters are old and large groups of stars that are gravitationally bound. These systems can reach central stellar densities as high as $10^6$ stars per cubic parsec, which makes them excellent laboratories to test theories of stellar dynamics and evolution. Due to their large concentration of stars, interactions among their members are common, giving birth to exotic systems \citep[see][for a review]{Maccarone2007}, several of which have not been found to exist in the field so far (e.g. yellow and red stragglers). Furthermore, the old age of globular clusters guarantees a sizeable sample of evolved stars and stellar remnants, which can be found either in isolation, in binaries or in triples, many of which are variable over different time scales. The variability and/or transient behavior of these exotic systems can be due to accretion, pulsations or to the binary’s configuration (e.g. high inclination systems). 

\begin{itemize}
\item Low hanging fruits

\begin{enumerate}

\item [a)]{\bf Stellar population studies}\\
Despite their predicted abundances, the known populations of variable stars in globular clusters are incomplete. Rubin LSST will help uncover the relative sizes of different stellar populations by combining variability and multicolor photometry with additional follow-up observations in other bands (e.g. UV, X-rays, radio). The large (time and spatial) coverage of Rubin LSST will open the possibility to identify for the first time ``missing" binaries such as symbiotic stars expected to reside in the outskirts of nearby low-density globular clusters. While the stars at the cores of clusters will not be individually resolved by Rubin LSST, techniques such as differential photometry will allow us to search for transients and variables also in the inner regions of globular clusters. The comparison to population models will be fundamental in order to obtain insights about the initial parameters of globular clusters that affect the creation and evolution of the different populations. 

\item [b)]{\bf Identification of dwarf novae}\\ 
The most abundant stellar remnants that have been identified in globular clusters are white dwarfs. Due to stellar interactions and their primordial formation, several of them are thought to be in cataclysmic variables (CVs). While many CVs are expected to be in the cores of clusters due to mass segregation, around $50\%$ of the predicted detectable CVs are believed to be outside the half-light radii, particularly in those globular clusters with half-mass relaxation times longer than a few Gyrs \citep[see][for a review on CVs in globular clusters]{Belloni-RS2021}. The wide-field coverage and 10 year duration of Rubin LSST will allow to us identify dwarf novae in the outskirts of globular clusters through their outbursts. Photometry in multiple filters will be helpful to corroborate variability detection and to place the identified systems in the color-magnitude diagram of the host cluster. This will be useful to further investigate the two-population problem of CVs in globular clusters. 

\item [c)]{\bf Characterization of exotic binaries and pulsating stars}\\
Recent studies have shown the presence of several X-ray sources in the outer parts of globular clusters, which  indicates binary interactions. Therefore, besides detecting accreting white dwarfs, the 10 year duration and multicolor photometry of Rubin LSST will help to characterize binaries with neutron stars or even black holes. For example, identifying the orbital periods of radio or X-ray sources such as binary pulsars \citep[including spider millisecond pulsars with long periods,][]{Pichardo-Marcano2021} or low mass X-ray binaries. We can also obtain information about the period and luminosity distributions of CVs and other exotic objects, which can then be compared to evolutionary models and to systems in the field. These comparisons will provide clues about the role and impact of stellar interactions, and cluster parameters such as the initial binary number and initial cluster mass on the currently detectable populations. Besides this, Rubin LSST will enable us to obtain the periods of pulsating stars such as SX-Phoenicis and RR-Lyrae. For example, combining the analysis of the light curves of RR Lyrae with semi-empirical relations, it will be possible to calculate parameters such as the effective temperature, luminosity, distance, metallicity ([Fe/H]), and mass of these pulsating stars \citep[e.g.][]{Arellano-Ferro2013}. See \autoref{sec:cepheids} for further discussion for more information on RR Lyrae variables. These studies additionally hold value because the characterization of exotica and pulsators gives information about the host globular clusters themselves. 

\item [d)]{\bf Effects of galactic tides}\\
By studying the kinematics, position and the structure of the outermost regions of globular clusters located towards the Galactic Bulge and away from it, we can obtain information about the effects of galactic tides \citep[e.g.][]{Piatti2020}. Since a significant fraction of the known globular clusters are on the Galactic plane, the monitoring of these region by Rubin LSST will be important to carry out such studies.

\end{enumerate}
\end{itemize}

\subsubsection{Preparations for Galactic Globular Cluster Science}
\begin{itemize}
\item {\bf{Follow-up observations/archival data}}
To further investigate the nature of transients and variable stars in globular clusters, multiwavelength observations in other bands such as the X-rays, UV, IR and radio with facilities like the Chandra Observatory, HST, JWST and the VLA will be desirable. For exotic/interesting systems identified in the outer regions of globular clusters, follow-up spectra taken from the ground with mid or large size telescopes could also be relatively easy to obtain, since crowding will not be a major problem in these parts. Additional photometry and/or spectroscopic follow-up will be likely required to determine membership to the host cluster. Archival observations from space telescopes or ground based telescopes with adaptive optics will also enable the investigation of systems in the inner parts of clusters. 

\item{\bf{Software Requirements}}

Pipelines to detect transient and variable sources in very crowded environments will be needed. Specially those including differential photometry techniques. Codes to determine the variable star periods will also be required. Up to date population models based on N-body or Monte Carlo methods are needed to make a fair comparison with the obtained observations. 

\end{itemize}

\subsection{Brown Dwarfs}\label{subsection:VBDs:Non-time}
\textsl{Authors: Markus Rabus}

\bigskip

\noindent Brown dwarfs (BDs) are objects with masses between the deuterium and helium burning limit. Mature field BDs cover a surface temperature range from 2000 K down to 400K. Their faintness makes them hard to detect, especially in the optical. Due to their temperature, BDs are brighter in the near-IR wavelength range and most studies have been done from 1 micron onwards.

\begin{figure}[ht]
\begin{center}
\includegraphics[width=0.65\textwidth]{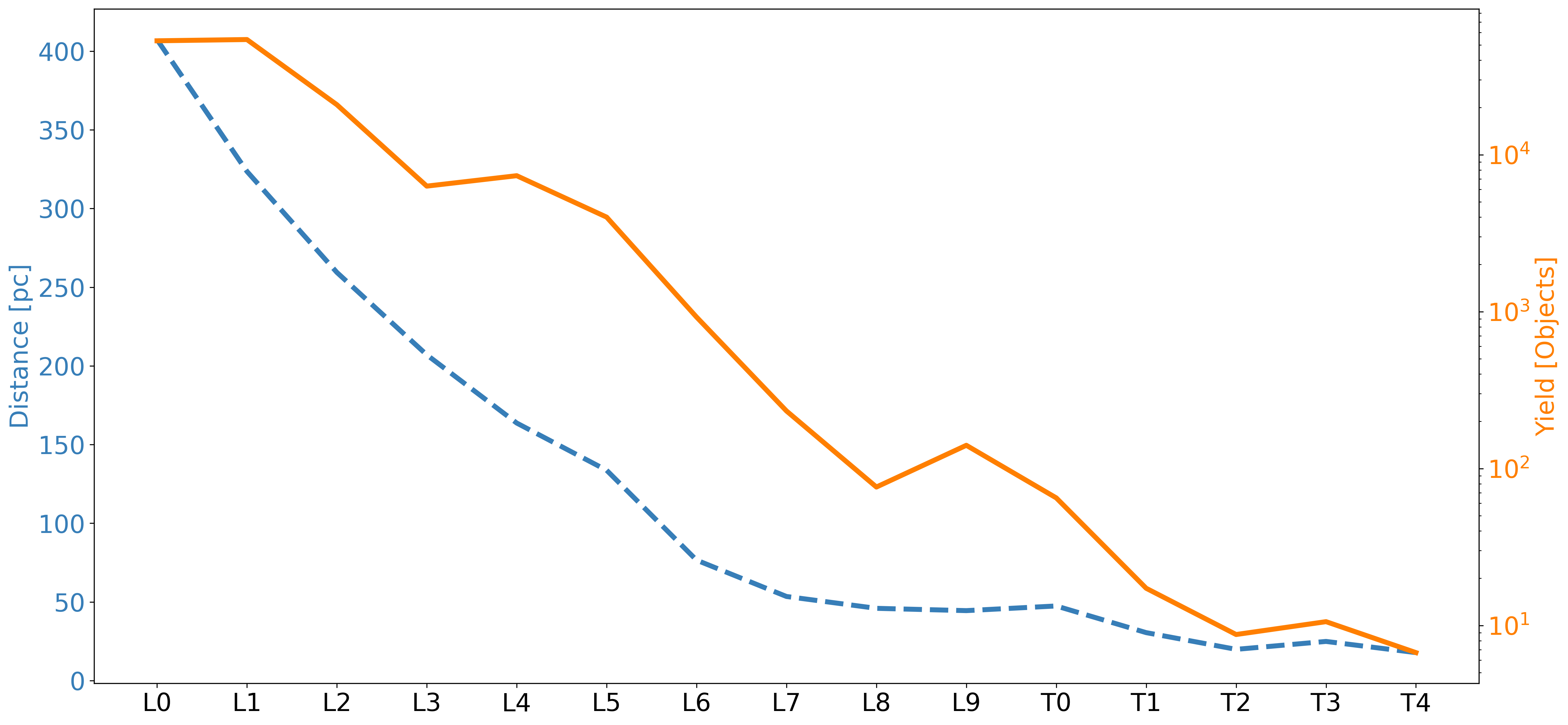}
\caption{{Expected distances and yield of BDs from spectral type L1 to T4 assuming the main survey's coverage of 25,000 square degrees. Based on magnitudes estimates from \cite{Hawley2002} and space densities from \cite{Kirkpatrick2021}.
{\label{BD_yield}}%
}}
\end{center}
\end{figure}

A study about the space density of brown dwarfs from \cite{Kirkpatrick2021} showed that BDs are well distributed across the sky. The Rubin Observatory, with its 8.4\,m mirror combined with a large etendue, will probe a large sample of BDs in the optical wavelength range. In \autoref{BD_yield} we show the expected yield of BDs for the LSSTCam's field-of-view. We will not only be able to detect them, but also get photometric time-series observations.

Photometric monitoring has shown that BDs show a rotational modulated variability. Almost all BDs show variability at a certain level \citep{Metchev2015} and $\sim$40\% of BDs at the L/T transition show amplitude variabilities larger than 2\% \citep{Radigan2014,Eriksson2019}. It is predicted that this variability is due to a global weather phenomena in the atmosphere. This prediction has initiated research about weather patterns in substellar objects outside our Solar System. While cloudy atmospheres could explain the observed variability, see e.g.\ \cite{Marley_2010}, some problems can only partially be explained, A resurgence of FeH absorption. As such, additional hypotheses were proposed, i.e. tne idea that the brightness fluctuations in BDs might emanate from non-uniform temperature profiles or perturbations in the atmosphere's temperature structure\citep{Robinson_2014}. An additional alternative, proposed by \cite{Tremblin_2016}, claims that non-uniform surface opacities caused by chemical abundance variations are the origin of the observed variabilities.

In \autoref{subsection:VBDs:DDF} we will show how deep drilling field observarions can contribute towards the variability study of BDs and in \autoref{subsection:VBDs:Mini-survey} we discuss the opportunities presented by the commissioning tests, which have high cadence observations. Regions overlapping with other missions, like e.g. NASA's Roman Space Telescope, can provide additional information in a different wavelength range. Mini or Microsurveys in crowded fields will probe regions neglected in the past. Here we discuss the science goals for the main survey.

\newpage
    \begin{itemize}
    \item Low hanging fruits
    
        \begin{enumerate}
        
        \item [a)]{\bf Enlarging the sample of BDs}\\ 
        Our current sample of BDs is limited to objects in our close Solar Neighbourhood, e.g. the faintness of BDs make measurements difficult and detection of early type L-dwarfs have only been achieved within a distance of 24 pc with Gaia \citep{Smart2017}. Also, \cite{Kirkpatrick2021} presented a volume limited sample of BDs in the celestial sphere with a radius of 20 pc. The Rubin Observatory will allow the detections of fainter objects and as such can produce a volume limited sample which will be complete to a larger distance.
        
        \item [b)]{\bf Detection of weather patterns in BDs}\\ 
        The sample of substellar objects monitored in the optical wavelength range is still too small to draw a meaningful conclusion. In addition, it is difficult to detect and monitor these objects, as shown by the example of Luhman-16 AB, which is the closest substellar object not belonging to our Solar System. Despite being very close to us, it was discovered only recently in 2013, but on the other hand, it has also become a benchmark system \citep[see e.g.][for the study of variability in brown dwarfs]{Street_2015, Buenzli_2015, Karalidi_2016}. A major problem concerning these objects is their low intrinsic brightness. However, only a few variability surveys have been conducted, most of them in the NIR bands. Unfortunately, the lack of high-precision light curves in the NIR wavelength range, combined with short observation spans, have hampered any robust conclusions. So far, the longest observing span has been obtained by \cite{Street_2015} and \cite{Apai2021}. \cite{Street_2015} observed Luhmann-16 in the optical band with the LCO 1-m telescope network for 42\,d and found a variabilty with an amplitude 0f 0.05-0.1 magnitudes and periodicities between 4.46-5.84 hours. \cite{Apai2021} produced a TESS light curve spanning several rotation periods to study the long-term variability seen in Luhman-16 and found changing periodicites, which they attributed to planetary wave beat patterns. Long-term optical observations have also been obtained of the $\epsilon$ Indi Ba/Bb BD binary. The detection of long-term variations, but the lack of variation on short time scale (a few hours, corresponding to the rotation period) argues for a system probably seen pole-on and with a detected long-term cloud evolution. Instead of observing confirmed BDs one-by-one, we can take advantage of Rubin LSST's extensive sky coverage. Given the faintness of these objects and the planned photometric precision of Rubin LSST, this facility will be the perfect way to photometrically monitor many BDs at once. 
        
        \item [c)]{\bf Distance and parallax measurements}\\
       The long baseline of the mission, the large field-of-view and the astrometric precision of the observations with the Rubin Observatory will provide us important astrometric parameters, like parallaxes and proper motions, of those faint objects.
         The distance measurements are an important additional information to provide a robust classification of BD candidates through luminosity estimates and without the immediate need for follow-up observations for classification. We also note that Gaia can only probe the bright end of the BD population, whereas the depth of Rubin observations will provide measurements of fainter BDs.

        \end{enumerate}

    \item Pie in the sky
    
        \begin{enumerate}

        \item [a)]{\bf Astrometric detection of sub-stellar companions}\\
        Astrometric variability can be indicative of an unseen companion. The astrometric signal can be divided into 2--3 main components: the parallax ellipse, the object's proper motion, and in the presence of a possible gravitationally bound companion, the orbital signature. The latter is only part of the astrometric signal if a companion is gravitationally bound. The companions' orbital signature can be estimated by unraveling the proper motion and parallax effects from astrometric measurements. Rubin LSST will provide a multi-year time-series of high precision astrometric measurements for stars and brown dwarfs, which will allow the disentanglement of the three major astrometric components and enable us to infer possible limits on gravitationally bound companions.
     As the differential chromatic refraction will likely be the determining systematic limiting the astrometric error budget, the data has to be carefully calibrated in each filter.

        \item [b)]{\bf Extremely short scale variability in BDs}\\ 
        While atmospheric variability due to rotation can be measured over a long time span, further short-term transient effects might be introduced through lightning and Aurora activity. Several works by \citet{Helling_2013}, \citet{Bailey_2014} and \citet{Hodos_n_2016} amongst others, discuss the possibility and detection of lightning effects in BD atmospheres. While the possible detection of lightning has been discussed theoretically, Aurora activity in BDs has already been detected by  \cite{Hallinan_2015}. The nature of these transient effects might be detected as very short brightening effects, usually only one exposure. The difficult task will be to identify these effects as anomaly in the light curves.

        \end{enumerate}
    
    \end{itemize}

    \subsubsection{Preparations for Brown Dwarf Science}
    
\begin{itemize}
\item{\bf Follow-up observations/archival data}\\
Follow-up observations will be necessary to identify objects that might have been misidentified as BDs as well as for spectral typing the objects of interests. These follow-up observations are essential in the first year as Rubin LSST parallaxes are not available, which will lead to a high rate of misidentified BDs. Follow-up observations to better characterize these objects are most efficient in the near-infrared wavelength range, because in this range, BDs are brightest, and their rich molecular features are best identified via near-infrared spectroscopy. For spectral typing one high SNR spectrum covering the near-IR wavelength range is necessary. As we go to cooler temperatures, the atmospheric chemistry of these objects changes, L-dwarfs lose the vanadium and titanium oxide features, which are seen in M-dwarfs, but they exhibit hydroxides. Even cooler T-dwarfs show a methane rich spectrum, while the ultra-cool Y-dwarfs exhibit ammonia absorption features. Therefore, spectra in the NIR are essential to classify these objects. Besides estimating the spectral types from spectroscopy, we can estimate the rotation period independently through the measurement of $v\sin i$ from spectral lines and subsequently, this can be compared with the rotational modulated photometric observations. Ideal facilities are the ARCoIRIS spectrograph mounted at the 4\,m SOAR telescope and the Fire spectrograph at the 6.5\,m Baade telescope. Brighter BD candidates can be cross-matched with existing BD catalogs and near-IR catalogs as well as with Gaia DR3.

As the parallaxes become available, we will be able to identify BDs better through distance measurements and luminosity estimates. The parallaxes, alongside proper motions, further help to detect possible companions.  
Near-infrared photometry can provide further information by increasing the cadence and supplying additional color information. From these multi-color observations we will be able to extract differences between amplitudes and phases at different wavelengths, which will help us to understand the possible mechanisms causing this variability. 

Brown dwarf candidates will be identified from Rubin LSST Data Release photometric catalogs, accessed via Data Access Centers. Cross-matching Rubin LSST data against the Gaia, 2MASS and WISE catalogs will provide the colors and proper motion information necessary to accurately select candidates.

\item{\bf{Facilities/Software Requirements}}\\
Existing procedures as outlined in e.g.\ \cite{Kelly2014}, \cite{celerite}, \cite{Feigelson2018}, \cite{Vos2020}, \cite{Apai2021} will be applied to Rubin LSST photometric and astrometric timeseries via an interactive Jupyter notebook environments and plotting tools. Multi-core or GPU computing facilities will be necessary to process the large data sets in a reasonable amount of time using Monte-Carlo Markov Chains \citep{emcee} or a Nested Sampling Monte Carlo \citep{pymultinest}. 

\end{itemize}

\subsection{Young Erruptive Variables}
\textsl{Authors: Teresa Giannini, Rosaria (Sara Bonito), Simone Antoniucci}

\bigskip

\noindent We have described EXor/FUor objects extensively in \autoref{sec:exorfuor} and so refer the reader to this Section for a basic overview of the objects. Here we focus on non time-critical EXor and FUor science.
    \begin{itemize}
    \item Low hanging fruits
    \begin{enumerate}
\item [a)]{\bf Define the statistical impact of eruptive vs. non-eruptive mass accretion in pre-main sequence stars (PMS)}\\ 
It is presently unknown whether EXors and FUors are peculiar objects or rather if they represent a short and recurring phase that all pre-main sequence stars (PMS) experience during their evolution. To answer this question, we need to compute the percentage of eruptive vs. non-eruptive PMS in different star forming regions. In particular, at least two outburst episodes should be seen to classify a source as an EXor-type variable. Present observations indicate that only about 2\% of PMS are eruptive variables, mostly identified through  serendipitous observations during campaigns carried out to pursue different scientific aims. We expect, therefore, that the percentage of eruptive vs. non-eruptive PMS is largely underestimated. A statistical study based on Spitzer/WISE data was already performed by our group \citep{Antoniucci_2014}. But now, in preparation for Rubin LSST observations, we intend to exploit the present public surveys to explore the topic in the optical domain, where bursts are far more intense.
With Rubin LSST, we expect to reach a statistically significant sample of eruptive protostars within 3 years.

\item [b)]{\bf Discriminate between intrinsic (EXor) and extrinsic (UXor) pre-main sequence variables}\\ 
This can be achieved through understanding relationships between accretion and extinction. The classification of a source as an EXor is often uncertain. Indeed, the observed properties (burst amplitude, cadence, optical/IR colors) are often attributable to accretion 
as well as to eclipse events related to orbiting bodies, the evaporation/condensation of circumstellar dust or ejected material by powerful outflows that 
move along the line of sight \citep[UXor variables,][]{vp1988}. Also, many young variables undergo both mechanisms, whose mutual prevalence depends on the level of activity. Hence, it is becoming clear that it is crucial to infer the variation of the mass accretion rate, the time dependence of visual extinction A$_V$ and more importantly, the possible existence of (inter-)relationships between accretion and extinction events.
A secure discrimination between EXor and UXor can be achieved through monitoring the flux over $1-3$\,yr. 
This is needed to separate light curves showing long periods of quiescence with superposed accretion bursts (EXors) from light curves that show quasi-periodic dimmings from a constant (high) brightness level (UXors). 

\item [c)]{\bf Obtain clues on the mechanism triggering the outbursts through lightcurve comparison}\\ 
Through the comparison of light curves in a statistically significant sample of objects, we aim to identify observational clues about the mechanism(s) triggering outbursts. Rubin LSST observations would provide an invaluable test-bed for existing models. For example, periodicities in the light curves (see \autoref{fig:audard_models}), burst durations, or asymmetries in the rising/declining phases, would represent observational constraints that we could use to discriminate thermal or gravitational instabilities inside the disk (slow rising, non-periodic light variations), from perturbations induced by external bodies (fast rising and periodic light variations). To reach this goal, we need extensive monitoring of these objects, and therefore we hope to obtain significant results within 10 years.
    \end{enumerate}
    
    \begin{figure}[ht]
    \centering
    \includegraphics[width=0.65\textwidth]{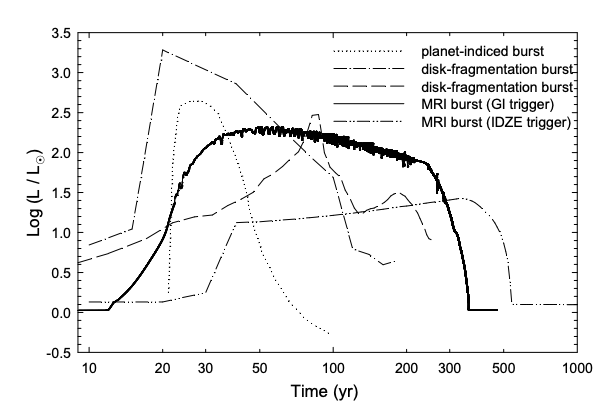}
    \caption{Time evolution of individual luminosity outbursts in
different burst-triggering models. The zero-time is chosen arbitrarily to highlight distinct models from \citet{Audard_2014}}
    \label{fig:audard_models}
\end{figure}    
    \item Pie in the sky
    \begin{enumerate}
\item [a)]{\bf Understand whether and how eruptive accretion can solve the ``luminosity problem'' } \\ The identification of a statistically significant sample of EXors (and possibly FUors) will allow us to understand the role of eruptive mass accretion in a more general context of star formation studies. In a classical star-formation scenario for low-mass objects, about 90\% of the final mass is accreted onto the star in about $10^5$\,yr, with typical mass accretion rates of $10^{-7}-10^{-5}$\,M$_\odot$/yr. Then, the accretion progressively fades to rates of 10$^{-9}-10^{-10}$ M$_\odot$/yr.
    \begin{figure}[ht]
    \centering
    \includegraphics[width=0.65\textwidth]{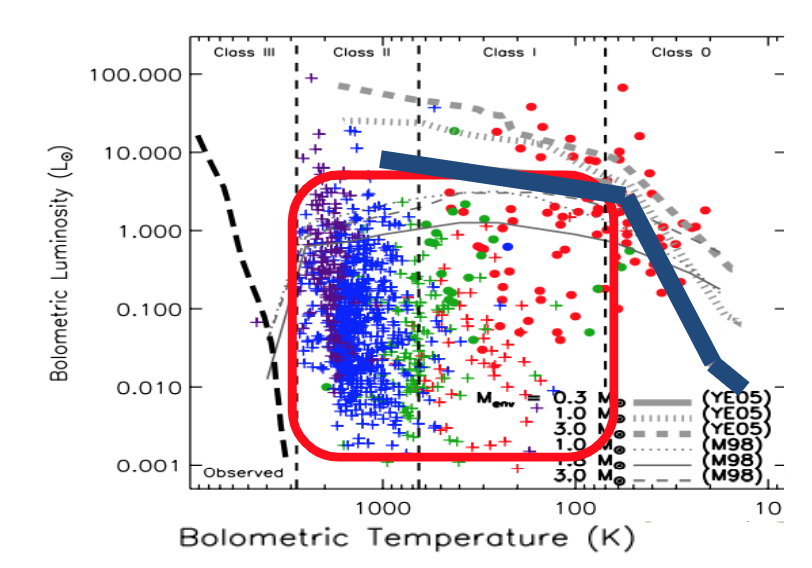}
   \caption{Bolometric luminosity (Lbol) is plotted vs. the bolometric temperature (Tbol). The bulk of the observational points (coloured dots) are located {\it under} the model predictions (black lines), indicating the ``luminosity problem". Adapted from \citep{Evans_2009}. }
    \label{fig:luminosity_problem}
\end{figure}    
In this quasi-stationary scenario, however, protostellar luminosities should be largely higher than observed the so-called ``luminosity problem'' (\autoref{fig:luminosity_problem}). Variable (and possibly eruptive) accretion has been proposed as a way of reconciling the observed star formation time and the mean protostellar luminosity \citep{Dunham_2015}. \citet{Offner_2011} also found that episodic accretion can contribute a significant fraction of the stellar mass. Rubin LSST will provide the opportunity to systematically cover and monitor most of the known star forming regions. These are the ideal conditions to determine the {\it true} percentage of eruptive variables, at least in regions where the extinction is not prohibitive. To solve the luminosity problem, however, we would need to demonstrate not only that eruptive PMS variables are much more common than presently expected, but also that burst frequency and amplitude produce an increase in the accretion luminosity high enough to compensate for the low values observed during the much longer fainter stages. The complete survey lifetime is needed to reach such an ambitious goal.
      \end{enumerate}
    \end{itemize}

\subsubsection{Preparations for EXor/FUor Science}
\begin{itemize}
\item {\bf{Follow-up observations/archival data}}
Optical/near-infrared spectroscopic follow-up is needed in order to confirm the indications retrieved from the light curves, since the shape of the continuum differs for EXors and UXors and the emission/absorption lines present in the spectrum also differ. Systematic monitoring of known objects is presently on-going at the Large Binocular Telescope \citep{Antoniucci_2014a,Giannini_2016,Giannini_2016a,Giannini_2018}, with the aim, among others, to construct template spectra of EXor and UXor sources,  which will be a reference for the spectroscopic follow-up of Rubin LSST observations. We intend to use our guarenteed telescope observations using SoXS (Son Of X-Shooter) on the New Technology Telescope for the same purpose.

Due to the nature of emissions during accretion events, it is highly desirable to cross-match the Rubin LSST Data Release products against catalogs at other wavelengths, especially UV and IR products of the Vista surveys, 2MASS, Gaia, Gaia-ESO survey (GES), WISE, Spitzer, Chandra, Herschel and eROSITA (when the data are made public). A database search interface that enables the user to select candidates based on these criteria will be essential to target selection, including Application Programmable Interfaces (APIs) that allow other data services to interact with Rubin LSST data products.

\item {\bf{Facilities/software requirements}}\\ 
Further work is needed to develop a suitable algorithm capable of identifying these targets and cross-matching them against the Gaia catalog in particular will be essential. Once identified, however, a dedicated software package will be used to analyse the photometry. If one is not available in the Science Platform, our group will develop a suitable tool for this kind of analysis. 
\end{itemize}

\subsection{Compact Binaries: Cataclysmic Variables}
\textsl{Authors: Paula Szkody and David Buckley}

\bigskip

\begin{figure}[ht]
    \centering
    \includegraphics[width=0.65\textwidth]{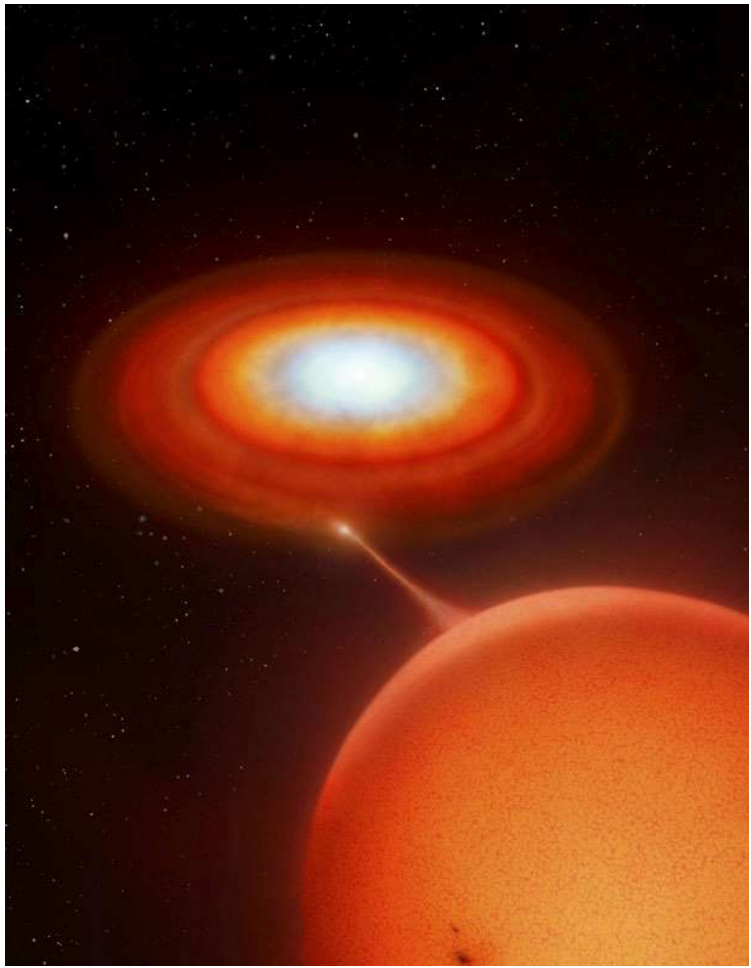}
   \caption{Artist's impression of a Cataclysmic Variable consisting of white dwarf accreting from a late-type (typically K$-$M) companion. Most of the optical luminosity and transient outbursts typically come from the surrounding accretion disk, which is continuously fed by an accretion stream from the Roche-lobe filling companion star.}
    \label{fig:CV}
\end{figure} 

\newpage
\noindent The variability of cataclysmic variables (CVs) involves timescales ranging from minutes for AM CVns and other ultra-compact binaries to 30 years or more for low mass accretion rate dwarf novae, and amplitudes from one magnitude to 15 magnitudes. While the time-critical studies will address some science goals related to immediate discoveries of novae and other outbursting systems (see \autoref{sec:CVs}, which describes the time-critical science cases for CVs), there are many aspects that need the 10-yr coverage afforded by Rubin LSST. Long term studies of CVs from the Rubin LSST archive, starting from year one and continuing to the full ten years of the survey, will involve the following science goals:

    \begin{itemize}
    \item Low hanging fruits:
    \begin{enumerate}
    \item [a)]{\bf Discover and classify new dwarf novae}\\ 
    This will be achieved using light and color curves, amplitudes and outburst recurrence timescales derived from the survey data. Their spatial distribution and distances will determine their number density in the galaxy as a function of galactic latitude. The newly identified distribution of objects will be used to compare with population models. This will require follow-up spectroscopic observations on a large telescope (4 to 8\,m, depending on brightness) to confirm their classifications. The full 10 years of the survey will provide the best number density as the majority of objects are expected to have long (decades) timescales between outbursts. Better population models are needed to provide a comparison of the resulting number densities with expectations from evolution.
    \item [b)]{\bf Monitor the known nova-like variables}\\
    Since nova-likes are known to sometimes decrease their mass transfer and at times to totally stop mass transfer, by observing the length of time that known nova-like variables spend in low versus high accretion states, we can begin to understand the total mass accreted and the angular momentum losses in these systems. This will require the long timebase of 10 years.
    \item [c)]{\bf Find candidate magnetic white dwarf systems (polars and intermediate polars)}\\ 
    Using the shape of the light and color curves and detection of relatively long-lived (weeks, months, years) bright and faint states, we are able identify new examples, which will lead to a better understanding of the ultimate evolution of high magnetic field systems after common envelope evolution. Confirmation of the magnetic nature will require circular polarimetry and/or medium resolution spectroscopy. Faint, low accretion rate systems are also expected to be discovered, which are important in understanding their ultimate evolutionary fate. While results will be found on a yearly basis, the full 10 years is needed to determine final numbers of magnetic vs. non-magnetic systems and how they are distributed in the galaxy. Such data will also help characterize the duty cycles and morphology of long-term light curves in terms of high, low and intermediate accretion states. 
    \item [d)]{\bf Find candidate eclipsing systems}\\ 
    Eclipsing systems enable inclinations (from light curve modeling) and the stellar masses (from radial velocity studies) to be determined. This will require follow-up observations to provide high cadence light curves and we will further need time-resolved spectra to obtain radial velocity curves.
    \end{enumerate}
    \item Pie in the sky:
    \begin{enumerate}
    \item [a)]{\bf Determination of the orbital period distribution of candidate dwarf novae}\\ 
    By obtaining follow-up photometric and spectroscopic studies we will be able to compare our results with population models. To do this we need to determine the mass of the secondary component for the shortest period systems to identify those after the ``period bounce" (when the secondary has become degenerate and the period of the system increases rather than decreases) as predicted by models. This will require time-resolved spectra from large (at least 10m) telescopes.
    \item [b)]{\bf Determine the spin and orbital periods of candidate intermediate polars}\\ 
    Despite the non-uniform temporal distribution of the survey data and the typical few day cadence, in some cases it may still be possible to detect short periods (minutes) in power spectra associated with the WD spin in intermediate polars (IPs).
    Time-resolved (5--15\,min) spectroscopic follow-up data for objects showing light curves resembling IPs will enable the determination of their orbital periods. Using high cadence (sec-min) photometric observations from follow-up data, we can determine the spin period which will confirm an IP classification. Obtaining the number of systems of this type will help us understand the magnetic binary evolution.
    \item [c)]{\bf Accomplish follow-up polarization studies of candidate polars}\\ 
    This will enable us to estimate the magnetic fields of the white dwarf components in order to confirm their classification and further understand the evolution of white dwarfs in binaries (compared to isolated single white dwarfs) that have the strongest magnetic fields.
    \end{enumerate}
    
    \end{itemize}

\subsubsection{Preparations for Compact Binary (Cataclysmic Variables) Science}
\begin{itemize}
\item {\bf{Follow-up observations/archival data}}\\
Due to the faint magnitudes of Rubin LSST, large (at least 8 to 10\,m) telescopes will be required for many of the follow-up observations. For the purpose of classification, spectroscopic follow-up is needed. To confirm the polar candidates, we require circular polarimetry and/or medium resolution spectroscopy. For intermediate polar candidates, high cadence light curves and time-resolved spectra for radial velocities will be needed. 

\item {\bf{Facilities/software requirements}}\\ 
Machine learning techniques will need to be developed to recognize the various types of dwarf novae and nova-likes that need follow-up spectra. To obtain the follow-up detailed classifications, large telescope facilities equipped with medium resolution spectrographs as well as spectropolarimeters will be needed, as well as continuous blocks of time in order to obtain the orbital periods.
\end{itemize}

\subsection{Compact Binaries: Neutron Star Binaries\label{sec:nsbinaries_ntc} }
\textsl{Authors: Elena Mason and Domitilla de Martino}

\bigskip

\noindent The study of transitional milli-second pulsars (tMSPs), milli-second pulsars (MSPs) and low mass x-ray binaries (LMXBs), in other words, accreting neutron stars is not limited to time-critical science, which requires the Rubin LSST alerts. Instead it will greatly benefit from the all-sky survey, long-term monitoring characteristics of the project. For a detailed introduction to Neutron Star Binaries see \autoref{sec:nsbinaries_tc}. Here we detail the science cases for non-time critical neutron star binary science.

\begin{itemize}
    \item Low hanging fruits
    
    \begin{enumerate} 
        \item [a)]{\bf A census of tMSPs, MSPs and LMXBs will be obtained}\\ 
        The multi-band deep Rubin LSST observations will be combined/cross-matched with data from other next generation X-ray surveys, such as THESEUS or eROSITA, allowing for the discovery of new accreting neutron stars (NS). The resulting census will also allow population analysis of complete sample once the Rubin LSST data are cross matched with GAIA parallaxes and proper motions.  
        
        \item [b)]{\bf Long term monitoring of all known tMSPs, MSPs and LMXBs}\\ 
        The Rubin LSST light curves of known objects will enable us to assess the long term variability of each system. 
        Known systems (MSPs, tMSPs and LMXBs) will be part of the monitoring program starting with the commissioning observations. New systems will enter the monitoring program once early Rubin LSST data sets are combined and cross-correlated with candidate MSPs, tMSPs and LMXBs from the coming radio and X-ray observations from the SKA and X-ray surveys. From the beginning to the end of the Rubin LSST survey mission, the database of accreting NSs will gradually increase in sample size, eventually culminating in a 10-yr long multi-color light curve for each system.

        \item [c)]{\bf {Mapping of the variability patterns displayed by interacting compact binaries}}\\
        Understanding the variability patterns will be invaluable to tackle the history of the mass accretion rate of each object and to frame the correct evolutionary scenario for each subclass of compact binaries. This is particularly compelling for the tMSPs \citep{demartino2010, papitto2013}, which are MSPs that have recently been discovered to change from a radio pulsar propeller state to that of an accreting LMXB, or vice-versa, also displaying intermediate states where a subluminous accretion disks maybe associated with the launching of jets. 
        By properly framing the MSP population in the context of LMXB, we will be able to answer the questions: (1) do all MSPs become transitional and (2) whether the tMSP phase is an evolutionary phase that precedes the total exhaustion of the donor star. 

        \item [d)]{\bf {Answer important questions about NS state changes}}\\
        While the change of state of an accreting NS might trigger follow-up observations (see \autoref{sec:nsbinaries_tc}, which describes NS binaries from a time-critical perspective), the long-term (multi-color) light curve of each system will allow the determination of the time scale of changes as well as the duration of each state and the driving factor for the change of state. Does the change of state depend solely on the change of the mass transfer rate? Does the mass transfer rate depend on the nature of the secondary star or other system parameters? Are tMSPs a specific evolutionary phase between the LMXBs and the MSPs or can any system turn into a tMSP? These questions are now becoming more important, given that the number of known MSPs has more than tripled since 2010 (thanks to both radio surveys and  blind searches in the {\it Fermi-LAT} sky scans that discovered a number of pulsar-like $\gamma$-ray sources that were eventually confirmed to be MSPs). The number of such systems is expected to further increase even more dramatically once SKA is operational (enabling much improved timing precision of the binary pulse). Furthermore, next generation X-ray missions will permit the discovery and high S/N modeling of X-ray pulsed signals. 
\end{enumerate}
\end{itemize}

\subsubsection{Preparations for Compact Binary (Neutron Star Binary) Science}
\begin{itemize}
\item {\bf{Follow-up observations/archival data}}\\
For the new MSPs discovered via radio surveys (e.g. ngVLA, ASKAP, MeerKAT, SKA), finding the X-ray and/or optical counterpart is a crucial first step to properly characterizing the population of MSP binaries in their various subclasses (e.g. black-widows with degenerate donors, and redbacks with non-degenerate donors). Rubin LSST will provde the optical light curve data to enable the classification of new MSPs.
\end{itemize}

\subsection{Compact Binaries: Black Hole Binaries (BHBs)}
\textsl{Authors: Michael Johnson, Poshak Gandhi}

\bigskip

\noindent Of the millions of stellar-mass black holes formed through the collapse of massive stars over the lifetime of the Milky Way, only $\sim20$ have masses that have been dynamically confirmed through spectroscopic measurements (e.g. \citealt{Corral_Santana_2016}). Expanding this sample of black hole masses could aid in answering many questions central to modern astrophysics.
 
There is expected to be a large population of black hole binaries (BHBs) in quiescence with low X-ray luminosities from $\sim 10^{30} - 10^{33}$\,erg\,s$^{-1}$. Such systems can be identified in the optical as variables that show unique, double-humped ellipsoidal variations of typical peak-to-peak amplitude $\sim0.2$\,mag due to the tidal deformation of the secondary star, which can be a giant or main sequence star. In some cases, analysis of the light curve alone can point to a high mass ratio between the components, suggesting a black hole primary; in other cases, the accretion disk will make a large contribution to the optical light which results in intrinsic, random, and fast variations in the light curve. The disk contribution to optical light can vary over time, and several years of data are required to properly subtract the accretion disk contribution in order to properly fit ellipsoidal variations \citep{Cantrell_2010}. As a result, data products generated by Rubin LSST for studying the ellipsoidal variability of BHBs will begin to become useful around the 3 year mark, allowing for some probe of the variability, however the results will be much improved after the full 10-year survey. The characteristic ellipsoidal modulation of X-ray binaries can also be observed in other classes of binary systems. Therefore, X-ray follow-up observations are required in order to classify X-ray binaries.

BHBs typically spend most of their time in quiescence, where the optical emission is dominated by the companion star. However, many systems undergo periods of outburst, thought to be triggered by instabilities in the accretion disc \citep{Lasota_2001}. These outbursts are characterised by very rapid and dramatic increases in optical luminosity and during outburst the disc becomes the dominant component in the optical luminosity and the characteristic ellipsoidal modulation can no longer be observed. 

Rubin LSST is expected to discover many unknown BHBs, therefore we expect archival pre-Rubin LSST data (i.e. prior to discovery) will be superfluous.

The brighter sources will be spectroscopically observable with the current generation of 4-m to 10-m telescopes to dynamically confirm new black holes; spectroscopy of all candidates should be possible with the forthcoming generation of large telescopes. Thus, Rubin LSST will trigger a rich variety of observational investigations of the accretion/outflow process through studies of this large, dark population.\\

\begin{figure}[ht!]
\begin{center}
\includegraphics[width=0.70\columnwidth]{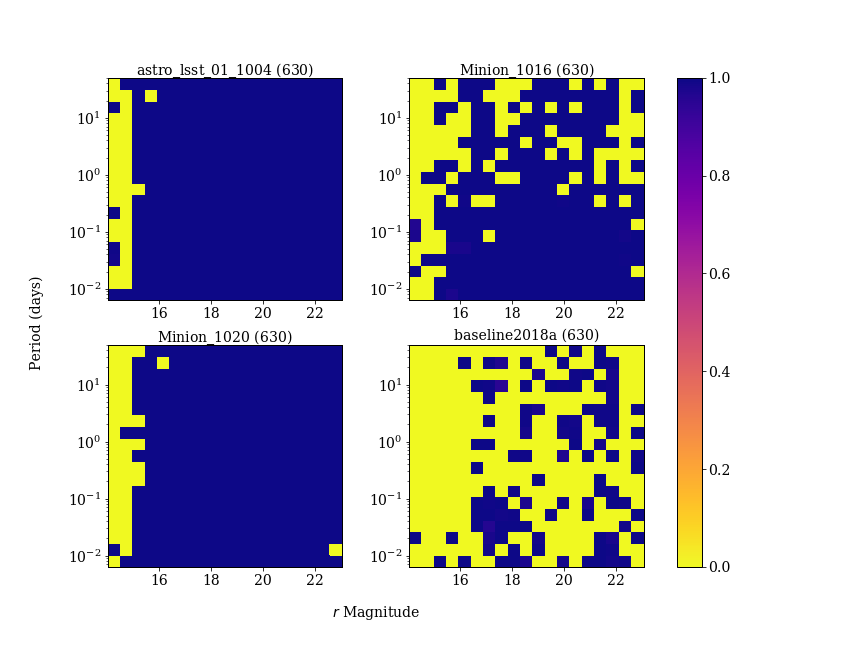}
\caption{{Color maps displaying the period determination of (low-mass X-ray binaries) LMXBs possible with
observing strategies astro\_sim\_01\_1004, Minion\_1016, Minion\_1020
and baseline2018a.~ All observations used were of Rubin LSST field 630 which
is in the Galactic Plane. The y-axis denotes the orbital period in days, the x-axis denotes the $r$ mag before adding contributions from ellipsoidal modulation,
flaring and noise. The color denotes the significance of the period detected. If the measured period differed from the actual period by more than 5\%, then the significance was set to zero.
{\label{935192}}%
}}
\end{center}
\end{figure}

    \begin{itemize}
    \item [a)] Low hanging fruits
        \begin{enumerate} 
        \item {\bf Period Determination and Quiescent Magnitude Observations}\\ 
        Many BHB candidates have currently only been observed during outburst as they are too faint to be observed with sufficient sensitivity in quiescence. Observations during quiescence have only been possible for a few of these systems to date. Rubin LSST should be able to observe a large fraction of the population of these systems in quiescence, for the first time, during the first year of its operations. As the companion star cannot always be observed during outburst, this will allow for counterpart identification for many of these systems.  
    
        \item [b)] {\bf {Period determination}}\\ 
        We will obtain the periods of a meaningful fraction of the black hole binary population. The majority of Galactic BHBs will likely be too faint to be observed by Rubin LSST. However, the fraction that Rubin LSST should be able to observe through ellipsoidal variability (approximately 1/3; \citealt{Johnson_2018}) should prove to be statistically significant with regards to population studies of BHBs. The current computational techniques are capable of determining the period through ellipsoidal variability. \autoref{935192} shows results for the period determination of BHB with Rubin LSST over a broad range of parameter space when using different observing strategies from \citet{Johnson_2018}.
        
        \end{enumerate}
    
    \item Pie in the sky
        \begin{enumerate} 
        \item [a)] {\bf Binary Component Mass Measurements}\\ Through knowledge of the binary component masses, we are able to learn about the history and evolution of black hole binaries. \citet{cherepashchuk1997} found correlations between the parameters of BHB optical light curves and the physical parameters of the system, such as binary component mass or orbital inclination. Therefore, Rubin LSST could be used to place constrains on these physical parameters in BHBs that are observed in quiescence. Potentially, this information could then be used to answer questions such as: ``which stars produce black holes and which neutron stars?''; ``whether there is a true gap in mass between these two types of compact object?''; and ``whether supernova explosions result in large black hole kicks?". Current constraints on black hole samples detected by Gaia are limited by small number statistics and non-negligible systematic uncertainties on the masses \citep{gandhi19, atri19}. However, these techniques will have to be investigated further, while using simulated Rubin LSST data in order to determine their applicability. 
        \end{enumerate}
    
    \end{itemize}

\subsubsection{Preparations for Compact Binary (Black Hole Binary) Science}
\begin{itemize}
\item {\bf{Follow-up observations/archival data}}\\
For classification purposes, X-ray follow-up is required. Follow-up for the majority of the discovered systems should be possible with the next generation of X-ray telescopes, e.g. at the deep sensitivity limits expected with Athena \citep{Barret_2018}, though the nearby population should also be detectable with eROSITA \citep{erosita}.

\item {\bf{Facilities/software requirements}}\\ 
Due to the expected total population of LMXBs (low-mass X-ray binaries) in the Milky Way ($\sim$\,1300; e.g. \citealt{Corral_Santana_2016}), period determination of a fraction of this population would require enough storage to host all images for each of the objects (or at least the regions in which they are contained) and computing resources to generate $\sim$~\(10^{5}\) power spectra per system.
\end{itemize}

\subsection{Luminous Blue Variables (LBVs)}
\textsl{Authors: Jorick Vink, Federica B. Bianco, Andrea Reguitti}

\bigskip

\noindent Luminous Blue Variables (LBVs) are very bright sources, and among the most massive stars detectable in galaxies. Famous examples in our Galaxy are AG Car, HR Car, and $\eta$ Carinae.
In the nearby Universe, we should also mention the famous S Doradus \citep{Leitherer85} in the large Magellanic cloud (LMC), the prototype of the homonym variability typical of this class of stars, AE And and AF And in M31, Var C and Romano's Star (GR 290) in M33 \citep{Richardson18}, along with several well-studied LBVs in NGC 2403 \citep{Tammann1968}. Collectively, they are known as the Hubble-Sandage variables \citep{Hubble1953}.
During S Dor-like outbursts, LBVs experience erratic brightness variability over timescales of several months to a few years or decades, with $\Delta$M of a couple of magnitudes, typically $\Delta$M$ = 1 - 2$ mag. Traditionally, it had been assumed there was no significant change in the bolometric luminosity \citep{Humphreys_1994}, but this has more recently been challenged \citep{Clark_2009,Groh_2009}. During outburst, S Doradus LBVs move to the right of the Hertzprung-Russell diagram (HRD), becoming redder and cooler, following which they come back in quiescence to the left side of the HRD. The reason for the HRD excursions is as yet unknown although envelope inflation due to the proximity to the Eddington limit is the main contender \citep{Gr_fener_2012,Vink12,Gras21}. \\ 

The full range of variability of a classical LBVs can be comfortably monitored with Rubin LSST observations up to about 30\,Mpc, but with periodic stacks, we can largely exceed this distance limit. An $\eta$ Carinae Great Eruption-like event \citep{smith2011revised}, instead, could be detected at tens or even hundreds of Mpc. With Rubin LSST, we plan to observe known LBVs and expect to find new LBV candidates in outburst.
\newpage
    \begin{itemize}
    \item Low hanging fruits
    \begin{enumerate} 
    \item [a)]{\bf Observations of ordinary luminous blue variable (LBV) outbursts (S Dor-like)}\\ 
    Known S Dor variables will automatically be observed and this will greatly enhance the baseline and quality of reference objects. In addition, some LBVs are known to undergo multiple giant outbursts. An example of such a restless LBV is SN 2000ch the LBV in NGC 3432 \citep{Wagner2004,Pastorello_2010}. Some LBVs may be identified in the first year of Rubin LSST observations, but given that the average time-scale of S Dor variations is about a decade, the sample will only start to become complete after 10 years. 
    \item [b)]{\bf Direct association of LBVs identified in their eruptive states with their quiescent counterparts}\\ 
    The depth and resolution of Rubin LSST will allow identification of stars in the local volume. With a single-image magnitude limit near magnitude 24 and a corresponding coadded depth near magnitude 27, the quiescent counterpart of outbursting LBVs can be detected in the 10-year Rubin LSST stacks. The unveiling of detailed progenitor information with deep, pre-eruption characterization of sources at the positions of LBVs will be a major contribution of LSST to LBV science.
    \item [c)] {\bf{Identifying SN precursors}}\\ 
    Precursor eruptions preceding the  explosions of interacting SNe, often associated with LBV-like progenitors \citep{GalYam2007} have been conclusively observed (SN 2009ip being the most important one (\citealt{Margutti2013,Fraser2013,Pastorello2013}), but see also e.g. \citealt{Reguitti2019,jacobson2022final}). Rubin LSST can provide a sufficiently large sample of such events to distinguish which of these optical transients are direct SN precursors, and which are not.
    \end{enumerate} 
    
    \item Pie in the sky
    \begin{enumerate} 
    \item [a)]{\bf Determination of the Evolutionary State of LBVs}\\ 
    Using the ``reference frame'' of known objects, new S Dor variables will be discovered with Rubin LSST in a variety of environments. Ultimately, it is the comparison of LBVs with respect to ordinary blue supergiants that will enable the determination of the duration of the LBV phase and thus the 
    evolutionary state of LBVs \citep{Kalari_2018}. This is the main science goal.\\
    In addition to relying on observed reference objects, theoretical development of envelope inflation and atmosphere modelling is needed to predict a range of LBV colors as a function of metal content. \\
    The need for modeling is particularly clear when considering that known transients show a rather wide range of behaviour \citep{Smith_2011}, which makes the simple use of templates rather limited.
    \end{enumerate} 
    \end{itemize}

\subsubsection{Preparations for LBV Science}
\begin{itemize}
\item {\bf{Follow-up observations/archival data}}\\
For known LBVs, using literature and archive data, for example, data from AAVSO, PanStarrs and Atlas, we aim to obtain light curves with baselines of many decades. This multiple survey strategy may reveal different types of variability, which develop on different timescales. Well-studied LBVs will become reference objects. 

\item {\bf{Facilities/software requirements}}\\ 
Software to generate periodic stacks will be needed to detect LBVs farther than 30Mpc. Likely, this can be done using the Rubin open software pipeline in-between data releases, possibly leveraging infrastructure such as LINCC\footnote{\url{https://www.lsstcorporation.org/lincc/node/1}}. Software to identify the quiescent state of LBVs in LSST stacks (whether Annual Dara Releases or custom made stacks) by co-location will be required. Depending on the location of the star, this operation may be complicated by crowdedness and blending. 
\end{itemize}

\subsection{Light Echoes of eruptions and explosions}
\textsl{Authors: Xiaolong Li \& Federica B. Bianco}
\selectlanguage{english}

\bigskip
 
\noindent Light Echoes (LEs) are the reflection of stellar explosions on interstellar material.  As the light from a transient propagates into space, it may be reflected towards Earth and our telescopes, if it encounters a sheet of interstellar material, it will generate a LE.  The geometry of LEs is straightforward, a transient and the Earth are at two foci of a 3D ellipsoid; light from the transient reflected off dust that intersects the same ellipsoid surface reaches Earth at the same time.  As time goes by, the ellipsoidal surface expands.  Due to the extra travel time, a light echo reaches the observer at a later time than the light directly detected from its source.

LEs provide unique insight into the transients that originated them and the dust that produced them. However, they are very challenging to detect. LEs appear as faint, extended, time-varying features~\citep{rest2005light, rest2013light}.  The complexity of their shape is inferred by the complexity of the underlying structure of the reflecting medium, while the time-changing aspect is due to the traveling of light across the dust sheet. 
\begin{figure}[h!]
    \centering
    \includegraphics[width=0.7\columnwidth]{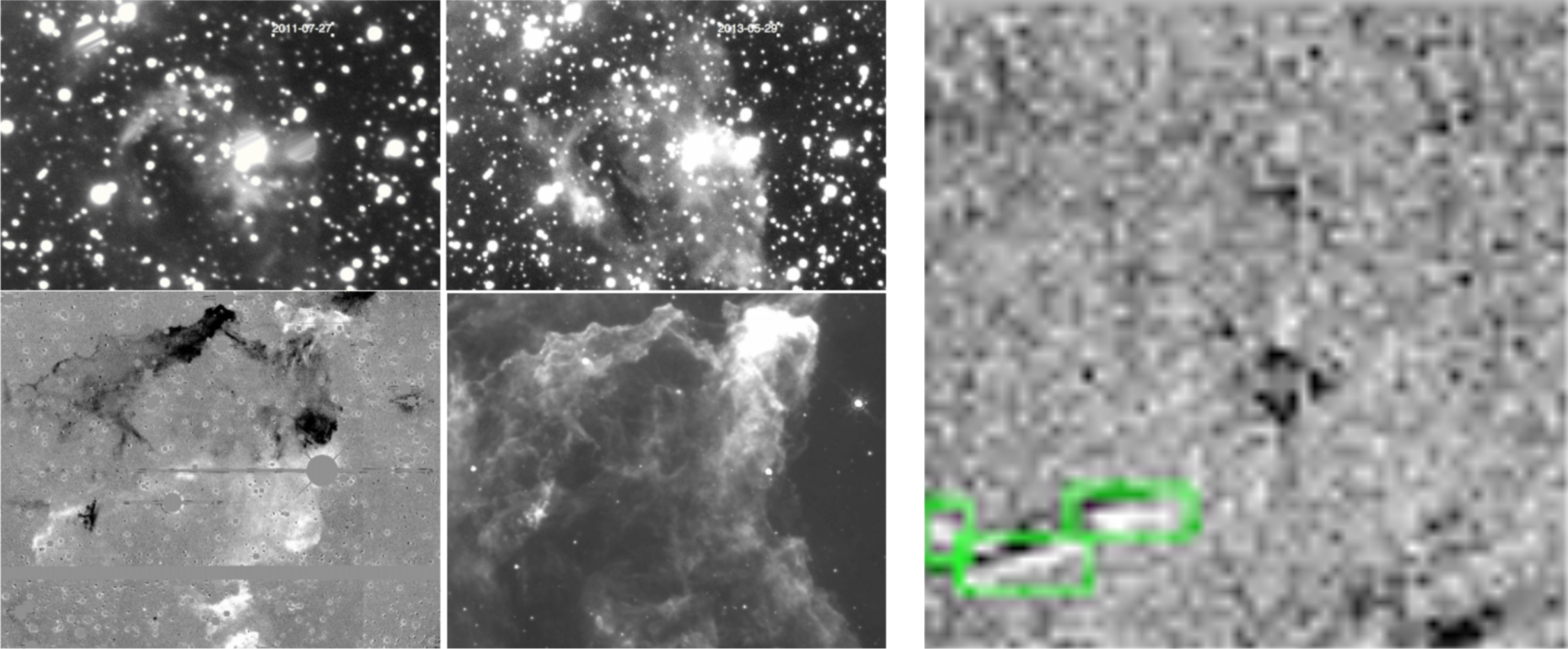}
    \caption{Left: The top two images show LEs of Eta Carinae’s Great Eruption lighting up the dust structure about 30 arcminutes away from the star, adapted from \citet{rest2012light}. The images, taken about 2 years apart, reveal the time-evolving structure of the LE phenomenon. On the bottom left, a LE of Eta Carinae in difference imaging. To its right, a Spitzer infrared image of the same dust structure illuminated by the LEs. The reflected light reveals an astounding level of detail in the dust structure. Right: a difference image of an ATLAS \citep{tonry2018atlas} field with a LE complex, labeled for ingestion by a Regional Convolutional neural network~\citep[RCNN,][]{faster2015towards}. This image shows the difficulty of the labeling exercise: the light echoes complex is composed of different echoes (different dust sheets) that are blending.  Several artifacts are present (saturated stars, star streaks, etc...) which are not labeled in this example. In reality, because light echoes diffuse until they blend into the noise, so that they form a continuum in detectability space, the RCNNs cannot rely on the assumption that the labels are complete (i.e. that they identify each instance of light echoes in the training set) and artifacts have to be labeled as negative examples as well (Li et al. in preparation).}
    \label{fig:les}
\end{figure}
LEs help characterize stellar explosions by offering the view of an explosion from different lines of sight, a unique opportunity in astrophysics~\citep{rest2011direct, finn2016comparison} and allow us to revisit events, even when the transient was originally not detected~\citep{rest2011direct}. The detection of LEs across a large portion of the sky, which Rubin LSST can enable, could lead to the detection of an unknown galactic supernova and enable an alternative way of tracing of the Milky Way dust structure, alternative to stellar extinction and direct detection surveys. Improved dust maps can improve inference on extragalactic sources by constraining extinction. A large scale census of LEs inform the evolutionary history of the Galaxy constraining the rate of massive star eruptions in the Galaxy and Local Volume.  

With its unique combination of deep, high-resolution imaging repeated at $\sim$day cadence across the whole southern sky, Rubin LSST is the ideal survey to detect LE. However, LEs are extremely hard to detect. As part of its federally funded operations, Rubin LSST will produce $10^6$ nightly alerts~\autoref{sec:lsst_prompt_dp}, each one announcing a changing or moving source. It will detect thousands of transients and variable “point sources” in each image, but the diffuse LEs will not reach 5-$\sigma$ significance over the size of the Rubin PSF and will be entirely missed.   

To date, LEs are still discovered chiefly by visual inspection, a method that obviously does not scale to the Rubin LSST data volume.  Even crowd-sourcing cannot help this science in the Rubin LSST era: simple scaling from the Galaxy Zoo \citep{fortson2012galaxy, rest2008scattered} project indicates that the entire population of the Earth would be insufficient to study the full dataset using the same method. Feature based approaches fail to separate the rare true positives from the many false positives, which include all sorts of artifacts in sky images. 

 \begin{itemize}
 \item Low hanging fruits:
 \begin{enumerate}
     \item [a)] {\bf Observations of known LEs, which will form a training set for furture detections}\\
     Several historical supernovae and stellar eruptions have known LEs \citep{rest2008scattered, rest2012light} and their detection is trivial with Rubin LSST. These light echoes can serve to train and tune models to detect lower signal-to-noise, unknown LEs. Their monitoring will trace interstellar dust filaments and provide detailed light curves of the explosions and eruptions. 
  \end{enumerate}

 \item Pie in the sky:
\begin{enumerate}
\item [a)]{\bf{Setting constraints on the Galaxy's explosion history}}\\ 
The detection of LEs across the southern hemisphere can be linked to stellar explosion models for the Galaxy. This will enable a more detailed understanding of the Galaxy explosion history.
\item [b)]{\bf The discovery of unknown supernovae}\\
The detection of "orphan" LE could lead to the discovery of an unknown Galactic supernova, which could be classified based on LE spectra collected from large telescopes. 
\end{enumerate}
\end{itemize}

\subsubsection{Preparations for Light Echo Science}
\begin{itemize}
\item {\bf{Follow-up observations/archival data}}\\
Once detected, LEs require repeated images to confirm their nature and to monitor their evolution and, if bright enough, spectroscopic follow-up with large telescopes (8+ meter class) will be applied to characterize the transient. The large aperture is required because the phenomena are faint (typically x10 fainter than the original transient). 

The current dataset of observed light echoes is extremely small (10s of examples, mostly from telescopes and surveys with vastly different properties than Rubin Observatory) and in particular the current sample is extremely small to train automated machine learning models for the detection of new LEs. These data should be augmented with simulations, either entirely ML based simulations (e.g. Generative Adversarial NN, \citealt{Goodfellow14}) or based on forward modeling of dust maps.

\item {\bf{Facilities/software requirements}}\\ 
A pipeline for the detection of LEs needs to be developed as a generated data product that reprocesses the Rubin LSST's difference and/or possibly the original images. This pipeline should run at a Data Access Centers (DACs) to avoid transferring large image datasets and would produce large databases (however, small compared to the image database) to be cross-referenced with explosion and eruption historical and contemporary datasets. 
Neural Networks (NN), including Deep Convolutional NN \citep{faster2015towards}, Generative Adversarial NN \citep{Goodfellow14} and Bayesian NN \citep{2020arXiv200601490C} to quantify uncertainties, and other methods in the NN family are a promising to address this low signal-to-noise regime computer-vision challenge and create the first pipeline for automated detection of LEs at scale.
\end{itemize}

\section{Extragalactic transients and variables} 
\textsl{Editor: Claudia M. Raiteri}

\subsection{Blazars}\label{blz2}
\textsl{Authors: Claudia M. Raiteri, Barbara Balmaverde, Maria Isabel Carnerero, Filippo D'Ammando, Chiara Righi}

\bigskip

\noindent While the detailed study of the jet emission variability in blazars is time-critical (see \autoref{blz1}) because it requires simultaneous multiwavelength observations, especially during active states, other studies on blazars are non time-critical. These include the statistical analysis of the properties of the whole blazar class or of blazar subclasses (BL Lacs and FSRQs). Of particular interest is the analysis of the blazar cosmological properties and their relationship with that of radio galaxies, which represent the unbeamed parent population of blazars according to unified models of AGN \citep{capetti15b}. Another important topic that is also connected to the understanding of the blazar parent population is the study of the environment of the blazar host galaxies. Blazars are commonly hosted by giant elliptical galaxies mostly located in dense environments, but the picture needs to be investigated further with a large sample of objects covering a wide range of redshifts.
With Rubin LSST data we plan to research the following science drivers:
    \begin{itemize}
    \item Low hanging fruits
    \begin{enumerate}
\item [a)] {\bf Perform a statistical study of long-term blazar variability}\\ 
Statistical studies will allow the duty cycles of the various blazar classes to be analysed. Further, we will search for peculiar behaviours that can shed light on the underlying variability mechanisms, such as periodicities, or strongly chromatic flares, i.e.\ flares characterized by noticeable colour changes. 
\item [b)]{\bf Study the environment of the blazar host galaxies}\\ 
Several radio-loud AGN are located in rich environments \citep{kotyla16}. The most powerful of them lie in galaxy clusters and those with the most massive hosts are found in the central cluster regions \citep{magliocchetti18}. If radio galaxies are the parent population of blazars, both classes should share the same clustering properties. However, this has recently been questioned by an analysis with a relatively small sample of objects \citep{sandrinelli19}. The matter needs to be further investigated with deep images of blazar and radio galaxy environments that Rubin LSST will provide.

The presence of a galaxy cluster around a blazar can be ascertained by looking for a “red sequence” of early-type galaxies in the colour versus magnitude diagram, where the colour and slope will depend on the blazar redshift. The higher the redshift we want to consider the deeper the observations must go.
We estimated \citep{raiteri2018} that magnitudes of 24, 25, and 26.5 must be reached in the $z$ band to extend the study to redshifts of 0.5, 1 and 2, respectively.
Therefore, as time goes on, Rubin LSST will allow us to explore the environment of our sources farther and farther away, starting from the local Universe in the first year, up to cosmological distances at the end of the project. 
At the same time, it will also sample fainter and fainter galaxies at low redshift, increasing the environment richness. 
    \end{enumerate}
    \item Pie in the sky
    \begin{enumerate}
\item [a)] {\bf Explore the cosmological properties of a much wider blazar population}\\
The identification of new blazars, together with the availability of their radio fluxes, will allow us to build luminosity functions (LFs) and to determine the cosmological evolution of the various blazar classes. \cite{wolter94} derived radio, optical and X-ray LFs for two small samples of radio-selected and X-ray-selected blazars. Through the comparison between LFs, they investigated the evolution of the blazar population(s). They found that the X-ray selected sources have a negative evolution, i.e.\ they were fewer or fainter in the past, while the radio-selected objects have a marginal positive evolution in the radio band. A number of studies can be found dealing with the radio LF of blazars \citep[see, e.g.][and reference therein]{capetti15b,capetti15a}. The 1.4 GHz radio LF shows a break at $\log L_{\rm radio} \sim 40.6$ [erg/s], which implies an abrupt decrease of the number of sources at low radio powers. The break is likely connected to the minimum power needed to launch a relativistic jet. For this kind of study, redshift is a necessary ingredient, so supporting spectroscopic observations will be needed to enlarge the blazar sample that can contribute to LF studies. 
 \end{enumerate}
 \end{itemize}

\subsubsection{Preparations for Blazar Science}
\begin{figure}[ht!]
\begin{center}
\includegraphics[width=0.7\columnwidth]{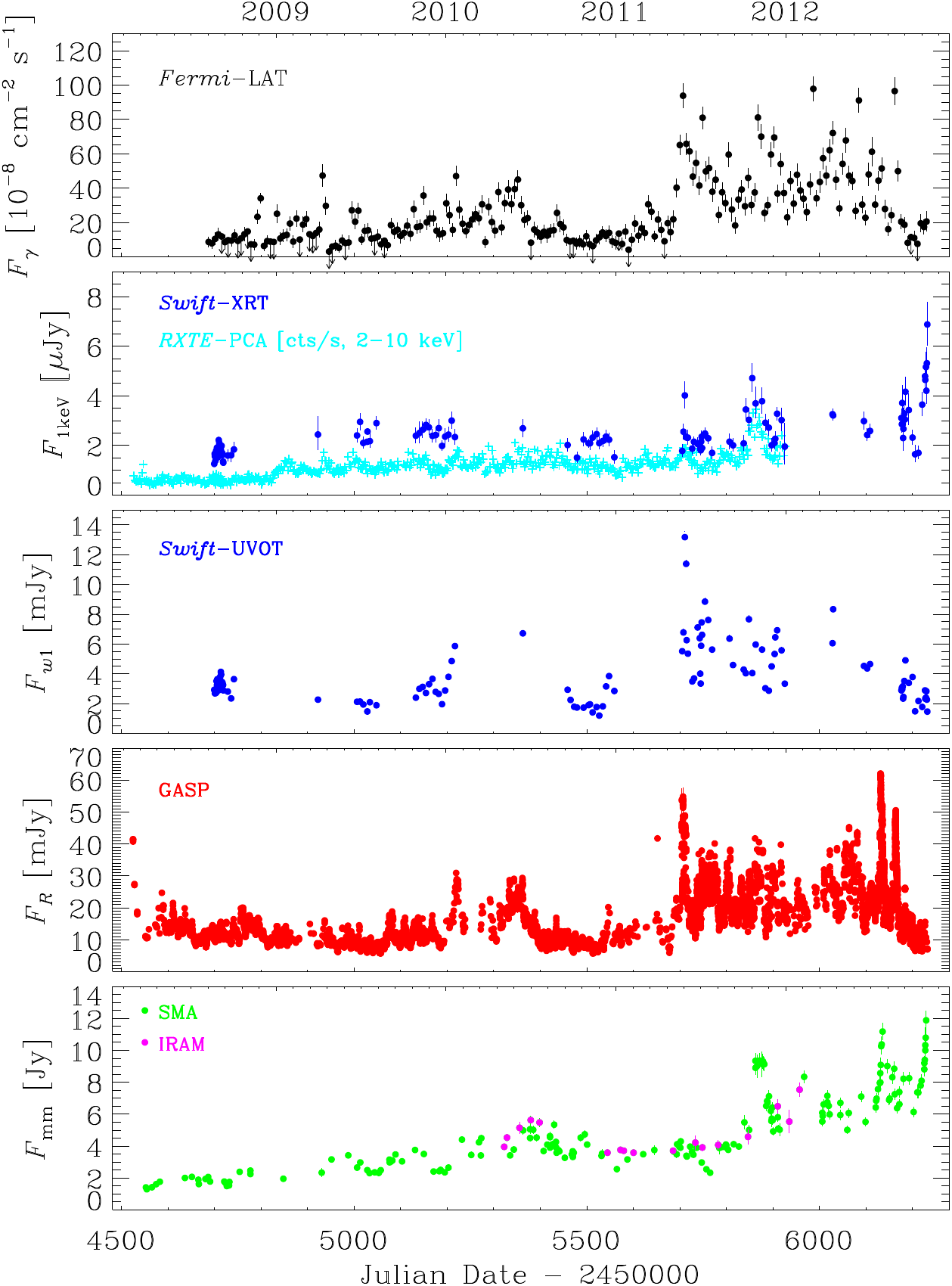}
\caption{Multiwavelength light curves of BL Lacertae. From top to bottom showing data in $\gamma$-rays (data from the {\it Fermi} satellite), X-rays (data from the {\it Swift} and RXTE satellites), in the ultraviolet (data from the {\it Swift} satellite), in the optical (data from the GASP Project of the Whole Earth Blazar Telescope, http://www.oato.inaf.it/blazars/webt/), and millimeter bands (bottom, data from the SMA and IRAM facilities). Adapted from \citet{raiteri2013}.}
\label{bllac}
\end{center}
\end{figure}

\begin{itemize}
\item {\bf{Follow-up observations/archival data}}\\
To identify new blazars we need multiwavelength follow-up, in particular in the radio band, because blazars are radio loud objects characterized by a flat radio spectrum. To further study the blazar population, it is important to know the associated redshift, and so spectroscopic follow-up is necessary.

\item {\bf{Facilities/software requirements}}\\ 
As in the time critical case (see \autoref{blz1}), methods to analyse time series will be required to identify characteristic variability time scales and possible (quasi)periodicities. Classical methods are the Structure Function \citep{simonetti1985}, the Discrete Correlation Function \citep{edelson1988,hufnagel1992} and the Lomb-Scargle Periodogram \citep{press1992,vanderplas2018}. 
The robustness of the results will increase as the Rubin LSST monitoring continues. These methods are expected to be applied to a wide variety of science cases both in the Transiting and Variable Star Science Collaboration (TVS SC) and in other SCs.

While diagnostic methods based on variability and colours to separate quasars from variable stars do exist \citep[e.g.][]{butler11}, we lack a similar method to identify blazars. Only the most violently variable blazars are clearly recognizable through their optical light curves without further spectroscopic or multiwavelength information. 

Radio spectral information is crucial for the detection of new objects, because a distinctive feature of blazars is to have flat radio spectra. Finding diagnostic methods for blazar detection is a critical issue that should be tackled ahead of Rubin LSST. Its solution would also allow Alert Brokers to characterize sources that are ideal for gamma-ray follow-up, since blazars are the most abundant population in the gamma-ray sky (see \autoref{blz1}). 

The Discrete Correlation Function can also check cross-correlations between flux variations in different bands and possible time delays. As mentioned in \autoref{blz1}, this gives information on the emission processes and on the location of the emitting zones inside the jet. Therefore, although Rubin LSST data alone will allow us to investigate several blazar science cases, the availability of multiwavelength time series is of great importance for the study of other blazar topics.
\autoref{bllac} shows a comparison of light curves of BL Lacertae in different bands \citep{raiteri2013}. It is clearly visible that the behaviour in $\gamma$-rays is well correlated to that in the optical band, though with some differences, while the trend in X-rays is more similar to that observed at millimeter wavelengths.
\end{itemize}

\subsection{Supernovae (SNe)}
\textsl{Authors: Maria Teresa Botticella, Laura Greggio, Enrico Cappellaro, Vincenzo Petrecca, Fabio Ragosta, Anais M\"oller}

\bigskip

\noindent The identification of the progenitors of both type Ia (Ia~SNe) and Core collapse SNe (CC~SNe) is of great importance in several astrophysical contexts, including constraining the evolutionary paths of close binary systems, the measurement of cosmological parameters, the description of the chemical evolution of galaxies and of the intergalactic medium, the evolution of galactic winds and the gravitational waves emission from merging binaries in galaxies.

In spite of the great efforts over the last decades on both theoretical and observational sides, the question about the nature of SN progenitors is still far from settled. 

The measurements of the SN rates in different stellar populations and their correlation with the properties of their parent galaxies provides an important tool to understanding the different types of SNe and their progenitors. 

Due to the short lifetime of progenitor stars, the rate of CC~SNe directly traces the current star formation rate (SFR) of the host galaxy. Therefore, the mass range for CC~SNe progenitors can be probed by comparing the birth rate of stars and the rate of CC~SNe occurring in the host galaxy, assuming the distribution of the masses with which stars are born, that is, the initial mass function.
\begin{figure}[ht]
\begin{center}
$
\begin{array}{c@{\hspace{.05in}}c@{\hspace{.05in}}c}
\includegraphics[width=7.5cm]{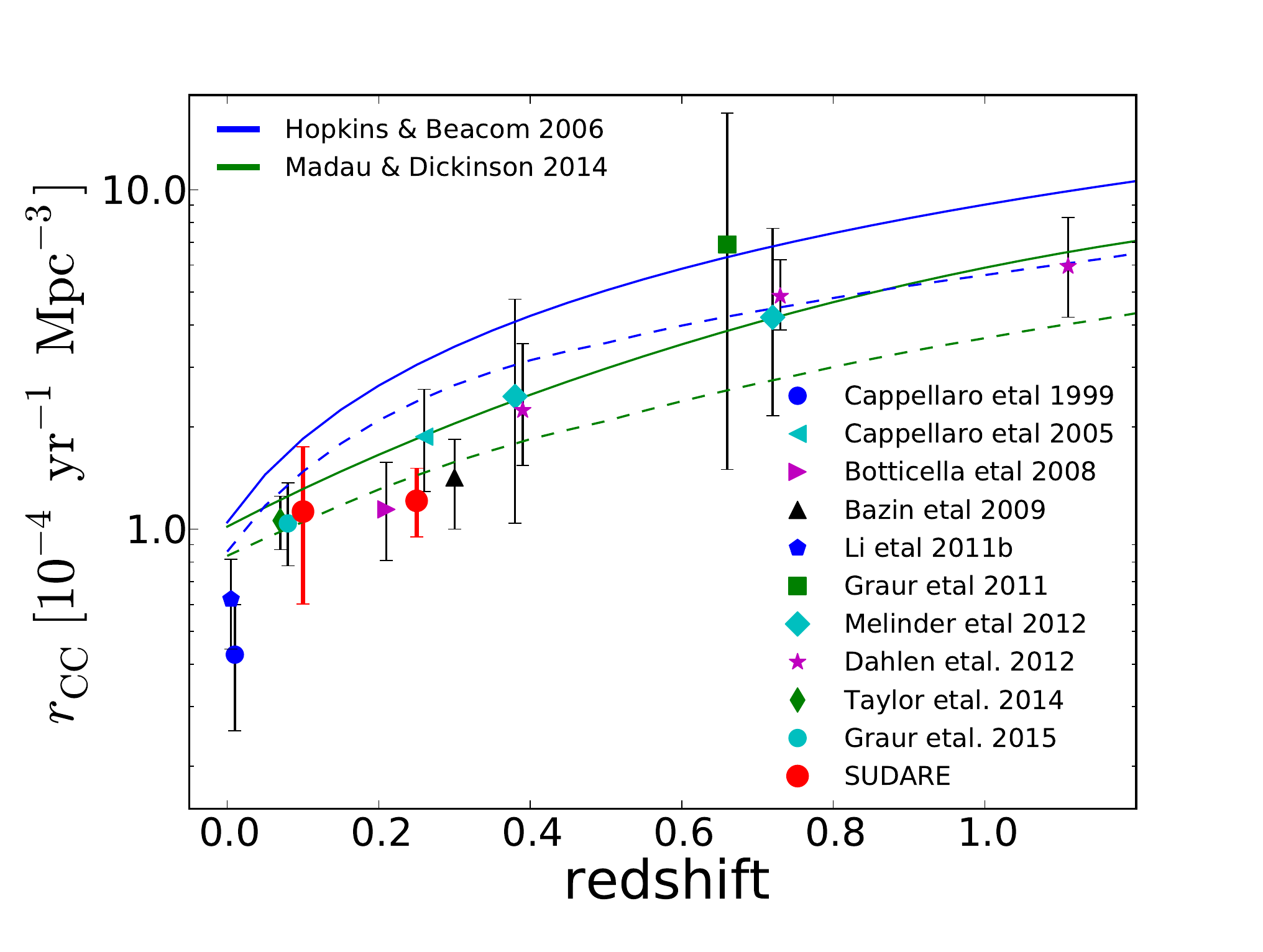} &
\includegraphics[width=7cm]{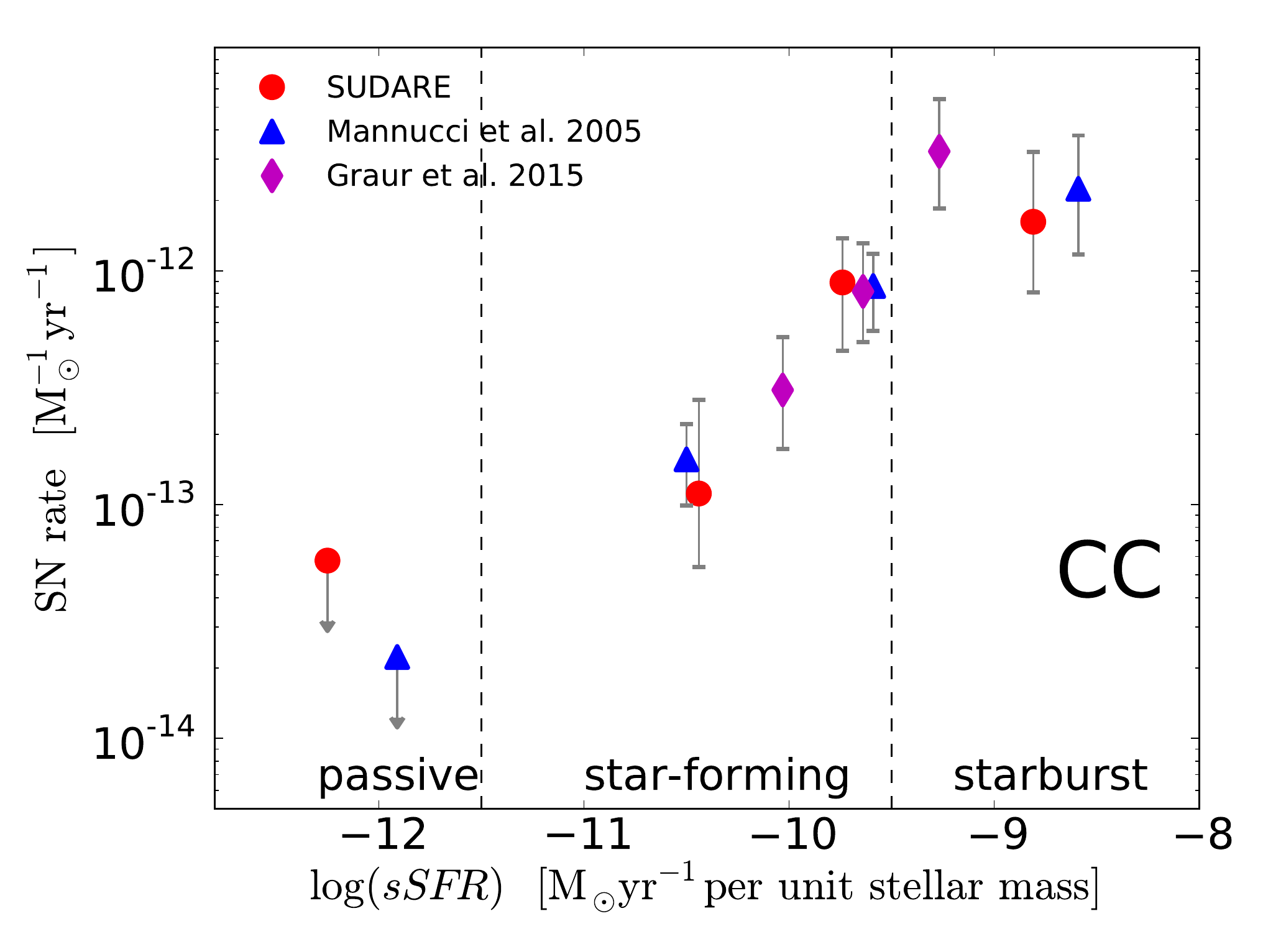}
\end{array}
$
\end{center}
\caption{Left panel: Volumetric SN rates vs redshift for CC~SNe from \citet{Cappellaro2015}. The filled symbols show measurements from literature (with no correction for hidden SNe).
Lines show the predicted CC~SN rate from two different SFHs (green for \citealt{Madau2014} and blue for \citealt{Hopkins2006}) adopting 8\,$M_\odot$ and 40\,$M_\odot$ as the lower and upper mass limits for CC~SN progenitors. The dashed lines show the predicted SN rates assuming the fraction of hidden SNes given in \cite{Mattila2012}.
Right panel: SN rates per unit mass vs specific SFR (sSFR) from \citet{Botticella2017} in three different groups of galaxies based on  their sSFR: the first group of passive galaxies with a zero mean SFR; the second group of galaxies with $-12.0 < \rm log(sSFR)<-9.5$; the third group of galaxies with $ \rm log(sSFR)> -9.5$.}
\label{fig_CCSNe}
\end{figure}

On the other hand, the rate of type Ia~SNe echoes the whole star formation history (SFH) of the host galaxy due to the time elapsed from the birth of the binary system to the final explosion, referred to as delay-time.
Type Ia~SNe are observed to explode both in young and old stellar populations and the age distribution of their progenitors is still a considerable matter of debate e.g. \citet{Maoz2014}.
 The cosmic evolution of the volumetric type Ia~SN rate and the measurements of the type Ia~SN rate per unit mass in the less massive and younger galaxies suggest a distribution of the delay times that are less populated at long delay times than at short delays e.g. \citet{Cappellaro2015,Botticella2017}.
By comparing the observed type Ia~SNe rate with what is expected for the SFH of the parent stellar population, it is possible to constrain the progenitor scenario and the fraction of the binaries exploding as type Ia~SNe \citep{Greggio2005,Greggio2009,Greggio2010}.
The success of the analysis of transient population demographics with respect to the stellar parent population depends on the selection of unbiased transient and galaxy samples, and a statistically significant number of transients.

\begin{figure}[ht]
\begin{center}
$
\begin{array}{c@{\hspace{.05in}}c@{\hspace{.05in}}c}
\includegraphics[width=0.45\textwidth]{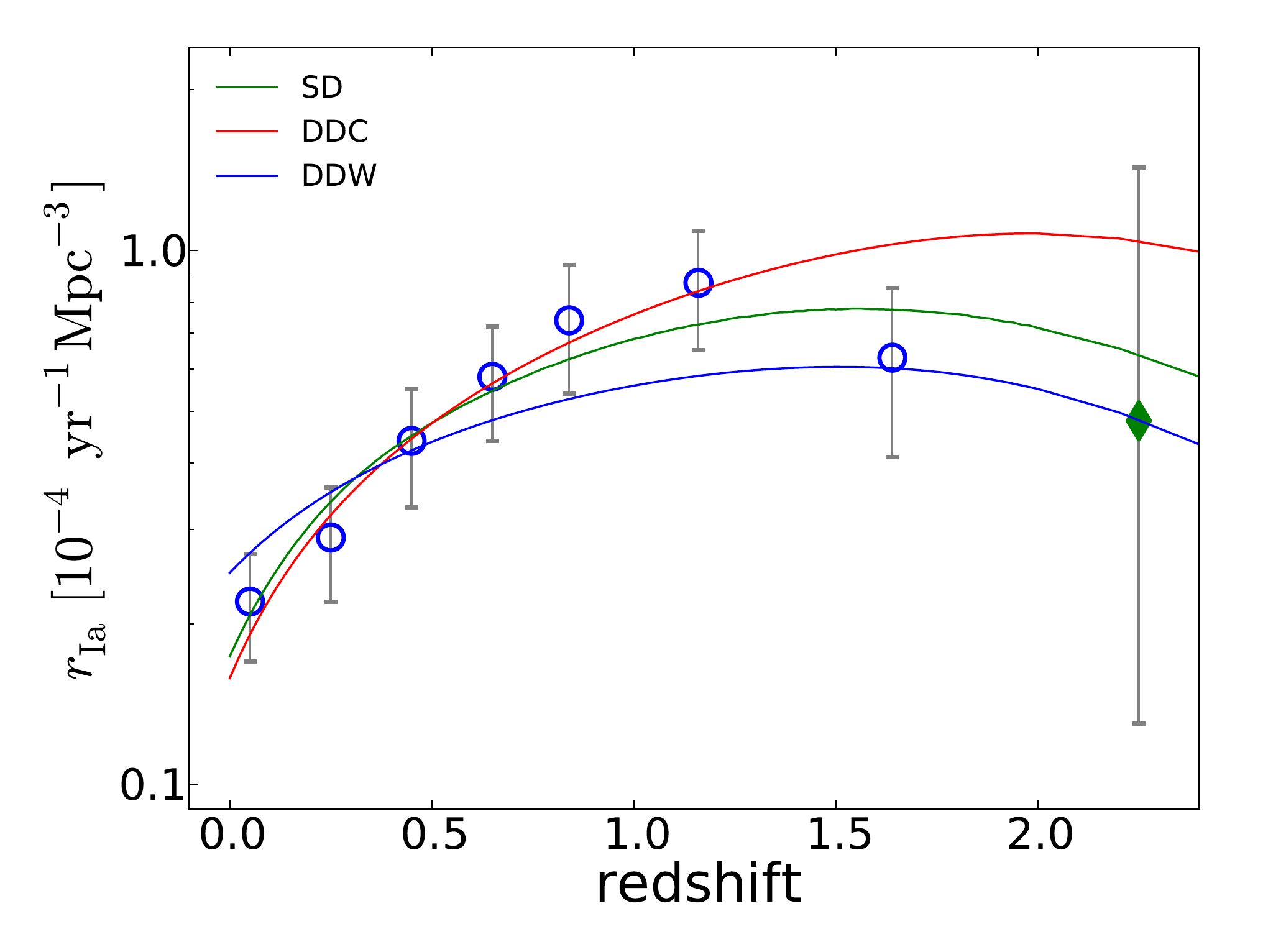} &
\includegraphics[width=0.45\textwidth]{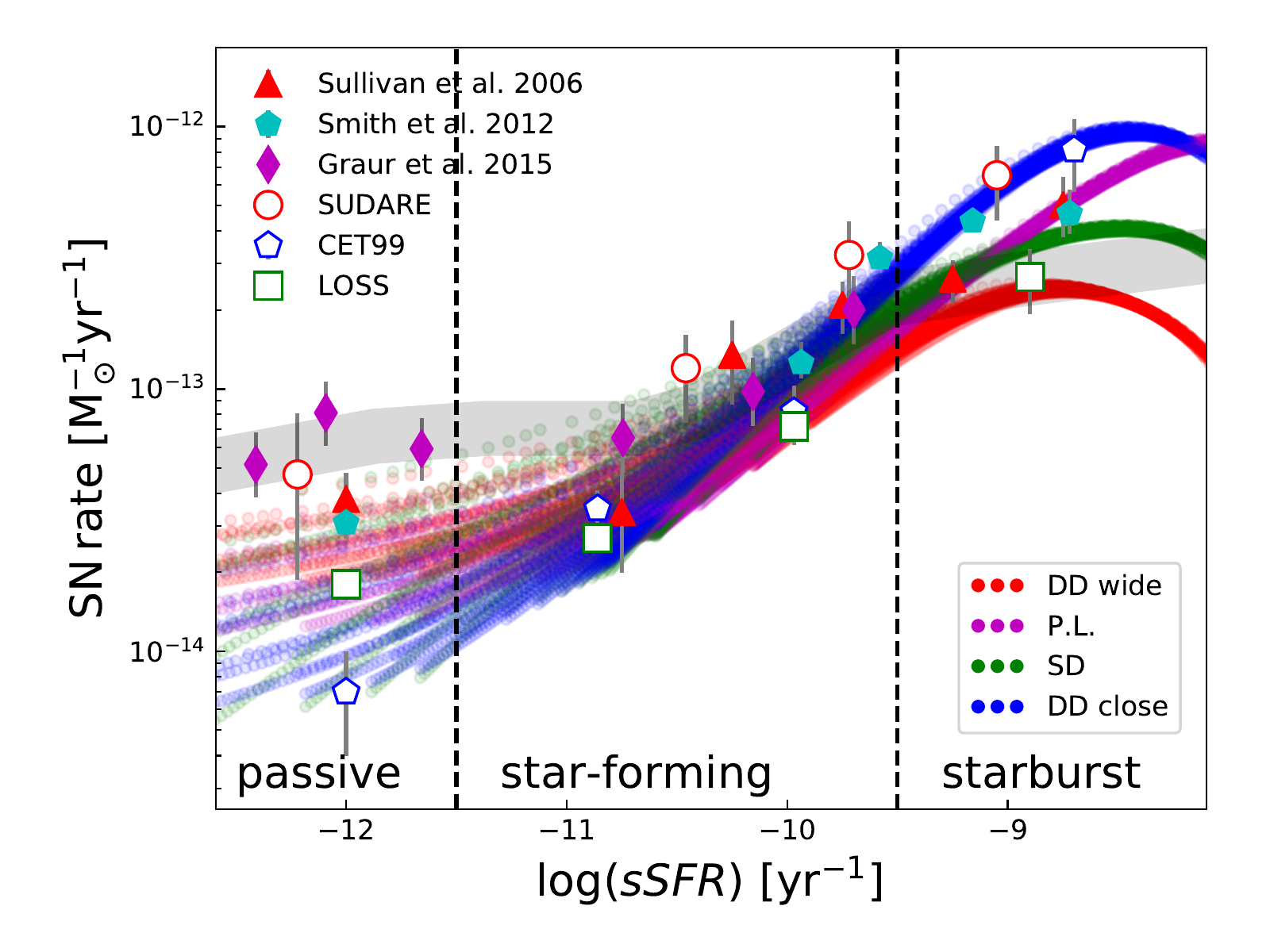}
\end{array}
$
\end{center}
\caption{Left panel: volumetric SN rates vs redshift for Type Ia  from \citet{Cappellaro2015}. The open circles are the averages of values from literature. The lines show the predicted rates as a function of redshift for three different Greggio's models for the distribution function of the delay times for single degenerates (SDs), double degenerates close (DDC) and wide (DDW), assuming the Madau and Dickinson cosmic SFH.
Right panel: SN rates per unit mass vs sSFR in three different groups of galaxies based on  their sSFR as shown for CC~SNe from \citet{Greggio2019}. Filled circles show Greggio's models computed with a log-normal SFH and four different Delay Time Distributions plotted with different colours as labelled. Open symbols show literature data (see legend in the top left corner). The grey stripe shows the result of the simulations in \citet{Graur2015}.}
\label{fig_Ia}
\end{figure}
 
The currently available surveys do not allow us to draw strong conclusions on the SN progenitor problem, due to both insufficient statistics and systematic effects, most notably related to the difficulty of pinpointing the SFH in the sample galaxies (\autoref{fig_CCSNe} and \autoref{fig_Ia}).

The Rubin LSST survey will provide a large number of events in galaxies with a large range of ages, strongly improving both the statistical uncertainty and the systematic uncertainty on the determination of the intrinsic properties of the parent stellar populations. Important constraints on the progenitors of all SN types will be derived by studying  the rates as a function of both the cosmic time and the parent galaxy properties including mass, colours, SFR, metallicity, e.g. \citet{Greggio2019}.
 
 Given the cosmic SFR  and the SFR in a galaxy sample monitored for SNe from Rubin LSST, the trend of the type Ia~SN rate with redshift and with galaxy colours will then be used to derive information on the distribution of the delay times and on the efficiency of type Ia~SN production from stellar populations. 
 
The measurement of the rate of CC~SNe as a function of the cosmic time and galaxy mass from the Rubin LSST survey will allow us to check how well this rate traces the global SFR in the Universe, as derived from other measurements of the production of massive stars. 
The proportionality constant will instead yield information on the mass range of the CC~SN progenitors \citep{Cappellaro2015,Botticella2017}. 
With the improved statistics from Rubin LSST, it will be possible to constrain the mass ranges of the different CC~SN subtypes, and derive the dependence on metallicity. 

The primary goals of our project are: 
 \begin{itemize}
    \item Low hanging fruits
    \begin{enumerate}
    
       \item [a)] {\bf Understanding the evolutionary scenarios, light curve differences and progenitors for all supernovae}\\  
       For SNIa’s, this will involve constraining the evolutionary scenario for their progenitors and  understanding the origin of the diversity of SN Ia light curves. For CC~SNe, this will involve constraining the mass range of the progenitors of the different CC~SNe subtypes and checking for trends with properties of the parent galaxy like, e.g. metallicity.

        The goals will be achieved by measuring the rates of the different SN types(Ia, Ib, Ic, IIP, IIL, IIn, SuperLuminous) as function of redshift and in galaxies of all types, and examining the dependence of these rates on the SFH of the parent galaxy. 

        To measure unbiased SN rates we need (1) to collect the light curves of all transients, (2) to identify the host galaxy for each transient (3) to obtain the photometric classification of each transient and the type for each SN, and (4) to estimate of the detection efficiency of the SN search as a function of the SN magnitudes and light curves and of the survey strategy.

        We aim to exploit the data from Wide Fast Deep Survey to obtain the detailed rates for all the SN sub-types in the local Universe (z < 0.1) (the Deep Drilling Field surveys will enable the exploration of the evolution of the rate of SNe (both type Ia and CC) at high redshifts).

The project will progress incrementally. Preliminary results can be achieved after the first three years of the survey. The whole ten-year lifetime of the survey is needed to reduce statistical errors on the rates and to obtain a more accurate galaxy characterization. 
 \end{enumerate}
    \item Pie in the sky
    
    \begin{enumerate}
    
       \item [a)]{\bf Comparison of the SN rate measurements with our theoretical predictions}\\ 
       Direct comparison will allow us to discriminate between different progenitor scenarios. The detailed analysis of the observed rates and of the host galaxies' properties will also help to shed light on the origin of the SN diversity.
       \newpage
       \item[b)]{\bf Measure intrinsic SN properties and their evolution through time}\\
       With the increase in the number of known supernovae at different redshifts, there a is new opportunity to model the intrinsic properties of SNe (e.g. decay rates, colour, etc) and their evolution through cosmic time. This will provide insights into SN progenitors and their diversity.
       
       \item [c)] {\bf Obtain light curves of all SN types}\\
       A large number of light curves will become available for the various SN types, enriching the template database, especially for the rarest types. The analysis of the light curve properties for a sample of high statistical significance will enable studies aimed at understanding the origin of the SN diversity. 
With this large database in hand, the accuracy of the photometric classification will improve, given the opportunity to include rare events in the sample, and to produce well sampled templates for all SN types.

     \end{enumerate}
        \end{itemize}

\subsubsection{Preparations for Supernova Science}
\begin{itemize}
 
\item {\bf{Follow-up observations/archival data}}\\
For the SN detection and typing we rely entirely on Rubin LSST data. 
This study will be based on an accurate characterization of the stellar populations in all galaxy types, which will be achieved through a detailed modelling of the galaxies Spectral Energy Distributions (SEDs) to estimate the galaxy colors and photometric redshifts and the SFH.
We aim to take full advantage of the  GALAXIES Science Collaboration determination of the photometric redshift and of the SFH of the galaxies in the survey. Ancillary data on the latter, where available, may be used to better constrain the SFH.

\item {\bf{Facilities/software requirements}}\\ 
A pipeline needs to be developed to deal with transient multi-band light curves and galaxy photometric redshift catalogues and to combine different wavelength data from other surveys (e.g. Euclid).
A method to associate transients with host galaxies and to classify transients and to determine the type of SNe will be required.

We will perform simulations to understand the anticipated supernova occurance rates. We will use current SN progenitor models and estimates of galaxy mixture evolution with cosmic time to predict the number of SNe for the different survey strategies and their expected light curves. 
We will compute the expected SN rates up to z\,=\,1  to construct various scenarios of the trends of the SN rates under different hypothesis on the progenitors.
We will perform tests to assess the detectability of the various SN types in the galaxies populating the field. 
This will be done by adding artificial SNe on the images and their light curves in order to evaluate both the detection efficiency and the classification efficiency.
The simulation tools developed for Rubin LSST survey simulation are essential for this task.
 
\end{itemize}


\chapterimage{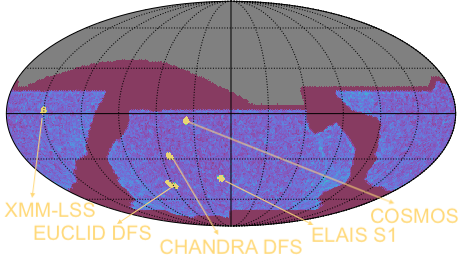} 

\chapter{Deep Drilling Fields science cases}
\section{Introduction} \label{sec:ddf}
\textsl{Author: Rachel Street}

\bigskip

\noindent Complementing\blfootnote{Image: A simulation of the LSST survey strategy} the regular cadence of the Wide-Fast-Deep wide-area survey, Rubin will pay significantly more visits to a small number of selected regions known as the Deep Drilling Fields (DDFs).  These are small spatial regions, typically consisting of one to a few Rubin pointings, centered on areas of special scientific interest. The scientific goals of each DDF varies with its location, but usually requires repeated imaging either to reach a fainter limiting magnitude and/or to achieve a high temporal cadence during the survey.  The science motivations were presented in the 2018 Rubin survey cadence White Papers\footnote{\url{https://www.lsst.org/submitted-whitepaper-2018}} and range from the discovery of Kuiper Belt objects to studying the origins of Dark Energy, galactic structure, large-scale structures and cosmology, to mapping the Magellanic system.  The current set of selected DDFs is presented in \autoref{tab:DDFs}. The original four DDFs selected were recently extended to include a proposed field covering the Euclid Deep Field South region \citep{Guy2022}, based on work by a joint Euclid-Rubin Working Group.  In this section we explore the benefits these intensive observations will bring to time-domain science. 

\newpage

\section{Extrinsic transients and variables} 
\textsl{Editor: Andrej Prsa}

\subsection{Transiting Exoplanets} \label{sec:ddexo}
\textsl{Authors: Mike Lund, Joshua Pepper, Keivan Stassun}

\bigskip

\noindent As discussed in \autoref{sec:ddf}, the deep-drilling fields constitute a small fraction of the Rubin LSST survey, but have a comparable cadence and number of observations to ground-based planet searches. 

\begin{figure}[ht!]
   \begin{center}
   \includegraphics[width=0.70\columnwidth]{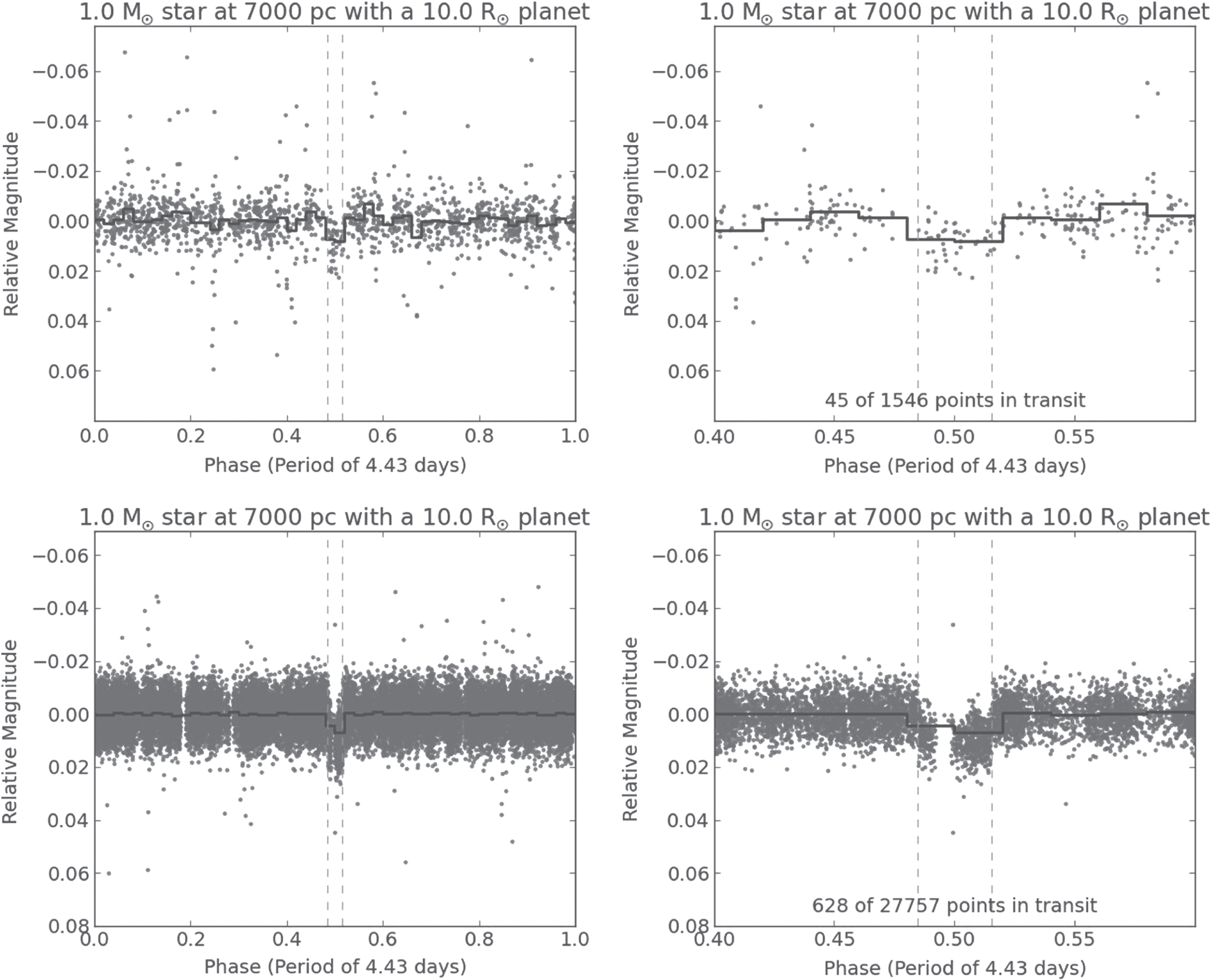}
   \caption{Light curve for a 10 R$_\oplus$ planet in a 4.43 day period around a 1.0 M$_\odot$ star at 7000 pc. The top two plots show a regular LSST field and the bottom two plots show an LSST deep-drilling field. The plots on the left show the full phase, and the plots on the right zoom in on the transit. Black lines are binned data of the light curve. First published in \cite{Lund_2015}.}
   \label{fig:dd6bandtransit}
   \end{center}
\end{figure}

\begin{itemize}
    \item Low hanging fruits
    \begin{enumerate}
        \item [a)] {\bf Detect planets orbiting stellar populations not usually observed}\\
        Exoplanet transits will be detectable in these light curves, as shown in \autoref{fig:dd6bandtransit}. The result is that for these fields the detection efficiency will be greater than for the wide-fast-deep fields, and the range of stars being searched will include populations not normally prioritized or accessible for most transiting planet searches, such as very late type stars, white dwarfs, stars in the galactic bulge, and cooler main sequence stars in clusters. This will enable Rubin LSST to provide insight into planet occurrence and formation rates around stars of varying masses, metallicities, and stages of stellar evolution.
\end{enumerate}
\end{itemize}

\begin{table}[ht!]
    \centering
    \begin{tabular}{c|c|c|c|c|c}
         & ELAIS S1 &  XMM-LSS & Extended Chandra & COSMOS  & Euclid$^{\dagger}$ \\
         &          &          & Deep Field-South &   & Deep Field-South \\
\hline
RA J2000.0   & 00 37 48 & 02 22 50 & 03 32 30 & 10 00 24 & 61 14 24\\
DEC J2000.0  & -44 00 00 & -04 45 00 & -28 06 00 & +02 10 55 & -48 25 12 \\
Galactic l [$^{\circ}$] & 311.30 & 171.20 & 224.07 & 236.83 & 256.060572\\
Galactic b [$^{\circ}$] & -72.90 & -58.77 & -54.47 & 42.09 & -47.17137\\
Ecliptic l [$^{\circ}$] & 346.68 & 31.74 & 40.99 & 151.40 & 36.49\\
Ecliptic b [$^{\circ}$] & -43.18 & -17.90 & -45.47 & -9.39 & -66.60\\
    \end{tabular}
    \caption{The Deep Drilling Field pointings selected for Rubin LSST. $\dagger$: proposed field, endorsed for selection in 2022. Barycentric True Ecliptic coordinates are quoted in each case, though the exact coordinates are to be finalized.}
    \label{tab:DDFs}
\end{table}

\subsection{Eclipsing binary stars} \label{sec:ddebs}
\textsl{Authors: Andrej Pr\v sa}

\bigskip

\noindent In \autoref{sec:tcebs} and \ref{sec:ntcebs} we addressed the time-critical and the non-time-critical aspects of eclipsing binary science, respectively; here we discuss the impact of deep drilling fields.
\begin{figure}[ht!]
    \centering
    \includegraphics[width=0.9\textwidth]{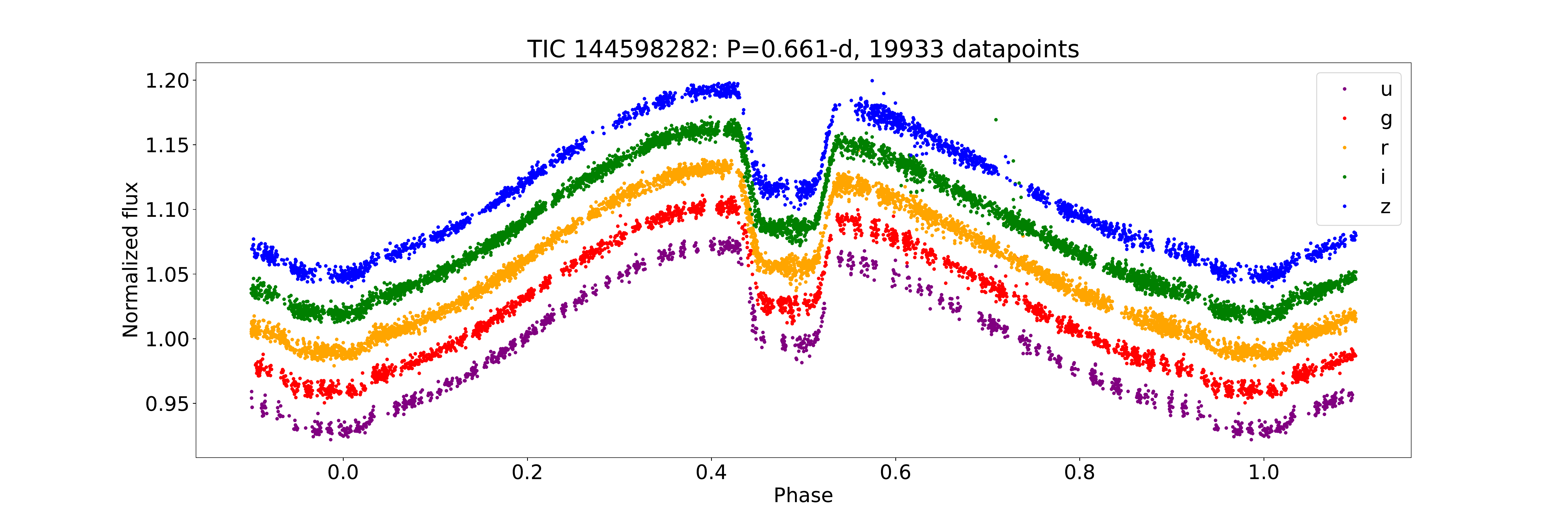} \\
    \includegraphics[width=0.9\textwidth]{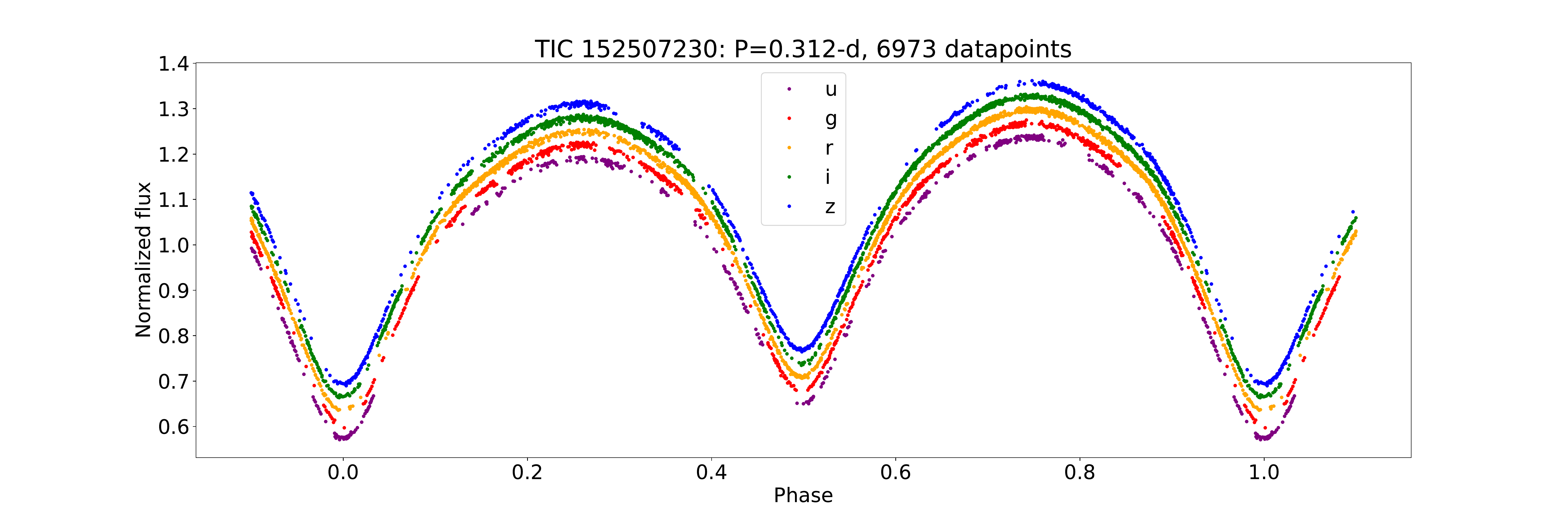} \\
    \includegraphics[width=0.9\textwidth]{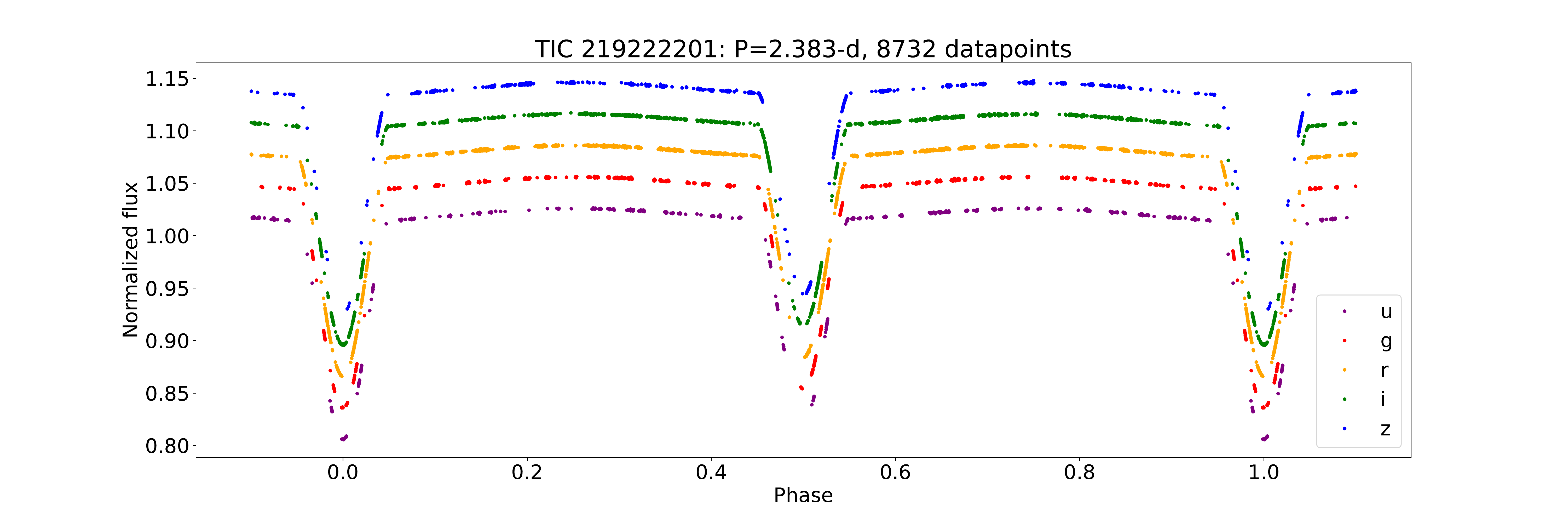} \\
    \includegraphics[width=0.9\textwidth]{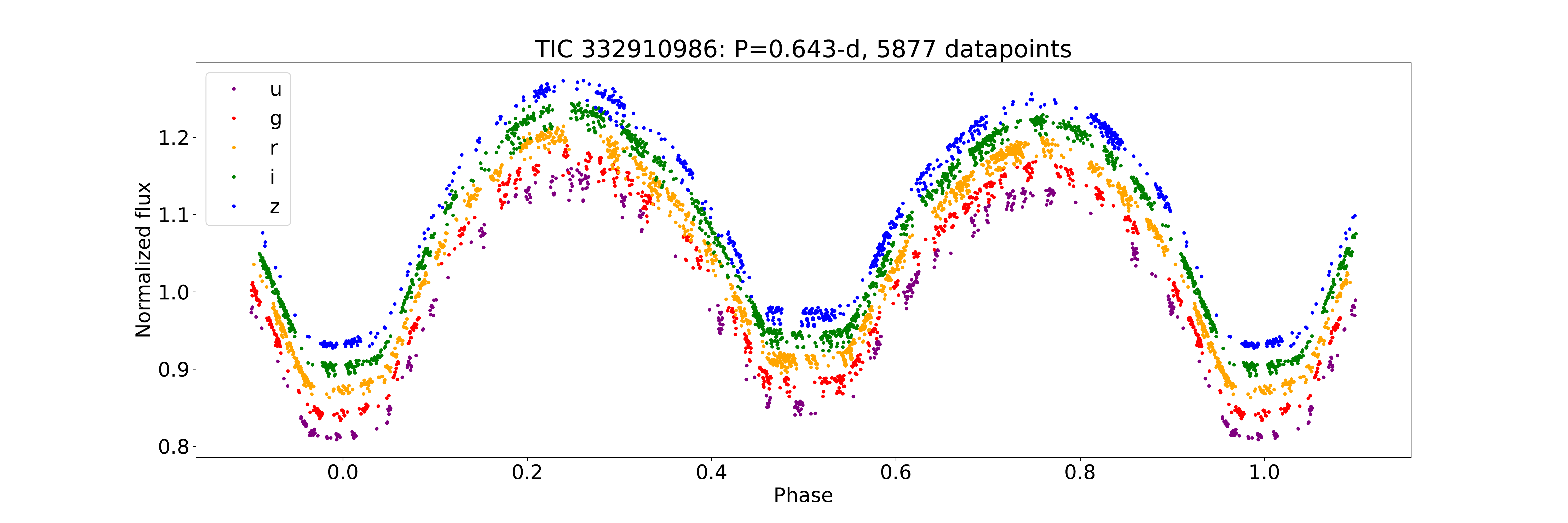} \\
    \caption{Simulated deep drilling light curves of eclipsing binaries, top to bottom: TIC 144598282, TIC 152507230, TIC 219222201 and TIC 332910986. Individual passbands are offset for clarity.}
    \label{fig:ddebs}
\end{figure}

Naturally, as cadence increases, so does the sensitivity to time-variable phenomena on shorter timescales. \autoref{fig:ddebs} depicts 4 TESS light curves sampled in Rubin LSST's deep drilling cadence, depending on their location. Fluxes are normalized and offset for better visibility; no actual color effects have been taken into account, only cadence sampling. While it is clear that the increased cadence helps a lot (cf. top panel), it can still fall short of full phase coverage for longer period systems (cf. third panel). The 20k+ visits certainly improve on the $\sim$800 visits of the main survey, but spread over 10 years still implies fewer than 6 observations per night, distributed across the 6 passbands.

\newpage
\begin{itemize}
    \item Low hanging fruits
    \begin{enumerate}
        \item [a)] {\bf Extending the completeness to longer orbital periods}\\ 
        While diurnal cycle certainly prevents us from reaching true completeness of short period systems, an increased number of visits over the same baseline will significantly improve the detection of longer period systems. A $\sim$20$\times$ increase in the number of visits will increase the phase coverage (barring the diurnal cycle) by the same factor; of course, the longer the orbital periods, the narrower the eclipses, which works to limit the benefits of the increased number of visits. Because of this interplay, we expect the detection sensitivity to increase from $\sim$2.5 days of the main survey to about 10 days for deep drilling fields \citep{kirk2016}.
        
        \item [b)] {\bf Increased sensitivity to intrinsic phenomena}\\
        In addition to extending the orbital period sensitivity, increased cadence will also extend the sensitivity to intrinsic time-variable phenomena such as spots, pulsations and other surface prominences, extrinsic phenomena such as extraneous body interactions (stars, circumbinary planets), and component interactions such as mass transfer or mass loss. Of course, we should maintain realistic expectations: the cadence of the deep drilling fields do not provide the same orbital coverage as Kepler or TESS.
    \end{enumerate}
    \item Pie in the sky
    \begin{enumerate}
        \item [a)] {\bf Increase in precision for all eclipsing binary science cases}\\
        Everything mentioned in \autoref{sec:ntcebs}, but to a higher precision -- thanks to the increased cadence. The number of eclipsing systems in deep drilling fields will of course be vastly smaller than from the main survey, but it will give us an opportunity to assess the amount of information loss due to limited cadence.
    \end{enumerate}
\end{itemize}

\newpage

\section{Intrinsic Galactic and Local Universe transients and variables} 
\textsl{Editor: Markus Rabus}

\subsection{Pulsating stars}
\textsl{Authors: Kelly Hambleton, Keaton Bell, Massimo Dall'Ora, Ilaria Musella, Maria Ida Moretti, Robert Szabo, Marcella Di Criscienzo}

\bigskip

\noindent The measurement of stellar pulsations enables the study of stellar interior structures and the determination of global stellar properties as described in \autoref{sec:pulse}. DDFs will deliver different science results as their observations are obtained with different observing strategies that change the locations and strengths of aliases or signal detection thresholds compared to the main survey.

\noindent A detailed introduction to pulsating stars science cases can be found in \autoref{sec:pulse}. Here we discuss pulsating star science in the deep drilling fields.
 
    \begin{itemize}
    \item Low hanging fruits
    \begin{enumerate}
    \item [a)]{\bf Observations of short-period/multi-periodic pulsators}\\
    If the additional observations of the deep drilling fields are carried out at a higher (possibly continuous; \cite{hermes2018}) cadence, accurate period solutions could be obtained for many shorter-period and multi-periodic variables (including $\delta$ Scuti stars amd $\gamma$ Dor stars, to name a few). The internal structures of these objects could be constrained by full asteroseismic analyses, and these well characterized variables could provide a reliable sample for training machine learning models to classify variables in the main survey.
    
    For the purpose of asteroseismology, radial velocity follow-up observations will provide a more detailed look into the pulsational nature of the objects and will allow for thorough pulsational modeling. The required spectroscopy must be of high resolution with high cadence (to obtain several spectra per pulsational cycle). This becomes significantly more difficult for higher frequency pulsations, as shorter exposures are required and thus the largest telescopes will be needed to reach the faint magnitude limits of Rubin LSST. For known pulsating stars in the field, spectroscopic observations can begin prior to the commencement of the survey observations.

    \end{enumerate}
    \item Pie in the sky
    \begin{enumerate}
  
    \item [a)] {\bf Observations of the Blazhko effect}\\
    The properties of RR Lyrae stars, including their period, color, period variations, and the detection of modulations in their light curves may be visible in the deep drilling fields. The Blazhko effect is the cyclical period and amplitude modulation of RR Lyrae pulsations. Given the large-area and depth of the deep drilling fields, we expect to obtain statistics on the occurrence rate of Blazhko modulations, which would be highly beneficial, since their origin is still heavily debated. The identification of multi-mode pulsations (RRd, double- and triple-mode Cepheids) will also be relatively easy in the deep drilling fields. Additionally, other dynamical phenomena (like period doubling) my be observable depending on the selected cadence.
    
    \item [b)] {\bf Pulsating stars in binary systems}\\
    The discovery of Cepheids and RR Lyrae stars in binary or even eclipsing systems would be of particular importance to derive their masses, radii and other physical parameters of these classical pulsators. The chance of finding them is low, especially for RR Lyrae stars because of binary evolutionary constraints, but the potential gain is high. To obtain the fundamental parameters from such objects, follow-up spectroscopy would be required so that the radial velocities of the stellar components could be determined.
    
    \item [c)] {\bf The asteroseismology of red giant stars}\\
    Red giant stars are stochastic pulsators. They pulsate with periods on the order of days to hours. Studies will have to evaluate how reliably the global seismic parameters $\nu_\mathrm{max}$ (frequency of maximum oscillation power) and $\delta\nu$ (large separation between radial overtones) can be determined given the aliasing caused by sparse and structured observing times.
    
    The Rubin LSST color measurements of red giants will provide temperatures which are necessary for the determination of radii and masses from the use of scaling relations. By considering red giants from several mini-surveys, population studies of red giants can be performed, including the consideration of their fundamental parameters. An assessment of the scaling relations as a function of location and thus metallicity can also be performed. If this is possible, it will only be possible at the end of the data acquisition period because of the need for a long observational baseline. With a 10-year baseline, it may be possible to detect solar-like oscillators with lower frequencies of maximum oscillation power than were detected from Kepler data.
    
    \end{enumerate}
    \end{itemize}

    Studying the potential problems caused by aliasing by the observational window function of different survey strategies will inform the realistic limitations for asteroseismology for these better studied fields. Ideally, the DDFs will be designed to minimize aliasing, possibly even using the past history of field visits to strategically time revisits to reduce existing aliases \citep{Bell2018}.  
    
    Information on adopted DDF strategies will enable more specific anticipation of science results and the associated tools and resources that they will require. The classification will commence with the first data release and will continue incrementally as the data are released. The infrastructure to store and host the catalog is required. Furthermore, the creation of an online catalog to provide the data to the community in a user-friendly format is desirable. 

\subsection{Brown Dwarfs}\label{subsection:VBDs:DDF}
\textsl{Authors: Markus Rabus}

\bigskip

\noindent The possibilities for brown dwarf science are similar to those for the main survey, as stated in \autoref{subsection:VBDs:Non-time}. However, the higher cadence for deep-drilling fields will allow us to better characterize the photometric variabilities in BDs.
 
    \begin{itemize}
    \item Low hanging fruits
    \begin{enumerate}
    \item [a)] {\bf The evolution of weather on BDs}\\
    High cadence long-term observations of BDs will help us to probe the atmosphere dynamics over several BD rotation periods, detect long term weather evolution \cite{Hitchcock2020} and possible planetary wave patterns \cite{Apai2021}. See \autoref{subsection:VBDs:Non-time} for more details.
    
    \end{enumerate}
    
    \item Pie in the sky
    \begin{enumerate}
    
    \item  [a)]{\bf Observe extremely short scale variability in BDs}\\ 
    This science case is outline in \autoref{subsection:VBDs:Non-time}. However, studying possible short-term variability effects through lightning and auroral activities may be achivable with the deep drilling field high-cadence data. 
    
    \item [b)]{\bf Identification of transiting planets around BDs}\\
        A favorable alignment between a potential planetary orbit and the observer's line of sight could also cause a periodic transit-shaped signal in the BD host star (details on transiting planets can be found in \autoref{subsection:Transiting exoplanets}).
    \end{enumerate}
    
    \end{itemize}

In \autoref{subsection:VBDs:Non-time} we have outlined the possible telescope and computational resources, which will be helpful for studying variabilities in BDs.

\subsection{Compact Binaries: AM CVn and Ultracompact Binaries}\label{sec:cbddf}
\textsl{Authors: David Buckley and Paula Szkody}

\bigskip

\noindent Ultra-compact binary systems (UCBs) have orbital periods less than $\sim$80\,min, which implies they have non-main sequence degenerate or semi-degenerate secondaries. Although UCBs are predicted to have a high Galactic spatial density \citep[e.g.][]{Nel2001, Nel2004}, there are only ~60 of these systems currently known \citep{Ramsay2018}, with periods ranging from 5 – 65 min. It is therefore not clear if the models seriously over estimate their number, or if many more systems await discovery. The Zwicky Transient Facility has so far discovered several new UCBs \citep{Cough2020}. Rubin LSST DDFs with sufficient cadence (ideally 15\,s), would result in the discovery of many new examples and therefore constrain their spatial density. Rubin LSST DDFs will enable the following science:

\begin{itemize}
    \item Low Hanging fruits
    \begin{enumerate}
\item [a)]{\bf Testing binary evolution models and the common envelope phase (CE)}\\ 
This will be performed through the comparison of the observed population with CE evolution models \citep[e.g. ][]{Neltout2005}. This will require a large homogeneous sample, as only Rubin LSST can provide, and will allow a more complete determination of their chemical compositions, masses, and orbital variability. These parameters are critical inputs to test theoretical predictions, such as temperatures and disk instability properties, \citep[e.g.][]{deloye2005, Bild2006}.

\item [b)]{\bf Discovering all of the UCBs in the solar neighbourhood}\\ 
With a significant number of UCBs, we will determine their orbital periods and space densities, which is crucial for developing algorithms for detecting sources in LISA data. UCBs and short period detached white dwarf binaries are predicted to be the dominant source of low frequency (mHz) persistent gravitational wave sources detectable by the new space-based gravitational wave observatories of the future, such as LISA  \citep{Strvecc2006}. The significance of UCBs is highlighted in the LISA science case\footnote{\url{https://www.elisascience.org/articles/lisa-mission/gravitational-universe-science-case-lisa/ultra-compact-binaries-milky-way}}.

    \end{enumerate}
\end{itemize}

Many identified UCBs will have very short orbital periods ($\sim$10 min), and are typically very faint ($g > 21$). Depending on the cadence, they may be detectable in the Rubin LSST deep drilling fields, where observations of $\sim$1 hour or more with a 10--30\,s cadence in a single filter would lead to the detection of eclipses or orbital modulations. Rubin LSST, with its huge etendue and ability to undertake high cadence observations continuously for up to several hours, will be crucial in discovering faint UCBs, including short period ($<80$\,min) detached binaries with white dwarfs. There is no doubt that Rubin LSST will succeed in detecting these intrinsically faint objects. As an example, the Faint Sky Variability Survey undertaken with the 2.5-m INT/WFC, was easily sensitive to variations on timescales of 10 min and longer \citep{Groot2003}. 

\subsection{Compact Binaries: Neutron Star binaries}\label{sec:nsbinariesddf}
\textsl{Authors: Elena Mason }

\bigskip

\noindent Deep drilling fields in the context of low mass x-ray binaries (LMXBs), milli-second pulsars (MSPs) and transitional milli-second pulsars (tMSPs) will work similarly to the Rubin LSST main survey, although the cadence, filter choice and sequence will be different. Hence, DDF data could produce additional alerts (see Time Critical Science \autoref{sec:nsbinaries_tc}) and, more importantly, light curves and colors (see non-Time-Critical Science \autoref{sec:nsbinaries_ntc}). Here we detail the science that we hope to achieve given the higher cadence of the deep drilling fields.

    \begin{itemize}
    \item Low hanging fruits
\begin{enumerate}
\item  [a)] {\bf Period search for known and newly discovered accreting neutron stars (NS)}\\
The higher cadence foreseen for DDFs will enable complementary science/analysis with respect to the Rubin LSST main survey. In particular, they will allow period searches to be performed (especially for the orbital periods) of the monitored systems, once a sufficient number of data points will have been acquired. 
\item  [b)] {\bf A census of accreting NSs in crowded fields}\\
The different stellar environments explored by the DDF surveys will enable comparison of the accreting NS populations and how they are affected by their environment (including star density and metallicity etc).
\item  [c)] {\bf Change of states and outburst alerts:}\\
Alerts from DDF surveys are in principle possible and will enable us to trigger follow-up observations of accreting NS binaries that enter a state which allow us to probe the binary parameters and/or physical mechanisms responsible for their observed phenomenology. 
\end{enumerate}
    \end{itemize}

\subsection{Intermediate Luminosity Optical Transients (ILOTs)}
\textsl{Authors: Andrea Pastorello \& Elena  Mason}

\bigskip

\noindent For a general introduction to ILOTs, see \autoref{sec:ilot-time}. In the investigation of the physical mechanisms responsible for the different types of ILOTs, detection of the earliest stages of their outburst is priceless. This is especially true because ILOTs differ from each other in the early development of their light curves and theoretical models predict different light curves depending on the progenitor. In view of their higher cadence, deep drilling fields should offer better monitoring than that of the main Rubin LSST survey. For the case of extra-galactic deep drilling fields, ILOTs will be too faint to allow for proper follow-up. Specifically, deep drilling fields in the Galactic Plane and nearby galaxies in the Local Universe will allow us to characterize the ILOT variability with good cadence and higher precision than that provided by the main survey. This may also constrain both shorter duration variability, produced by erratic stellar flares and (quasi-) periodic variability due to close binary interaction, possibly associated with the ILOT's progenitor.

    \begin{itemize}
    \item Low hanging fruits
    \begin{enumerate}
\item [a)] {\bf The discovery of ILOTs in their early phases}\\
Deep drilling fields will enable the discovery of ILOTs at very early phases. A higher, multi-band cadence allows us to characterize the stellar variability and (including the data obtained with the main survey) constrain the variability history and the nature of the progenitor system. This strategy allowed the OGLE survey to detect the famous Red Nova V1309 Sco, a stellar merger event \citep{Tylenda2011,Mason2010}. 
    \end{enumerate}
    \newpage
    \item Pie in the sky
    \begin{enumerate}
\item [a)] {\bf The search for luminous red nova (LRN) precursor candidates}\\ 
The photometric monitoring of new and/or known contact binaries allows us to find new LRN precursor candidates \citep[e.g. the claimed KIC 9832227;][]{Molnar2017}. The identification of binary systems, whose photometric period declines with time, indicates inspiraling motions that can lead to a common envelope ejection. The common envelope may rearrange the geometry of the stellar system or cause the two stars to merge. The monitoring of such stellar systems would require a cadence of one shot every several hours.

\item [b)] {\bf The identification of smaller-amplitude/shorter timescale variability}\\
Dedicated surveys on nearby galaxies (within 30 Mpc) and with a higher cadence can provide high-quality light curves. High-precision photometry of bright objects is crucial to identify lower-contrast modulations superimposed on the well-known larger variability of LBVs. This is a key step to constraining the presence of binary LBV companions.

\item [c)] {\bf Observations of fast evolving transients}\\
High-cadence observations reveal fast-evolving transients, including new species of ILOTs and faint, fast-evolving supernovae. In particular, studying the rapid photometric evolution of some types of ILOTs (e.g. putative supernovae impostors or intermediate luminosity red transients) during the very early phases may reveal shock-breakout signatures, which may enable us to unequivocally discriminate terminal (faint) SN explosions from non-terminal outbursts.
    \end{enumerate}
    \end{itemize}

\section{Intrinsic/Extrinsic Extragalactic Transients} 
\textsl{Editor: Claudia M. Raiteri}

\subsection{Blazars}\label{blz3}
\textsl{Authors: Claudia M. Raiteri, Barbara Balmaverde,  Maria Isabel Carnerero Martin, Filippo D'Ammando, Chiara Righi}

\bigskip

\begin{figure}[ht]
\begin{center}
\includegraphics[width=0.70\columnwidth]{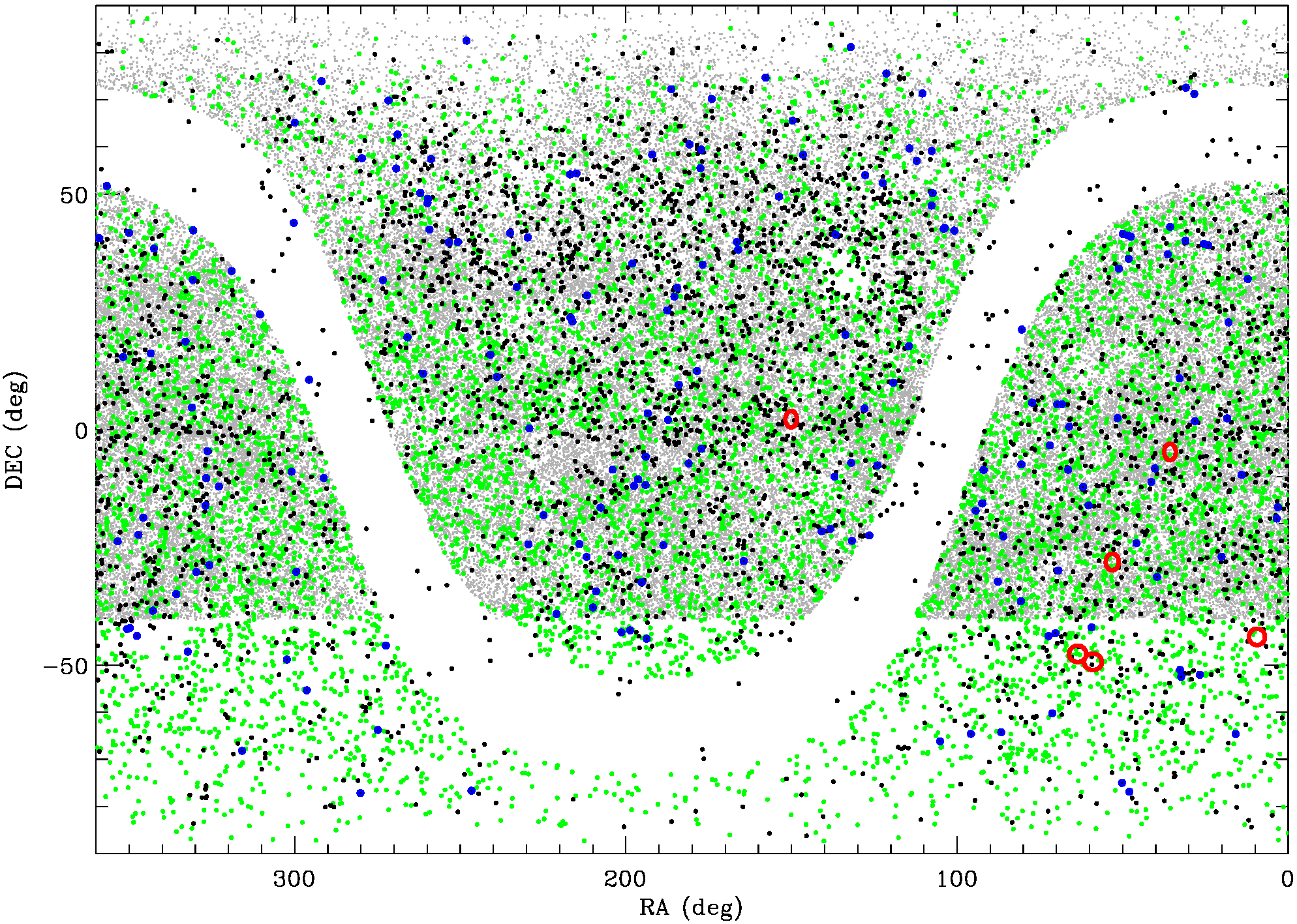}
\caption{The distribution of blazars in the sky according to the BZCAT5 (black symbols), BROS (grey), and CRATES (green) blazar catalogues. Blue circles mark the BZCAT5 objects whose $R$-band catalogue magnitude is brighter than 15.5 and can thus be affected by saturation problems. Red lines represent the 
planned DDFs. From North to South: COSMOS, XMM-LSS, ECDFS, ELAIS S1, and the double field EDFS. From \citet{raiteri2022}.}
\label{mappa}
\end{center}
\end{figure}

\noindent With their denser sampling, DDFs would allow us to study blazar variability in more detail, in particular at shorter time scales than the Main Survey. 
As discussed in \citet{raiteri2022}, we may expect that Rubin LSST will observe about 150 blazars in the six planned DDFs (see \autoref{mappa}). The majority of them  will likely be detected with single exposures. These sources will constitute a small golden sample, whose monitoring will be highly valuable, especially if multiband intra-night observations are performed to follow flux and spectral variability on hour-long time scales.

    \begin{itemize}
    \item Low hanging fruits
    \begin{enumerate}
    \item [a)] {\bf Analyse the fast variability of blazars}\\
    Variability at higher cadences can reveal different emission mechanisms with respect to those that are seen on longer time scales. Fast variability is a distinctive signature of the non-thermal jet emission contribution, and is likely due to intrinsic, energetic processes. On longer time scales, we may see the result of changes in the jet orientation (see \autoref{blz1}). Moreover, in flat-spectrum radio quasars (FSRQs), time scales of the order of weeks/months also characterize the variability due to the thermal nuclear emission, i.e. due to the quasar core \citep[e.g.][]{raiteri2019}.
    \end{enumerate}
    \newpage
    \item Pie in the sky
    \begin{enumerate}
   \item [a)] {\bf Push the study of the environment of the blazar host galaxies to higher redshifts}\\ 
   As mentioned in \autoref{blz2}, deeper and deeper photometry is required as we proceed to higher redshifts. This will be provided by the DDFs. Moreover, the DDF observations would increase the environment richness at low redshifts, through a better sampling of the faintest galaxies.
     \end{enumerate}
   \end{itemize}

Precursor optical monitoring of DDFs would improve detection of unknown sources (blazars, but also active galactic nuclei in general) based on variability at the beginning of the survey, creating a longer monitoring baseline. Moreover, it would extend the timeline on which blazar variability studies can be carried out and diagnostics for blazar identification can be tested.

\chapterimage{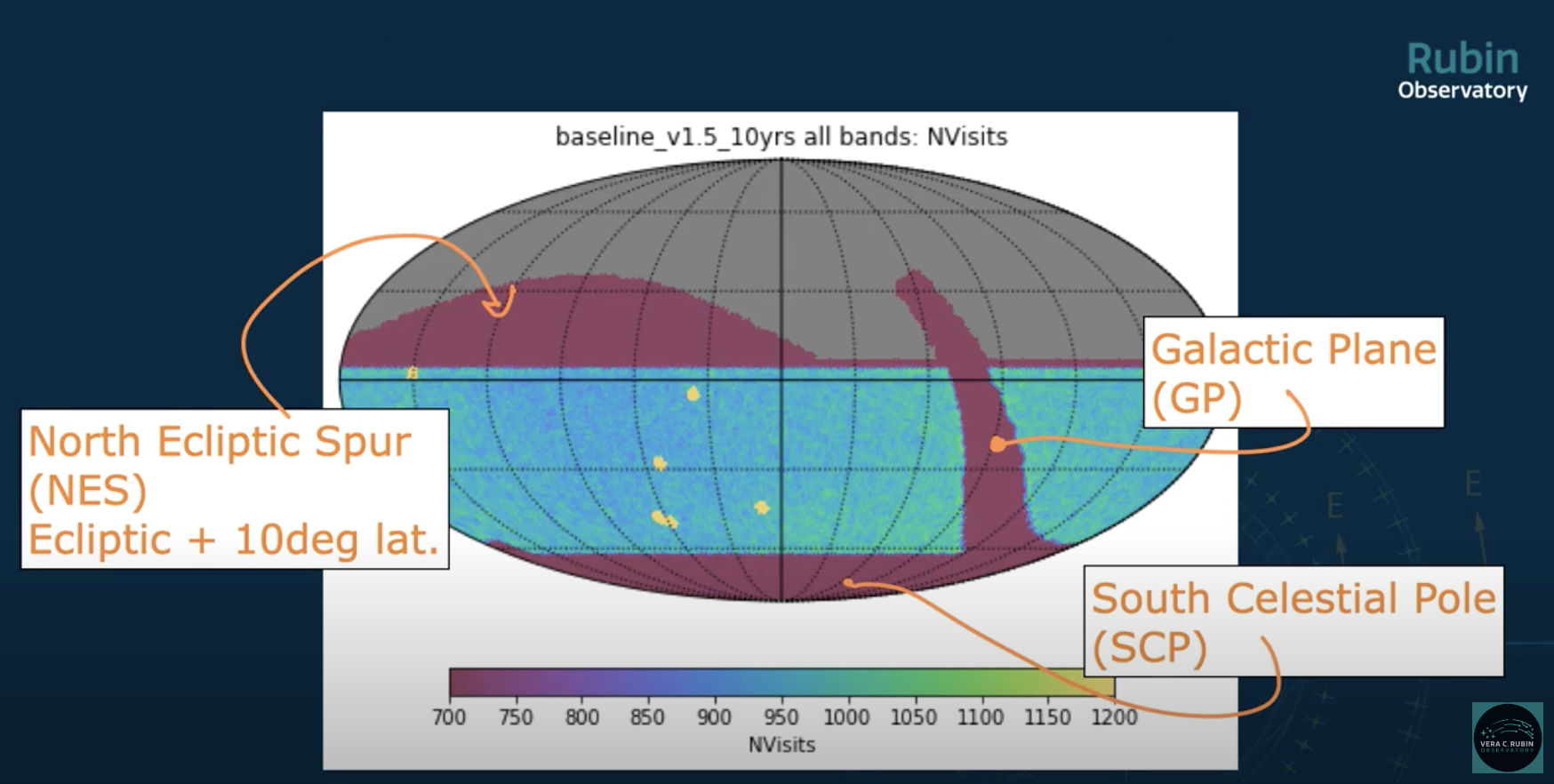} 


\chapter{Minisurvey and Microsurvey Science Cases}

\section{Introduction}\label{chap:minisurveys}
\textsl{Author: Rachel Street}

\bigskip
\noindent The\blfootnote{Image: A simulation of the LSST survey strategy \url{https://community.lsst.org/c/sci/survey-strategy/37}} third category of survey strategy proposed for Rubin LSST are the so-called ``minisurveys" and ``microsurveys".  These represent proposals for novel survey strategies of small regions, though typically larger than that of a Deep Drilling Field, and often located outside the main Wide-Fast-Deep footprint.  The survey cadence and filter selection for these surveys is sometimes specific to the science case that motivated it, as described in 2018 survey strategy White Papers\footnote\url{{https://www.lsst.org/submitted-whitepaper-2018}}, and in the following sections.  These strategies are referred to as ``microsurveys'', as they require a relatively small telescope time commitment ($<$3\% of total survey time) to generate a significant scientific return.  However, numerous White Paper authors presented scientific motivations for increasing the number of visits Rubin dedicates to the Galactic Plane \citep{Gonzalez2018, Lund2018, Prisinzano2018, Strader2018, Street2018} relative to the original implementation.  The survey regions and cadence strategies proposed by these authors overlapped sufficiently well to motivate the exploration of a common strategy for the time-series monitoring of a significant area in the Galactic Plane. As this region will serve a range of science goals, it is referred to as a ``minisurvey``. Additional minisurveys include the North Ecliptic Spur and the Southern Celestial Pole. Currently, the minisurveys proposed have either similar cadence to the main survey (Ecliptic Spur) or 5 times fewer observations with respect to the main survey (Galactic Plane and Southern Ecliptic Pole). In this chapter, we describe the science anticipated from each category of survey.  
\newpage

\section{Extrinsic transients and variables} 
\textsl{Editor: Marc Moniez}

\subsection{Microlensing}
\textsl{Author: Rachel Street}

A general introduction to microlensing can be found in the time-critical section, \autoref{sec:micro-time} and in the non-time critical section, \autoref{sec:micro-non-time}. Here we discuss microlensing science in the minisurveys.

One of our most powerful tools for understanding planetary formation is to compare the actual planet population with that predicted by simulations, but there remain important gaps in our planet census. Low-mass planets in orbits between $\sim$1--10 AU are of particular interest, because the core accretion mechanism predicts a population of icy bodies (e.g. \citealt{IdaLin2013}) in this region. Evolutionary models further predict that gravitational interactions between migrating protoplanets should result in some being ejected from their systems \citep{Mustill2015, Chatterjee2008}. However, this parameter space coincides with a gap in the sensitivity of the planet-hunting techniques used to date, leading to it being sparsely sampled. 
Microlensing offers a way to test both of these predictions, being capable of detecting planets down to $\sim$0.1\,$M_{\bigoplus}$ at orbital separations of $\sim$1--10\,AU. It is also capable of detecting free-floating planets (FFPs) that have been ejected from their parent star systems. The effectiveness of this technique has now been demonstrated by the discovery of three candidate FFP events \citep{Mroz2018}. 
    \begin{itemize}
    \item Low hanging fruit
    \begin{enumerate}
    \item[a)] {\bf Exoplanets in the Bulge}\\
        A survey of the Galactic Bulge region, where the rate of microlensing events is highest, is one of the main goals of NASA's Roman Space Telescope (previously WFIRST) Mission \citep{Spergel2015}. The spacecraft will discover $\sim$1400 bound planets and will provide a dataset ideal for detecting FFPs \citep{Penny2019}.  The physical properties of bound planets should be constrained by direct measurement of the light from the lensing system, but this technique cannot be applied for FFPs. Microlensing models suffer from a number of degeneracies, while the physical properties of the lens are extremely hard to measure without an additional constraint on the event parallax.  For long timescale events ($>$30\,d) this can be derived from a single lightcurve thanks to the orbital motion of the observer, but Roman data alone cannot measure the parallax for short timescale ($t_{E}$\,$\leqslant$30\,d) events.  Thanks to the $\sim$0.01\,AU separation between Rubin and Roman (at L2), the observatories will measure different magnifications and times of maximum, enabling us to derive the physical and dynamical properties of short timescale events. 
        
        Rubin LSST will substantially improve constraints on the lens properties, particularly the FFP mass function, distances and kinematics, by performing regular multi-band imaging of the Roman survey of the Bulge region during the periods when the field is visible to both Rubin and Roman observatories simultaneously.  Rubin LSST will also complete event lightcurves that remain partially sampled by Roman by continuing to monitor the Bulge during the multi-month gaps in the Roman survey.  Roman will observe $i\sim$ 19--25\,mag stars in the Bulge for a total of $\sim$432\,d spread over 6 `seasons'.  But the spacecraft can only monitor the field for $\sim$72\,d at a time, due to pointing constraints. Since microlensing events can peak at any time, and have durations $t_{E}\sim $1--100\,d, many Roman lightcurves will be incompletely sampled.  This will make it difficult to measure the parallax (and hence physical properties) for even long timescale events, and raises the probability that anomalous features (and lens companions) will be missed in the inter-season gaps. 
    \end{enumerate}    
    \end{itemize}

   Originally, the Roman Galactic Exoplanet Survey region was proposed as a Deep Drilling Field in a 2018 white paper. While this was not selected, it was later incorporated into a Minisurvey that covers the Galactic Plane and Magellanic Clouds. To prepare for this science, it will be necessary to cross-match all Rubin LSST DIASources with the Roman catalog.  This will require access to the full DIASource catalogs for an extremely star-rich field - \cite{Penny2019} predicts that Roman will detect $\sim$240$\times$10$^{6}$ stars with $W149<$25\,mag - and thus this substantial task will necessarily be accomplished within the Rubin Science Platform\footnote{\url{https://data.lsst.cloud}} (RSP). For brighter objects, it will be advantageous to cross-match against additional catalogs, including the OGLE Variable Star catalogs (e.g. \citealt{Soszynski2018, Soszynski2019}), and the VISTA Variables in the V{\'i}a Lactea (VVV) Survey \citep{Medina2019} as well as the Gaia catalog, where available.  The combination of optical and NIR photometry will enable us to  determine the spectral types source stars, estimate their distances, and in some cases place constraints on blending.  
   The astrometric information will also help to constrain microlensing models.  
   
   Although NASA are developing a pipeline for Roman that will independently detect microlensing events within that dataset, it is highly desirable to combine the full timeseries photometry for all stars detected by both Roman and Rubin LSST within this Minisurvey.  Since Rubin LSST will fill in gaps in Roman's lightcurves, this combined analysis is likely to reveal additional events as well as to substantially improve the classification of all other events and to detect previously-unseen anomalies.  Since this will require the combination of two very substantial datasets, this analysis is likely to take place within the Rubin LSST Science Platform and/or a NASA Archive.  The resulting data products will then be used to select subsets of events for more detailed analysis, which could be done either within the Rubin Science Platform\footnote{\url{https://data.lsst.cloud/}} (RSP) or at the researcher's home institution.

\subsection{Eclipsing binary stars}
\textsl{Authors: Andrej Prsa}

\bigskip

\noindent The impact of minisurveys on eclipsing binary science is unlikely to be fundamentally different from the main survey: minisurveys feature a different cadence to reach their respective science goals, but that does not mean deeper/more visits. While the details of the impact will depend on the minisurvey at hand, the main driver for EB science remains overall cadence. If there is no appreciable increase in field visits, science yield will be similar to that of the main survey discussed in \autoref{sec:ntcebs}.

\begin{itemize}
\item Pie in the sky
    \begin{enumerate}
        \item [a)] {\bf Binary stars in the Magellanic Clouds}\\
        Magellanic Clouds have been a target of many variability studies, so we have a healthy set of (mostly early-type) eclipsing binaries known to date. The main power of Rubin LSST LMC/SMC minisurvey will be to follow-up and calibrate those systems. We can certainly expect color calibration and, in turn, improved spectral type/luminosity class determinations, and any changes to light curve shape and time of arrival compared to archival data.
    \end{enumerate}
\end{itemize}

\subsection{Interstellar scintillation towards LMC and SMC}
\textsl{Author: Marc Moniez}

Stars twinkle because their light propagates through the Earth's turbulent atmosphere.
On the order of one percent light modulation due to scintillation
is also expected to happen on the order of a few minute time scales when remote stars are observed through an interstellar turbulent cloud, although, it has never been
observed at optical wavelengths.
The time scales of the weak optical intensity fluctuations resulting from the
wave distortions induced by a turbulent medium (visible nebulae or
hidden molecular gas) are accessible only now to the current technology.
Rubin LSST is the ideal setup to search for this signature of gas,
thanks to the fast readout and to the wide and deep field.
As a first result, the detection of such a signal would provide a new tool to
measure the inhomogeneities and the dynamics of nebulae.
Our long term objective is to search for
{\it cold transparent} molecular $H_2$ {\it dust-free} clouds, which are the last possible
candidates for the missing baryons \citep{f1994, Pfenniger_2005} representing $\sim 50\%$ of the Milky Way baryons \citep{McGaugh_2009}.

We propose to take a long series of consecutive images of the same field towards the LMC or SMC during two nights through the ``moving mode". This
``micro-survey" could be done during the commissioning of the camera, but does not need all the mechanical functionalities of the telescope mount,
since the telescope should point in the same direction during each full night.

It has been suggested that a hierarchical structure of cold $\mathrm{H_2}$ could fill the Galactic thick disk \citep{f1994} or halo, 
providing a solution for the Galactic dark matter problem.
This gas should form transparent (dust-free)
``clumpuscules'' of $10\,\mathrm{AU}$ size,
with a column density of $10^{24-25}\,\mathrm{cm^{-2}}$, and a surface filling factor smaller than 1\%.
Such clumpuscules are invisible to the direct observations
since they do not emit or absorb light, but only increase the total optical path of the light by $5-50$ cm;
as a consequence, the diffractive and refractive scintillation caused by their turbulence
- similarly to the well known radio scintillation - is the only way to detect them. 

\begin{figure}[ht]
\begin{center}
\includegraphics[width=0.70\columnwidth]{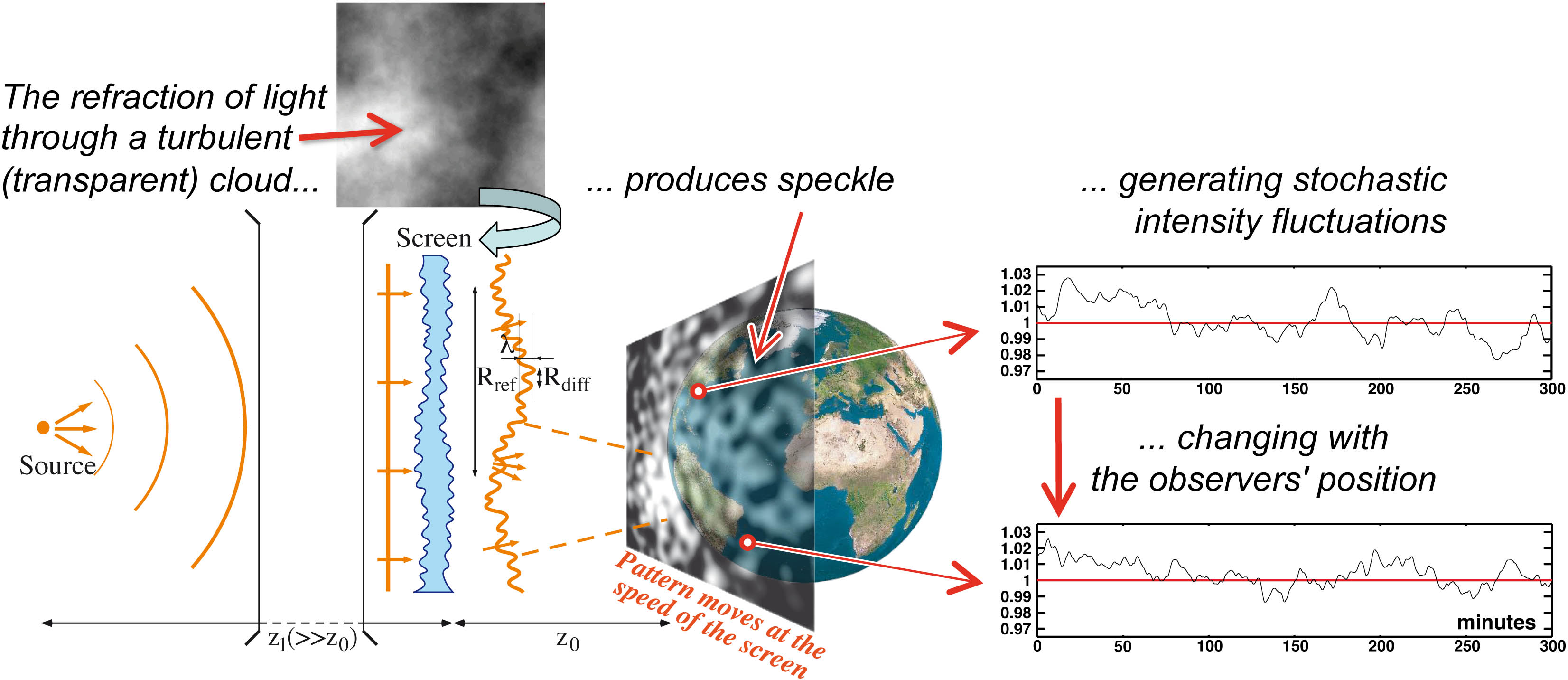}
\caption{{Upper region: A simulation of a 2D stochastic phase delay (grey scale) caused by the
column of gas affected by Kolmogorov-type turbulence. Lower region: The propagation of light from a stellar source (left) after crossing the cloud (represented as a phase screen) and the resulting illumination pattern on Earth. The distorted wavefront produces structures on the scale of: \(R_{ref}=3086km\times\left[\lambda/1\mu m\right]\left[z_{0}/100\,pc\right]\left[R_{diff}/1000\,km\right]^{-1}\). 
As a consequence, two telescopes separated by a few thousand kilometers are differently illuminated at a given time. These structures sweep the Earth at the transverse speed of the screen (typically a few \(10\,^{\prime}s\) of km/s), producing uncorrelated illumination fluctuations over time scales of minutes. The configuration simulated here corresponds to a scintillating star that is half the size of the Sun, located at 1\,kpc, seen through a turbulent cloud at 160\,pc with \(R_{diff}=1000\)\,km translating at\(V_{T}\sim 17\)\,km/s with respect to the line of sight. The two telescopes are 10,000\,km away (the GEMINI telescopes are at 9430\,km linear distance).
{\label{947729}}%
}}
\end{center}
\end{figure}

\autoref{947729} shows how refraction through an inhomogeneous transparent cloud
produces an irregular illumination on Earth \citep{Moniez_2003, Habibi_2010}.
The turbulence strength of the refractive medium is quantified by the
diffraction radius $R_{diff}(\lambda)$, defined as the transverse separation
for which the rms of the phase difference is 1 radian at $\lambda$.
Assuming that the cloud turbulence is isotropic and is described
by the Kolmogorov theory up to the largest cloud's scale $L_z$,
$R_{diff}$ can be expressed as:
\begin{equation}
R_{diff}(\lambda)=263\, km\times
\left[\frac{\lambda}{1\mu m}\right]^{\frac{6}{5}}
\left[\frac{L_z}{10\ AU}\right]^{-\frac{1}{5}}
\left[\frac{\sigma_{3n}}{10^{9}\, cm^{-3}}\right]^{-\frac{6}{5}}\!\! ,
\label{expression-rdiff}
\end{equation}
where $\sigma_{3n}$ is the cloud's molecular number density dispersion. Here, we assume that the cloud is a mixing of $76\%$ of ${\rm H_2}$ and $24\%$ of He by mass. The values of the denominators are typical of the parameters of the already mentioned hidden molecular clouds.
The refractive medium, located at distance $z_0$ from Earth,
and moving with transverse velocity $V_T$
relative to the line of sight, is responsible for stochastic
intensity fluctuations of the light received from the star
at a typical characteristic time scale of a few minutes, scaling as:
\begin{equation}
t_{ref}(\lambda) = 
5.2\, {\rm minutes}\left[\frac{\lambda}{1\mu m}\right]\left[\frac{z_0}{100\, pc}\right]\left[\frac{R_{diff}(\lambda)}{1000\, km}\right]^{-1}\left[\frac{V_T}{10\, km/s}\right]^{-1},
\label{dureescint}
\end{equation}
with a typical intensity modulation index $m_{scint.}=\sigma_I/<I>$ of a
few $\%$,
limited by the source spatial coherence, thus decreasing when
the angular stellar radius $\theta_r$ increases, according to:
\begin{equation}
m_{scint.} = 0.05 \, \left[\frac{\lambda}{1 \mu m}\right] \left[\frac{z_0}{100\, pc}\right]^{-1/6} 
                      \left[\frac{R_{diff}(\lambda)}{1000
                          km}\right]^{-5/6}
                      \left[\frac{\theta_r}{\theta(Sun\  at\  10 kpc)}\right]^{-7/6}.
\label{xparam}
\end{equation}
Since the illumination on Earth depends on the position (\autoref{947729}-right),
we expect variations of the light-curves observed from two telescopes
to decorrelate when their distance increases. This signature
--  incompatible with an intrinsic source variability --
points to a propagation effect. \\
One should notice that the signal cannot be confused with atmospheric scintillation
- that induces fast PSF variations but negligible intensity variations within a large aperture \citep{Dravins_1997a, Dravins_1997b, Dravins_1998} - or with atmospheric absorption fluctuations, that can be precisely taken into account in the analysis, by the simultaneous monitoring of all stars in the field.

\begin{figure}[ht]
\begin{center}
\includegraphics[width=0.5\columnwidth]{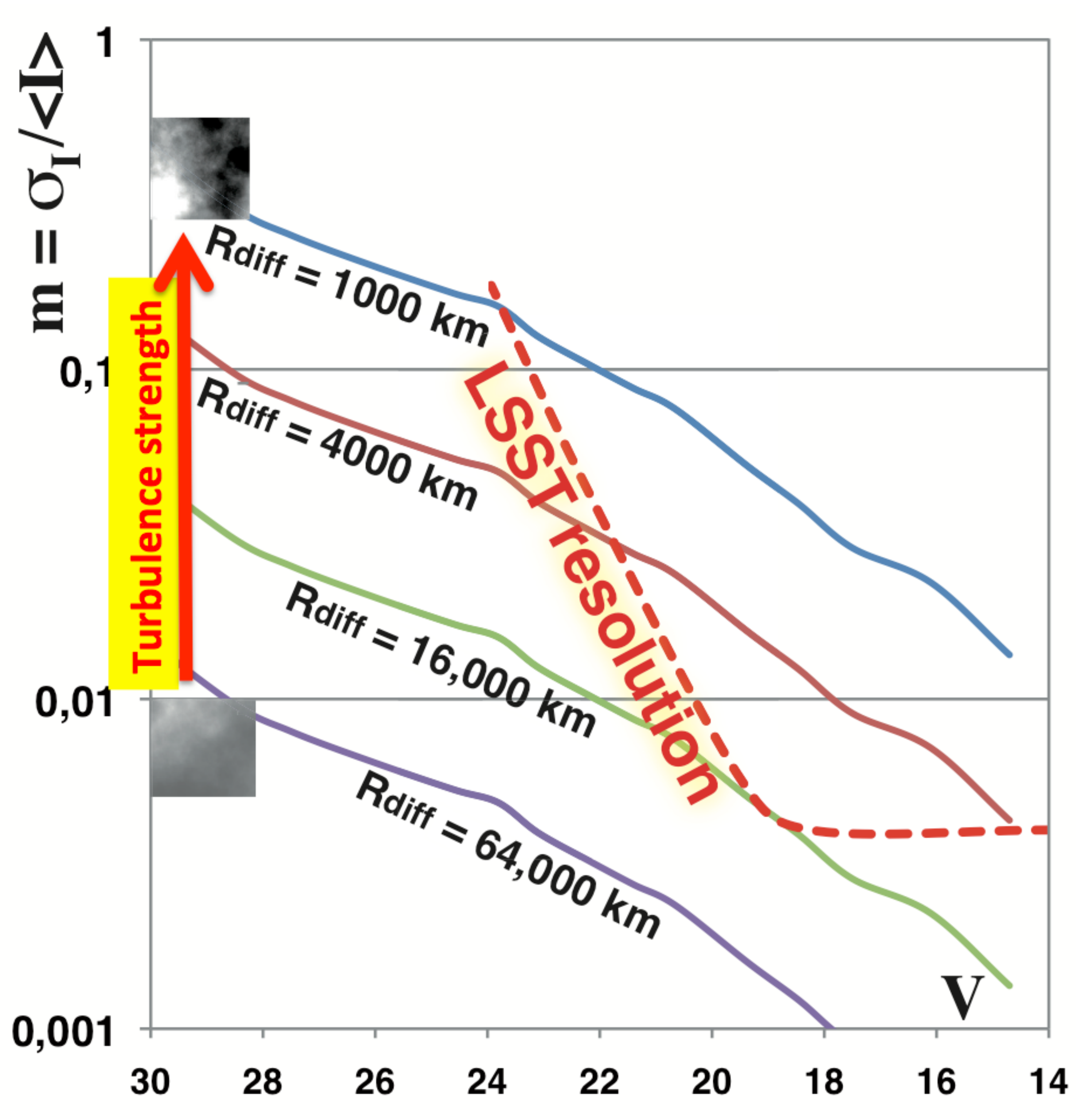}
\caption{{The expected modulation index m in~\(I\), as a function of
the source's apparent magnitude~\(V\), for 4 different
turbulence strength parameters~\(R_{diff}\)
(smaller~\(R_{diff}\) corresponds to stronger turbulence). The
screen is at~\(z_{0}=1\)Kpc (invisible halo clumpuscule typical
case), the source is in the LMC. The
instrumental photometric precision (dashed line) is taken from the Rubin LSST
science book~\protect\citep{Abell_2009}. We see that few percent precision
observations already allow a sensitivity to a medium with relatively low
strength turbulence, characterized by ($R_{diff}<16,000$\,km), that should
induce scintillation detectable for stars with~\(16<V<22\).
{\label{471833}}%
}}
\end{center}
\end{figure}

\begin{itemize}
    
    \item Low hanging fruits
    
    \begin{enumerate}
    \item [a)] {\bf Searching for scintillation signals}\\
    Using Rubin LSST,
two runs (either with neutral filter, to benefit from a maximum of light, or
with g filter) of a few hours towards a given
direction (LMC or SMC to benefit from the wide field),
taking series of 15\,s consecutive exposures, will allow
to obtain a few tens of millions of light-curves with 1--2 thousand measurements each
at the requested high rate ($\sim3$ per minute) and the requested
photometric precision (better than $1\%$ on stars with $M_V=20.5$).
Such a harvest of data would allow to efficiently search for scintillation signals down to an optical depth of order of $10^{-6}$.
\begin{itemize}
\item
If no scintillation is discovered, a strong upper limit of the molecular gas contribution 
to the Galactic mass will be established (following the analysis of \citealt{Habibi_2010}).
\item
If a scintillation signal is suspected, it could be confirmed thanks to the repetition of the observations of stochastic variations
during two different nights (which do not need to be consecutive).
Moreover, if a trigger is available at the time of observations (based on simple peak-to-peak variation
threshold), complementary observations could be simultaneously done with remote telescopes, allowing to test the decorrelation of the stochastic fluctuations between distant locations.
\item
If enough twinkling stars are discovered,
the decrease of the modulation index when the size of the source
increases could also be measured, as a check of the properties
of the scintillation process.
\end{itemize}
\autoref{471833} shows the configurations (source magnitude and
turbulence strength expressed in $R_{diff}$) that should produce
detectable scintillation with Rubin LSST observations (upper-right region).
The ultimate sensitivity corresponds to $R_{diff} \sim 16,000$ km, which is typically associated to a medium
with density fluctuations as low as $2\times 10^7\ cm^{-3}$.
   
The database produced by the suggested ``movie mode" will certainly be useful for many other
science subjects like the searches for planetary transits, the detailed study of eclipsing
binaries \citep[e.g. to refine the knowledge of the LMC distance,][]{Muraveva_2014}, the microlensing
searches for hidden very low mass compact objects and for microlensing caustic crossing if observations are coordinated with the microlensing networks.
\end{enumerate}
\end{itemize}

\section{Intrinsic Galactic and Local Universe transients and variables} 
\textsl{Editor: Keaton Bell }

\subsection{RR Lyrae Stars}
\textsl{Authors: Massimo Dall'Ora, Giuseppe Bono, Michele Fabrizio, Giuliana Fiorentino, Alessia Garofalo, Kelly Hambleton, Davide Magurno, Silvia Marinoni, Tatiana Muraveva}

\bigskip

\noindent Despite all the efforts so far, we still lack a clear knowledge of the old stellar population in the Galactic inner bulge, mainly because of the strong reddening. However, RR Lyrae stars are well-known pulsating stars typical of the old stellar populations and sound standard candles, since they follow well-defined near infrared period-luminosity relations. As standard candles, in the GAIA DR2 era they can provide individual distances with an accuracy better of 3\%. Moreover, they follow reddening-free Weseheneit functions, of great interest in environments affected by strong and/or differential reddening. With these tools we can measure the density profile of the old population, the 3D structure of the bulge and of the bar, and get fundamental observables to constrain the Milky Way formation models. The use of three bands \textit{iZy}, together with \textit{JHK} (VVV survey) measurements, will allow us to provide new individual estimates of distance, reddening and metallicity.

\begin{itemize}

\item Low hanging fruits
\begin{enumerate}
\item [a)]{\bf Individual distances of the RR Lyrae stars, of the SX Phe stars and of the Anomalous Cepheids in the Galactic Bulge}\\
As in \autoref{sec:cepheids}, we plan to obtain distances to the RR Lyrae, SX Phe and Anomalous Cepheids in the Bulge. The crowding in the bulge will require the extraction of accurate photometry from point sources. While a data product of Rubin LSST will be differential photometry, in the bulge, PSF photometry will be necessary.
\end{enumerate}

\item{Pie in the sky}
\begin{enumerate}
\item [a)] {\bf The comparison of results with theory}\\
The collected database of individual distances, reddening and metallicity estimates of the RR Lyrae stars in the inner bluge will allow us to: (1) Put constraints on the 3D model and density profile of the old population of the inner bulge and (2) Compare the derived profiles with the results from other stellar population tracers (red clump stars, etc.), and with the theoretical models of the bulge formation
\end{enumerate}

\end{itemize}

The effective detection and robust characterization of the variables will be essential. To obtain these goals, we plan to use variability criteria that have been widely tested in the literature, most of them being implemented in the VaST software (http://scan.sai.msu.ru/vast/). This software is freely available, and it does not have special hardware requirements or software dependencies. The characterization is the evaluation of all the relevant pulsational properties of the candidate variables: periods, mean magnitudes and amplitudes. The characterization of the candidate variables will be accomplished with two different approaches: during the first year, when only a few data points per source will be available, the characterization will be performed by comparing the observations with available templates. When at least 15-20 data points are available, such evaluation can be performed directly on the observed data, with well tested methods.

As previously mentioned, we will only require Rubin LSST data for our scientific goals. However, where possible, cross-matching our objects with Gaia (for the positions, proper motions and parallaxes), OGLE-IV (for the classification and very-long term characterization) and VVV (for the NIR measurements) will provide useful, complementary information.

\subsection{Brown Dwarfs}\label{subsection:VBDs:Mini-survey}
\textsl{Authors: Markus Rabus}
 
\bigskip

\noindent A general introduction to brown dwarf (BD) science can be found in \autoref{subsection:VBDs:Non-time}. Already during the commissioning phase, several test surveys will be conducted to assess the performance of the system. Those test surveys are designed to simulate 10 and 20 years of survey operation within a few months. For this propose a small region of the sky will be observed with high cadence and will provide an excellent data set to study the variability of brown dwarfs. Another possible microsurvey proposes to observe fields which will allow us to combine data from the Rubin Observatory and NASA’s Roman Space Telescope (aka WFIRST). Also, DDFs are usually sparse fields, choosing crowded fields, e.g. in the galactic plane, for mini/microsurvyes will allow us to probe those denser regions for BDs. 
 
    \begin{itemize}
    
    \item Pie in the sky
    \begin{enumerate}
    
    \item [a)]{\bf Probe the space density and sub-stellar mass function in crowded fields}\\
    We hope to extend and improve our understanding of the initial mass function (IMF) of BDs in dense regions. These regions have been avoided in the past due to crowding and increased confusion with other objects, e.g. `O'-rich and `C'-rich Long Period Variables (LPV), asymptotic giant branch (AGB) stars, distant highly reddened luminous early-type main-sequence/giant branch stars and Young Stellar Objects (YSOs). Therefore, only few limited searches for BDs in the galactic plane have been conducted, see e.g.\ \cite{Reid_2003}, \cite{Phan_Bao_2003} and \cite{Folkes_2012}. While the minisurvey of the galactic plane will have a reduced number of observations with respect to the main survey ($\sim1/5$), we hope to obtain a first glimpse of BD science through these observations.

    \end{enumerate}
    
    \end{itemize}

In \autoref{subsection:VBDs:Non-time} we have outlined the possible telescope and computational resources, which can be helpful for studying variabilities in BDs. The main survey WFD will provide additional information on proper motion and parallaxes which will help us to place our objects on the color-magnitude diagram and better characterize their spectral types.

\subsection{Variables in Large Magellanic Cloud and Small Magellanic Cloud}
\textsl{Authors: Paula Szkody}

\bigskip

\noindent The Magellanic Clouds (MCs) contain a large range of variables and transients, all at the same distance, with low extinction and low metallicity. Thus, the minisurvey of the Clouds provides the ability to collect a large statistical sample of both known and unknown variability. To obtain light curves covering the full range of periodic variability from 30 sec to 10 years, as well as transient and eruptive objects, would require observations ranging from continuous 15 second exposures in a single filter (to catch short period variables such as $\delta$ Scuti and pulsating white dwarfs) to multiple (30-300) visits in \emph{gri} filters, depending on the type of variable, and to catch transient or eruptive variables. Current simulations of the proposed cadence include a minisurvey of the LMC and SMC with five times fewer observations than the main survey, however, a microsurvey has been proposed to observe 10 local volume galaxies in a higher cadence in g-band. It has been recommended by the Survey Cadence Optimization Committee that the cadence for the minisurveys and microsurveys should be finalized during the first year of operations. As some variables will benefit from multiwavelength (X-ray, UV, IR) observations, the dates of Rubin LSST observations of the Clouds should be advertised to facilitate simultaneous or contemporaneous observations of the Rubin LSST fields. During the first survey year, most of the short period variables should be discovered,
at 3 yrs, new LPVs and novae, while the end of the survey should have a complete list of variability in the Clouds.
    \begin{itemize}
    \item Low hanging fruits
    \begin{enumerate}
    \item [a)] {\bf Study the effect of metallicity on novae occurrences}\\ 
    Use novae to determine how the differences in metallicity affect the rate and location of novae compared to those in the Milky Way.
    \item [b)] {\bf Search for supernova type 1a progenitors}\\
    Find the numbers of short period recurrent novae which are the best candidates for Type Ia SNe progenitors.
    \item [c)] {\bf Fully characterize variable stars in the MCs}\\ 
    Much higher cadence of observations in \emph{g} and/or \emph{r} filters will enable us to determine the ranges of variability between 15 sec and 3 days that are missed in the main survey.
    \item [d)] {\bf Absolution magnitude limited period determination of all variables in the clouds}\\
    Determine accurate periods for all regularly variable objects down to approximately absolute magnitude of 6.5 (assuming a limit for variability of apparent magnitude of 25).
    \end{enumerate}
    
    \item Pie in the sky
    \begin{enumerate}
        \item [a)] {\bf Analyse all non-periodic variables in the Clouds}\\ 
        By identifying non-periodic variables in the Clouds, we will be able to probe the cause of their variability. This will require machine-learning
    to sort out known variables and theoretical work to understand the causes of previously
    unknown types of variables.
    \end{enumerate}
    \end{itemize}

\subsection{Compact Binaries: Neutron Star Binaries}
\textsl{Authors: Elena Mason }

\bigskip

\noindent Mini surveys will support and serve our studies of binary NS of the type LMXBs, MSP and tMSP in an identical way to the main Rubin LSST and deep drilling survey but in different patches of sky. Again, they potentially probe different stellar environments compared to the Rubin LSST main survey and the DDFs (see \autoref{sec:nsbinaries_tc}, \ref{sec:nsbinaries_ntc} and \ref{sec:nsbinariesddf}). 
The smaller total number of visits foreseen for the mini surveys, will however produce somewhat more poorly sampled light curves and less timely alerts. 
    \begin{itemize}

    \item Low hanging fruits
    
        \begin{enumerate}
        \item [a)] {\bf Change of states and outbursts alerts}\\
        Alerts from a Minisurvey are in principle possible and will enable us to trigger follow-up observations of accreting Neutron Star (NS) binaries that enter a state that allows probing the binary parameters and/or physical mechanisms responsible for the observed phenomenology. This is identical to the aforementioned sections, however in a different region of the sky.
        
        \end{enumerate}

    \end{itemize}

\newpage
\subsection{Intermediate Luminosity Optical Transients (ILOTs)}
\textsl{Authors: Andrea Pastorello \& Elena  Mason}

\bigskip

\noindent In the investigation of the physical mechanisms responsible for the different types of ILOTs, to detect the earliest stages of their outburst is priceless, since ILOTs differ from each other in the early developments of their light curves and theoretical models predict different light curves,  depending on the progenitor. Hence, in view of their likely higher cadence,  if selected, some of the microsurveys might offer a better monitoring than the main Rubin LSST survey.  Unfortunately, DDFs are expected to monitor variability in the distant Universe where ILOTs appear as too feeble sources for a proper follow-up.  Instead,  the cadence of the microsurveys (which is somewhat lower than that of DDFs but higher than that of the main Rubin LSST survey)  is an important tool to study nearby ILOTs. Specifically, the microsurveys of individual galaxies and the Local Universe will allow us to characterize the ILOT variability with good cadence and higher precision than that provided by the main survey. This may also constrain both shorter duration variability produced by erratic stellar flares and (quasi)-periodic variability due to close binary interaction possibly associated to the ILOTs progenitors.

    \begin{itemize}
    \item Low hanging fruits
    \begin{enumerate}
\item [a )] {\bf ILOTs observed in their early phases}\\
Microsurveys with high cadence targeting the Local Universe, LMC/SMC and crowded fields in the Milky Way can discover ILOTs caught at very early phases. A higher, multi-band cadence allows us to characterize the stellar variability and (including the data obtained with the main survey) constrain the variability history and the nature of the progenitor system. This strategy allowed the OGLE survey to detect the famous Red Nova V1309Sco \citep{Tylenda2011}. 

\item [b)] {\bf Determination of occurrence rates}\\ 
Reliable rate estimates for the different ILOT types are still incomplete, or even missing. With the minisurvey, we aim to precisely constrain the frequency with which the different families of transients may occur in the surveyed galaxies (in particular, Milky Way, SMC, LMC and M31).
    \end{enumerate}
    \item Pie in the sky
    \begin{enumerate}
\item [a)] {\bf The search for Luminous Red Nova precursors}\\
The photometric monitoring of new and/or known contact binaries allows us to find new Luminous Red Nova precursor candidates \citep[e.g. KIC 9832227,][]{Molnar2017}. The identification of binary systems whose photometric period declines with time indicates inspiraling motions that can lead to a common envelope ejection. The common envelope may rearrange the geometry of the stellar system or lead the two stars to merge. The monitoring of such stellar systems would require a cadence of one shot every several hours.

\item [b)] {\bf Understanding LBV companions}\\ 
Dedicated microsurveys of nearby galaxies (within 30 Mpc) and with a higher cadence can provide high-quality light curves. High-precision photometry of bright objects is crucial to identify lower-contrast modulations superimposed on the well-known larger variability of LBVs. This is a key step to constrain the presence of binary LBV companions.

\item [c)] {\bf ILOTS in higher cadence}\\
High-cadence observations will reveal fast-evolving transients, including new species of ILOTs and faint, fast-evolving supernovae. In particular, studying the rapid photometric evolution of some types of ILOTs (e.g. putative SNe impostors or ILRTs) during the very early phases may reveal shock-breakout signatures, which may enable us to unequivocally discriminate terminal (faint) SNe explosions from non-terminal outbursts.
    \end{enumerate}
    \end{itemize}  
    
\subsection{Young Stellar Objects}
\textsl{Authors: Rosaria (Sara) Bonito, Laura Venuti, Patrick Hartigan}

\bigskip

For an introduction to young stars, particularly EXor and FUor stars, see \S\ref{sec:exorfuor}. Here we discuss young stars in the context of minisurveys and microsurveys. We aim to investigate stellar variability of single objects or statistically in stellar clusters. We plan to analyze the variability induced by several mechanisms, including stellar activity, the disk accretion process (which can also occur in eruptive bursts, i.e. FUors and EXors), and rotation. Our goals include identifying specific regions in the sky, selecting suitable targets, and determining the appropriate cadence to pursue the study of variability for young stars in general. We plan to take advantage of data collected in existing surveys and previous programs (i.e. the Gaia-ESO Survey, Chandra, etc.) to characterize interesting fields and objects, also using a multi-wavelength approach. Indeed, stellar variability is a panchromatic phenomenon, with distinctive features on all timescales from hours to years \citep[e.g.][]{Fischer2022}. While long-term variability can be optimally traced with the Wide-Fast-Deep approach, studies of shorter-term variability, driven by inner disk dynamics, will benefit from a denser cadence, which will also enable exploring the impact of different properties, i.e. ages, metallicity, and environment. The Galactic Plane, and in particular, select star forming regions, should therefore be investigated with a cadence higher than that of the main survey to follow the variability of stars on short time scales (hours and days; see \autoref{fig:YSO-LC-dip}). Analysis of available data from previous surveys, as well as the development of diagnostic tools, will be undertaken to lay the groundwork for the proposed science cases.

    \begin{itemize}
    \item Low hanging fruits
    \begin{enumerate}
\item [a)]{\bf Analyzing short term variability in young stellar objects}\\
Photometric variability, on short-term (hours), mid-term (days, months), and long-term (years) timescales, is part of the definition of classical T Tauri stars \citep[CTTSs,][]{Joy1945}. Young stellar objects (YSOs) are characterized by photometric variability caused by several distinct physical processes: mass accretion events from circumstellar disks, the presence of warps in envelopes and disks, the creation of new knots in stellar jets, stellar rotation, starspots, magnetic cycles, and flares. We can study all these phenomena if we acquire both short-term and long-term light curves of a statistically significant sample of YSOs. The analysis of ``static'' color-magnitude diagrams (CMDs) in Rubin LSST filters, such as $r$ vs. $g-r$ or $r$ vs. $u-r$, will allow us to identify weak-line T Tauri stars (WTTS; non-accreting) within the observed clusters. Although, the bulk of CTTS members are typically spread out at bluer colors than the cluster sequence traced by WTTSs, as a result of the short-wavelength color excess related to the accretion activity that is only present in CTTSs \citep[e.g.][]{Venuti2014}. For this reason, it is important that the selected fields be observed in both bluer and redder bands so that we are able to discriminate between WTTSs and CTTSs.

We have data from previous programs, such as DECam observations of Carina at the CTIO 4-m, that reached depths similar to those expected for Rubin LSST. We will also take advantage of data collected in existing surveys and data from previous programs (for instance, many team members are involved in the Gaia-ESO Survey, Chandra, etc.) to characterize fields and objects of interest, using a multi-wavelength approach. Minisurveys or microsurveys that overlap with these cluster observations will provide an in-depth look at young stars in clusters, as well as a significantly extended the baseline over which the inner disk processes can be tracked and characterized.

For those stars identified as young stars, classifications based on photometric colors can be confirmed spectroscopically with, for example, FLAMES observations of the H$\alpha$ emission line \citep[see also][]{Bonito2013, Bonito2020}.

\begin{figure}[ht]
\begin{center}
\includegraphics[width=0.7\columnwidth]{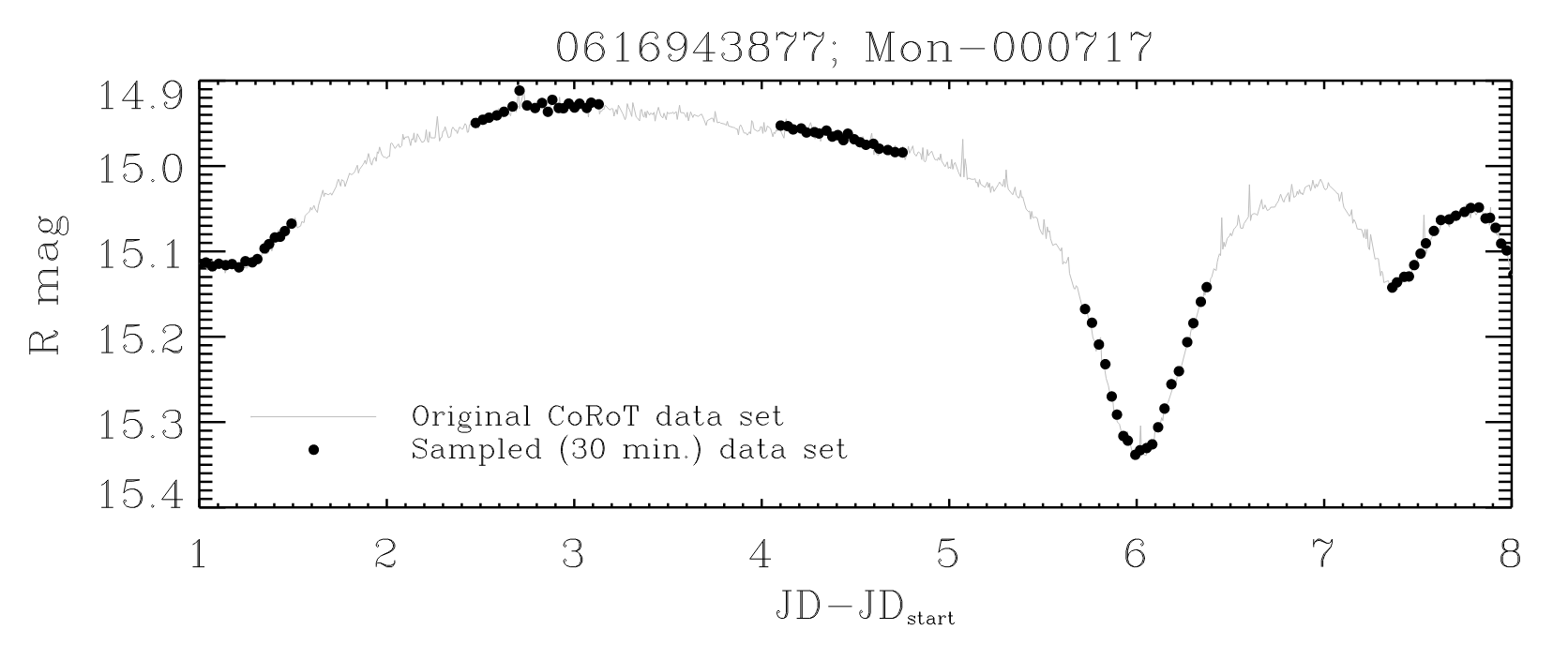}
\caption{The light curve of a young star showing short-term variability manifested as a flux dip due to the presence of 
a warped inner disk, as observed with CoRoT (see also \citealp{McGinnis2015}). Black dots mark periods of 10 hours per night for 7 days with 30
minutes cadence (as discussed in the text). Figure adopted from \citet{Bonito2018}. }
\label{fig:YSO-LC-dip}
\end{center}
\end{figure}
 
\end{enumerate}
    \item Pie in the sky
   \begin{enumerate} 
\item [a)]{\bf Dense coverage of star forming regions}\\
As mentioned above, young stars exhibit short-term photometric variability caused by mass accretion events from circumstellar disks; the presence of dusty warps within the inner disks; heterogeneous starspot coverage across the stellar surface; and flares. Some of these processes (e.g. accretion bursts and stellar flares) develop on timescales as short as a few hours or less \citep[e.g.][]{Stauffer2014, Cody2017, Getman2021}, while others exhibit characteristic timescales of variability comparable to the stellar rotation rates ($\lesssim$\,7~d; e.g. \citealt{Venuti2017, Venuti2021}).
Therefore, adopting a denser observing cadence (i.e. one datapoint every 30\,min each night for one week per year) to target different star forming regions is necessary to follow the evolution of each process and clarify the nature of short-term variability for thousands of young stars, thus complementing the long-term information that will be accumulated in the course of the main survey (more details can be found in the White Paper submitted in response to the call for Rubin LSST Cadence Optimization; in \citealp{Bonito2018}; in the Cadence Note Bonito \& Venuti et al., 2021; and in Bonito \& Venuti et al., in preparation).

This approach will allow us to relate the observed variability to stellar properties, such as mass, age, binarity, and to environmental properties such as location within or exterior to the H II region, and to the presence or absence of a circumstellar disk.
Large samples are needed to quantify how the various physical processes depend upon stellar properties, environmental conditions, and the evolutionary stages of the stars.
Rubin LSST will allow us to survey an outstanding collection of star forming regions (SFRs) in the Southern hemisphere: the closest low-mass populations, the intermediate-mass populations, and massive (like Carina) SFRs.
Therefore, we propose to target one major SFR every year.
We plan to start with observations of the Carina Nebula, which is well-placed for observations from Chile with Rubin LSST, and guarantees a large number of sources (11,000 members identified; \citealt{Townsley2011}).

Different causes of variability (e.g. stellar flares, accretion bursts, absorption due to warped disks, rotational modulation due to spots) can
be discriminated on the basis of their significantly differing observational characteristics. 
A dense coverage of SFRs with Rubin LSST will allow us to characterize different classes of light curves \citep[see Figure~1 of][]{Bonito2018}, including light curves dominated by accretion bursts \citep{Stauffer2014}, or periodic or quasi-periodic flux dips \citep[associated with rotating inner disk warps partly occulting the stellar photosphere,][]{Bouvier2007A&A,Alencar2010}.

As young stars with variability undergo significant and rapid color changes owing to both accretion processes and extinction variations, it is important to include multiple filters in any dense coverage campaign, i.e. $g$-, $r$-, and $i$-bands.
Data in each band (changing every 30 minutes between $g$-, $r$-, and $i$-filters, and possibly also the $u$-band) will provide their own light curves, making it possible to follow how the stellar colors vary with phase: 140 photometric points in each filter (corresponding to a 30-minute observing cadence implemented for one week, assuming 10-hour-long observing nights) should be collected in order to populate the phases well enough.
Variability in different colors helps to discriminate between hot spots, cold spots, and circumstellar extinction \citep[e.g.][]{Venuti_2015}.
Flaring in WTTSs can also be monitored, though the rapid decline of chromospheric flares requires a rapid cadence to capture correctly. Monitoring the accretion events will be pivotal to trigger an alert to observe the same objects with other instruments and in different bands (from X-rays to IR).

At the beginning of Rubin LSST operations we argue that a targeted test field (Carina Nebula) should be observed in the above manner to illustrate what can be done with Rubin LSST in this mode. In subsequent years, we will either choose a different region or possibly return to the same regions to monitor slow changes in periods or amplitudes that may arise from differential rotation or starspot cycles. Combining a densely-packed short-interval dataset with a sparse but long-baseline study maximizes the scientific return for both methods, and allows Rubin LSST to address all of the accretion and rotational variability associated with young stars, and to bridge the knowledge gap between short-term and long-term behaviors documented for young stellar objects.
\end{enumerate}

    \end{itemize}

The analysis of variability in YSOs described here will be based on Rubin LSST data, but for a complete description of these complex systems (consisting of a central young star, the surrounding disk, the accretion streams, the jets, and the shocks formed at the stellar surface and at the intersection with the ambient medium) it will be crucial to also take advantage of additional data from existing surveys (e.g. Gaia-ESO Survey) or through synergies with new observations and surveys (e.g. with 4MOST, WEAVE, eROSITA).

In particular, spectroscopic follow-up data will be used to investigate variable YSOs showing accretion/ejection activity. FLAMES, 4MOST, WEAVE, eROSITA, SoXS are a few examples of facilities that we plan to use for follow-up and additional new observations.

Detailed 3D MHD models of the infalling material have been developed to investigate the accretion process in young stars. These models account also for the observed variability in the inverse P-Cygni line profiles as we view accretion streams along the line of sight to the star
\citep{Kurosawa2013,Bonito2014,Revet_2017}.
Models suggest accretion cooling timescales of 30\,min to several hours in accordance with observations of the shortest bursts in BP Tau \citep[0.6 h; see discussion in][]{Siwak2018}.

This study will additionally require the development of software dedicated to the analysis of variability related to accretion/ejection activity in YSOs. In more detail, this software will be necessary to characterize the light curves, discriminate the physical processes behind the light curves and to investigate the CMDs to distinguish CTTSs from WTTSs. This will be developed in collaboration with the Italian National Institute of Astrophysics - Observatory of Palermo and the team already working on this topic of YSOs and young clusters. 

Our current hardware resources are not suited to handle the Rubin LSST data volume. Therefore, updated computational resources are required to keep up with the increased data-flow rate (already seen with Gaia) that will become available in the near future with Rubin LSST.
With its unprecedented sensitivity, spatial coverage, and observing cadence, Rubin LSST will allow us to employ a statistical approach for the first time to achieve a comprehensive view of the process of star formation.


\chapterimage{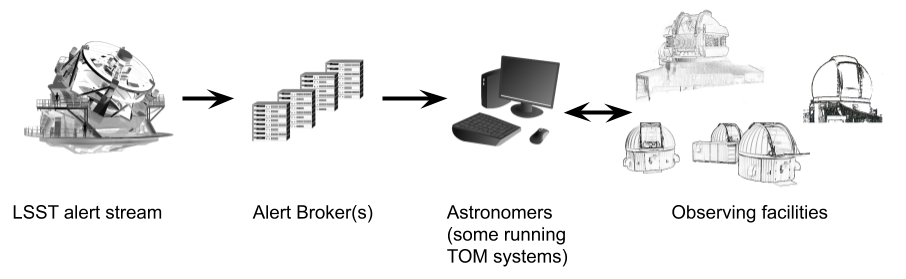} 

\chapter{Methodology and Infrastructure}

\bigskip

\section{Infrastructure Required for Time-Critical Science}
\textsl{Editor and Author: Rachel Street}
\bigskip

\begin{figure}[ht]
    \centering
    \includegraphics[width=14cm]{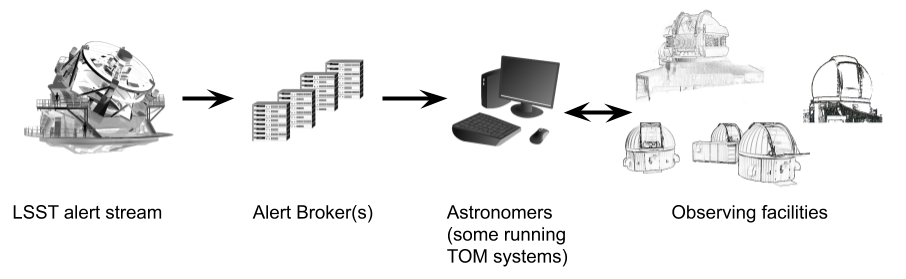}
    \caption{Overview of the Rubin LSST follow-up ecosystem; the chain of services and facilities necessary for Rubin LSST discoveries to be categorized, selected for further study and additional characterization observations made where necessary.}
    \label{fig:followup_ecosystem}
\end{figure}

\noindent The discovery of new transient phenomena occurring on short timescales  motivates a rapid response to fully characterize the objects in question while we have the chance.  In some cases, the window of opportunity may be $<$1\,day.  This drives the most stringent technical demands of any program following-up on Rubin LSST discoveries, particularly since the high data volume and data rate produced by Rubin LSST effectively mandate a reliance on software infrastructure at all stages.  In this section, the requirements that these science use-case will place on all astronomical facilities involved are examined within the context of this infrastructure, which is presented in \autoref{fig:followup_ecosystem}.  

\newpage
\subsection{Rubin LSST Prompt Data Products and Alert Stream}
\textsl{Author: Rachel Street}

\begin{figure}[hb!]
    \centering
    \includegraphics[width=0.9\textwidth]{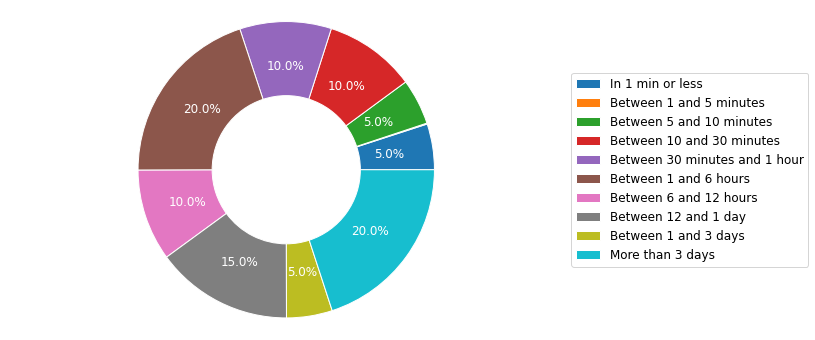}
    \caption{The breakdown of TVS scientists' responses to the question: "What is the maximum time delay tolerable between new alerts being produced by the Rubin LSST data reduction and the alert information becoming available through a broker service? In this context "tolerable" means that you would still be able to complete your scientific goals."}
    \label{fig:max_alert_delay}
\end{figure}

\noindent An informal survey of TVS members was used to gauge the maximum tolerable delay between new alerts being produced by the Rubin LSST Data Management system and the alert information being made available to the community (responses are presented in \autoref{fig:max_alert_delay}).  30\% of respondents preferred to have access to that information within 1\,hr of alert production, while 75\% indicated a maximum delay of 24\,hrs.  The primary scientific drivers behind this requirement were the follow-up of gravitational wave detections and early-time supernovae.  

It is expected that the Rubin LSST Data Management system will meet this requirement by producing real-time photometry and alerts for every image within $\sim$1\,min of shutter-close.  Current designs for the alert handling infrastructure will immediately serve the alert data publicly in the form of a Kafka event stream. 

\subsection{Rubin LSST Alert Brokers}\label{ssec:brokers}
\textsl{Authors: Rachel Street,  R\'{o}bert Szab\'{o}, Nina Hernitschek, Anais M{\"o}ller, Andreja Gomboc, Sjoert van Velzen, Federica B. Bianco, Konstantin Malanchev}

\noindent Owing to the high data rate and volume of the full Rubin LSST alert stream (approximately 10 million alerts per night), it is not envisioned that individual science users will subscribe to it directly, but rather that it will be sent to several Community Brokers.  These services are expected to collate the available alert data and serve it to their communities online.  Brokers also provide value-added services such as aggregating data from multiple survey alert streams and existing data catalogs, and attempt to classify the alerts by astrophysical type.  Some also provide additional science-specific modeling such as orbit-linking for Solar System objects.   


Due to the anticipated high bandwidth of the Rubin Observatory alert stream, only a limited number of brokers will receive the alert stream directly from Rubin Observatory.
In May 2019, the broker selection process was started with Letters of Intent (LOI) submitted from 15 teams, after which in August 2019 the Science Advisory Committee (SAC)\footnote{\url{https://project.lsst.org/groups/sac/welcome}} announced that all LOI teams were encouraged to submit full proposals. From the nine full proposals submitted by December 2020, the SAC recommended that seven brokers should receive direct access to the full stream, with the remaining two brokers acting as downstream brokers, getting their input stream from another broker.

Currently, these brokers are under development, being tested on an alert stream delivered by ZTF \citep{patterson2019}. Also, for testing purposes, an archive of ZTF alerts is available from the University of Washington.

To facilitate the development of the brokers, the alert broker teams held a two-part virtual workshop - two LSSTC Enabling Science Brokers Workshops, in 2020 and 2021, for which the recordings, slides and collaborative products are available online.

These services are expected to be the primary point of contact for astronomers accessing the Rubin LSST Prompt Data Products in real-time. This is especially important for TVS science where rapid access to alert data is critical.

\subsubsection{Overview of existing Brokers}

The list of Rubin Observatory Full-Stream Alert Brokers is as follows:

\begin{itemize}
    \item ALeRCE (\url{http://alerce.science/})
    \item AMPEL (\url{https://ampelproject.github.io/})
    \item ANTARES (\url{https://antares.noirlab.edu/})
    \item Baba-Mul
    \item Fink (\url{https://fink-broker.org/})
    \item Lasair (\url{https://lasair-ztf.lsst.ac.uk/})
    \item Pitt-Google (\url{https://pitt-broker.readthedocs.io})
\end{itemize}

The list of Rubin Observatory Downstream Alert Brokers is as follows:

\begin{itemize}
    \item SNAPS 
    \item POI: Variables (a downstream broker from ANTARES)
\end{itemize}

In the following, we give an overview of the technical specifications, especially regarding the classification capacities, of these brokers.

\subsubsection{ALeRCE}

The ALeRCE broker \citep{carrasco-davis2021, sanchez-saez2021, forster2020} is a Chilean-led initiative to build a community broker for ZTF, Rubin LSST, and other large etendue survey telescopes. Its goal is to facilitate the study of variable and transients objects, while using its classification to connect survey and follow-up resources in Chile and abroad.

ALeRCE's classification is implemented by both a stamp classifier and a light curve classifier. The stamp classifier is implemented with a convolutional neural network \citep{carrasco-davis2021}. The light curve classifier utilizes a Hierarchical Random Forest Classifier for light curves with at least 6 observations \citep{sanchez-saez2021}.

ALeRCE will help with answering science questions regarding transients (especially progenitors of stellar explosion and explosion physics), variable stars (including low-mass microlensing events, changing-mode stellar pulsators, eclipsing events and eruptive events) and supermassive black holes (including changing state AGNs, the detection of intermediate mass black holes, tidal disruption events and reverberation mapping studies). 

The developers demonstrate use cases in a series of Jupyter notebooks available at \url{https://github.com/alercebroker/usecases}. The ALeRCE broker web interface is available at \url{http://alerce.science/} with the ZTF Explorer (\url{http://alerce.online}) and the SN Hunter (\url{http://snhunter.alerce.online}).

\subsubsection{AMPEL}
 
 AMPEL \citep{nordin2019} is a modular and scalable platform with explicit provenance tracking, suited for real-time processing of large astronomical datasets, but also potentially other heterogeneous data streams.
 
 The goal of the AMPEL broker is to answer science questions regarding multi-messenger science (such as real-time comparisons between optical and
gravitational wave events, neutrino and GRB alerts), autonomous transient selection (triggering immediate follow-up observations, for example for supernovae) and complex light curve evaluation (with the application of domain-specific algorithms that allow e.g. the detection of tidal disruption events, \citealt{velzen2021}). The AMPEL repository is available at \url{https://ampelproject.github.io/}.

\subsubsection{ANTARES}

 ANTARES \citep{matheson2021} is a real-time broker system under development at NOIRLab. ANTARES applies filters to the alert stream which it ingests in real-time. Filters are Python functions that flag loci (each \texttt{locus} is a compilation of alerts, possibly from different surveys) for distribution by various output streams depending on their properties (called \textit{tags}), as well as produce classifications. Users can write their own filters in Python and submit them to run them in ANTARES. ANTARES client can be used by the community to communicate with the broker for both listening of an alert sub-stream, and archival loci requesting. ANTARES also provides cross-matching results with external catalogs and is open to add new catalogs by a science community request.
 
 ANTARES provides build-in light-curve feature extraction filter as well as a number of filters contributed by the broker team, including anomaly detection filters for the weird Galactic and extra-galactic transients.

 The ANTARES broker web interface is available at
 \url{https://antares.noirlab.edu/}

\subsubsection{Baba-Mul}

Baba-Mul is a broker developed at Caltech (PI: Matthew Graham).  In contrast to the other brokers, which are designed around a centralized service operated by the broker developers, Baba-Mul proposes a decentralized network of smaller brokers.  The Baba-Mul team proposed to make available a public repository with the code for the broker packaged as a containerized instance, enabling users to deploy their own instance of the software, and customize it for their own science case.   The software, which makes use of the TensorFlow\footnote{https://www.tensorflow.org/} machine learning tools, is optimized to run on inexpensive Coral Edge TPUs, as a means of making these powerful tools easily available to users. 

\subsubsection{Fink}

Fink \citep{moller2021} is a community-driven broker that processes time-domain alert streams and connects them with follow-up facilities and science teams. The goal of Fink is to enable discovery in many areas of time-domain astronomy. It has been designed to be flexible and can be used for a variety of science cases, from stellar microlensing, to extra-galactic transients. It currently processes the public alert stream of ZTF and has been extensively tested for deployment with Rubin LSST.


Fink is built on high-end technology that enables real-time selection of transients and variable sources in big data. To achieve this, it enriches alert data with existing catalogues, multi-wavelength and messenger detections, as well as machine learning algorithms to select promising candidates for a variety of science cases. Selected events are communicated in real-time for follow-up coordination using customizable filtering; as well as through access to the data through a web-portal and a REST API (\url{https://fink-protal.org/}). Example user cases and tutorials can be found in a series of Jupyter Notebooks at \url{https://github.com/astrolabsoftware/fink-tutorials}.

Fink aims to provide classification scores for a wide-range of science cases. These classification scores are and can be used for customizable selection of events with partial and complete data. Current algorithms include filters for: early supernova \citep{Leoni:2022}, early and complete supernova \citep{Moller:2020}, kilonovae (Biswas et al. in prep), microlensing \citep{Godines2019}, and satellite glints \citep{Karpov:2022}. Algorithms are based in many techniques ranging from Active Learning to Supervised Learning algorithms such as Random Forests and Recurrent Neural Networks. Further information on Fink, can be found at \url{https://fink-broker.org/}.


\subsubsection{Lasair}
    
The Lasair broker \citep{smartt2021} will provide a cross-match between observations and objects from astronomical catalogs, such as stars, galaxies, active galactic nuclei and cataclysmic variables from photometric as well as spectroscopic redshift catalogues. In addition, the Lasiar borker will cross-match sources with gravitational waves, gamma rays, and neutrino observations. Further information on Lasair can be found at \url{https://lasair-ztf.lsst.ac.uk/}.
    
\subsubsection{Pitt-Google}

The Pitt-Google Broker is a scalable broker system designed to maximize the availability and usefulness of the Rubin LSST alert data by combining cloud-based analysis opportunities with value-added data products. It utilizes publicly available classifiers and a Bayesian belief network meta-classifier. Further information on Pitt-Google can be found at \url{https://pitt-broker.readthedocs.io}.

\subsubsection{SNAPS}

SNAPS, The Solar System Notification Alert Processing System, is a broker targeted towards asteroid detection and classification. At present, nearly 800,000 asteroids are known, with the number expected to increase with Rubin LSST, e.g. more than 5 million main-belt asteroids are expected to be observed. Asteroids are tracers of the Solar System's dynamical and physical evolution, they contain the intrinsic material properties of primitive Solar System bodies, and finally they provide information about the origin of life on Earth as they brought the water and organic material to Earth when it was in it's earlier stages. To fulfil the goal of asteroid detection, astronomical data, such as those gathered by the Vera C. Rubin Observatory, must be converted into physical properties of Solar System objects.
SNAPS will provide the tools necessary to derive those physical properties.

\subsubsection{POI:Variables}

The Point of Interest broker (PI: Nina Hernitschek) is tailored towards the needs of astronomers looking for updated observations of variable stars in specific on-sky regions. Stellar streams, as well as their progenitors -- dwarf galaxies and globular clusters -- are of great interest because their orbits are sensitive tracers of a galaxy formation history and the gravitational potential. Many of such ``points'' or regions of interests can be traced by periodic variable stars, i.e. RR Lyrae and Cepheid stars. Additional attributes of these objects include that they are quite easily detectable due to their periodicity and their distances can be constrained from rather simple period-luminosity-metallicity relations.

The POI alert broker will provide users with updates on variable star observations within regions of interest. The output consists mainly of the light curves of machine-learning identified pulsating variable stars (RR Lyrae and Cepheid stars), and their derived light curve features such as periods and phase offsets 
(once LSST has reached the point of sufficient revisits). Depending on the classification result, further information, such as the modulation of the pulsation (Blazhko effect) shown by some RR Lyrae stars, or a distance estimate, can be additionally quantified.

The Point of Interest broker is a downstream broker from ANTARES.

\subsection{Managing Follow-up Programs}
\textsl{Author: Rachel Street}

\noindent Studying astrophysical phenomena in real-time can be an extremely demanding task.  For many of the science cases above, once targets are selected, astronomers will need to compile and analyze data from Rubin LSST and a number of other sources, often involving science-specific modeling.  Additional, follow-up observations may also be required, and for time-critical science these will often need to be taken very rapidly (within days at most) of discovery, sometimes from multiple facilities.  This creates a need to share data efficiently between team members and to coordinate their efforts, as well as to manage the observations themselves and the data products and analyses they entail.  These issues become especially acute when subject to the data rate of Rubin LSST.  

Target and Observation Manager systems (TOMs) are often used to manage the workload of follow-up programs, particularly in time-critical science areas.  These database-driven systems harvest information on targets of interest, and provide display and visualization tools for teams to share and discuss the information online.  They also offer programmatic interfaces to observing facilities so that observations can be planned and requested and the data products shared.  TOMs often make extensive use of APIs to automate many or even all aspects of their operation  and ensure the most rapid response possible to an alert.  

Though TOMs generally need to be customized to the needs of each specific science use-case, open-source packages now exist \citep{Street2018b} to make them easier to develop and maintain.  

\subsubsection{Observing Facilities}
Perhaps counter-intuitively, time-critical science often requires that follow-up observations be made across a range of timescales.  In many cases, such as for a gravitational wave events, there is the obvious need to observe as soon as possible, often with many different ground- and space-based facilities operating at different wavelengths and with a range of instrumentation.  With this in mind, many observing facilities offer a Target-of-Opportunity override mode of operation.   However, the way targets behave over the longer term (days, months or even years) can be equally diagnostic, for example to establish supernovae type, or to monitor for stellar accretion events or outbursts.  The traditional block-scheduling of telescope facilities, where consecutive hours or nights of time are allocated months in advance, does not adequately support the needs of these `monitoring' observations, and queue-scheduling is preferred.  

Though technically challenging, programmatic access to a follow-up telescope facilitates' rapid follow-up observations strongly enhances an astronomer's ability to conduct effective follow-up observations of time-critical targets.   By requesting observations via a software interface, and receiving information about the status of those observations and the facilities themselves, astronomers are able to respond quickly and efficiently to new alerts, which in-turn leads to more efficient use of limited follow-up resources.  

\subsubsection{Data Archives}
It is often necessary to access and analyse the data obtained from follow-up observations as rapidly as that from the initial survey discovery.  Since the purpose of follow-up data is usually to characterize a target, the results of that analysis serve to update the assessment of that target, and hence it's priority for further observations.  It may also decide what form subsequent observations may take.  

For this reason, it is a priority for time-critical science programs to be able to access data products from follow-up facilities rapidly via an online data archive.  Making these data products accessible via API is the final step in the `follow-up chain', as it enables astronomers to fully automate a time-critical science program, from discovery to characterization, and thus represents the most-efficient possible response to the Rubin LSST alert stream.  

\section{Classification}
\textsl{Authors and Editors:  R\'{o}bert Szab\'{o}, Nina Hernitschek, Anais M{\"o}ller, Andreja Gomboc, Sjoert van Velzen, Konstantin Malanchev}

\bigskip


\noindent We aim to classify variable stars, or variable objects in general, as well as other transients. The classification of transient objects is vital for answering the science questions that Rubin LSST endeavours to address: understanding the nature of Dark Matter and Dark Energy (for which tracers, such as quasars and supernovae, are vital), cataloging the Solar System (which incorporates the detection of asteroids and other small bodies as well as calculating accurate ephemerides), exploring the changing sky (which refers to the detection of transients and variable stars in general), and understanding the structure and formation of our Milky Way (which draws heavily on distance calculations and population studies relying on variable stars such as Cepheids and RR Lyrae).

Given the large amount of transients and variable stars that will be detected by Rubin LSST, up to 10 million detections per night, it is imperative to develop  classification algorithms using only photometric data from Rubin LSST, as spectroscopic classification will be available only for a very small fraction of events.

Classification algorithms are crucial for TVS science to: (i) select new and richer samples of known transients and variables, (ii) select samples of unknowns or barely characterized objects, and (iii) early (with partial information) selection of promising transients and variables for follow-up coordination. The latter is a particular focus of the classification efforts by Rubin LSST Community Brokers that will analyze Rubin LSST detections in real-time.

In the following, we give more detailed descriptions of recent projects regarding the classification of sources in the LSST main-survey data.

\subsection{The Classification of Periodic Variable Stars}
\textsl{Authors and Editors:  R\'{o}bert Szab\'{o}, Nina Hernitschek, Anais M{\"o}ller, Andreja Gomboc, Sjoert van Velzen, Konstantin Malanchev}

\noindent  A classification algorithm to classify periodic variables is currently being constructed. Assuming several groups will select and investigate transient phenomena, the primary goal of this project is to identify and classify non-transient variable stars. In order to classify periodic variables (pulsating, eclipsing), first we need to identify them and have an estimated period. We suppose that these pieces of information are available. By classification we mean assigning major variable star classes to individual objects, which later (as more and more data come in) can be further refined into subclasses.

\subsubsection{Methods} 

For classification, Convolutional Neural Networks (CNN) are used, which can identify the features of several variable classes. Convolution is a powerful tool to identify high- and low-level features. We made extensive tests based on the OGLE-III and OGLE-IV variable star catalogs (several hundred thousand variable stars with several thousand individual photometric measurements), and these light curves proved to be suitable as a training set for these methods.

Data augmentation is a key process before the training phase, because the number of stars within each class has to be balanced. With this step and with a judiciously planned neural network we are able to avoid over-fitting the model. 

Our preliminary tests show a high level of precision \citep[above 80\%[]{Szklenar2020}, but these tests were based on well-sampled OGLE light curves with several thousands of data points with known periods as opposed to new variable stars discovered and monitored by Rubin LSST. Clearly, more works is needed to test these methods with Rubin LSST-like datasets.

\subsubsection{Future developments} 

One future goal is to apply the method to light curves sampled by the Rubin LSST observing strategy. The OGLE database might be ideal for benchmarking, because it features high quality light curve coverage in the $I$-band filter with significantly less coverage in the $V$-band filter. The latter might be used as a proxy for Rubin LSST data.

Another direction is to exploit the inherently multi-color Rubin LSST observations. Multi-color light curves and color curves will provide an extra source of information that can be fed into the neural network to balance the lower sampling rate and to improve the accuracy of the classification.

\subsection{Transient and Anomalies Classification Algorithms}
\textsl{Authors and Editors:  R\'{o}bert Szab\'{o}, Nina Hernitschek, Anais M{\"o}ller, Andreja Gomboc, Sjoert van Velzen, Konstantin Malanchev}


\noindent Transients provide information on the extreme and fundamental physics of the Universe. However, for many of them, their progenitors, explosion mechanisms and diversity remain unknown. These transient phenomena include supernovae and the recently detected (for the first time) kilonovae. Additionally, perhaps the greatest promise of Rubin LSST, is its potential to discover entirely new phenomena, never seen before or even predicted from theory. The ability of Rubin LSST to deliver on this promise and the details of the algorithms that enable these discovery cannot be finalized until the Rubin LSST survey strategy itself is finalized (see \citep{li2022} for a study of the effectiveness of different Rubin LSST strategies in enabling anomaly detection). 

There already existing classification algorithms that can be applied to select transients to address LSST science questions. Here we discuss some of the classification approaches and software that we intend to apply.
\begin{itemize}
\item{Low hanging fruit:} 
\begin{enumerate}
    \item [a)]{\bf Exploring supernovae diversity}\\ 
    There are already a significant number of classification algorithms for supernovae using their light-curves \citep{Muthukrishna:2019,Moller:2020}. These algorithms allow us to obtain the largest number and most diverse sample of supernovae across cosmic time. Additionally early classifiers can allow us to trigger follow-up to obtain additional follow-up observations such as spectra \citep{Leoni:2022}. This will allow us to characterize supernovae rates, population properties and determine sub-classes boundaries.
    \item [b)] {\bf Classification of transients}\\
    Most of the existing transient classification algorithms assume some pre-filtering of objects to be done to select transient-like light-curves before they consume the light curves and contextual information. If no relevant contextual data is available (such as a close galaxy or variability catalog match) a relatively fast classification algorithm can be used for this purpose, for example a stamp classifier or light-weight variability time-series feature classifier. The successful usage of both approaches was presented by the ALeRCE broker team \citep{carrasco-davis2021,sanchez-saez2021}. A set of light-curve features to be broadcasted as a part of alert package (see \href{https://dmtn-118.lsst.io}{DMTN-118}) whith community driven solutions, such as a light-curve toolkit \citep{light-curve-toolkit}, are utilized by some ZTF/LSST community brokers.
    \item [c)] {\bf Anomalies}\\
    With billions of transients to be detected by Rubin LSST, new classes are essentially guaranteed. Some classification algorithms based on Isolation Forests and other methods are already designed for anomaly detection and have been applied to current datasets \citep{Pruzhinskaya:2019,Malanchev:2021,martinez2021method} which can be adapted to Rubin LSST. A combination of anomaly detection and active learning could be used for expert-driven anomaly detection and active classification tasks \citep{Ishida:2021,Lochner:2021}.
    \item [d)] {\bf Multi-survey astronomy}\\
    The early identification of transients such as supernovae or potential afterglows, will allow joint analyses in multiple wavelengths as well as follow-up optimisation. Later classification can incorporate these additional information to further classify objects into their relative sub-types.
\end{enumerate}
\end{itemize}

\subsubsection{Challenges for Classification Algorithms in the Rubin LSST Era}

Here we present the known challenges for classification the given the characteristics of Rubin LSST data and the challenges of classification in general.
\begin{itemize}
\item {\bf Event-dependent baselines}\\ 
Different transients and variables require different light-curve spans for accurate classification, e.g. yearly or multi-yearly for microlensing or RR Lyrae; weeks to months for kilonovae and supernovae. It will be useful to continue developing algorithms that can accurately classify all different classes using the same data and/or designing hierarchical mechanisms that allow a distinction between events such as the ones participating in the PLAsTiCC challenge \citep{Kessler:2019} and the current LSST alert stream challenge ELaSTiCC.
\item {\bf Incomplete data}\\ 
Rubin LSST's observing strategy will provide non-homogeneous light-curve sampling. Classification methods are being developed to tackle this, which use Recurrent Neural Networks with uneven time sampling \citep{Moller:2020}, model or gaussian process extrapolation \citep{Boone:2019} and wavelength/sampling-agnostic classifiers \citep{Qu:2021}. It is imperative to make these classifiers more robust to all transient classes to eliminate biases.
\item {\bf Training sets for supervised learning algorithms}\\ 
Current training sets, whether simulated or survey data, are an incomplete representation of our transient and variable Universe. We need to explore methods to improve training sets such as augmentation and optimisation of spectroscopic follow-up using Active Learning \citep{Ishida:2019, Leoni:2022,Boone:2019}.
\end{itemize}

\subsection{Tidal Disruptive Events Filtering}
\textsl{Authors and Editors:  R\'{o}bert Szab\'{o}, Nina Hernitschek, Anais M{\"o}ller, Andreja Gomboc, Sjoert van Velzen, Konstantin Malanchev}

\noindent Rubin LSST, with its large field of view, depth and image quality has the potential to detect many rare transients including Tidal Disruption Events (TDEs). 
TDEs are occurring (mostly) in the galactic centers and are currently discovered at a rate of about 10 TDEs per year. This number will substantially increase with Rubin LSST. Simulations show that we could expect $\sim$ 10 TDEs detected per night \citep[depending on their rate and Rubin LSST observing strategy,][]{Bricman:2019mcg}. The reliable identification of TDEs detected with Rubin LSST is important to enable sample studies, probe the supermassive black hole mass distribution, and to probe emission processes etc.

The main challenges involved in identifying TDEs are: 
\begin{enumerate}
\item [a)]{How to reliably identify TDEs based solely on Rubin LSST's photometric data, which may not have ideal time and multi-band coverage (in particular in the $u$-band filter)}
\item [b)]{How to measure the purity of the filtered TDEs using realistic light curves of the most frequent contaminants (i.e. supernovae and AGNs)}
\item [c)]{How to identify a TDE before the peak in the light curve, i.e. on the order of days to weeks, depending on the time of the first detection}
\end{enumerate}

\noindent We propose a dedicated TDE filter to run on the Rubin LSST Alert stream data on one or more of the Rubin LSST Community Broker(s). We propose the following stages for development: 
\begin{enumerate}
\item[a)] {Extract nuclear flares (from the centers of galaxies)},
\item [b)] {Extract the photometric features (e.g. rise-time, color, color-evolution, fade timescale),}
\item [c)]{Photometric typing, including machine learning.} 
\end{enumerate}
Required data would be the information included in the Rubin LSST Alerts: the history of an object, full photometric light curve, astrometric data (galaxy cross-match, off-set from the galactic center), galaxy photo-z, and galaxy color/type. The TDE filter output would be a stream of nuclear flares with light curve features, including classification labels or probabilities.

\chapterimage{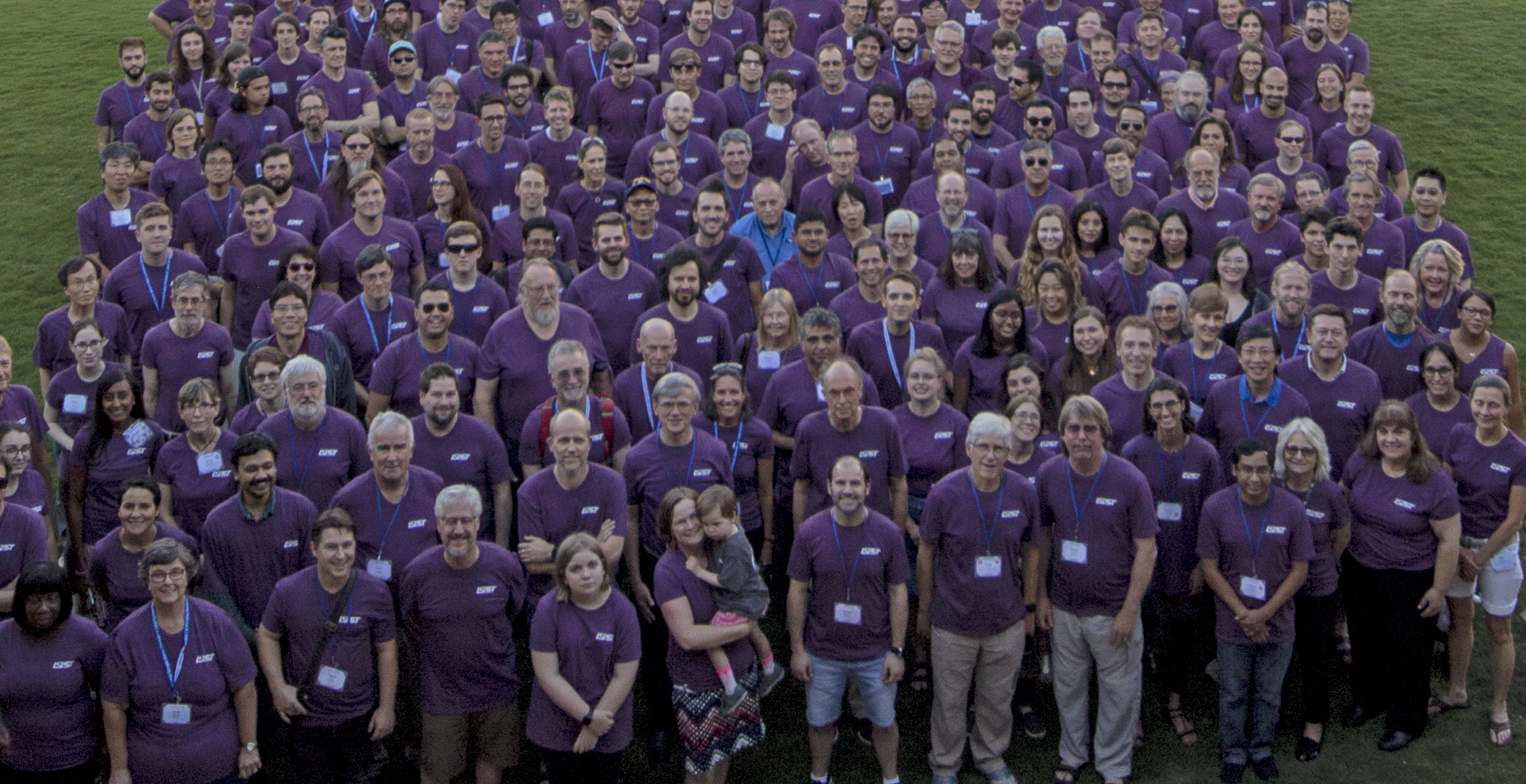} 

\chapter{Equity and Inclusivity in the Rubin LSST Transient and Variable Stars Science Collaboration}
\textsl{Authors and Editors: Federica B. Bianco, Rosaria (Sara) Bonito, Rachel Street}

\bigskip
\section{Introduction and current state}\label{chapter:DEI}
\textsl{Author: Federica B. Bianco, Rachel Street, Rosaria (Sara) Bonito}\\
\bigskip

\noindent The\blfootnote{Image: Rubin Observatory Project Community Workshop 2019} TVS SC is designed to be a collaborative environment to advance the science potential of the Rubin Legacy Survey of Space and Time. Through the TVS, the community can prepare to turn the Rubin LSST data into discoveries about the Universe and advocate for science-driven decisions that can enhance the power of Rubin LSST to explore and advance our knowledge of the transient and variable Universe.

The TVS members and leadership believe that the maximization of the potential of Rubin LSST will only happen if we create a supportive and inclusive environment for all our members and we strive to make our community inclusive. 

In the wake of the COVID-19 pandemic, which brought to light and amplified established social inequities, and of the racial reckoning following the murder of George Floyd, Breonna Taylor, and many other people of color, killed by authorities in the USA, we have renewed our commitment to advance equity and justice within our organization and to promote equitable and fair scientific practices. In June 2020, the TVS released its first statement of values, and a Code of Conduct. The Statement of Values was open for collaborative editing by all TVS members and was adopted by consensus by all TVS members. The original statement of value is reported here in full:

\bigskip

\emph{The goal of the LSST TVS SC is to advance our understanding of the Universe through science and
to create and sustain a research environment in which all members can thrive. Supporting an equitable space, free of discrimination is first and foremost a matter of social justice. We recognize that academia is embedded within, and takes advantage of, systemic racism that perpetuate white supremacy and suppresses non-white voices and the voices of diverse scholars along other axes or privilege. We, as an organization, renew our commitment to be proactive in order to fix and stop this “status quo” in our corner of the Universe. The TVS SC is inherently a multinational organization with members from a wide range of cultures and this diversity strengthens our research by bringing in different perspectives and expertise. We seek to enhance our diversity of membership and to create an educational and research culture that is welcoming and supportive for all members. We encourage people to apply and be an active part of TVS, regardless of their race, color, country of origin, sex, age, national origin, religion, sexual orientation, gender identity and/or expression, disability or veteran status. We have an ongoing commitment to a range of initiatives to make our organization more equitable and just, undertaken with the guidance of the Justice, Equity, Diversity, and Inclusion (JEDI) group of the TVS SC. More information on these activities can be found in our TVS call to action. The leadership of the TVS SC and JEDI are responsible for, and all members are empowered to, ensure full support for these activities. Accountability metrics are explicitly included in the same document.}

\bigskip

A specific roadmap to fulfill our commitment to equity should be created in the framework of Change Management to ensure its feasibility and sustainability \citep{james2019time}. For now, TVS established a ``Call to Action'' with several equity focused items which are discussed in the next section. 

A note on western-centric bias is in order: while we are a highly international organization, we work alongside a US/Chilean National facility, and our composition reflects the statistics of the STEM population in general. Our members' composition is primarily white, and primarily from Western countries. 
It is natural for us to focus on US and Europe-based rights issues and sociopolitical events in the US, Europe, and other ``developed'' countries that have traditionally been involved in large international STEM collaborations. For example, as discussed below, we strove to collect resources in support of members impacted by the Ukraine war, and by police violence, gun violence, and the right to health-care services in the US. However, we recognize we are likely to overlook international events that are just as impactful, if not involving directly as many of our members. As part of our path to truly becoming an equitable organization, and as we accrue more international members from different countries, we recognize we need to educate ourselves and pay closer attention to international and global events and learn how to support members across cultures. This will require self-education and working closely with members from different communities, while striving not to burden them with the obligation to help those of us in dominant cultures to expand our perspectives.   

We emphasize that equity-focused work is time consuming and emotionally draining. Unfortunately, most of this work in academic and STEM-focused organizations is also done on a volunteer basis and with little to no academic recognition. While several TVS member routinely engage in equity-focused work, to truly ensure the sustainability of this effort and the continuing progress of TVS toward equity and justice, support for these activities needs to be secured. In 2021 the Heising-Simons Foundation generously awarded \$900,000 to three science collaborations, the TVS SC, the Stars, Milky Way, and Local Volume Science Collaboration, and the Solar System Science Collaborations.\footnote{\url{https://lco.global/news/heising-simons-foundation-grant-will-fund-equity-and-excellence-in-science}} This grant supported may equity-focused activities, as described below. This grant will end in August 2022.  
 
\subsection{TVS Call for Action: Current Progress and Future Commitments}
  
  \noindent In June 2020 TVS released a ``Call to Action''\footnote{\url{https://lsst-tvssc.github.io/calltoaction.html}} and established the Justice, Equity, Diversity, and Inclusion (JEDI) group. The Call to Action included several areas of activities, the current status of which is reported here:
 \begin{itemize}
     \item {\bf Create a Justice, Equity, Diversity, and Inclusion group}. Status: completed.
     \item {\bf Collect, retain, and publicly release anonymized TVS demographic data}. Status: ongoing. A demographic assessment requires: (1) formulation of a survey; (2) distribution of the survey; (3) aggregation of the responses to ensure anonymization; (4) analysis of the responses and formulation of an action plan to address concerns. At the time of writing, a census of the TVS SC has been formulated and released (phase 1). Collection of demographic data is ongoing  (phase 2). An analysis of this document will be released to all TVS members on the TVS website in December 2022. We project the census will be reissued every two years.

     \item {\bf Establish formal, structured mentoring relationships to pair senior members with new members, particularly students, and ensure that all members have the support and preparation they need to be successful.} Status: while steps to improve on-boarding procedures have been taken, a formal mentoring structure is yet to be established. The volunteer nature of our organization, which is not supported by any stable source of funding at this time, makes recruitment of members for these kind of activities difficult. One project, which focused on mentoring, involved pairing 4 faculty members with students from minority serving institutions. For this purpose, a grant has been awarded by a 2021-2022 Heising-Simons Foundation\footnote{\url{https://lco.global/news/heising-simons-foundation-grant-will-fund-equity-and-excellence-in-science}}. \emph{Action: fundraising to support this activity may be necessary.}
     \item {\bf Recruit diverse cohorts of members, especially junior members}. Status: the process to expand our membership through targeted recruitment is ongoing. This process was actively supported by a 2021-2022 Heising-Simons Foundation grant awarded to the TVS SC, the Stars Milky Way Science Collaboration, and Local Volume, and the Solar System Science Collaboration\footnote{\url{https://lco.global/news/heising-simons-foundation-grant-will-fund-equity-and-excellence-in-science}}. A fraction of the funds from this grant was used to support kickstarter programs that focused on partnerships between research-focused institutions (with ongoing Rubin LSST programs and access to research funds), and primarily teaching and/or minority serving institutions to establish Rubin-related research programs that cater to a broader and more diverse community.\footnote{\url{https://lsst-sci-prep.github.io/kickstarter_grants.html}} \emph{Action: The sustainability of this effort will depend on our ability to secure additional funds.}
     \item {\bf Fundraise to advertise our organization at meetings and also at conferences directed to specific underrepresented groups and identities}. Status: TVS has honored its commitment to increase its visibility by participating in STEM URM-focused conferences/organizations. In the US, TVS purchased booths at the National Society of Black Physicist annual conference (2020), the Blacks in Physics conference (2020), and has advertised TVS opportunities by supporting participation of its members at relevant meetings (e.g. SACNAS). At this stage, however, there is no clear throughput from these activities. It is possible that engaging in these activities while the meetings were held remotely was ineffective. We are hopeful that in person participation will be more effective. A more comprehensive strategy to increase our visibility in the broader (more diverse) community needs to be developed. 
     
     The TVS has recently been successful at involving communities in Africa. Notoably, South African astronomers have long been members of TVS (and contributors to this roadmap). But until recently South Africa was the only country in Africa represented in our membership. In March 2022, as a members of TVS SC, SER-SAG in-kind team organized a one day “Student intro training on Python for data processing of AGN variability within the LSST” for students from several Ethiopian and Serbian universities and research institutions\footnote{\url{https://github.com/LSST-sersag/dle/tree/main/activities/workshop}}. The workshop was aimed to raise staff and student awareness of scientific and cooperation opportunities within the LSST. Riding on the tail wind of this initiative, TVS has recruited members of the Ethiopian community to become member of our SC and encouraged participation in an intermediate software development skills training workshop held in July 2022.  This workshop, supported through the Heising-Simons Foundation grant, provided access to training materials written by the Software Carpentries and presented by the Software Sustainability Institute.  It was deliberately designed to be fully virtual and largely asynchronous, structured to enable participation from different timezones and from different levels of connectivity. This is a promising beginning to expand our presence and offer opportunities in areas not traditionally connected with large astrophysical programs. \emph{Action: As mentioned earlier, we acknowledge and regret that our structure and composition leads naturally to a western-centric bias. As we continue to work to mitigate this, we are discovering the many ways in which the opportunities we offer are not suitable for everyone. For example, when Rubin LSST begins operations, the accessibility of computing resources may very well be one of the most important disparities that will exist. While Rubin will offer a platform for processing data, connectivity may be a bottleneck. Groups within the TVS SC ---including groups offering in-kind contributions from the Astronomical Observatory – Belgrade (AOB) and University of Belgrade - Faculty of Mathematics (UB-MatF)--- are working to address this by offering to execute a pipeline on available HPC platforms on behalf of users. More solutions to secure missing resources and enable the participation of under-resourced communities undertake scientific discovery through Rubin LSST clearly need to be conceptualized and realized. Fund-raising through agencies and foundations, as well as leveraging in-kind resources from international communities seeking access to Rubin LSST data\footnote{See \url{https://community.lsst.org/t/international-in-kind-contribution-evaluation-committee-cec-update-charge-and-science-collaboration-representation/3998}} are all viable paths that our SC should commit to.}

    \item {\bf Keep our webpage up to date and make it more accessible to all, taking into consideration the needs of differently-abled members and prospective members of our organization.} Status: The TVS website\footnote{The TVS website is accessible at \url{https://lsst-tvssc.github.io/}} was redesigned in 2020 to enhance accessibility, information retrieval, transparency, and have a space to elevate the work and achievements of TVS members. In the wake of the Ukraine crisis, the JEDI group has also collected information to support our colleagues suffering due to the ongoing war, such as opportunities for scientists from Ukraine and Russian dissidents to be hosted at international locations. This information has been shared on the TVS website.  \emph{Action:  The sustainability of this work will only be ensured if TVS can recruit members to serve as web-masters and perform website maintenance on an ongoing basis. Ensuring service work such as this is recognized will be a necessary step to recruiting a web-master. Furthermore, TVS intends to provide support to its community through the national and international crisis leveraging of its large international network of members. This will be done in the future through the TVS JEDI as well as a new (at the time of writing) Diversity Equity and Inclusion (DEI) Council of the Science Collaborations. This action requires ongoing involvement for all TVS members to support each other which, in turn, requires the continue reinforcement of ethically-focused collaborative practices, a responsibility of the TVS leadership and JEDI.}
    
    \item{\bf Create a Slack channel where questions about TVS can be asked. Assign primary members to the channel to make sure that the questions are answered.} A SLACK channel  \texttt{\#tvs-whos-in-the-what-now} has been added to the LSSTC SLACK Workspace, the primary communication venue for the Rubin TVS SC and and all SCs (together with Rubin Community\footnote{\url{community.lsst.org}}). \emph{Action: The usage has been limited since its creation. This could be due to poor advertising of this venue, or to inherent ineffectiveness. A review of the usefulness of this feature should be initiated.}

    \item{\bf Create a speaker’s bureau that ensures the representation of TVS at conferences and meetings is diverse.} Status: On hold. The TVS Chairs, however, have committed to increase the visibility of junior members by directly recruiting speakers for talks and speaking engagements. 

    \item{\bf Develop and collect bystander intervention training resources for our members}. The TVS has shared information and supported the participation of its members at bystander virtual training workshops. The TVS leadership and JEDI have been and continue to be involved in the organization of facilitated events toward DEI for TVS and the Rubin community, including anti-racism and bystander training (at the 2021 Rubin Project Community Workshop), other Rubin meetings, and in ad-hoc TVS-specific sessions. The JEDI team has been awarded a \$9,000 grant (as part of the 2021-2022 Heising-Simons Foundation grant to support the Science Collaborations\footnote{\url{https://lco.global/news/heising-simons-foundation-grant-will-fund-equity-and-excellence-in-science}}) to provide training on DEI topics to members of the SCs \emph{Action: the JEDI will continue to engage in the organization of equity-focused training. Fundraising to support our training will be necessary.}

    \item{\bf Enable and reward mentoring training targeted to mentor minority students for senior members.} Status: on hold. \emph{Action: appropriate training opportunities need to be identified or created.}
    
    \item{\bf Commit to hosting meetings only in places that are near minority serving institutes and in areas where police practices are progressive and don’t make our members and guests unsafe.} Status: this action has been on hold since the creation of our Call for Action as no in-person meetings have been organized due to COVID-19. \emph{Action: in the changing political and social landscape, further considerations should be included in the selection of venues. These include: the access to and criminalization of pregnancy-related medical procedures that may put pregnant meeting members at risk;  concealed-carrying laws; ability to impose COVID-related precautions such as mask-wearing. Furthermore, the TVS leadership has committed to provide remote access to all its meetings going forward, explicitly working to ensure a quality experience for the remote attendants, in order to better support the needs of TVS members that cannot or prefer not to engage in person. Currently, this may be a common choice because of the ongoing pandemic, but TVS commits to continue to enable remote meeting attendance past the pandemic. Circumstances that will continue to make remote attendance the preferred mode of participation, identified by the TVS, include family and care-taking responsibilities, lack of flexibility in work engagements, or neuro-diversities. The JEDI should compile information on all potential venues and, on an ongoing basis, continue to keep statistics up-to-date on all sites where members of the SC are located, including statistics on policing, weapons laws, abortion laws. This is time consuming and complex work. Structures within the Rubin larger ecosystem, such as the Diversity Equity and Inclusion Council of the Science Collaborations, may be able to support this work.}
 \end{itemize}

\section{Increasing Accessibility to Rubin Data}
\textsl{Editor and Author: Federica B. Bianco}

The TVS has recently instantiated a subgroup for Data Visualizations and Representations dedicated to conceptualize, foster, and prototype ways to access Rubin LSST data that will enable all members, including members with vision impairments, to participate in LSST-driven discovery. 
 
Astronomy is a highly visual science. However,  Rubin’s commitment to equity would not be complete if it did not enable Blind and Visually Impaired (BVI) users to explore its rich, scientific data. Ways to enable access to the BVI community include (see Astrobite article\footnote{\url{https://astrobites.org/2022/07/17/}}: enabling \emph{tactile access}: ``using the sense of touch to explore a model. One example of a tactile model is a skymap that can be touched.''
\emph{Enabling auditory access}: 
Through ``sonifications'' by mapping Rubin LSST data features to sounds, and ``Using audio descriptions for images or videos''. 

As the Rubin data will be accessed through the Rubin Science Platform (RSP), BVI users will not have the opportunity to use tools for its analysis on their own devices if the only data representation enabled is visualization-based. 
Furthermore, the jupyter notebook environment requires specific tweeks to enable integration with a Smart Reader \citep{EasyChair:3936}, so the integration of sonification has to happen at the RSP level. Without such efforts, Rubin runs the risk of inadvertently locking BVI users out of the process of scientific discovery. 

\subsection{Sonification}
\textsl{Authors: Federica Bianco, Riley Clarke}

\bigskip

\noindent Sonification is the practice of giving an audible representation of information and processes \citep{kramer1994introduction}. While visualizations are the traditional means of making data accessible to scientists as well as to the public, sonification is a less common but powerful alternative.
In scientific visualizations, specific data properties are mapped to visual elements such as color, shape, or position in a plot. Similarly, data properties can be matched to sound properties, such as pitch, volume, timbre, etc. 
In order to successfully convey information, both visualization and sonification need to be systematic, reproducible, and avoid distorting the data. 

Today, more and more examples of discovery through sound are emerging across disciplines\footnote{\url{https://sonification.design/}, \url{https://eos.org/articles/set-to-music-exoplanets-reveal-insight-on-their-formation}. See also the TED talk by Wanda Diaz Merced ``How a blind astronomer found a way to hear the stars''  \url{https://www.ted.com/talks/wanda_diaz_merced_how_a_blind_astronomer_found_a_way_to_hear_the_stars?language=en}}.
While sonification is generally considered a ``potentially useful alternative and complement to visual approaches, [it] has not reached the same level of acceptance” \citep{de2009science}. Nevertheless, \cite{Zanella_2022} report an exponential growth of sonification over the last 10 years and discuss the application of sonification to research, education, and public engagement. Compared to visualization, sonification of scientific data are scarce, despite the fact that the differences in the processing of visual and auditory stimuli imply that sonification, by emphasizing complementary relationships in the data, can reveal properties that are missed in visualizations. Compared to the human eye the ear is orders of magnitudes more sensitive to changes in signal intensity and frequency, can pick out low signal-to-noise sources better than computer code, can respond and identify aspects of the data faster (e.g., alarms), can determine variable source periods fast and accurately, and can reach new regimes using sound quality and higher-order harmonics to convey and interpret a large number of parameters at once \citep{treasure_2011,cooke2017exploring}. Data plotted visually in more than 3 dimensions are too difficult and much harder to interpret compared to sound.
 
We argue that integrating sonification in the workflow of Rubin LSST data analysis serves three important purposes. Firstly, it is an issue of equity and research inclusion. Rubin's commitment to equity would not be complete if it did not enable Blind and Visually Impaired (BVI) users to explore its rich, scientific data. As the Rubin data will be accessed through the Rubin Science Platform\footnote{\url{https://data.lsst.cloud}} (RSP), BVI users will not have the opportunity to use tools for its analysis on their own devices, so the integration of sonification has to happen at the RSP level. Access to such data also enables real BVI scientific contribution and will promote STEMM careers for the large BVI community that would otherwise be missed. Secondly, we argue that sonification will enable alternative and complementary ways to explore Rubin data in all its complexity and high dimensionality, and that will provide a tremendous benefit to the Rubin community from the extension of the representation of the data beyond visual elements. Just as Rubin will open up the parameter space for astrophysical discovery with its rich dataset, investment in sonification strategies has the potential to expand the representation space, ultimately benefiting the entire scientific community. Finally, sonification provides a means of verificatiion for the detections and discoveries identified in the data via software or visual methods, in particular the low signal-to-noise sources and events that are often the most important and impactful. We believe the integration of sonification into the Rubin data access and analysis workflow is achievable, and even may be considered a ``low hanging fruit'', although dedicated resources will be required, including funding support for the community activities required to establish a sonification standard for Rubin LSST data and for Rubin Observatory to implement technical solutions that are out of scope at the moment.
\begin{figure}[ht]
\begin{center}
\includegraphics[width=0.6\columnwidth]{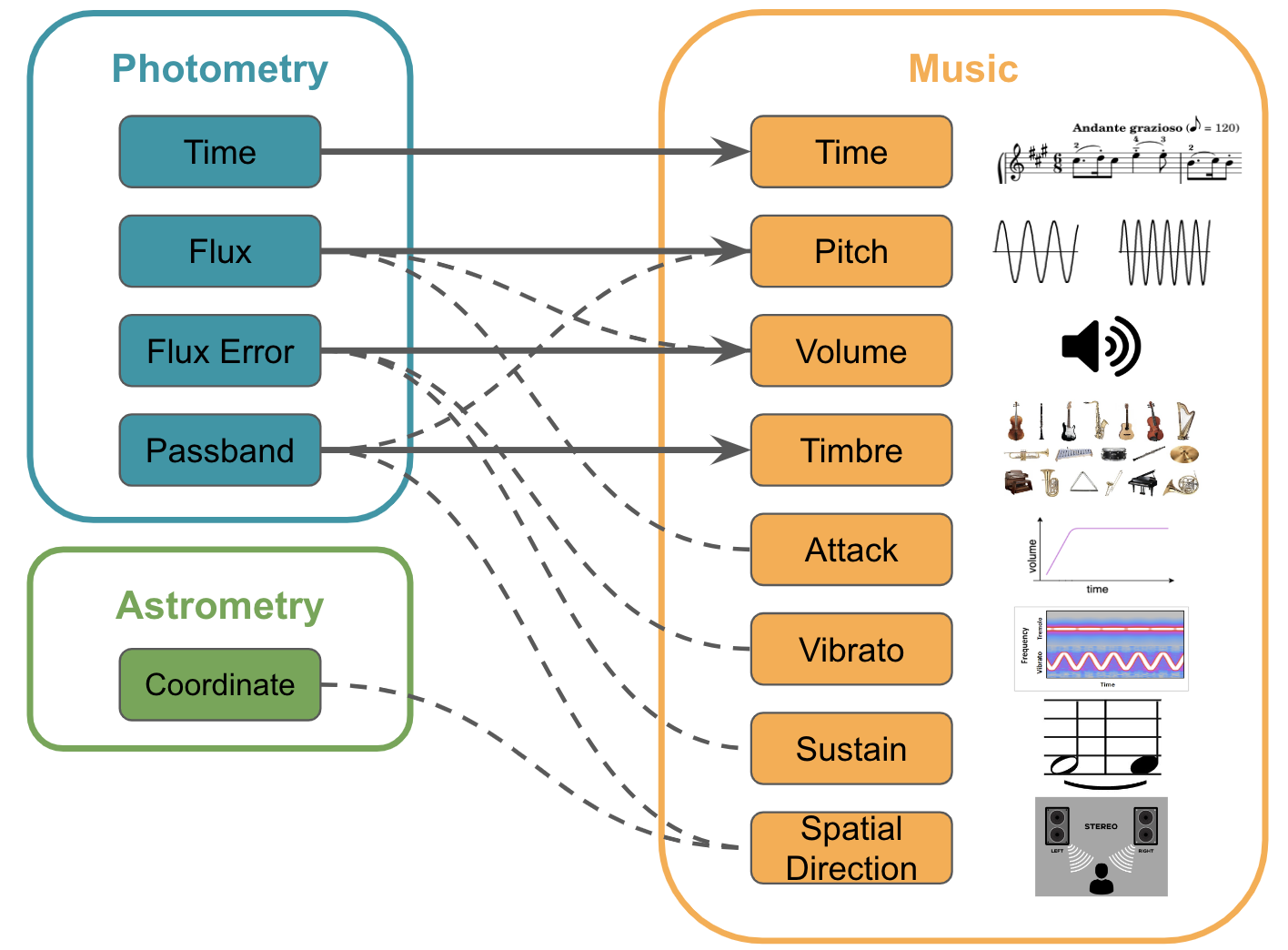}
\caption{Example of potential parameter mapping schemes for Rubin LSST data sonification: on the left are LSST time-domain data features, on the right sound parameters (and a graphical representation of their meaning). Arrows represent potential mapping, in gray are the  mappings that have been tested (results from preliminary interdisciplinary studies involving musicians and TVS members can be found on the TVS website).}

{\label{fig:sonification}}
\end{center}
\end{figure}

Considerations when working towards the sonification of Rubin LSST data include:
    \begin{itemize}
    \item Because LSST data are complex, they can benefit tremendously from the extension of the representation of the data beyond visual elements. But the sound representation of a source is only scientifically effective if a deliberate sonification design scheme is created for the complexity and richness of the Rubin dataset, with perceptive effectiveness and clarity \citep{walker2005mappings}. 
    Sonification is relatively common in astronomy to represent (1) 1D time-series, typically evenly sampled (e.g. Kepler data\footnote{\url{https://starsounder.space}}) and spectra, and (2) 2D images (e.g. Chandra images\footnote{\url{https://chandra.si.edu/sound/}}) and 3D spectroscopic intergral field unit (IFU) collapsed datacubes. Yet, the effective sonification of 6-bands of unevenly spaced lightcurves \emph{and the corresponding metadata} is essentially an uncharted territory (however, see also \citealt{cooke2017exploring}). The very first step to proceed in the process of enabling the sonification of Rubin LSST data is to establish an effective standard for the mapping of Rubin LSST features to sound properties (\autoref{fig:sonification})\footnote{Early work in this direction was supported by the Preparing for Astrophysics with LSST Program, funded by the Heising-Simons Foundation and is available at \url{https://lsst-tvssc.github.io/RubinRhapsodies}}. The proposed mapping will have to be vetted by user communities in controlled experiments, likely with separated experiments for members of the BVI community. This effort will require time and resources.
    \item Sonification software tools, including tools specifically created for the sonification of astrophysical data exist and are proliferating (e.g.  \texttt{astronify}\footnote{\url{https://astronify.readthedocs.io}}, \texttt{SonoUno}\footnote{\citealt{cooke2017exploring}}, \texttt{soni-py}\footnote{\url{http://www.sonification.com.au/sonipy}}, \texttt{StarSound and VoxMagellan}\footnote{\url{https://www.jeffreyhannam.com/starsound}}). However, it is likely that the complexity of the Rubin data will require additional capabilities if we truly want to enhance data interaction through sound. For example: the sonification of the data (numerical data such as time series) without the metadata (e.g. categorical information such as galaxy association) would provide an incomplete picture. Interaction with engineers building sonification software will be required.
    \item Issues of accessibility need to be considered early in the design of the Rubin Science Platform\footnote{\url{https://data.lsst.cloud}} (RSP) to be embedded as natural solutions instead of ad-hoc workarounds. For example, while the emphasis RSP places on Jupyter notebooks is welcome as an effective and popular coding platform, these benefits must be weighted against accessibility issues: Jupyter notebooks do not work with screenreaders, for example. This issue of accessibility has been noted by the NASA Mikulski Archive for Space Telescopes\footnote{\url{https://archive.stsci.edu}} (MAST). The MAST archive is implementing a workaround to convert notebooks to HTML so they are accessible with screenreaders. Enabling similar solutions will require close collaboration with the Rubin Project and Operation teams.
    \end{itemize}

\subsection{3D Rendering of Rubin Data}
\textsl{Author: Rosaria (Sara) Bonito}

\noindent To increase accessibility to Rubin LSST's scientific results for Blind and Visually Impaired (BVI) students and researchers, we can develop 3D models of astrophysical objects studied by Rubin LSST and utilize these models to print 3D renderings. 

3D models can be useful in the research dissemination, increase accessibility for the BVI community, and support sighted astronomers in research: 3D printed models allow us to better understand the complexity of astrophysical systems.
For example, we aim at investigating the stellar variability on a wide range of timescales (from hours to years), exploiting the full capability that will be available with Rubin LSST data. We will analyse the light curves (LCs) of a very diverse range of physical processes leading to variability detections, as discussed in several sections of this document.
Producing 3D renderings of these astrophysical objects, and the associated 3D printing kits, will enhance Rubin LSST science accessibility and inclusiveness and help develop understanding of these systems which consist of multiple components.

We note that in addition to 3D renderings, Virtual Reality (VR) experiences relating to the physical processes we study can be produced based on the same 3D physical models \citep[see][]{2019RNAAS...3..176O}. While not aimed at increasing accessibility to differently-abled scientists and communities, this is also a method of dissemination and will build awareness and excitement about Rubin science. We note that while VR sets have historically been expensive, free platforms, such as Sketchfab\footnote{\url{https://sketchfab.com}} and low-cost visors (e.g. cardboard visors that can even be made at home) are becoming more common, mitigating equity issues and concerns in devoting resources to means of scientific communication accessible to only a subset of the population.

The 2021-2022 Heising-Simons Foundation grant to the Science Collaborations\footnote{\url{https://lco.global/news/heising-simons-foundation-grant-will-fund-equity-and-excellence-in-science/}} has supported a pilot project to produce 3D models and 3D printed kits as described here, focusing on young stellar objects (see \autoref{sec:exorfuor}).
To make our scientific results more accessible, we are creating publicly available interactive 3D graphics and 3D printed kits based on our interpretation of a diverse range of physical processes causing photometric stellar variability.

To make the 3D modeling and rendering of physical systems, whose study is aided by Rubin LSST data, from prototype to an integral element of the Rubin research and dissemination workflow, the following steps are needed:

\begin{itemize}
\item  A path for accessibility of VR renderings and experiences is clear: free and low cost resources to enable VR access are becoming more common and TVS will be able to take advantage of this increased accessibility
\item The transfer-of-knowledge to the scientifc community as well as to students to develop 3D models that can be both: (i) printed in 3D on one hand for a more accessible dissemination of Rubin LSST results aslo including the BVI community and (ii) converted into VR experience through repositories focused on asrophysical objects of interst of the TVS members and beyond on platforms as Sketchfab is forseen.
\end{itemize}

Members of TVS developing 3D models will make them available as printable files of the 3D models or 3D printed kits. Although the previous items are easily achievable, to truly advance this data representation method and make it common practice, additional resource will be needed. These include:

\begin{itemize}
\item Advanced laptops for 3D visualization equipped with VR visors. Further, 3D printers should be provided for every team engaging in 3D rendering work for a variety of physical phenomena whose modeling is aided by Rubin LSST data. 
\item Advanced laptops for 3D visualization equipped with VR visors would also be needed to enable the teams developing the 3D models to share the VR experience at meetings and conferences, engaging the community of researchers and students, as well as for outreach and education related to the exploitation of future Rubin LSST data.
\item Collaboration with experts to develop standalone apps\footnote{See for example \url{http://www.pharos.ice.csic.es/star-blast}, PI M. Miceli} to fully exploit the VR experience for an immersive investigation of 3D models of astrophysical objects that could allow an higher degree of engagement of the community.
\end{itemize}

\chapterimage{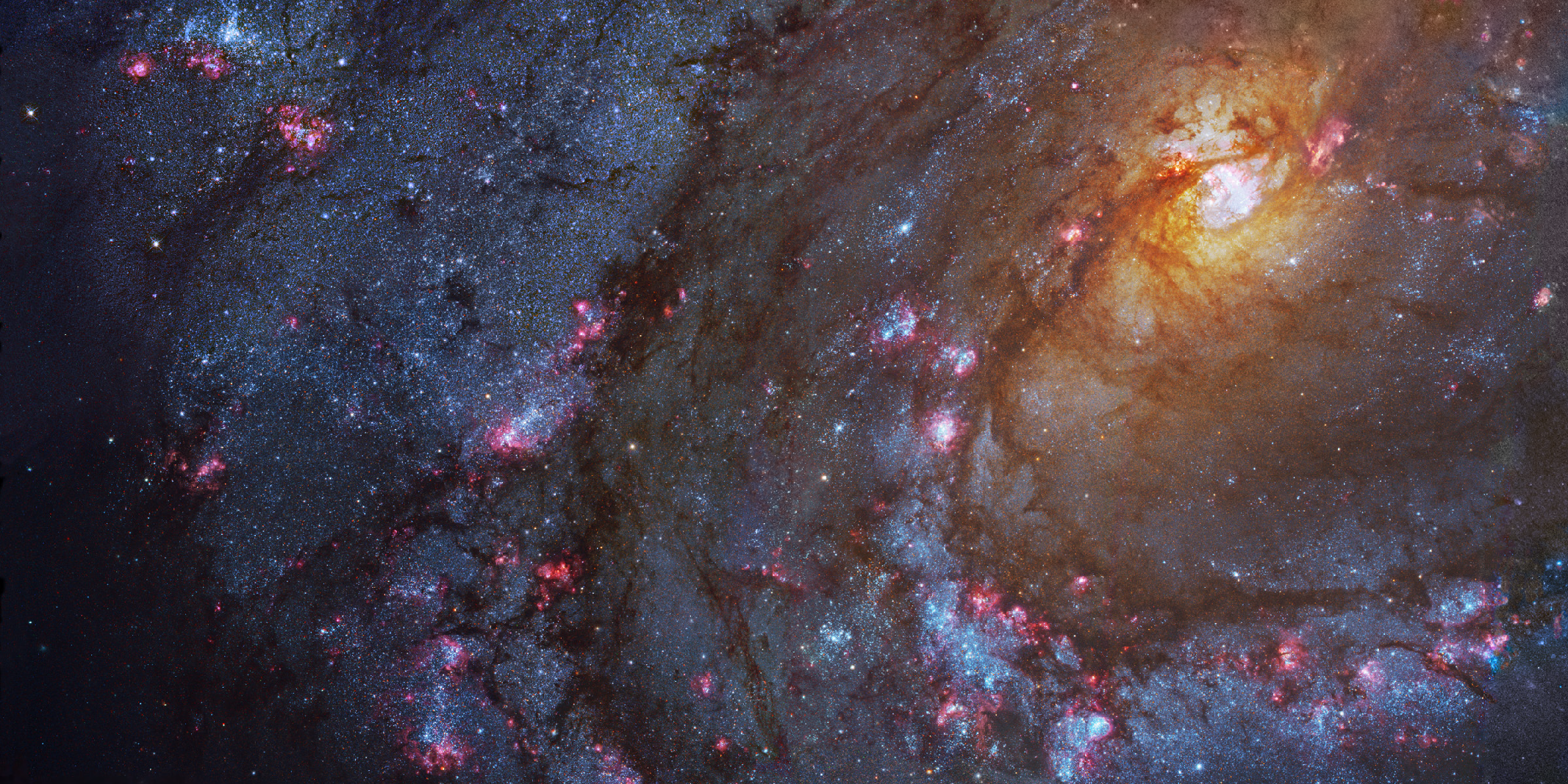} 

\chapter{Summary and next steps} 
This document has been in the making for several years, during which time the Rubin Legacy Survey of Space and Time (LSST) Transients and Variable Stars Science Collaboration (TVS SC) has doubled in size, and expanded the scope of its science. Additionally, time domain science itself has changed with discoveries of new classes of object, a growing emphasis on Multi-Messenger Astronomy, and a renewed energy as the start of the Rubin LSST approaches, with its potential to revolutionize our science. 

The plans for Rubin LSST have been refined, with growing knowledge of the system as built, a transition from constructions to operations, and changes the paths to acquire data rights, and, notably, with the survey strategy decisions still being finalized at the time of publication of this document. Rubin LSST cadence decisions are being made through a process that involves the community at large and the Science Collaborations in particular to an unprecedented degree \citep{bianco2022}. The TVS SC has contributed to the cadence optimization process with over 20 white papers and peer reviewed publications (most of them collected in the Rubin LSST Survey Strategy Optimization Special Issue of the Astrophysical Journal Supplements\footnote{\url{https://iopscience.iop.org/journal/0067-0049/page/rubin_cadence}} \citealt{Li_2021, Raiteri_2021, Hernitschek_2021, Andreoni_2021, Bellm_2022}).  Our roadmap to science depends inextricably on the final survey strategy that Rubin will implement.

As a result of this ever evolving landscape, we recognize that our roadmap will have to be responsive,  adapting to changes in our science domain and Science Collaboration composition. With this publication, we wish to benchmark our roadmap, roughly a year prior to the start of  Rubin LSST commissioning, and share with the world our plan to support and leverage Rubin Observatory in the discovery of the ever-changing sky. 

Updates to this document will be posted on the TVS website.\footnote{\url{https://lsst-tvssc.github.io/roadmap.html}}
    
\chapterimage{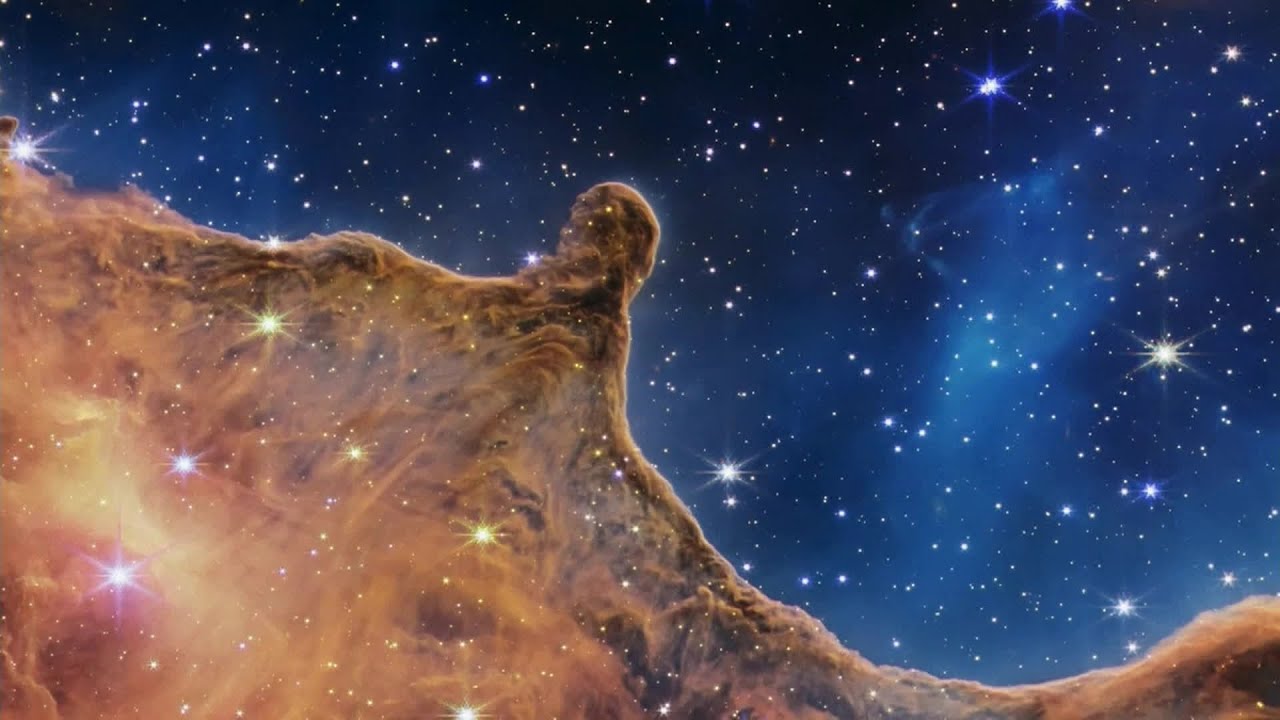} 

\chapter{Acknowledgements}

\bigskip 

This\blfootnote{Image: one of the first James Webb images Credits: NASA, ESA, CSA, and STScI} work was produced in the nursery of the Rubin LSST Transient and Variable Star Collaboration (TVS). TVS particularly acknowledges the support of the LSST Corporation\footnote{The LSST Corporationa not-for-profit 501(c)3 corporation formed to initiate the LSST Project and advance the science of astronomy and physics \url{https://www.lsstcorporation.org}}
and its commitment to securing and directing private funds toward activities that supported the work of the TVS over the years.

This work was supported by the Preparing for Astrophysics with Rubin LSST Program, funded by the Heising Simons Foundation through grant 2021-2975, and administered by Las Cumbres Observatory.

We thank the external reviewers: Maurizio Paolillo, Massimo Villata, Dan Holdsworth, Olivier Perdereau.

KMH acknowledges support through NASA ADAP
grant 80NSSC19K0594. KCD acknowledges funding from the Natural Sciences and Engineering Research Council of Canada (NSERC), and fellowship funding from the McGill Space Institute. This work was performed in part at Aspen Center for Physics, which is supported by National Science Foundation grant PHY-1607611. XL and FBB were partially supported by the National Science Foundation Grant No.2108841 and University of Delaware General University Research grant GUR20A00782. AR acknowledges support from ANID BECAS/DOCTORADO NACIONAL 21202412. KBB and AG acknowledge the financial support from the Slovenian Research Agency (P1-0031, I0-0033,  J1-8136, J1-2460). AG acknowledges the support from the Fulbright Visiting Scholars program. RB acknowledges financial support from the project PRIN-INAF 2019 ``Spectroscopically Tracing the Disk Dispersal Evolution".	

\chapterimage{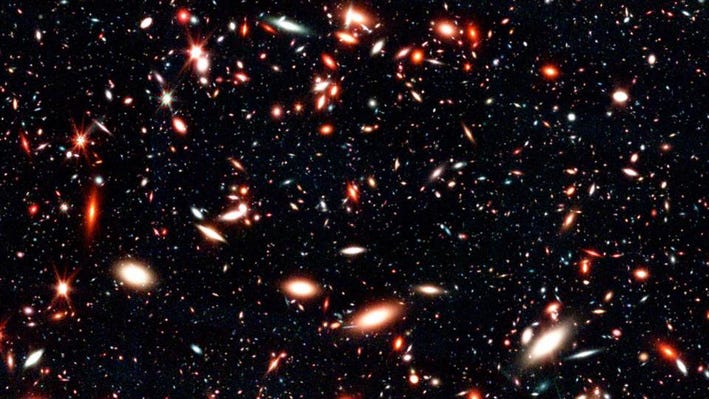} 

\bibliography{main}
\blfootnote{Image: one of the first James Webb images Credits: NASA, ESA, CSA, and STScI}
\end{document}